\documentclass[oneside,12pt,a4paper]{IB_tesis}

\usepackage{graphicx, color}
\usepackage{geometry}
\usepackage{makeidx}
\usepackage[pdftex,bookmarks=true]{hyperref}
\usepackage{setspace}
\usepackage[normalem]{ulem}
\usepackage{indentfirst}
\usepackage[square,sort,comma,numbers]{natbib}

\usepackage[latin1]{inputenc}

\usepackage[english,spanish]{babel}
\usepackage[latin1]{inputenc}

\usepackage[T1]{fontenc}

\usepackage{type1cm}
\usepackage{courier}
\usepackage{graphicx} 
\usepackage{color}

\usepackage{geometry}
\usepackage{palatino}

\usepackage[square,comma]{natbib}

\usepackage{fancyhdr}
\renewcommand{\chaptermark}[1]{\markboth{\thechapter.\ #1}{}}

\usepackage[small,normal,bf,up]{caption}

\usepackage{subfigure}
\usepackage{epsfig}

\usepackage{amsmath, amsthm, amssymb}
\usepackage{amsfonts}
\usepackage{mathrsfs}
\usepackage{latexsym} 
\usepackage{epic,eepic,color}
\usepackage{graphicx}  
\usepackage{lscape}

\usepackage{tocbibind}

\usepackage{verbatim}	
\usepackage[T1]{fontenc}
\usepackage[spanish]{babel}
\usepackage[latin1]{inputenc}
\usepackage{listings} 	
\usepackage{url} 			

\newcommand\be{\vspace{0.05cm}\begin{equation}}
\newcommand\ee{\end{equation} \vspace{0.05cm}}
\newcommand\bea{\vspace{0.05cm}\begin{eqnarray}}
\newcommand\eea{\end{eqnarray}\vspace{0.05cm}}
\newcommand\bdm{\vspace{0.05cm}\begin{displaymath}}
\newcommand\edm{\end{displaymath} \vspace{0.05cm}}

\usepackage{float}
\floatstyle{boxed}
\newfloat{bordepage}{h}{lop}[chapter]
\floatname{bordepage}{Microfabricación}

\newfloat{Bordepage}{H}{lop}[chapter]
\floatname{Bordepage}{Microfabricación2}

\definecolor{RedOrange}{cmyk}{0,0.77,0.87,0} 
\definecolor{Melon}{cmyk}{0,0.46,0.50,0}

\usepackage[absolute]{textpos}
\TPGrid[0.1mm,0.1mm]{19}{28}

\newcommand{\labell}[1]{\label{#1}}

\newcommand{\ba}{\begin{eqnarray}}
\newcommand{\ea}{\end{eqnarray}}

\newcommand{\beqar}{\begin{eqnarray*}}
\newcommand{\eeqar}{\end{eqnarray*}}

\newcommand{\reef}[1]{(\ref{#1})}

\newcommand{\mt}[1]{\textrm{\tiny #1}}
\newcommand{\veps}{\varepsilon}

\newcommand{\N}{\mathcal{N}}

\newcommand{\la}{\lambda}
\newcommand{\lp}{\ell_{\mt P}}

\newcommand\ra{\rangle}
\renewcommand\la{\langle}

\newcommand{\hr}{\hat\rho}

\newcommand{\mv}{v}

\newcommand{\tr}{{\rm tr}}
\newcommand{\Tr}{{\rm Tr}}

\newcommand{\tz}{\tilde{z}}
\newcommand{\zh}{z_\mt{h}}
\newcommand{\vep}{\alpha}

\def\({\left(} \def\){\right)}
\def\[{\left[} \def\]{\right]}

\graphicspath{{entropia/}{plano/}{introduccion/}{agujeros/}{extensividad/}}

\begin{document}

\encabezado{Thesis Submitted to UNCuyo\\for the Ph.D. Degree in Physics}

\title{Quantum information measures and their applications in quantum field theory}

\author{David D. Blanco}
\rolAutor{Author}

\directorIzq{Dr. Horacio Casini}
\rolIzq{Supervisor}

\location{San Carlos de Bariloche}
\date{January 2016}

\frontmatter
\pagestyle{empty}

\maketitle

\chapter*{\hfill\foreignlanguage{english}{Abstract}\hfill}

\begin{foreignlanguage}{english}
In the last decades, it has been understood that a wide variety of phenomena in quantum field theory (QFT) can be characterised using quantum information measures, such as the entanglement entropy of a state and the relative entropy between quantum states in the same Hilbert space. In this thesis, we use these and other tools from quantum information theory to study several interesting problems in quantum field theory. These topics range from the Aharonov-Bohm effect in QFT to the consistence of the Ryu-Takayanagi formula (proposed in the context of the AdS/CFT duality) with some properties of relative entropy. We also have derived new interesting quantum energy inequalities using properties of relative entropy, that constrain the spatial distribution of negative energy density.
\end{foreignlanguage}
\vspace{\fill}

\begin{foreignlanguage}{english}
\noindent\textsf{\textbf{Keywords:}
entanglement entropy, quantum information, relative entropy, holography, quantum entanglement, holographic entanglement entropy, negative energy, modular hamiltonian.}
\end{foreignlanguage}

\renewcommand\tablename{Tabla}
\renewcommand\contentsname{Contenidos}




\renewcommand{\baselinestretch}{1.1}
\mainmatter
\pagestyle{fancy}
\tableofcontents 

\setlength{\parskip}{3mm}

\cleardoublepage
\addcontentsline{toc}{chapter}{Resumen}
\thispagestyle{empty}
\chapter*{\hfill Resumen \hfill}

En las últimas décadas, se ha comprendido que una gran variedad de fenómenos en teoría de campos puede entenderse y caracterizarse utilizando medidas de información cuántica, tales como la entropía de entrelazamiento, la información mutua y la entropía relativa, entre muchas otras. Este trabajo de investigación apunta a utilizar estas y otras herramientas de la teoría de la información cuántica, para estudiar diversos problemas relacionados con la física de altas energías y las teorías de gravedad.

Desde el punto de vista de la teoría cuántica de campos, la entropía de entrelazamiento es una cantidad no local con interesantes propiedades no perturbativas, que puede ser definida para cualquier teoría, independientemente del contenido de campos. Si bien el cálculo preciso de esta cantidad suele ser bastante dificultoso (aún para teorías libres), en los últimos años se ha logrado un gran progreso analítico y numérico en este sentido. El abanico de aplicaciones de los resultados es muy variado, y abarca aspectos formales de la teoría de campos como también algunos más prácticos (búsqueda de teoremas C en dimensiones superiores, relación con efectos topológicos, cotas para la distribución espacial de energía negativa, etc.). En este trabajo, una de las cosas que mostramos es que, como una de sus aplicaciones, la entropía de entrelazamiento de campos libres exhibe una dependencia en la fase de Aharonov-Bohm.

En el contexto de la dualidad de J. M. Maldacena, en 2006 se ha propuesto una fórmula para calcular la entropía de entrelazamiento entre una región y su complemento en una teoría de campos conforme, a partir del cómputo del área de una superficie extremal, homóloga al borde de la región, en la teoría dual gravitatoria. En este trabajo, demostramos que esta prescripción satisface una desigualdad muy restrictiva derivada de la positividad de la entropía relativa, lo que le brinda un gran soporte a esta prescripción holográfica. La entropía relativa es una cantidad de la teoría de la información que da una medida de la distancia entre dos estados cuánticos. Un análisis más profundo de la desigualdad para estados similares nos permite arribar también a interesantes conclusiones relacionadas con la localización en teoría de campos.

Otra de las propiedades de la entropía relativa, la monotonicidad ante la inclusión, nos permite también derivar una cota cuántica reminiscente a la cota de Bekenstein, que pone severas restricciones a la distribución espacial de energía negativa en una teoría de campos conforme.

En las aplicaciones mencionadas, se manifiesta la importancia que tiene un objeto de la teoría de campos denominado hamiltoniano modular. En esta tesis, también nos centramos en el estudio de este objeto, analizando en particular la forma que tienen los términos locales de los hamiltonianos modulares de campos libres.

Esta tesis, en su conjunto, además de proveer interesantes y novedosos resultados en varios ámbitos de la teoría de campos, demuestra lo amplio que resulta el espectro de aplicaciones de herramientas de la teoría de la información cuántica al estudio de problemas en física de altas energías y en teorías gravitatorias.

\vspace{\fill}

\noindent\textsf{\textbf{Palabras clave:}
entropía de entrelazamiento, información cuántica, entrelazamiento cuántico, entropía relativa, holografía, entropía holográfica, energía negativa, hamiltoniano modular
}

\thispagestyle{empty}

\cleardoublepage

\chapter{\label{ch:introduccion}Introducción}

El tema de esta tesis se enmarca en un área interdisciplinaria de gran interés actual, donde confluyen nuevas investigaciones en materias tan diversas como la teoría de la información cuántica, los modelos de sistemas cuánticos extendidos, la teoría de campos relativista, las teorías de la gravedad y la holografía.

Para hacer cuantitativo el grado de entrelazamiento entre dos subsistemas se han propuesto varias medidas, entre las que tiene especial importancia la llamada entropía de entrelazamiento, correspondiente a la entropía de von Neumann de la matriz densidad reducida del subsistema en consideración. En sistemas cuánticos extendidos la entropía de entrelazamiento correspondiente al estado reducido a una región del espacio suele denominarse entropía geométrica. Esta mide la información compartida por distintas regiones debida a las fluctuaciones de vacío.

En teoría de campos la entropía en una región $V$ del espacio contiene varios términos divergentes, el más relevante de los cuales es proporcional al área de la frontera de $V$ \cite{review}. Las divergencias se deben a la creación de pares de partículas como consecuencia de la localización en $V$. Este es el punto de partida de varios trabajos que intentan explicar la entropía de los agujeros negros, proporcional al área del horizonte, como entropía geométrica \cite{bh,sred,casinid}. Las divergencias de la entropía localizada aparecen también en cualquier estado de energía finita. El cálculo tradicional de la entropía de gases u otros sistemas pasa por alto este problema imponiendo condiciones de borde artificiales en la región considerada (por ejemplo una caja con paredes reflectantes), y luego definiendo el sistema y el estado en el interior, eliminando de esta forma las correlaciones con el exterior. Luego, las cantidades intensivas, como la entropía por unidad de volumen, se calculan en el límite de volumen infinito. De esta forma, se obtienen resultados independientes de las condiciones de borde particulares. Sin embargo, no siempre es posible tomar el límite de volumen infinito, como sucede en general cuando se considera la interacción gravitatoria. Casos notables incluyen la entropía de un agujero negro, o la que corresponde al interior del horizonte cosmológico. Estos sistemas nos llevan a las sutilezas del problema de localización en relatividad. De hecho, la presencia de horizontes no es excepcional, sino que constituye la regla cuando la gravedad está presente y se intenta tomar el límite de volumen infinito de un sistema. Hablando estrictamente, no existe entonces una noción de entropía localizada en gravedad semiclásica. Sin embargo, la noción de entrelazamiento entre regiones puede definirse de modo preciso, en teoría de campos, aún en presencia de gravitación. Las medidas de entrelazamiento entre regiones disjuntas, como la información mutua, no contienen divergencias.

La entropía de entrelazamiento se ha consolidado como una herramienta fundamental para estudiar diversos problemas de la física. En materia condensada, por ejemplo, puede ser utilizada para distinguir nuevas fases topológicas y puntos críticos \cite{wenx,hamma,cacardy}. En el contexto de la teoría cuántica de campos, la entropía de entrelazamiento juega un papel relevante al estudiar transiciones de fase en teorías de calibre \cite{igor0,buivi} y ha provisto nuevas ideas acerca de la estructura del flujo del grupo de renormalización \cite{twoD,balas} (siendo fundamental para establecer teoremas C en 2+1 y dimensiones superiores \cite{threeD,msinha1,msinha2}). A un nivel más fundamental, se ha sugerido que la entropía de entrelazamiento tiene un rol relevante para entender la estructura cuántica del espacio-tiempo \cite{mvr,mvr1,bianchi}.

Otra de las aplicaciones interesantes se da en términos de la dualidad de Maldacena. En el marco de esta dualidad, se conjetura que la entropía de entrelazamiento en la teoría de campos del borde se determina a partir de un elegante cálculo en la teoría dual gravitatoria \cite{rt1,rt2}. A la entropía calculada de esta forma se la conoce como entropía holográfica de entrelazamiento. Esta prescripción para el cálculo de la entropía ha superado exitósamente diverdas pruebas de consistencia y ha sido demostrada para ciertos casos particulares (ver por ejemplo \cite{rt1,rt2,hee,hms,hmm,lewko}). En particular, en \cite{blanco2} hemos demostrado que satisface una desigualdad muy restrictiva derivada de la positividad de la entropía relativa. Este trabajo fue pionero en la utilización de la entropía relativa en el contexto de la holografía y las consecuencias de la validez de esa desigualdad son variadas e interesantes. Por un lado, ha puesto de manifiesto que existe un análogo a la primera ley de la termodinámica para la entropía de entrelazamiento. También ha permitido idear un procedimiento para reconstruir el operador densidad del vacío sólo a partir del funcional entropía (``tomografía del vacío a partir del entrelazamiento'').

La entropía relativa también tiene interesantes aplicaciones dentro de la teoría de campos. Es la herramienta fundamental utilizada para probar la versión cuántica de la cota de Bekenstein \cite{beke0,beke00}, que relaciona la energía de una región y su entropía. La demostración de esta desigualdad en \cite{beke1} pone de manifiesto que esta cota no impone nuevas restricciones provenientes de la física de agujeros negros para la física en espacio plano. También soluciona el llamado ``problema de las especies''. En \cite{blanco3}, demostramos que la monotonicidad de la entropía relativa provee una nueva desigualdad similar a la cota de Bekenstein. De esta desigualdad, se desprenden interesantes relaciones que restringen la distribución espacial de energía negativa para teorías de campos conformes. Una interesante línea de aplicación y exploración futura de este trabajo consiste en analizar la validez de alguna cota similar en espacio curvo y, eventualmente, en alguna geometría similar a la de un agujero de gusano (ya que se entiende que es necesaria una cantidad suficiente de energía negativa para sostener la formación de un agujero de gusano \cite{wormhole}).

En conexión con estos temas que mencionamos resulta relevante conocer un objeto llamado hamiltoniano modular, que aparece por ejemplo en las formulaciones cuánticas de la cota de Bekenstein. El hamiltoniano modular es generalmente un objeto no local y, desafortunadamente, se conoce explícitamente sólo en unos pocos casos. En esta tesis encontramos la forma que tiene el hamiltoniano modular para un estado térmico en distintas regiones y también abordamos el estudio de la forma que tienen los términos locales que aparecen en un hamiltoniano modular (genérico) de campos libres. 

Esta tesis está dividida en 10 capítulos. El objetivo del autor es brindar una presentación bastante autocontenida, haciendo énfasis en detenerse a presentar los conceptos más relevantes para el entendimiento del trabajo de investigación realizado.

El capítulo \ref{ch:entropia} es una introducción general sobre la entropía de entrelazamiento en teorías cuánticas y sus aplicaciones. Se discuten las definiciones básicas y se presentan algunas de las herramientas de cálculo más relevantes en el área. Luego, en el capítulo \ref{ch:abe}, se aborda un análisis del efecto Aharonov-Bohm sobre las fluctuaciones de vacío utilizando la entropía de entrelazamiento. Este capítulo, además de brindar resultados novedosos que indican que la entropía de entrelazamiento `mide' el efecto Aharonov-Bohm, sirve como un buen ejercicio para que el lector se meta de lleno en un cálculo explícito de la entropía en teoría de campos.

El estudio de hamiltonianos modulares está dividido en dos capítulos. Comienza en el capítulo \ref{ch:modular}, donde se presenta su definición y propiedades fundamentales. En ese capítulo, se deriva además la forma que tiene el hamiltoniano modular en distintas situaciones, algunas de las cuales resultan relevantes para el resto de este trabajo.

El capítulo \ref{ch:relativa} presenta los conceptos de entropía relativa y holografía. Se exploran sus aplicaciones y se sientan las bases para el desarrollo de los dos capítulos siguientes. La entropía relativa es una cantidad estadística fundamental que mide la ``distancia'' entre dos estados en el mismo espacio de Hilbert. Una de sus propiedades más relevantes es que resulta positiva (salvo para estados iguales, en cuyo caso es nula). La propiedad de positividad se traduce en una desigualdad muy interesante: la diferencia entre la entropía de entrelazamiento de dos estados está acotada por la diferencia del valor de expectación del hamiltoniano modular en esos estados. Esta desigualdad se utiliza en el capítulo \ref{ch:hee} para realizar pruebas de consistencia no triviales para la entropía holográfica de entrelazamiento (conjetura de Ryu-Takayanagi), mientras que en el capítulo \ref{ch:aplicaciones} se discuten algunas paradojas aparentes que resultan del análisis de la desigualdad para estados similares y se mencionan algunas aplicaciones a las que dio lugar parte de este trabajo.

En el capítulo \ref{ch:negativa} utilizamos la propiedad de monotonicidad de la entropía relativa para derivar una relación entre la entropía y energía, reminiscente a la cota de Bekenstein. En este capítulo discutimos las diferencias y similitudes entre esta nueva cota, la versión original propuesta por Bekenstein y la formulación cuántica previa que se conocía de esta desigualdad. Este capítulo comienza motivado por algunas conclusiones obtenidas en el capítulo \ref{ch:aplicaciones}, que sugieren la existencia de restricciones a la localización de energía negativa. En efecto, demostramos en particular que la nueva cota cuántica que obtenemos pone severas restricciones a la distribución espacial de energía negativa en teorías de campos conformes.

A lo largo de este trabajo se manifiesta la relevancia que tiene conocer el hamiltoniano modular para distintos problemas. En vistas de esto, el capítulo \ref{ch:local} es un análisis de la forma general que tienen los términos locales del hamiltoniano modular de campos libres. Este avance representa un puntapié inicial en el estudio de estos objetos en las teorías de campos.

Para finalizar, en el capítulo \ref{ch:conclusiones} presentamos un resumen de los resultados más relevantes obtenidos en este trabajo, así como también las perspectivas futuras en esta área de investigación.
\chapter{\label{ch:entropia}Entropía de entrelazamiento}

\section{Entropía en teorías cuánticas}

El entendimiento moderno de la entropía como la cantidad que conecta las interpretaciones microscópica y macroscópica de un sistema es debido principalmente a L. Boltzmann y J. Gibbs. La entropía estadística de un sistema termodinámico se define como
\begin{equation}\label{ent-est}
S=k_B \ln(\Omega)\,,
\end{equation}
donde $k_B=1.38066 \times 10^{-23} J K^{-1}$ es la constante de Boltzmann y $\Omega$ es el número de microestados posibles para un dado macroestado y se interpreta como el ``grado de desorden'' que hay en el sistema.

En mecánica estadística clásica, el número de microestados de un sistema es infinito, dado que las variables que se utilizan para describir a los sistemas clásicos son continuas. Para poder contar el número de microestados en mecánica clásica se suele realizar una discretización en el espacio de fases, definiendo una ``celda unidad'' en forma arbitraria. El número de microestados se define entonces como el cociente entre el volumen accesible del espacio de fases y el volumen de la celda unidad. Nótese que dada la arbitrariedad en la definición de la celda unidad, la entropía queda definida a menos de una constante. En mecánica cuántica no existe este tipo de ambigüedad ya que la discretización viene dada por la misma teoría.

En mecánica cuántica uno trata con \emph{observables} y con \emph{estados}. Los observables, como el momento, la posición, etc., se describen matemáticamente por operadores autoadjuntos en un espacio de Hilbert. La entropía no es un observable (es decir, no existe un operador con la propiedad de que su valor de expectación en algún estado sea la entropía) sino que es una función del estado.

\subsection{Operador densidad}
El estado de un sistema cuántico se conoce completamente si se puede representar mediante un cierto vector $\left|\psi\right\rangle$ en el espacio de Hilbert (se dice entonces que el estado es puro). Si la información que se posee acerca del sistema es incompleta, el estado del mismo no puede representarse por un vector único sino por una mezcla estadística de vectores $\left|\psi_1\right\rangle$, $\left|\psi_2\right\rangle$,..., $\left|\psi_n\right\rangle$ y decimos que el sistema tiene probabilidades $p_1$, $p_2$,..., $p_n$ ($p_i\geq0$) de encontrarse en alguno de estos estados.

Supongamos que se realiza la medición de una magnitud física del sistema representada por el operador $A$. El valor de expectación de $A$ viene dado por
\begin{equation}\label{valorexp}
\left\langle A\right\rangle=\sum_{n}p_n\left\langle \psi_n\left|A\right|\psi_n\right\rangle\,,
\end{equation}
donde hemos supuesto que cada $\left|\psi_i\right\rangle$ tiene norma uno.

Es común describir la mezcla estadística de los vectores $\left|\psi_i\right\rangle$, mediante el llamado \emph{operador densidad}
\begin{equation}\label{densidad}
\rho=\sum_{n}p_n\left|\psi_n\right\rangle\left\langle \psi_n\right|\,.
\end{equation}
El valor de expectación de un observable $A$ dado por (\ref{valorexp}) se reescribe en términos del operador densidad
\begin{equation}\label{expectacion}
\left\langle A\right\rangle=\textrm{tr}(\rho A)\,.
\end{equation}
Si se toma $A=Id$, se obtiene la condición
\begin{equation}\label{norm}
\textrm{tr}(\rho)=1\,,
\end{equation}
que está asociada a que los números $p_i$ son \emph{pesos estadísticos} y por lo tanto debe resultar
\begin{equation}\label{pesos}
\sum_{n}p_n=1\,.
\end{equation}
En general entonces, el estado de un sistema queda descripto por un operador autoadjunto, de autovalores positivos y traza igual a la unidad. Estas propiedades, permiten probar que el espectro de $\rho$ es discreto y sus autovalores están comprendidos entre 0 y 1.

El formalismo del operador densidad permite estudiar los casos puros como casos particulares de mezcla estadística. Si el sistema se encuentra en un estado puro $\left|\alpha\right\rangle$, se puede escribir su operador densidad como
\begin{equation}\label{puro}
\rho=\left|\alpha\right\rangle\left\langle \alpha\right|\,,
\end{equation}
que no es más que un proyector sobre el estado $\left|\alpha\right\rangle$. En este caso, es sencillo comprobar que
\begin{equation}\label{rhocuadrado}
\rho^2=\rho\,.
\end{equation}
En general, todo operador hermítico definido positivo de traza uno, posee la propiedad $\textrm{tr}\rho^2\leq1$. De las ecuaciones (\ref{norm}) y  (\ref{rhocuadrado}) se ve trivialmente que si el estado del sistema es puro, entonces debe ser
\begin{equation}\label{rhocuadrado1}
\textrm{tr}(\rho^2)=1\,.
\end{equation}
Un operador densidad genérico $\rho$ representa un estado puro si y sólo si vale la ecuación (\ref{rhocuadrado1}).

En conclusión, el estado de un sistema se representa por un operador densidad con las características mencionadas, y dado este operador, es posible determinar los valores de expectación de observables utilizando la ecuación (\ref{expectacion}).

\subsubsection{Operador densidad reducido}\label{red}
Consideremos dos sistemas $A$ y $B$ con espacios de Hilbert $\cal{H}_A$ y $\cal{H}_B$ respectivamente y supongamos que el estado del sistema completo es $\left|\psi\right\rangle\in \cal{H}_A\otimes \cal{H}_B$. Introducimos una base $\left|i,j\right\rangle \doteq \left|i\right\rangle _A \otimes \left|j\right\rangle _B$ del espacio $\cal{H}_A\otimes \cal{H}_B$ ($\lbrace\left| l \right\rangle _A \rbrace_l$ y $\lbrace\left| l \right\rangle _B \rbrace_l$ son bases de $\cal{H}_A$ y $\cal{H}_B$, respectivamente); con esa representación $\left|\psi\right\rangle$ se escribe como $\left|\psi\right\rangle=\sum\lambda_{kl}\left|k,l\right\rangle$, donde $\lambda_{kl}=\langle \left. k,l\right|\psi\rangle$ y $\sum\left|\lambda_{kl}\right|^2=1$. Para hallar los valores de expectación de observables que sólo actúan en $\cal{H}_A$ (nos referimos así a aquellos observables que actúan como la identidad en $\cal{H}_B$) se define el operador densidad reducido al subsistema $A$ como
\begin{eqnarray}\label{reducido}
\rho_A &=& \textrm{tr}_B\left|\psi\right\rangle \left\langle \psi\right| = \\ \nonumber
&=& \sum_l \left\langle l \right| _B \left( \left|\psi\right\rangle \left\langle \psi\right| \right) \left| l \right\rangle _B =\sum \lambda_{kl}\lambda_{il}^*\left|k\right\rangle _A \left\langle i\right| _A \,,
\end{eqnarray}
Es relevante remarcar que $\rho_A$ es independiente de la base $\left| l \right\rangle _B$ elegida.

Los valores de expectación de operadores $O:\cal{H}_A\rightarrow \cal{H}_A$ vendrán dados por
\begin{equation}\label{expectacionred}
\left\langle O\right\rangle=\textrm{tr}(\rho_A O)\,.
\end{equation}
Es interesante notar que cualquier operador densidad puede expresarse como estado reducido de un operador densidad puro. En efecto, consideremos un sistema caracterizado por el operador densidad $\rho$ y supongamos que los autovalores del mismo son $\lambda_i$, con lo que podemos escribir
\begin{equation}\label{ejreduccion}
\rho= \lambda_i\left|i\right\rangle\left\langle i\right|\,,
\end{equation}
para ciertos estados $\left|i\right\rangle$ en el espacio de Hilbert $\cal{H}$ del sistema. Considemos una base $\left|i,j\right\rangle=\left|i\right\rangle\otimes\left|j\right\rangle\in \cal{H} \otimes \cal{H}$. Se define
\begin{equation}\label{producto}
\rho_1=\sum \sqrt{\lambda_{i}\lambda_{j}}\left|i,i\right\rangle\left\langle j,j\right|=\left(\sum\sqrt{\lambda_i}\left|i,i\right\rangle\right)\left(\sum\sqrt{\lambda_j}\left\langle j,j\right|\right)\,,
\end{equation}
Es sencillo ver que $\textrm{tr}({\rho_1}^2)=1$, por lo que el estado que representa $\rho_1$ es puro (es un proyector al estado $\sum\sqrt{\lambda_i}\left|i,i\right\rangle$). Además, resulta
\begin{equation}\label{reduccion}
\rho= \textrm{tr}_B\rho_1\,.
\end{equation}
donde $\textrm{tr}_B$ indica la \textit{traza parcial} sobre el espacio de Hilbert $\cal{H}_B$, como se ha definido en (\ref{reducido}).

Una propiedad interesante surge de considerar un sistema para el cual el espacio de Hilbert se factoriza $\cal{H}_A \otimes \cal{H}_B$. Si el estado total del sistema $\rho$ es puro, entonces, los autovalores no nulos de los operadores densidad reducidos $\rho_A$ y $\rho_B$ son los mismos. Esto puede demostrarse viendo que vale la igualdad $\tr (\rho_A^n)=\tr (\rho_B^n)$ para toda potencia $n$ y consecuentemente, los autovalores no nulos de $\rho_A$ y $\rho_B$ y sus multiplicidades coinciden.

\subsection{Entropía de von Neumann}\label{neumann1}

Como se mencionó anteriormente, la entropía en mecánica cuántica es una función del estado del sistema. La forma más natural de definir la entropía de un sistema cuántico que se encuentra en un estado representado por el operador densidad $\rho$ es la siguiente
\begin{equation}\label{neumann}
S(\rho)=-k_B \textrm{tr}(\rho \ln \rho)\,,
\end{equation}
o sea, $S=-k_B \sum_{i}\lambda_i \ln \lambda_i$, donde $\lambda_i$ son los autovalores de $\rho$. Esta fórmula es debida a J. von Neumann (1927) y generaliza la expresión (\ref{ent-est}) de la entropía del caso clásico al cuántico.

Se comentó que la definición (\ref{ent-est}) establecía una relación entre la variable de estado $S$ y la ``cantidad de desorden'' del sistema. Este grado de desorden está asociado a la cantidad de microestados posibles que pueden generar un estado con las mismas propiedades macroscópicas. En mecánica cuántica, el ``número de microestados'' puede interpretarse como el número de estados puros en los que se puede encontrar al sistema. Para ver esto, considérese por ejemplo un sistema cuántico que puede hallarse con la misma probabilidad en un número $N$ de estados puros. El operador densidad del sistema es entonces $\rho=(1/N)P$, donde $P$ es un operador que proyecta vectores sobre cierto subespacio de dimensión $N$. La entropía resulta $S=-k_B \textrm{tr}(\rho \ln \rho)=k_B \ln(N)$ y se observa la similitud entre las ecuaciones (\ref{ent-est}) y (\ref{neumann}).

La entropía de von Neumann es la única cantidad para la cual aparece esta semejanza con (\ref{ent-est}) y para la cual se cumplen ciertas propiedades\footnote{Tales como la aditividad y subaditividad, y diversas propiedades de mezclado. En las siguientes secciones estudiamos con más detalle las propiedades generales de la entropía.} que se desea que ``herede'' una entropía definida en un contexto cuántico, de las propiedades conocidas para el caso clásico.

De la propia definición de la entropía de von Neumann, se sigue que la misma es nula sólo cuando el operador densidad que describe el estado tiene sólo un autovalor no nulo e igual a la unidad, es decir, cuando el estado es puro.

De ahora en adelante tomaremos $k_B=1$. Esta elección hace que la entropía sea una cantidad adimensional (y, correspondientemente, la temperatura pasa a tener unidades de energía).

\subsubsection{\label{sec:propentro}Propiedades generales de la entropía}

La entropía de von Neumann definida a través de (\ref{neumann}), para cualquier estado $\rho$ es una cantidad positiva. Para estados puros (y sólo para ellos) resulta $S=0$. Es sencillo verificar que el rango de $S\left(\rho\right)$ es $\left[ 0,+\infty \right]$.

Si el espacio de Hilbert del sistema tiene dimensión finita $N$, entonces, para cualquier estado $\rho$ se tiene $S\left(\rho\right) \leq \log N$. La igualdad sólo se da para los llamados estados maximalmente entrelazados.

La entropía es invariante ante transformaciones unitarias de $\rho$, es decir: $S\left(\rho\right)=S\left(U \rho U^{-1}\right)$ para cualquier operador $U$ unitario.

Entre las propiedades de la entropía de von Neumann, vamos a destacar la aditividad, subaditividad y subaditividad fuerte \footnote{Para un estudio detallado de estas y otras propiedades de la entropía ver \cite{wehrl}.}.

\begin{enumerate}

\item \textbf{Aditividad de la entropía}

La \textit{aditividad} de la entropía establece que, dados dos operadores densidad $\rho_A$ y $\rho_B$, endomorfismos de los espacios de Hilbert $\cal{H}_A$ y $\cal{H}_B$ respectivamente, la entropía del operador densidad $\rho_A \otimes \rho_B : \cal{H}_A \otimes \cal{H}_B \longrightarrow \cal{H}_A \otimes \cal{H}_B$ es
\begin{equation}
S\left(\rho_A \otimes \rho_B\right)= S\left(\rho_A \right) + S\left(\rho_B\right)\,.
\label{aditividad}
\end{equation}
La demostración de esta propiedad es bastante sencilla. Llamemos $\vert j\rangle_A$ y $\vert k\rangle_B$ a los autovectores de $\rho_A$ y $\rho_B$, asociados a los autovalores $\lambda^A_j$ y $\lambda^B_k$, respectivamente. Por lo tanto, $\vert j\rangle_A \otimes \vert k\rangle_B$ es autovector de $\rho_A \otimes \rho_B$ con autovalor $\lambda^A_j \lambda^B_k$ y se tiene entonces
\begin{eqnarray}
S\left(\rho_A \otimes \rho_B\right)&=& -\sum_{j,k}{\lambda^A_j \lambda^B_k \log\left(\lambda^A_j \lambda^B_k\right)}=\nonumber \\
&=& -\sum_{j}{\lambda^A_j \log\left(\lambda^A_j \right)} -\sum_{k}{\lambda^B_k \log\left(\lambda^B_k \right)} = \nonumber \\
&=& S\left(\rho_A \right) + S\left(\rho_B\right)\nonumber \,.
\end{eqnarray}
Desde el punto de vista de la teoría de la información esta propiedad es bastante clara: si tenemos dos sistemas independientes, entonces la información total sobre el sistema es la suma de la información de cada uno de sus constituyentes.

Es importante entender que cuando hay correlaciones (es decir, cuando el operador densidad del sistema no tiene la forma $\rho_A \otimes \rho_B$ en $\cal{H}_A \otimes \cal{H}_B$) esta propiedad no vale. Si las correlaciones entre los subsistemas $A$ y $B$ son importantes, la entropía total puede llegar a ser mucho más pequeña que la suma de las entropías $S\left(\rho_A \right) + S\left(\rho_B\right)$ (de hecho, puede llegar a anularse).
\bigskip
\item \textbf{Subaditividad de la entropía}

Consideremos un sistema cuyo espacio de Hilbert es de la forma $\cal{H}_A \otimes \cal{H}_B$. Un estado genérico $\rho$, endomorfismo de $\cal{H}_A \otimes \cal{H}_B$, no necesariamente es un producto tensorial de un endomorfismo de $\cal{H}_A$ por un endomorfismo de $\cal{H}_B$. La propiedad de aditividad demostrada anteriormente no vale para un estado general $\rho$ de este tipo, sino sólo para estados producto. Sin embargo, también pueden decirse cosas sobre la entropía de un estado general que no sea un producto en $\cal{H}_A \otimes \cal{H}_B$. Para ello, se recurre al concepto de operador densidad reducido, definido por la ecuación (\ref{reducido}).

Llamamos $\rho_A$ y $\rho_B$ a los operadores densidad reducidos a $\cal{H}_A$ y $\cal{H}_B$ provenientes del estado global $\rho$ de nuestro sistema en $\cal{H}=\cal{H}_A \otimes \cal{H}_B$. Es decir
\begin{eqnarray}
\rho_A &=& \tr _B \left(\rho \right)\,\,\,y \nonumber \\
\rho_B &=& \tr _A \left(\rho \right)\,. \nonumber
\end{eqnarray}
La propiedad de \textit{subaditividad} establece que
\begin{equation}
S\left(\rho\right) \leq S\left(\rho_A\right) + S\left(\rho_B\right) = S\left(\rho_A \otimes \rho_B\right)\,.
\label{subaditividad}
\end{equation}
Esta propiedad es plausible dado que al formar los estados $\rho_A$ y $\rho_B$ se pierde información acerca de las correlaciones (no es posible reconstruir $\rho$ a partir de $\rho_A$ y $\rho_B$). La propiedad de subaditividad es una consecuencia de la propiedad de subaditividad fuerte, que presentaremos más adelante.

A partir de la subaditividad (\ref{subaditividad}) se desprende la propiedad de \textit{concavidad} de la entropía
\begin{equation}
S\left(\lambda \rho _1 + \left(1-\lambda\right)\rho _2\right) \geq \lambda S\left(\rho _1\right) + \left(1-\lambda\right) S\left(\rho _2\right)\,,
\end{equation}
para todo $\lambda \in \left[0,1\right]$.

\item \textbf{Subaditividad fuerte de la entropía}

Una relación más fuerte que la subaditividad surge al considerar un estado $\rho$ endomorfismo de un espacio del Hilbert de la forma $\cal{H}_A \otimes \cal{H}_B \otimes \cal{H}_C$. La propiedad de \textit{subaditividad fuerte} dice que
\begin{equation}
S\left(\rho_{AB}\right)+S\left(\rho_{BC}\right)-S\left(\rho\right)-S\left(\rho_{B}\right) \geq 0\,,
\label{subfuerte}
\end{equation}
donde se introdujeron los operadores densidad reducidos a los espacios $\cal{H}_A \otimes \cal{H}_B$: $\rho_{AB}=\tr _C \rho$, $\cal{H}_B \otimes \cal{H}_C$: $\rho_{BC}=\tr _A \rho$, y $\cal{H}_B$: $\rho_{B}=\tr _A \tr _C \rho$.
Volveremos a hacer referencia varias veces a esta propiedad. En particular, daremos la demostración de la misma a partir de la propiedad de monotonicidad de la \textit{entropía relativa}, cantidad que se introduce en el capítulo \ref{ch:relativa} de este trabajo.

Notemos que si elegimos $\cal{H}_B$ tal que $\dim \left(\cal{H}_B\right) = 1$, la propiedad de subaditividad fuerte (\ref{subfuerte}) se reduce a la propiedad de subaditividad dada por la ecuación \reef{subaditividad}.

\end{enumerate}

\section{Entrelazamiento cuántico}

El \textit{entrelazamiento} es un fenómeno cuántico, sin equivalente clásico, en el cual los estados de dos o más subsistemas se deben describir haciendo referencia a los estados del sistema global, incluso cuando la separación entre los subsistemas es grande.

Las mediciones realizadas sobre algún observable de un sistema que está entrelazado con otros parecen influenciar instantáneamente a estos. Por ejemplo, si tenemos un sistema de dos electrones con espín total nulo y los separamos infinitamente, la medición del espín de uno de ellos determina de forma unívoca el espín del otro. Esto parece sugerir que existe alguna interacción instantánea entre los sistemas, a pesar de la separación entre ellos, lo que violaría el principio de relatividad.

El entrelazamiento cuántico fue en un principio utilizado por A. Einstein, B. Podolsky y N. Rosen (EPR)\cite{einstein} como un argumento para intentar probar la incompletitud de la mecánica cuántica como teoría física, alegando que las correlaciones predichas por la mecánica cuántica son inconsistentes con el Principio del Realismo Local \footnote{El Principio de Realismo Local es una conjunción entre el Principio de Localidad y la asunción de que todas las magnitudes físicas tienen valores preexistentes para cada medición incluso antes de que se realice la medición.}.

En 1964, un trabajo presentado por J. Bell \cite{bell} permitió cuantificar matemáticamente en forma de desigualdades las implicaciones teóricas planteadas en el trabajo de EPR. El hecho de que las desigualdades de Bell sean violadas\footnote{Hecho que se ha comprobado experimentalmente, en primera oportunidad por A. Aspect en 1982 \cite{aspect}.} provee evidencia empírica a favor de la mecánica cuántica y en contra del Realismo Local.

\subsection{El experimento EPR}

En el trabajo original de Einstein, Podolsky y Rosen se hace un análisis del entrelazamiento que presenta un sistema de dos partículas, que interactuaron durante un intervalo de tiempo determinado y luego son separadas espacialmente, en términos del momento lineal y las posiciones de las partículas. Una situación en la que se puede visualizar el fenómeno de entrelazamiento en forma más sencilla fue introducida por D. Bohm en 1951, y consiste en pensar en un sistema formado por dos partículas de espín 1/2 en un estado inicial de espín total igual a cero. En tal caso, el estado de espín inicial del sistema se escribe como
\begin{equation}\label{espinini}
\left|\chi\right\rangle=\frac{1}{\sqrt{2}}\left(\left|+ -\right\rangle-\left|- +\right\rangle\right)\,,
\end{equation}
donde $\left|+ -\right\rangle=\left|+\right\rangle_1\otimes\left|-\right\rangle_2$, y $\left|-\right\rangle_1$ y $\left|+\right\rangle_2$ son los estados con proyección de espín negativa y positiva en alguna dirección determinada. Una de las partículas se aleja significativamente de la otra para suprimir las interacciones entre ellas. Si un observador mide el espín de una de estas partículas y la encuentra, por ejemplo, en el estado de espín positivo, este observador sabe inmediatamente que la partícula que está alejada de él tiene que encontrarse en el estado de espín negativo. Esto está asociado al colapso de la función de onda total de las dos partículas después de la medición; si la partícula $1$ tiene un espín positivo, entonces el estado inicial colapsa a
\begin{equation}\label{espinfin}
\left|\chi\right\rangle=\left|+ -\right\rangle\,,
\end{equation}
lo que indica que si midiera la proyección del espín de la partícula $2$ (en la dirección en cuestión) el mismo va a ser negativo. A pesar de haber medido una propiedad de la partícula $1$, la partícula $2$ es ``afectada'' por la medición. En realidad, esto dice que los sistemas de muchas partículas que estuvieron interactuando durante un tiempo no pueden ser descriptos luego como sistemas desconectados o independientes, aún a tiempos posteriores de que la interacción se suprima. Esto es lo que constituye el fenómeno del entrelazamiento cuántico.

\subsection{Medidas de entrelazamiento}

Hasta ahora se ha introducido una idea de lo que representa el fenómeno de entrelazamiento. Ahora daremos la definición de un estado entrelazado\footnote{Al hablar de entrelazamiento nos estaremos refiriendo a entrelazamiento bipartito, es decir, entre dos subsistemas. El concepto de entrelazamiento se extiende al caso de un sistema compuesto por más de dos subsistemas y se lo estudia bajo el nombre de entrelazamiento multipartito \cite{horo}} y después presentaremos algunas cantidades que sirven para cuantificar el grado de entrelazamiento entre las partes de un sistema.

Consideremos un sistema cuyo espacio de Hibert es de la forma $\cal{H}_A \otimes \cal{H}_B$. Decimos que un estado $\rho : \cal{H}_A \otimes \cal{H}_B \longrightarrow \cal{H}_A \otimes \cal{H}_B$ es \textbf{separable} si existe un conjunto de números $\lbrace p_{\lambda}\rbrace$, con $\sum p_{\lambda}=1$ y $\rho_A^{\lambda}$, $\rho_B^{\lambda}$ endomorfismos de $\cal{H}_A$ y $\cal{H}_B$, respectivamente, tales que
\begin{equation}
\rho=\underset{\lambda}\sum p_{\lambda}\rho^{\lambda}_A\otimes\rho^{\lambda}_B\,.
\label{separable}
\end{equation}
Si $\rho$ no es separable, entonces decimos que $\rho$ es un estado \textbf{entrelazado} entre $\cal{H}_A$ y $\cal{H}_B$.

Notemos que el entrelazamiento es una propiedad que depende de la partición que uno hace del espacio de Hilbert del sistema completo. Por ejemplo, si consideramos un sistema de tres electrones no interactuantes fijos en el espacio, el espacio de Hilbert del sistema es $\cal{H}_A \otimes \cal{H}_B \otimes \cal{H}_B$, donde $\cal{H}_A=\cal{H}_B=\cal{H}_C=\cal{H}$ es un espacio de Hilbert separable de dimensión $2$ (el espacio generado por los autovectores del operador de espín de un electrón en alguna dirección). El siguiente estado puro de tres electrones
\begin{equation}
\vert + \rangle_A \otimes \left[ \frac{1}{\sqrt{2}} \left(\vert + \rangle_B \otimes \vert - \rangle_C - \vert - \rangle_B \otimes \vert + \rangle_C \right)\right]\,
\end{equation}
es separable entre $\cal{H}_A$ y $\cal{H}_B \otimes \cal{H}_C$, mientras que está entrelazado entre $\cal{H}_B$ y $\cal{H}_A \otimes \cal{H}_C$.

\subsection{Entropía de entrelazamiento}

Ya se comentó que el entrelazamiento (bipartito) da cuenta, en algún sentido, de las correlaciones existentes entre dos subsistemas aún luego de que los mismos no están interactuando. Una cantidad que permite cuantificar el grado de entrelazamiento entre dos subsistemas es la entropía de entrelazamiento.

Consideremos un sistema compuesto de dos subsistemas $A$ y $B$ (es decir, el espacio de Hilbert es $\cal{H}_A \otimes \cal{H}_B$), en un estado total puro. Los resultados de las mediciones que se realizan sobre observables en el subsistema $A$ ($B$) pueden predecirse utilizando el operador densidad reducido $\rho_A$ ($\rho_B$), como indica la ecuación (\ref{expectacionred}). Se define la \textbf{entropía de entrelazamiento} $S_A$ entre el subsistema A y el B como la entropía de von Neumann (\ref{neumann}) para la matriz densidad $\rho_A$
\begin{equation}\label{entrelazamiento}
S_A=S(\rho_A)\,.
\end{equation}
Es sencillo ver que la entropía de entrelazamiento $S_A$ coincide con la entropía de entrelazamiento $S_B$, definida a partir de la matriz densidad reducida $\rho_B$, si el estado total del sistema compuesto es puro\footnote{Esta igualdad sale como una consecuencia directa de la igualdad de los autovalores de $\rho_A$ y $\rho_B$ que se mencionó en la sección~\ref{red}.}. La igualdad entre $S_A$ y $S_B$ sólo es cierta cuando el estado del sistema compuesto es puro, en cuyo caso la entropía $S_{A\cup B}$ es nula. Esto pone de manifiesto el hecho de que, en general, la entropía no es aditiva (como se mencionó en la sección \ref{sec:propentro}). En este caso puro se tiene $S(A)=E(A,B)$ para toda medida de entrelazamiento $E$.

También es sencillo comprobar que para un estado separable, la entropía de entrelazamiento se anula.
%

En la siguiente subsección nos concentramos en calcular la entropía de entrelazamiento para sistemas de fermiones y bosones en la red. Esta derivación, además de representar un buen ejercicio, será de gran utilidad para realizar cálculos numéricos (en particular, utilizaremos estos resultados en el estudio de hamiltonianos modulares que se presenta en el capítulo \ref{ch:local}).

\subsubsection{Cálculo de entropía en la red}\label{discreto}

A continuación, presentamos el cálculo de la entropía de entrelazamiento para el estado fundamental de un sistema reducido a una región $V$ utilizando el formalismo de tiempo real, siguiendo la derivación dada en \cite{review}. Recordamos que el operador densidad reducido a una región $V$ del espacio representa al estado que reproduce los valores de expectación en el vacío de operadores $O_V$ localizados en $V$, es decir
\begin{equation}\label{valores}
\left\langle O_V\right\rangle =\textrm{tr}(\rho_V O_V)\,.
\end{equation}

\begin{itemize}

\item \textbf{Bosones}

Los operadores hermíticos coordenada $\phi_i$ y su momento conjugado $\pi_j$ satisfacen las siguientes relaciones de conmutación
\begin{equation}
[\phi_i,\pi_j]=i\delta_{ij}\, ,\,\,\,[\phi_i,\phi_j]=[\pi_i,\pi_j]=0\,.
\end{equation}
Supongamos que se conocen las funciones de correlación de dos puntos dentro de $V$
\begin{equation}\label{corres1}
\left\langle \phi _{i}\phi _{j}\right\rangle =X_{ij}\, ,\,\,\,\left\langle \pi _{i}\pi _{j}\right\rangle =P_{ij}\,,
\end{equation}
\begin{equation}\label{corres2}
\left\langle \phi _{i}\pi _{j}\right\rangle =\left\langle \pi _{j}\phi _{i}\right\rangle^{*}= \frac{i}{2}\delta_{ij}\,.
\end{equation}
La última ecuación podría parecer un poco arbitraria y en realidad puede generalizarse. Sin embargo, a los efectos de trabajar con el estado de vacío es suficiente pedir que valga (\ref{corres2}). De las ecuaciones en (\ref{corres1}) se deduce que las matrices $X$ y $P$ son hermíticas y tienen autovalores positivos. Además, es sencillo demostrar que el producto
\begin{equation}\label{mayor}
X\cdot P\geq \frac{1}{4}\,,
\end{equation}
donde debe entenderse que la condición es para los autovalores de las matrices $X$ y $P$. Los correladores de más de dos puntos pueden obtenerse a partir de estos, en virtud del teorema de Wick
\begin{equation}\label{wickboson}
\left\langle Of_{i_{1}}f_{i_{2}}...f_{i_{2k}}\right\rangle =\frac{1}{2^{k}k!
}\underset{\sigma }{\sum }\left\langle Of_{i_{\sigma (1)}}f_{i_{\sigma
(2)}}\right\rangle ...\left\langle Of_{i_{\sigma (2k-1)}}f_{i_{\sigma
(2k)}}\right\rangle \,,
\end{equation}
donde la suma es sobre todas las permutaciones $\sigma$ de índices, $f_i$ puede ser $\phi_i$ o $\pi_i$ y $O$ es un operador de orden específico, por ejemplo, podría indicar ordenar los productos dentro de los valores de expectación con las coordenadas a la izquierda y los momentos a la derecha.

Se consideran operadores de creación y destrucción $a_l$, $a_l^{\dagger}$, con $[a_i$, $a_j^{\dagger}]=\delta_{ij}$, que se expresan como combinaciones lineales de los $\phi_i$ y $\pi_j$
\begin{equation}\label{creadest1}
\phi_i=\alpha_{ij}^*a_j^{\dagger}+\alpha_{ij}a_j\,,
\end{equation}
\begin{equation}\label{creadest2}
\pi_i=-i\beta_{ij}^*a_j^{\dagger}+i\beta_{ij}a_j\,.
\end{equation}
Las relaciones de conmutación entre las coordenadas y los momentos dan
\begin{equation}
\alpha^*\beta^T+\alpha\beta^{\dagger}=-1\,.
\end{equation}
Proponemos la siguiente forma para el operador densidad reducido
\begin{equation}\label{matrizdens}
\rho_V=K\,e^{\cal{H}}=K\,e^{-\sum \epsilon_l a_l^{\dagger} a_l}\,,
\end{equation}
siendo $K=\prod\left(1-e^{-\epsilon_l}\right)$ una constante de normalización. Con esta forma del operador densidad, se satisface la propiedad (\ref{wickboson}). Utilizando la ecuación (\ref{valores}) y la forma explícita del operador densidad reducido, se puede hallar, a partir de las relaciones (\ref{corres1}) y (\ref{corres2})
\begin{equation}
\alpha^*n\beta^T-\alpha(n+1)\beta^{\dagger}=\frac{1}{2}\,,
\end{equation}
\begin{equation}
\alpha^*n\alpha^T+\alpha(n+1)\alpha^{\dagger}=X\,,
\end{equation}
\begin{equation}
\beta^*n\beta^T-\beta(n+1)\beta^{\dagger}=P\,,
\end{equation}
siendo $n$ la matriz diagonal dada por
\begin{equation}
n_{k}=\left\langle a_k^{\dagger}a_k\right\rangle=(e^{\epsilon_k}-1)^{-1}\,.
\end{equation}
De estas ecuaciones, se obtiene que $\alpha=\alpha_1 U$y $\beta=\beta_1 U$, siendo $U$ unitaria y diagonal, y $\alpha_1$ y $\beta_1$ reales. La matriz $U$ puede reabsorberse en la definición de los $a_i$, por lo que ponemos $U=1$. Así $\alpha=-\frac{1}{2}(\beta^T)^{-1}$ y
\begin{equation}\label{loqueda}
\alpha\frac{1}{4}(2n+1)^2\alpha^{-1}=XP\,.
\end{equation}
Al ser $\alpha\frac{1}{4}(2n+1)^2\alpha^{-1}$ semejante a $\frac{1}{4}(2n+1)^2$, de (\ref{loqueda}) se puede leer el espectro del operador densidad reducido en términos del espectro de $XP$
\begin{equation}
\frac{1}{2}\textrm{ctgh}(\epsilon_k /2)=\nu_k\,,
\end{equation}
donde $\nu_k$ son los autovalores de
\begin{equation}
C=\sqrt{XP}\,.
\end{equation}
Invirtiendo las relaciones (\ref{creadest1}) y (\ref{creadest2}) y reemplazando en (\ref{matrizdens}), el operador densidad reducido queda
\begin{equation}
\rho_V=K\,e^{-\sum(M_{ij}\phi_i\phi_j+N_{ij}\pi_i\pi_j})\,,
\end{equation}
donde
\begin{equation}
M=\frac{1}{4}\alpha^{-1\,T}\epsilon \alpha^{-1}=\frac{P}{2C}\ln\left( \frac{C+1/2}{C-1/2}\right)\,,
\end{equation}
\begin{equation}
N=\alpha\epsilon\alpha^T=\frac{1}{2C}\ln\left( \frac{C+1/2}{C-1/2}\right)X\,,
\end{equation}
siendo $\epsilon$ la matriz diagonal de los $\epsilon_k$. La entropía, dada por (\ref{neumann}), queda
\begin{equation}
S=\textrm{tr}\left[ \left( C+1/2\right) \ln \left( C+1/2\right) -\left( C-1/2\right)
\ln \left( C-1/2\right) \right]\,,
\end{equation}
que resulta positiva en virtud de la ecuación (\ref{mayor}).

Para el estado fundamental de un hamiltoniano de la forma $H=\frac{1}{2}\sum \pi_i^2+\frac{1}{2}\sum \phi_iK_{ij}\phi_j$, los correladores (\ref{corres1}) están dados por
\begin{equation}
X_{ij}=\frac{1}{2}(K^{-1/2})_{ij}\,,
\end{equation}
\begin{equation}
P_{ij}=\frac{1}{2}(K^{1/2})_{ij}\,.
\end{equation}
Para el estado global, se tiene $X\cdot P=1/4$, que tiene entropía nula, como corresponde para un estado puro.

Esta formulación pone en evidencia el hecho de que para hallar la matriz densidad reducida sólo es necesario conocer los correladores dentro de $V$.

\item \textbf{Fermiones}

Los operadores $\psi_i^{\dagger}$ y $\psi_j$ satisfacen las relaciones de anticonmutación
\begin{equation}
\left\{ \psi _{i},\psi _{j}^{\dagger}\right\} =\delta _{ij}\,.
\end{equation}
Supóngase que los correladores de dos puntos son
\begin{equation}\label{fermion1}
\left\langle \psi_{i}\psi_{j}^{\dagger}\right\rangle =C_{ij}\, ,\,\,\,\left\langle \psi_{i}^{\dagger}\psi_{j}\right\rangle =\delta_{ij}-C_{ij}\,,
\end{equation}
\begin{equation}\label{fermion2}
\left\langle \psi_{i}\psi_{j}\right\rangle =\left\langle \psi_{i}^{\dagger}\psi_{j}^{\dagger}\right\rangle=0\,.
\end{equation}
Se asume aquí también que el teorema de Wick es válido y todos los correladores de más de dos puntos que no se anulan se pueden obtener a partir de los correladores de dos puntos como
\begin{equation}
\left\langle \psi _{i_{1}}...\psi _{i_{k}}\psi _{j_{1}}^{\dagger}...\psi
_{j_{k}}^{\dagger}\right\rangle =(-1)^{k(k-1)/2}\underset{\sigma }{\sum }
\epsilon _{\sigma }\underset{q=1}{\overset{k}{\prod }}\left\langle \psi
_{i_{q}}\psi _{j_{\sigma (q)}}^{\dagger}\right\rangle\,,
\end{equation}
donde la suma se realiza sobre todas las permutaciones de los índices $j_1$, $j_2$, ..., $j_k$, y $\epsilon_\sigma$ es el signo de la permutación. De las ecuaciones (\ref{fermion1}) se ve que los autovalores de $C$ se encuentran en el intervalo $[0,1]$. Si $V$ es el espacio total, $C$ es un proyector y sus autovalores son $0$ y $1$.

Al igual que en el caso bosónico, los correladores dentro de $V$ calculados haciendo uso del operador densidad reducido satisfacen la propiedad de Wick si se tiene
\begin{equation}\label{latercera}
\rho_V=K\,e^{-\cal{H}}=K\,e^{-\sum H_{ij}\psi_i^{\dagger}\psi_j}\,.
\end{equation}
Como $\rho_V$ es hermítico, $H$ también debe ser hermítico. El exponente de la ecuación anterior se puede diagonalizar entonces utilizando la transformación de Bogoliuvov $d_l=U_{lm}\psi_m$, siendo $U$ unitaria (para que resulte ${d_i,d_j^{\dagger}}=\delta_{i,j}$. $U$ se elije de modo que $UHU^{\dagger}=\epsilon$, siendo $\epsilon$ una matriz diagonal de los autovalores de $H$. Se tiene
\begin{equation}\label{lacuarta}
\rho_V=\prod\frac{e^{-\epsilon_l d_l^{\dagger}d_l}}{(1+e^{-\epsilon_l})}\,;
\end{equation}
nótese que de la ecuación anterior se obtiene el valor de la constante $K=\textrm{det}=(1+e^{-H})^{-1}$. La relación entre los autovalores de $C$ y los de $H$ sale de las ecuaciones (\ref{valores}), (\ref{fermion1}), (\ref{latercera}) y (\ref{lacuarta}). Si se nota con $\nu_l$ a los autovalores de $C$ se obtiene la siguiente relación
\begin{equation}
e^{\epsilon_l}=\frac{\nu_l}{1-\nu_l}\,,
\end{equation}
que se puede poner en notación matricial como
\begin{equation}
H=-\ln(C^{-1}-1)\,.
\end{equation}
El cálculo de la entropía de von Neumann para este operador densidad reducido puede evaluarse como una suma sobre cada modo independiente de $H$ autovalor $\epsilon_l$, obteniéndose
\begin{equation}
S=-\textrm{tr}\left[ \left( 1-C\right) \ln \left( 1-C\right) +C\ln \left( C\right)\right]\,.
\end{equation}
\end{itemize}
\section{Entropía en teoría de campos}

El estado relevante para el cálculo de valores de expectación de observables en una región $V$ es el operador densidad reducido $\rho_V$. A partir del operador densidad reducido es posible evaluar la entropía de entrelazamiento en la región, definida a partir de la ecuación (\ref{neumann}) (cuando el estado global es el vacío, se la suele llamar \textbf{entropía geométrica}).

\begin{figure}[h]
\begin{center}
\includegraphics[width=5cm]{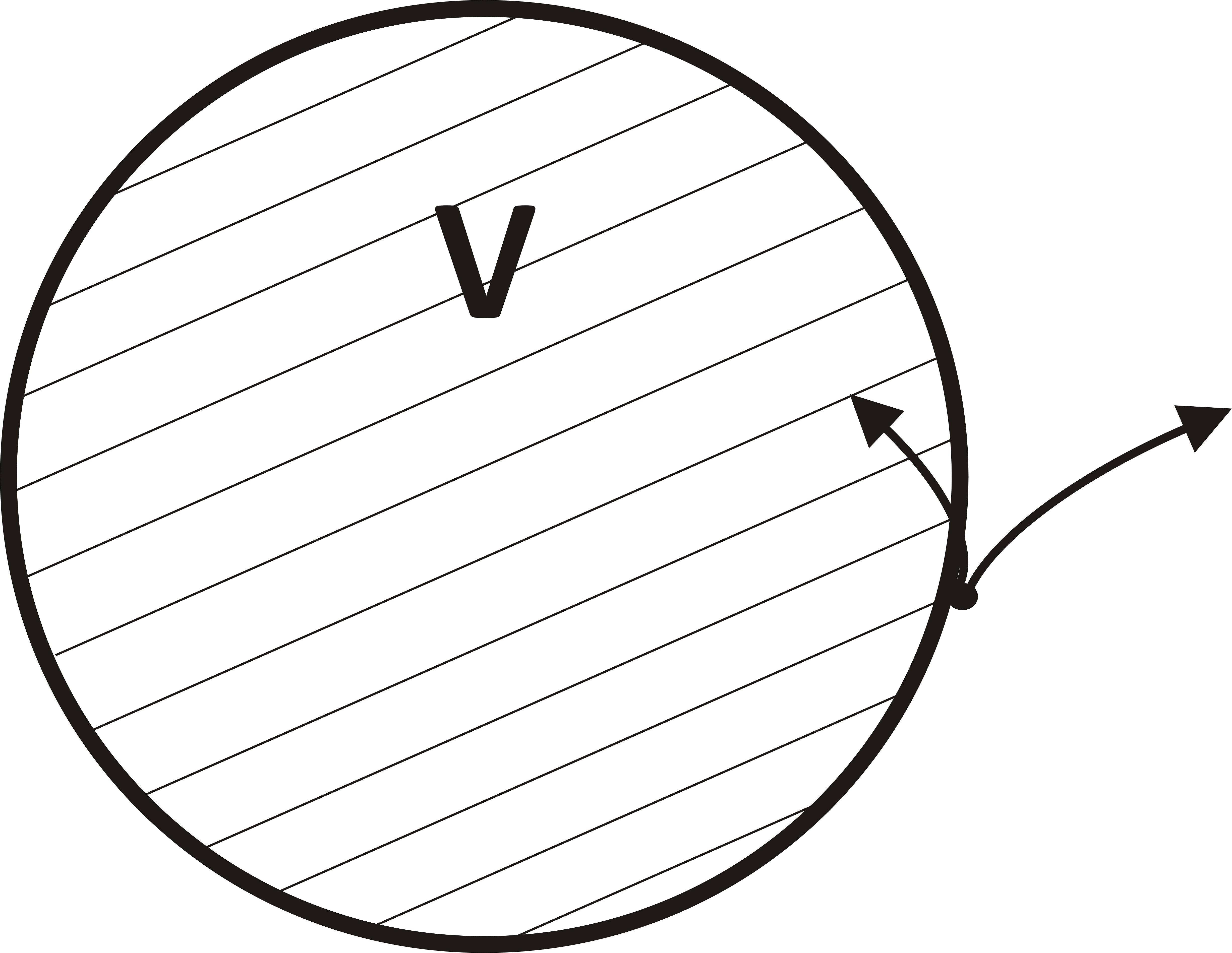}
\end{center}
\caption{\textbf{Fluctuaciones del vacío}. El proceso de localización en teorías relativistas lleva a la creación de pares de partícula-antipartícula. El entrelazamiento en estos pares es la causa de las contribuciones divergentes a la entropía geométrica.} \label{fluctuaciones}
\end{figure}

Al considerar teorías relativistas, la entropía geométrica resulta ser divergente \cite{bh,hooft}. Estas divergencias se pueden entender como consecuencia del entrelazamiento presente en las fluctuaciones del vacío alrededor de la frontera de $V$ (ver figura \ref{fluctuaciones}). El proceso de localización conlleva inevitablemente a la creación de pares de partícula-antipartícula. Si una de las partículas se encuentra dentro de $V$ y la otra fuera, para hallar el operador densidad reducido $\rho_V$, la partícula dentro de $V$ queda como una partícula real mientras los grados de libertad asociados a la otra se trazan.

Dada la naturaleza ultravioleta del problema, los términos divergentes aparecen en el cálculo de la entropía de entrelazamiento también para los estados excitados de energía finita reducidos a la región $V$.

En esta sección presentamos una introducción general al cálculo de entropía en teoría de campos. También comentamos algunas de las aplicaciones de esta cantidad y sus conexiones con diversos e interesantes problemas de la física.

\subsection{Estructura de divergencias para la entropía en teoría de campos}

Como comentamos anteriormente, el problema que se plantea es calcular la entropía de entrelazamiento en una región del espacio, para una teoría de campos. Una forma de entender a la teoría de campos es pensando primero en una teoría con un número discreto de grados de libertad y tomando luego el límite al continuo.

\begin{figure}[h]
\begin{center}
\includegraphics[width=7cm]{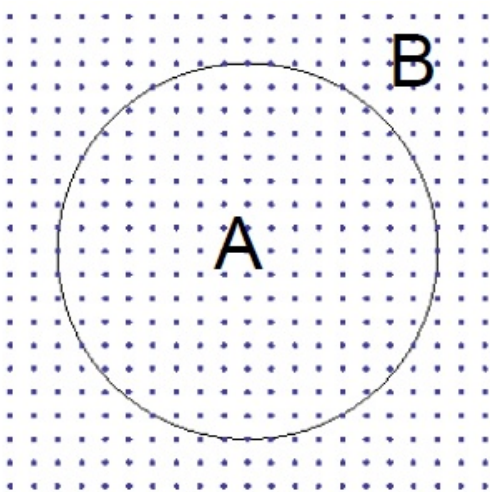}
\end{center}
\caption{\textbf{Teoría de campos en la red.} Una forma de entender a la teoría de campos es pensar en su formulación discretizada. Para ello, generamos una red de puntos en el espacio-tiempo. En este caso, ilustramos la región $A$, un círculo, y su complemento $B$, junto con la red de puntos que discretiza el espacio-tiempo.} \label{red1}
\end{figure}
En la figura \ref{red1} representamos una región del espacio dividida en dos partes: un círculo $A$ y su complemento $B$. Los puntos representan una red que discretiza el espacio. El espacio de Hilbert de nuestro sistema estará compuesto por dos partes $\cal{H} = \cal{H}_A \otimes \cal{H}_B $; supongamos que nos interesa estudiar la entropía para la región $A$, para un estado global $\rho$. Por lo comentado en el capítulo anterior, sabemos que el estado relevante para hacer este cálculo está dado por el operador densidad reducido $\rho_A=\tr_B \rho$, y que la entropía es la entropía de von Neumann para $\rho_A$: $S_A=-\tr \left(\rho_A \log \rho_A\right)$. Esta entropía se reduce a la entropía termodinámica cuando el estado es térmico. Cuando el estado no es térmico aparecen otras contribuciones. Para ver estas contribuciones presentamos el siguiente ejemplo.

\subsubsection{Un primer ejemplo}
\begin{figure}[h]
\begin{center}
\includegraphics[width=7cm]{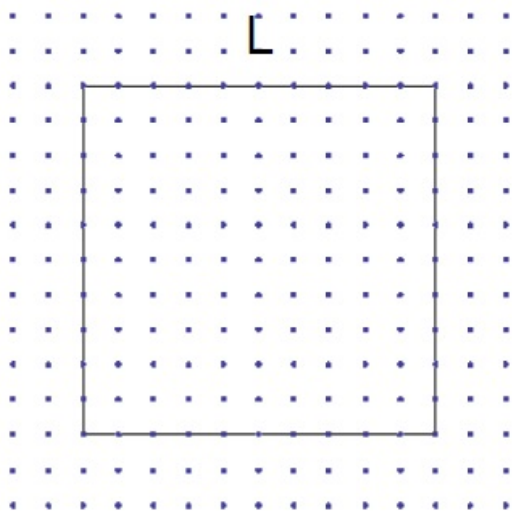}
\end{center}
\caption{\textbf{Discretización para el cálculo de entropía.} Para hallar la entropía de un campo escalar libre en una región cuadrada, discretizamos el espacio utilizando una red con espaciado $\epsilon$.} \label{red2}
\end{figure}

Consideremos como ejemplo el caso de una teoría escalar libre no masiva en dos dimensiones espaciales. Siguiendo las ideas presentadas en \cite{sred}, discretizamos el espacio como en la figura \ref{red2} para obtener versión de la teoría escalar libre en la red
\begin{eqnarray}
H &=&\frac{1}{2} \int d^2 x \left[ \dot{\phi}\left( x\right)^2+\left( \nabla\phi\left(x\right) \right)^2 \right] \longrightarrow \nonumber \\ 
\longrightarrow H &=& \frac{1}{2} \sum \epsilon ^2 \left[\dot{\phi}_i^2 + \sum \frac{\left(\phi_i-\phi_j\right)^2}{\epsilon ^2} \right]\,,
\end{eqnarray}
donde $\epsilon$ es la distancia entre dos primeros vecinos en la red.

Elegimos a $\rho$ como el estado de vacío (estado fundamental en la teoría en la red) para hacer el cálculo de entropía geométrica en la región cuadrada de lado $L$ de la figura \ref{red2}. Esta cuenta puede hacerse numéricamente y se obtiene
\begin{equation}
S=0.075 \frac{4L}{\epsilon}-0.047 \log\left(\frac{L}{\epsilon}\right)+cte = 0.075 \frac{\cal{A}}{\epsilon}-0.047 \log\left(\frac{L}{\epsilon}\right)+cte\,,\label{unaecmas}
\end{equation}
donde llamamos $\cal{A}$ al perímetro del cuadrado.

El término proporcional al perímetro se conoce como \textbf{término de área}, dado que es un término proporcional al área del borde de la región (recordar que en este caso las regiones son figuras en dimensión espacial $2$, por lo que el perímetro de la región es el `área' del borde de la región).  También vemos que hay una contribución logarítmica que, al igual que el término de área, diverge en el límite $\epsilon \longrightarrow 0$. Estas son el tipo de divergencias ultravioletas de las que hablamos al comienzo de este capítulo.
\begin{figure}[h]
\begin{center}
\includegraphics[width=7cm]{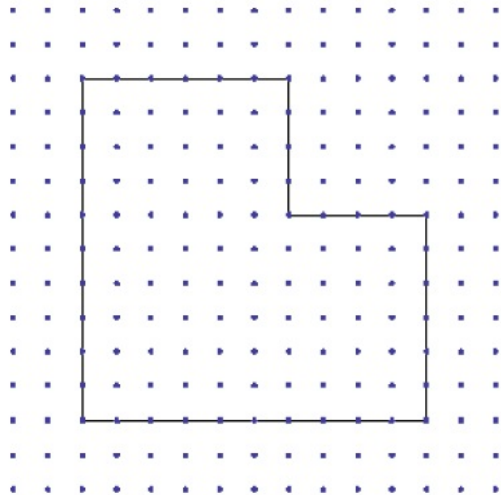}
\end{center}
\caption{\textbf{Dependencia de la entropía con la geometría.} Si mantenemos el perímetro de la región, pero cambiamos la forma de la misma vemos que la entropía mantiene igual el término de área. Sin embargo, el coeficiente logarítmico aumenta en forma proporcional al número de singularidades cónicas de la región.} \label{red3}
\end{figure}
Consideremos ahora la región marcada en la figura \ref{red3}, de igual perímetro que la anterior, pero diferente geometría. El cálculo para la entropía de esta región arroja el siguiente resultado
\begin{equation}
S = 0.075 \frac{\cal{A}}{\epsilon}- \frac{6}{4} 0.047\log\left(\frac{L}{\epsilon}\right)+cte\,,
\end{equation}
es decir, tenemos el mismo término de área y además vemos que el coeficiente del término logarítmico aumenta. De hecho, el coeficiente del término logarítmico aumenta con el número de vértices y en general se tiene
\begin{equation}
S = c_1 \frac{\cal{A}}{\epsilon}-\sum_{vert.} c_{\log}\left(\theta\right) \log\left(\frac{L}{\epsilon}\right)\,.
\end{equation}

\begin{figure}[h]
\begin{center}
\includegraphics[width=7cm]{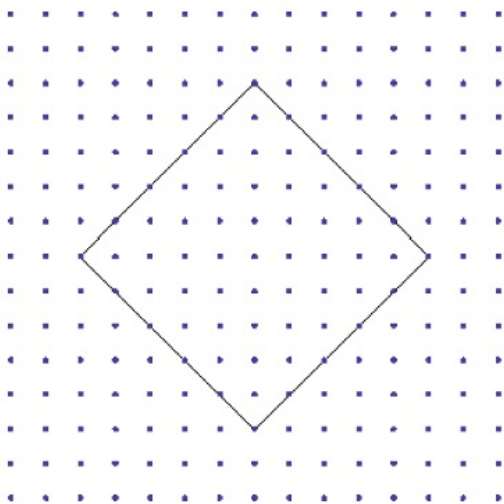}
\end{center}
\caption{\textbf{Independencia del término logarítmico de los detalles de la red.} Si cambiamos la forma de la red, en este caso, rotando la región considerada, vemos que el término logarítmico no cambia, lo que indica que contiene información universal sobre la teoría de campos en el continuo. La ley de área en cambio no es una propiedad del continuo.} \label{red4}
\end{figure}
Hasta aquí pareciera apreciarse cierta regularidad en el coeficiente del término de área; sin embargo, si tomamos una región como la diagramada en la figura \ref{red4} (es decir, también un cuadrado de lado $L$ pero rotado respecto al de la figura \ref{red2}, obtenemos
\begin{equation}
S = 0.085 \frac{\cal{A}}{\epsilon}- 0.047 \log\left(\frac{L}{\epsilon}\right)+cte\,.
\end{equation}
Esto nos demuestra que el término de área no tiene la simetría rotacional que debería tener en la teoría del continuo. El punto interesante es que el coeficiente del término logarítmico es igual al dado por la ecuación \reef{unaecmas}, lo que indica que no siente los efectos producidos por cambiar la forma de la red. En este sentido, diremos que el término logarítmico es un término universal, que se espera que contenga información relevante de la teoría de campos.

Con este ejemplo sencillo confirmamos el hecho de que la entropía para una región es divergente en la teoría de campos. Además, vimos que en general los términos divergentes no son universales (a excepción del término logarítmico). Esta universalidad del término logarítmico también se confirma con el resultado para la entropía de un intervalo de longitud $r$ en teorías conformes en $1+1$ dimensiones \cite{cacardy,wilcCFT,lakita,latorre}
\begin{equation}
S\left(r\right)=\frac{c}{3}\log\left(\frac{r}{\epsilon}\right)+c_0\,,\label{paracft}
\end{equation}
siendo $c$ la carga central de Virasoro de la teoría.

\subsubsection{Intervalo en 1+1 para el campo de Dirac a T$\neq$0}

Otro ejemplo interesante, para visualizar el rol de los términos divergentes y finitos en la entropía, es el del cálculo de la entropía de un intervalo espacial en una teoría de Dirac no masiva en $1+1$ dimensiones a temperatura $T$. En este caso, la entropía resulta\cite{cacardy}
\begin{equation}\label{tfinita}
S=\frac{1}{3}\ln\left[\frac{1}{\pi \epsilon T}\textrm{sinh}\left(\pi L T\right)\right]\,,
\end{equation}
donde $\epsilon$ es un cutoff ultravioleta con unidades de distancia. En el límite en que $L T\rightarrow \infty$, la ecuación (\ref{tfinita}) da
\begin{equation}\label{finnn}
S\approx\frac{1}{3}\ln(\epsilon)+\ln\left(\frac{1}{T}\right)+\frac{\pi}{3}L T\,,
\end{equation}
donde el último sumando de \reef{finnn} da una contribución finita y dice que la entropía crece en forma proporcional al volumen, como suele ocurrir para un estado térmico. Para $L T\rightarrow 0$ se obtiene
\begin{equation}
S\approx \frac{1}{3}\ln\left(\frac{L}{\epsilon}\right)\,,
\end{equation}
que es un resultado que también puede obtenerse independientemente utilizando el formalismo de tiempo real, para la misma teoría con temperatura nula (este cálculo, utilizando el formalismo de tiempo real se puede ver por ejemplo en \cite{fermion}). El término $\frac{\pi}{3}L T$ se obtendría también si uno pusiera al sistema en una caja con condiciones de borde a temperatura $T$, tomando el límite de $L$ grande.

\subsubsection{Estructura de divergencias y términos universales}

En $d$ dimensiones espaciales, para cualquier teoría de campos se tiene \cite{review}
\begin{equation}\label{ultimaec}
S(V)=g_{d-1}[\partial V]\epsilon^{-(d-1)}+...+g_1 [\partial V]\epsilon^{-1}+g_0[\partial V]\ln(\epsilon)+S_0(V)\,,
\end{equation}
siendo $S_0(V)$ una contribución finita, $\epsilon$ un \textit{cutoff} ultravioleta (proveniente por ejemplo de la discretización realizada para llevar la teoría de campos a la red) y cada $g_i$ una función local extensiva en el borde $\partial V$~\footnote{Esto está asociado a que los modos de corta longitud de onda que se encuentran entrelazados entre la región exterior y la interior a $V$ sólo pueden contribuir como una integral sobre el borde. Recordamos que para un estado puro, las entropías de una región arbitraria y su complemento son iguales, lo que indica naturalmente que la entropía dependa del borde, que es común a la región y su complemento.}.

El coeficiente del primer término $g_{d-1}[\partial V]$, como ya mencionamos, se conoce cómo término de área. Principalmente en relación a la entropía de un agujero negro se suele decir que la entropía satisface una \textit{ley de área} $ S \sim \frac{R^{d-2}}{\epsilon^{d-2}}$ \cite{sorviejo,bh,sred}. Los agujeros negros son clásicamente objetos cuya entropía $S$ es proporcional al área $A$ de su horizonte \cite{bekenstein}
\begin{equation}
S=\frac{A}{4G},
\end{equation}
siendo $G$ la constante de gravitación de Newton. Sin embargo, el origen de esta entropía no es evidente y aún no ha sido clarificado.  Una propuesta hacia el entendimiento de esta entropía es que su significado microscópico está relacionado con la entropía de entrelazamiento \cite{sorviejo,bh}, hecho motivado por la presencia del término de área que hemos visto que aparece en el cálculo de la entropía en teoría de campos. Es importante volver a remarcar que (a pesar de que resulta evidente después de estudiar el ejemplo presentado en la sección anterior) esta ley de área no es una propiedad general de la teoría del continuo. 

Los términos proporcionales a $g_i$ para $i>0$ no tienen una interpretación física en teoría de campos, dado que no están asociados a cantidades independientes de la discretización a partir de la cual se interpreta a la teoría de campos. Sin embargo, el coeficiente $g_0$ del término logarítmico sí es universal.

El término finito $S_0\left(V\right)$ contiene mucha información sobre el estado y la región. Por ejemplo, si el estado es un estado de equilibrio térmico a una temperatura $T$, se tiene $S_0\left(V\right) \sim \kappa T^{d-1} V$, que es la entropía termodinámica usual, extensiva en el volumen $V$.

\subsection{Información mutua}\label{secmutua}

Se ha comentado que la localización estricta de un estado en una región del espacio, en teorías relativitas, da lugar a la formación de pares partícula-antipartícula y que en la aparición de estos modos de alta energía se encuentra el origen de las divergencias en la entropía de entrelazamiento. Una contribución divergente similar aparecerá en cualquier cálculo de la entropía en una región separada de otra por un horizonte causal, como es el caso de los agujeros negros. Sin embargo, la noción de entrelazamiento entre dos regiones distintas puede definirse de modo preciso en teoría de campos, aún en presencia de gravitación. Las medidas de entrelazamiento entre regiones disjuntas no presentan divergencias y son, en principio, calculables en la teoría semiclásica.

\begin{figure}[h]
\begin{center}
\includegraphics[width=5cm]{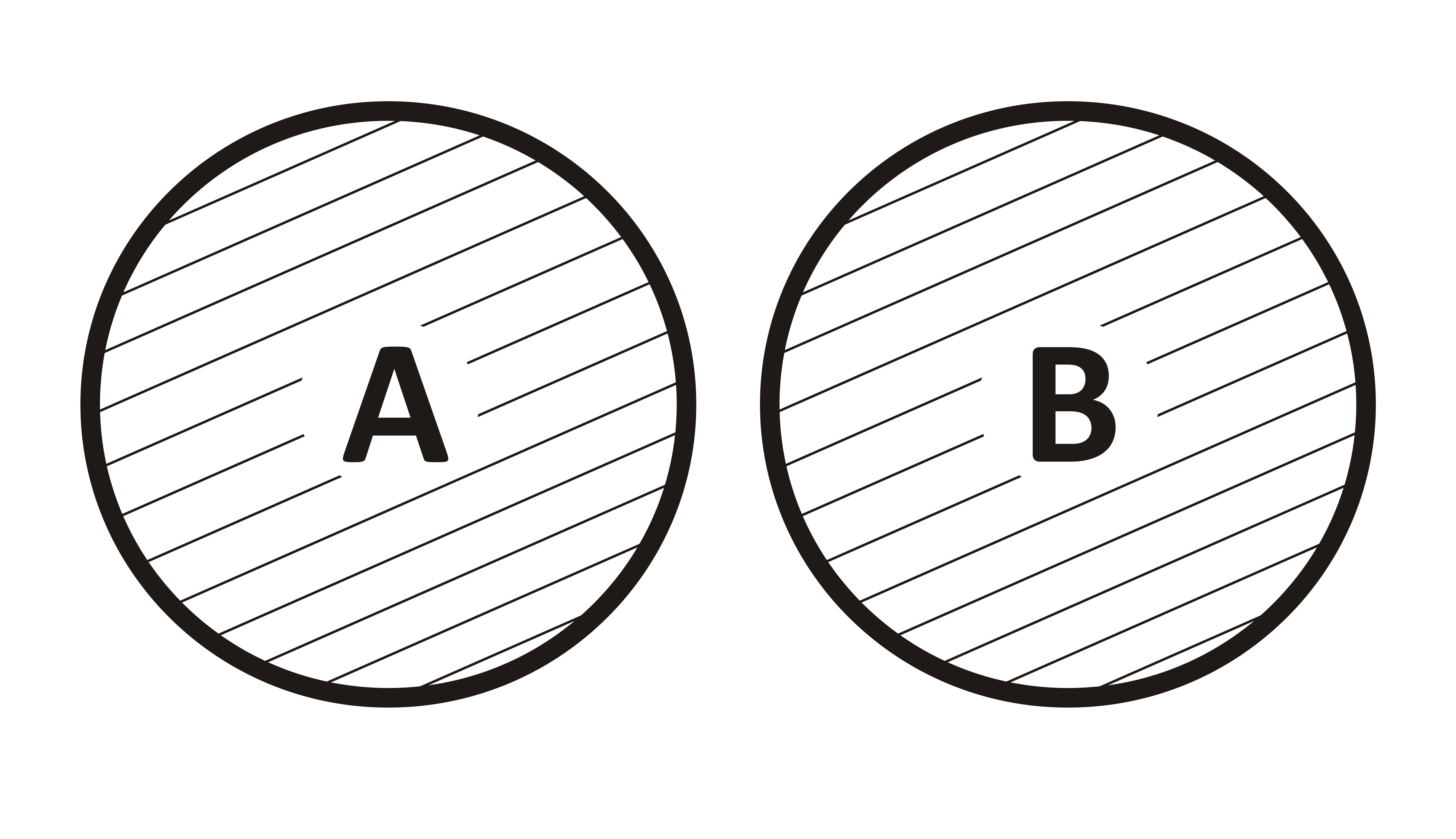}
\end{center}
\caption{\textbf{Información mutua}. Para dos regiones $A$ y $B$ disconexas, como las que se muestra en la figura, es posible definir una cantidad universal de la que puede inferirse la entropía de entrelazamiento.} \label{mutua}
\end{figure}

Es posible realizar una substracción en la entropía de entrelazamiento del vacío para eliminar las divergencias provenientes de las fluctuaciones del vacío. Sin embargo, no se puede simplemente llevar a cero el valor de esta entropía, dado que existen cantidades finitas universales (independientes de la regularización) que se pueden obtener a partir de ella. Una de esas cantidades es la \textbf{información mutua}
\begin{equation}
I(A,B)=S(A)+S(B)-S(A\cup B)\,,\label{infomutua}
\end{equation}
entre dos conjuntos disjuntos $A$ y $B$ (ver figura \ref{mutua}). Nótese que los términos divergentes, que están asociados a los bordes de las regiones se sustraen en (\ref{infomutua}). En estadística, la información mutua da una medida de la información compartida por dos sistemas. La información mutua es una cantidad bien definida en teoría de campos, con el único requerimiento de que exista una separación no nula entre las regiones $A$ y $B$, y mide el grado de no extensividad que tiene la entropía $S$.

Si la región $B$, por ejemplo, se extendiera a todo el espacio de modo de tender a cubrir la región $\bar{A}$, y el contorno de $B$ se aproximara al de $A$ desde afuera, se obtendría $S(A)\approx S(\bar{B})=S(B)$ y $S(A\cup B)\approx 0$ y, entonces
\begin{equation}\label{recupera}
I(A,B)\approx 2S(A)\,.
\end{equation}
Así, la información mutua diverge del mismo modo que la entropía cuando los conjuntos se acercan uno al otro y, de algún modo, se reproducen las divergencias presentes en $S(A)$, pero ahora en forma independiente de la regularización.

\subsubsection{Propiedades de la información mutua}\label{sectripartita}

Para un sistema cuántico general, la función $I(A,B)$ es simétrica y positiva. Si se considera un estado producto, dado por $\rho_{A\cup B}=\rho_A \otimes \rho_B$, la información mutua $I(A,B)$ se anula. Otra propiedad interesante es la de monotonía respecto al tamaño de los conjuntos
\begin{equation}\label{monoinfo}
I(A,B)\leq I(A,C)\,,\,\textrm{si}\,\,\,B\subseteq C\,.
\end{equation}
Esta propiedad, indica que la información mutua varía suavemente con el conjunto y es una consecuencia de la propiedad de monotonicidad ante inclusión de conjuntos de la entropía relativa (ver sección \ref{subcapsub}). También se tiene la siguiente desigualdad
\begin{equation}
I(A,B)\leq 2 \,\textrm{min}(S(A),S(B))\,.
\end{equation}
La \textbf{información tripartita}
\begin{equation}\label{tripartita}
I(A|B,C)=I(A,B)+I(A,C)-I(A,B\cup C)\,,
\end{equation}
es otra cantidad que mide la información compartida entre $B$ y $C$ respecto de $A$. Por la propia definición de la información mutua (\ref{infomutua}), la información tripartita tiene simetría de permutación total
{\small
\begin{eqnarray}\label{triparentro}
&&I_3(A,B,C)\equiv I(A|B,C)=\\
&&S(A\cup B\cup C)-S(A\cup B)-S(B\cup C)-S(C\cup A)+S(A)+S(B)+S(C)\,.\nonumber
\end{eqnarray}}
Así como la información mutua mide el grado de no extensividad de la entropía, la información tripartita, en virtud de la ecuación (\ref{tripartita}), mide el apartamiento de la extensividad de la información mutua. Nótese que a diferencia de $I(A,B)$, $I_3(A,B,C)$ puede ser tanto positiva o negativa, y es nula en particular cuando el estado total es puro.

\subsection{El método de réplicas}

En las siguientes páginas introducimos el método de réplicas, una herramienta poderosa mediante la cual se puede hallar la traza de potencias del operador densidad. Es particularmente útil para calcular las entropías de Renyi, a partir de las cuales se puede obtener la entropía de entrelazamiento por una continuación analítica.

\subsubsection{Entropías de Renyi}

Las denominadas entrop\'{\i}as de Renyi son una familia de cantidades de la teoría de la información dependientes de un número natural $n$
\begin{equation}
S_n\left(V\right)=\frac{1}{1-n}\log tr\rho_V^n\,.
\label{renyi}
\end{equation}
Las entropías de Renyi resultan particularmente útiles para hallar $S\left(V\right)$ como una continuación analítica hacia el valor de $n=1$
\begin{equation}
-\lim_{n\longrightarrow 1}\frac{\partial}{\partial n}tr\rho_V^n=\lim_{n\longrightarrow 1}S_n\left(V\right)=S\left(V\right)\,,
\end{equation}
donde la última igualdad se obtiene utilizando la regla de L'Hôpital.

Como veremos a continuación, es posible representar estas cantidades utilizando el formalismo de integrales de camino en tiempo euclídeo.

\subsubsection{Representación funcional del operador densidad reducido}

Consideramos un campo escalar $\hat{\phi}\left(x,t\right)$ y una base de autovectores de este operador a tiempo $t=0$: $\hat{\phi}\left(x,0\right)\left|\alpha\right\rangle=\alpha\left(x\right)\left|\alpha\right\rangle$, donde $\alpha$ es cualquier función real. Luego, el funcional de onda del vacío se escribe como \cite{pokorski}
\begin{equation}
\Phi\left(\alpha\right)=\left\langle 0\right|\left.\alpha\right\rangle=N^{-1/2}\int_{\phi\left(x,-\infty\right)=0}^{\phi\left(x,0\right)=\alpha\left(x\right)}{\textit{D}\phi\, e^{-S_E\left(\phi\right)}}\,.
\label{funcional}
\end{equation}
Para elegir el estado de vacío, la integral funcional se realiza sobre la mitad inferior del espacio (para ``suprimir'' los estados excitados) y en tiempo euclídeo. $N^{-1/2}$ es un factor de normalización. El operador densidad en esta base es $\rho\left(\alpha,\alpha '\right)=\left\langle \alpha\right|\left.0\right\rangle\left\langle 0\right|\left.\alpha '\right\rangle=\Phi\left(\alpha\right)^{\dagger}\Phi\left(\alpha '\right)$. Hallar la matriz densidad reducida a la región $V$ implica trazar sobre los grados de libertad en $\bar{V}$ (el complemento de $V$). Para realizar esto, consideramos funciones $\alpha=\beta\oplus\alpha_V$, $\alpha '=\beta\oplus\alpha'_V$, que coinciden en $\bar{V}$ (son iguales a $\beta$), y sumamos sobre todas las posibles funciones $\beta$. Utilizando la representación (\ref{funcional}), la matriz densidad reducida se obtiene tomando dos copias de la mitad del espacio, pegándolas en $\bar{V}$ y realizando la integral funcional en este nuevo espacio \cite{larsen}
\footnotesize
\begin{equation}
\rho_V\left(\alpha_V,\alpha'_V\right)=\int{\textit{D}\beta\,\Phi\left(\beta\oplus\alpha_V\right)^{*}\Phi\left(\beta\oplus\alpha'_V\right)}=N^{-1}\int_{\phi\left(x,0^-\right)=\alpha'_V\left(x\right),\,x\,\in V}^{\phi\left(x,0^+\right)=\alpha_V\left(x\right),\,x\in V}{\textit{D}\phi\, e^{-S_E\left(\phi\right)}}\,.
\end{equation}
\normalsize
Los argumentos de la matriz densidad reducida son las condiciones de contorno de la integral funcional a cada lado de los cortes en $V$.

\begin{figure}[h]
	\centering
		\includegraphics{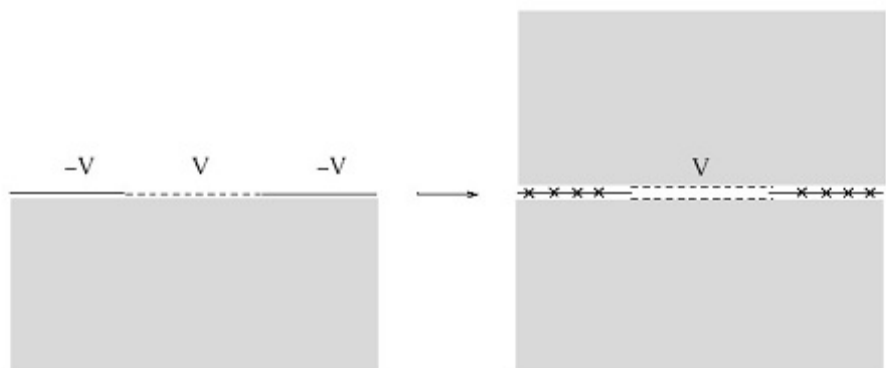}
	\caption{\textbf{Método de réplicas.} La integral funcional sobre la mitad inferior del espacio euclídeo nos permite obtener el funcional de onda del vacío. La matriz densidad reducida $\rho_V$ se obtiene ``pegando'' dos copias de este semiespacio a lo largo de $\bar{V}$, el conjunto complementario a $V$.}
	\label{densidad}
\end{figure}

\subsubsection{El método de réplicas aplicado al cálculo de la entropía}

Es posible escribir las entropías de Renyi (\ref{renyi}) en esta representación funcional. Para ello, hay que tomar $n$ copias del plano euclídeo con cortes a lo largo de $V$ y pegar el lado superior de la copia $k-$ésima con el lado inferior de la copia $\left(k+1\right)-$ésima, para $k=1,\,2,\,...,\,n$ (la copia $n+1$ se identifica con la primera copia). Esto nos deja con un espacio euclídeo de la misma dimensión que el espacio original, con singularidades cónicas de ángulo $2\pi n$ localizadas en el borde $\partial V$ de $V$. Con estas consideraciones, se tiene

\begin{equation}
tr \rho_V^{n}=\frac{Z\left[n\right]}{Z\left[1\right]^n}\,,
\end{equation}

\begin{equation}
S_n\left(V\right)=\frac{\log Z\left[n\right]-n\log Z\left[1\right]}{1-n}\,,
\label{renyi2}
\end{equation}

\begin{equation}
S\left(V\right)={\lim}_{n\longrightarrow 1} \left(1-n\frac{d}{dn}\right)Z\left[n\right]\,,
\end{equation}
donde $Z\left[n\right]$ es la integral funcional en la variedad con $n$ ``láminas'' (que llamaremos \textbf{función de partición}), y elegimos el factor de normalización $N=Z\left[1\right]$ para que se cumpla $tr\rho_V=1$. La ecuación (\ref{renyi2}) da una representación funcional de las entropías de Renyi para $n$ entero. La entropía de entrelazamiento se obtiene haciendo una continuación analítica de $S_n$ hacia $n=1$.
\begin{figure}[h]
\begin{center}
\includegraphics[width=10cm]{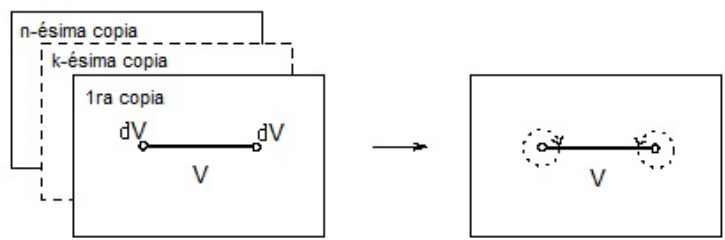}
\end{center}
\caption{\textbf{Cálculo de la traza de potencias de la matriz densidad utilizando el método de réplicas.} La $tr\rho_V^n$ está dada por una integral funcional en una variedad con $n$ láminas, construída pegando el espacio euclídeo replicado con un corte $V$.} \label{densidad}
\end{figure}

\subsubsection{Métodos para calcular la función de partición}

En general, el cálculo de la función de partición es complicado dado que aparecen involucradas variedades no triviales debido al método de réplicas. En las siguientes subsecciones presentamos muy brevemente algunos métodos existentes para hallar $Z$.

\begin{enumerate}
\item Diagonalización en el espacio replicado

Para el caso de campos libres, es posible trasladar el problema de la variedad con $n$ ``láminas'' a uno equivalente en el cual uno trata con $n$ campos libres multivaluados desacoplados. Este método se encuentra aplicado en \cite{fosco} para calcular las entropías de Renyi de un campo de Dirac libre.

\item Heat Kernel

Otra herramienta poderosa para realizar el cálculo de la función de partición, cuando la acción es cuadrática, es el método del \textit{heat kernel}. Para introducirlo, consideramos un campo escalar libre, y notamos que
\begin{equation}
W=-\log Z=\frac{1}{2}\log\det\left(M^2-\nabla^2\right)=\frac{1}{2}tr\log\left(M^2-\nabla^2\right)\,.
\label{elibre}
\end{equation}
La función de partición se obtiene entonces hallando la función espectral de un operador diferencial. El heat kernel se define como $K\left(x,y,t\right)=\left\langle x\right|e^{t\nabla^2}\left|y\right\rangle$, y su traza se escribe como
\begin{equation}
\zeta\left(t\right)=tr e^{t\nabla^2}=\int{dx\,K\left(x,x,t\right)}\,.
\end{equation}
La ``energía libre'' $W$ puede escribirse en términos de esta función espectral
\begin{equation}
W=-\frac{1}{2}\int_{\epsilon}^{\infty}{dt\,e^{-M^2 t}\frac{1}{t}\zeta\left(t\right)}\,,
\end{equation}
donde $\epsilon$ es un cutoff. La ventaja de esta ecuación es que, en contraste con (\ref{elibre}), está escrita como una función de la traza de un operador que satisface una ecuación del calor local
\begin{equation}
\frac{\partial K}{\partial t}=\nabla^2 K\,,\,\,\,\,\,K\left(x,y,0\right)=\delta\left(x-y\right)\,.
\end{equation}

\item Función de Green

Otra estrategia posible para hallar la función de partición consiste en estudiar la función de Green $G=\left(-\nabla^2+M^2\right)^{-1}$ en la variedad asociada al problema. $Z$ y $G$ están relacionadas a través de la identidad
\begin{equation}
\frac{d}{dM^2}\log Z=-\frac{1}{2}tr G\,.\label{greengo}
\end{equation}
Sin embargo, no existen métodos generales para hallar $G$ en este tipo de variedades con singularidades cónicas, por lo que hay que analizar caso por caso viendo si la geometría del problema ayuda a reducir la dificultad del cálculo. Este es el método que utilizaremos en la siguiente sección para mostrar el ejemplo de un cálculo sencillo de la entropía de un campo escalar libre utilizando el método de réplicas.
\end{enumerate}

\subsubsection{Ejemplo: Campo escalar libre en $1+1$}

Consideramos el ejemplo de un campo escalar libre en dimensión $1+1$, cuya acción euclídea es
\begin{equation}
S_E=\int d^2 r \frac{1}{2}\left[\partial_\mu\phi\partial^\mu\phi+M^2\phi^2\right]\,.
\end{equation}
Pretendemos calcular la entropía para una semirrecta, es decir, $V=\left[0,\infty\right]$ (seguiremos como referencia el cálculo realizado en \cite{cacardy}). Para hallar $tr\rho_V^n$ debemos calcular la función de partición $Z[n]$ de esta teoría en una variedad con $n$ ``láminas'' conectadas por un corte. Como ya comentamos, hay diferentes maneras de calcular la función de partición; en este caso, resultará conveniente recordar la relación con la función de Green $G_n\left(\bar{r},\bar{r}'\right)$ dada por la ecuación \reef{greengo}
\begin{equation}
\frac{\partial }{\partial M^2}\log Z\left[n\right]=-\frac{1}{2}\int{d^2 rG_n\left(\bar{r},\bar{r}\right)}\,.
\end{equation}
En este caso, la función de Green satisface la siguiente ecuación sobre la variedad
\begin{equation}
\left(-\bar{\nabla}_{\bar{r}}^2+M^2\right)G_n\left(\bar{r},\bar{r}'\right)=\delta^2\left(\bar{r}-\bar{r}'\right)\,,
\label{parti}
\end{equation}
y se puede obtener hallando una base de autofunciones $f_k\left(\bar{r}\right)$ del operador de Helmholtz $\left(-\bar{\nabla}_{\bar{r}}^2+M^2\right)$
\be
\left(-\bar{\nabla}_{\bar{r}}^2+M^2\right) f_k\left(\bar{r}\right) = \lambda_k f_k\left(\bar{r}\right)\,.
\ee
En términos de estas autofunciones, la función de Green se expresa como
\begin{equation}
G_n\left(\bar{r},\bar{r}'\right)=\sum{\frac{N_k f_k^*\left(\bar{r}\right) f_k\left(\bar{r}'\right)}{\lambda_k}}\,,
\label{helm}
\end{equation}
siendo $N_k$ una constante de normalización tal que
\begin{equation}
N_m \int{d^2 r f_m\left(\bar{r}\right)f_n\left(\bar{r}\right)}=\delta_{mn}\,.
\end{equation}
Por la simetría del problema, es conveniente escribir todo en coordenadas polares
\begin{equation}
x= r\cos \theta\,,\,\,y=r\sin \theta\,.
\end{equation}
El hecho de que la variedad tenga un corte se expresa en estas coordenadas diciendo que las autofunciones tienen que tener el mismo valor para diferencias de $2\pi n$ en $\theta$
\begin{equation}
f_k\left(r,\theta\right)=f_k\left(r,\theta+2\pi n\right)\,.
\label{contorno}
\end{equation}
Proponiendo una solución a la ecuación (\ref{helm}) de la forma $f\left(r,\theta\right)=A\left(r\right)B\left(\theta\right)$, utilizando la ecuación \reef{contorno} y pidiendo regularidad en $r=0$, obtenemos finalmente
\begin{equation}
G_n\left(r,\theta,r',\theta\right)=\frac{1}{2\pi n}\sum_{k=0}^{\infty}{d_k\int_{0}^{\infty}{\lambda}\frac{J_{k/n}\left(\lambda r\right)J_{k/n}\left(\lambda r'\right)}{\lambda^2+M^2}C_k\left(\theta,\theta'\right)d\lambda}\,,
\end{equation}
donde $d_k$ es $1$ si $k=0$ y $2$ si $k>0$, y $C_k\left(\theta,\theta'\right)=\cos\left(\frac{k\theta}{n}\right)\cos\left(\frac{k\theta'}{n}\right)+\sin\left(\frac{k\theta}{n}\right)\sin\left(\frac{k\theta'}{n}\right)$.
Para hallar la función de partición usando la ecuación (\ref{parti}) ponemos $\bar{r}=\bar{r}'$ e integramos la ecuación anterior en $\theta$ y $\lambda$, obteniendo
\begin{equation}
G_n\left(r\right)=G_n\left(\bar{r},\bar{r}\right)=\sum_{k=0}^{\infty}{d_k I_{k/n}\left(M r\right)K_{k/n}\left(M r\right)}\,,
\label{coin}
\end{equation}
donde $I_{k/n}$ y $K_{k/n}$ son las funciones de Bessel modificadas de primer y segundo tipo respectivamente.

La suma sobre $k$ en (\ref{coin}) es divergente y debe regularizarse. Sin embargo, si formalmente intercambiamos la integración con la suma, podemos hallar
\begin{equation}
-\frac{\partial }{\partial M^2}\log Z\left[n\right]=\frac{1}{2}\sum_{k=1}^{\infty}{d_k\int{d^2 r d_k I_{k/n}\left(M r\right)K_{k/n}\left(M r\right)}}=\frac{1}{4n M^2}\sum_{k=0}^{\infty}{d_k k}\,.
\end{equation}
Interpretando la última suma en términos de la función zeta de Riemann en el límite hacia el valor $-1$, tenemos $\sum_{k=1}^{\infty}{d_k k}=2\zeta\left(-1\right)=-\frac{1}{6}$. Esto da el resultado para la entropía del campo escalar libre en $1+1$
\begin{equation}
S\left(V\right)=-\frac{1}{6}\log\left(M \epsilon\right)\,.
\end{equation}
Aquí utilizamos la regularización por la función zeta de Riemann, aunque es posible realizar un procedimiento más limpio para regularizar las integrales, como se muestra en \cite{cacardy}. De todos modos, la regularización aquí utilizada da, sorprendentemente, el resultado correcto para la entropía del campo escalar.

Este capítulo introductorio concluye aquí. Naturalmente, muchas aplicaciones de la entropía de entrelazamiento no han sido expuestas en esta presentación, por lo que se sugiere al lector interesado revisar las referencias mencionadas en esta sección y (los muchos) otros artículos sobre el tema que se encuentran en la literatura.

En el siguiente capítulo, presentaremos otro cálculo explícito de la entropía de entrelazamiento para campos. En particular, analizaremos el efecto Aharonov-Bohm sobre las fluctuaciones de vacío utilizando la entropía de entrelazamiento. 
\chapter{\label{ch:abe}Entropía y el efecto Aharonov-Bohm}

En el capítulo anterior realizamos una presentación de la entropía de entrelazamiento en teoría de campos y también expusimos algunos ejemplos de su cálculo. En este capítulo, realizamos un cálculo que indica que la entropía de entrelazamiento mide el efecto Aharonov-Bohm. Este trabajo fue realizado en colaboración de Raúl E. Arias y Horacio Casini, y sus resultados fueron plasmados en \cite{abe}.

\section{Efecto Aharonov-Bohm}

El efecto Aharonov-Bohm (AB) es un fenómeno cuántico fundamental en el cual una partícula cargada eléctricamente es afectada por un potencial electromagnético $A_{\mu}$, incluso si los campos eléctrico y magnético son nulos en la región donde la partícula se encuentra confinada. Este efecto surge del hecho de que la circulación de $A_{\mu}$ alrededor de una curva {\cal C} ($\Phi:=\oint_C A_{\mu}dx^{\mu}$) cambia la función de onda $\psi\left(x\right)$ de la partícula cargada, adquiriendo un factor de fase adicional $e^{i e\Phi}\psi\left(x\right)$ (que es independiente de los valores que $A_{\mu}$ toma en la región donde la partícula está confinada). Este factor de fase puede observarse en experimentos de interferencia de partículas que recorren distintas trayectorias.

El efecto AB fue estudiado por primera vez por W. Ehrenberg y R. Siday en 1949 \cite{siday}, y Y. Aharonov y D. Bohm en 1959 \cite{bohm}, y ha sido observado experimentalmente \cite{exp,exp1}. Recientemente, se ha estudiado la respuesta al flujo magnético que presentan los valores de expectación de ciertos operadores en la teoría de campos (en geometrías cilíndricas) usando holografía \cite{Montull,Montull2}. En la literatura de materia condensada también se ha explorado el efecto del flujo magnético en superconductores no convencionales con geometría cilíndrica \cite{cmt,cmt1,cmt2}.\

En este capítulo, analizamos el efecto AB sobre las fluctuaciones de vacío utilizando la entropía de entrelazamiento. En particular, consideramos campos escalares y de Dirac libres con soporte en un cilindro 2-dimensional y estudiamos la dependencia de la entropía de entrelazamiento con el flujo magnético para una región tipo faja sobre la superficie del cilindro. Mostramos que la entropía de entrelazamiento exhibe una dependencia en la fase de Aharonov-Bohm $\Phi$; esto la convierte en una herramienta más que interesante para explorar fenómenos topológicos asociados.

\section{Efecto AB sobre la entropía de entrelazamiento}

\begin{figure}[h]
\begin{center}
\includegraphics[width=7cm]{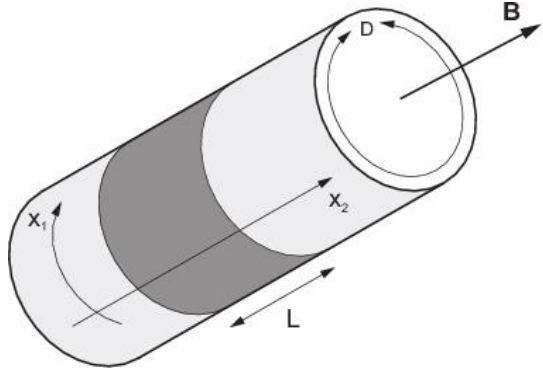}
\end{center}
\caption{\textbf{Configuración geométrica.} La coordenada $x^1$ está compactificada a un círculo de perímetro $D$. Se calcula la entropía de la región tipo faja anular de ancho $L$ sobre el cilindro. El campo de gauge externo induce una fase en la dirección $x^1$.} \label{figu1}
\end{figure}

Vamos a analizar el caso de un campo escalar libre $\phi$ de masa $m$ en $d$ dimensiones espaciales, cargado con respecto a un campo de calibre externo $A_{\mu}$, que es puro gauge en la región de interés. El lagrangiano es
\be
{\cal{L}}=-(\partial_\mu+i e A_\mu)\phi^* (\partial^\mu-i e A^\mu)\phi-m^2 \phi^* \phi\label{uno}
\ee
El espacio está compactificado a un círculo $S^1$ de perímetro $D$ en la dirección $x^1$ direction (ver figura \ref{figu1}), con condiciones de contorno periódicas para el campo: $\phi(x^0,0,x^2,...,x^{d})=\phi(x^0,D,x^2,...,x^{d})$. Elegimos al campo de gauge constante en la dirección $x^1$. La presencia de este campo de puro gauge $A_\mu=\partial_\mu \alpha(x)$ puede eliminarse haciendo una transformación de gauge
\be
\phi(x)\rightarrow  e^{-i e \int_{\tilde{x}}^x dy^\mu A_\mu(y)}\,\phi(x)\,,
\ee
donde $\tilde{x}$ es un punto arbitrario. Esto cambia la condición de contorno del campo escalar
\be
\phi(x^0,0,...,x^{d})=e^{-i e \oint A_1 dx^1 }\phi(x^0,D,...,x^{d})\,.
\ee
La integral
\be
e \oint A_1 dx^1=\varphi=e \Phi
\ee
es proporcional al flujo $\Phi$ de un campo magnético a través de $S^1$. Este campo magnético se encuentra completamente fuera del espacio y su efecto en el campo escalar se da sólo a través del efecto AB (da una fase $e^{-i\varphi}$ en la condición de contorno del campo $\phi$ que queda desacoplado de fuentes externas \footnote{El efecto de las condiciones de contorno en la entropía de entrelazamiento ha sido estudiado para algunos modelos en $1+1$ dimensiones, ver por ejemplo\cite{spin,spin1,cacardy}.}).

Veamos ahora que es posible calcular la entropía en la configuración ilustrada haciendo una reducción dimensional del lagrangiano. Partiendo del lagrangiano (\ref{uno}), elegimos el campo electromagnético de modo que $e A_1=\varphi/D$ sea constante (y todas sus otras componentes nulas), y descomponiendo en modos de Fourier en la dirección $x^1$, $\phi=\sum_{n} e^{-i 2 \pi n x/D} \phi^{(n)}$, tenemos 
\be
{\cal{L}}=\sum_{n=-\infty}^\infty  \left(-\partial_\mu \phi^{(n)*} \partial^\mu \phi^{(n)}-\left(m^2+ \frac{(2\pi n+\varphi)^2}{D^2}\right) \phi^{(n)*} \phi^{(n)}\right)\,,
\ee
donde cada campo $\phi^{(n)}$ depende del tiempo y de $d-1$ dimensiones espaciales $x^2,...,x^{d}$.

Luego, la entropía de la faja de ancho $L$ alrededor del cilindro (ver figura \ref{figu1} estará dada por una suma sobre estos modos de entropías de los campos masivos en dimensión $d-1$.

A partir de ahora, nos concentraremos en el caso $d=2$ (dos dimensiones espaciales). El cálculo de la entropía de la región anular mencionada se reduce a hallar la entropía de un intervalo unidimensional de longitud $L$, para una torre infinita de campos cuyas masas están dadas por
\be
M(n,\varphi)=\sqrt{m^2+ \frac{(2\pi n+\varphi)^2}{D^2}}\,.\label{masses}
\ee
La misma reducción dimensional se aplica para el caso en el que se consideran campos de Dirac con lagrangiano
\be
{\cal L}=i\bar{\Psi}\gamma^{\mu}\left(\partial_{\mu}-i e A_{\mu}\right)\Psi-m\bar{\Psi}\Psi\,,
\ee
donde las masas efectivas para las campos unidimensionales están nuevamente dadas por (\ref{masses}). Por lo tanto, la entropía de la región anular es
\be
S(L,m,\varphi)=\sum_n S_1(L,M(n,\varphi))\,,\label{sum}
\ee
donde $S_1(L,M)$ es la entropía geométrica del campo de Dirac unidimensional de masa $M$ para un intervalo de longitud $L$ (utilizamos la letra $M$ para referirnos a la función $M\left(n,\varphi\right)$).

Estas entropías unidimensionales han sido calculadas en \cite{review}. Se tiene
\be
S_1(L,M)=-\int_{L M}^\infty dy \, \frac{C(y)}{y} -C(0) \log(M \epsilon)\,,\label{pp}
\ee
siendo $\epsilon$ un regulador ultravioleta, $M$ la masa efectiva del campo y
\be
C(ML)=L \frac{dS_1(L,M)}{dL}\label{cccc}
\ee
la función entrópica C \cite{twoD}. Esta función es positiva y monótonamente decreciente. Para $ML=0$ toma el valor $C(0)$ dado por un tercio de la carga central conforme en el límite $M\rightarrow 0$. Este valor es $C(0)=1/3$ para el campo de Dirac y $C(0)=2/3$ para el campo escalar complejo. Para valores grandes de la masa, $C(ML)$ es exponencialmente decreciente. Precisamente, los  límites para argumentos pequeños y grandes de esta función están dados por \cite{review}
\be
C(y)\simeq \frac{2}{3}+\frac{1}{\log(y)}+...\,\,\textrm{para}\,\, y\ll 1\,, \,\,\,\, C(y)\simeq \frac{1}{2} y K_1(2 y)\,\, \textrm{para}\,\, y\gg 1\,,\label{dixi}
\ee
para el campo escalar complejo, y
\be
C(y)\simeq\frac{1}{3}-\frac{1}{3}\,y^2 \,\log^2(y)+... \,\, \textrm{para}\,\, y\ll 1\,, \,\,\,\, C(y)\simeq \frac{1}{2}\, y\, K_1(2 y) \,\,\textrm{para}\,\, y\gg 1\,,\label{dixit}
\ee
para el campo de Dirac. Las funciones C pueden calcularse numéricamente con alta precisión integrando las soluciones de una ecuación diferencial ordinaria \cite{review}.

La dependencia de la entropía con el regulador ultravioleta en (\ref{pp}) no juega un rol relevante, dado que nos interesa evaluar cómo cambia la entropía con el flujo magnético (termina siendo en una constante independiente de la masa y de $L$). Para el campo escalar, la entropía incluye un término adicional dependiente de la masa
\be
\log(\log(- M \epsilon))\,.\label{vuelta}
\ee
Este término es debido a las divergencias infrarrojas que tienen los campos escalares no masivos en dos dimensiones \cite{review}. Sin embargo, este término dependiente de la masa debe pensarse como una constante global infrarroja dado que sus derivadas con respecto a la masa se anulan en el límite ultravioleta. Por este motivo, obviaremos este término en el resto del desarrollo.

Para concentrarnos en la parte universal de la variación de la entropía con el flujo magnético podemos calcular la cantidad
\be
S(\varphi)=\int_0^\varphi d\varphi^\prime\frac{d}{d\varphi^\prime} S(L,m,\varphi^\prime) =S(L,m,\varphi)-S(L,m,\varphi=0)\,.\label{subtract}
\ee
La contribución a $S(\varphi)$ del segundo término en (\ref{pp}) está dada por
\bea
&&\sum_{n=-\infty}^\infty \int_0^\varphi d\varphi^\prime\frac{d}{d\varphi^{\prime}}\left(-C(0)\log\left(M(n,\varphi') \epsilon\right)\right)=\nonumber\\&&=-\int_0^\varphi d\varphi^\prime \sum_{n=-\infty}^\infty C(0)\,\,\frac{2\pi n+\varphi^\prime}{m^2 D^2+(2 \pi n+\varphi^\prime)^2}=\nonumber\\&&=-\int_0^\varphi d\varphi'\int_0^{\varphi'} d\varphi'' C(0)\sum_{n=-\infty}^\infty \frac{(mD)^2-(2\pi n+\varphi'')^2}{\left[(mD)^2+(2\pi n+\varphi'')^2\right]^2}=\nonumber\\&&=-\int_0^\varphi d\varphi'\int_0^{\varphi'} d\varphi'' C(0) \frac{\cosh(mD)\cos(\varphi'')-1}{2\left[\cos(\varphi'')-\cosh(mD)\right]^2}= \nonumber\\&&=-
\int_0^\varphi d\varphi^\prime \frac{C(0)\,\sin{\varphi^\prime}}{2 (\cosh(m D)-\cos(\varphi^\prime))}
=\nonumber\\&&=-\frac{C(0)}{2}\log\left(\frac{\cosh(m D)-\cos(\varphi)}{\cosh(m D)-1}\right)\,.\label{eq1234}
\eea
Afortunadamente, la suma en la segunda línea de (\ref{eq1234}) se puede hacer. La contribución dada por (\ref{eq1234}) es independiente del ancho $L$ de la faja $L$, y es siempre negativa.

Por lo tanto, de las ecuaciones (\ref{sum}), (\ref{pp}) y (\ref{subtract}) se tiene
\begin{eqnarray}
S(\varphi)&=&-\sum_{n=-\infty}^\infty  \int_{L M(n,\varphi)}^\infty dy \, \frac{C(y)}{y}
+\\
&+&\sum_{n=-\infty}^\infty  \int_{L M(n,0)}^\infty dy \, \frac{C(y)}{y}
-\frac{C(0)}{2}\log\left(\frac{\cosh(m D)-\cos(\varphi)}{\cosh(m D)-1}\right)\,.\nonumber
\label{complete}
\end{eqnarray}
Esta expresión es finita, mostrando que la dependencia con $\varphi$ de la entropía es independiente de la regularización. Algunos aspectos generales de$S(\varphi)$ pueden obtenerse de (\ref{complete}) sin necesidad de hacer más cuentas. Evidentemente, de (\ref{complete}) la entropía $S(\varphi)$ será una función periódica de la fase $\varphi$ con período $2 \pi$. Cuando el cuanto de flujo a través del cilindro es $\Phi=e/2\pi$ tenemos $S(\varphi)=0$ y por lo tanto no hay efecto en la entropía geométrica. De (\ref{masses}) ve que el efecto es simétrico ante $\varphi\rightarrow -\varphi$ y $\varphi\rightarrow \pi-\varphi$. $S(\varphi)$ puede calcularse numéricamente a partir del conocimiento numérico de la función C. El resultado muestra que $S(\varphi)$ es siempre negativa; el máximo de $|S(\varphi)|$ se alcanza para $\varphi=\pi$. Esto indica que el efecto AB siempre hace decrecer la entropía de entrelazamiento respecto a la de vacío sin campo magnético.

\subsection{Varios límites}

Para estudiar el caso no masivo comenzamos considerando la ecuación (\ref{complete}) para $m D\ll 1$ y $mL\ll 1$. Hasta primer orden en $m D$ el tercer término en (\ref{complete}) da
\be
C(0)\log\left(m D\right)-\frac{C(0)}{2}\log\left(2-2\cos\left(\varphi\right)\right)\,.\label{loga}
\ee
Podemos tomar $m=0$ en la primera suma infinita de (\ref{complete}) sin que aparezcan divergencias (a menos que $\varphi$ sea un múltiplo entero de $2\pi$). La segunda suma porta una divergencia proveniente del modo $n=0$ cuando se toma $m=0$, pero se puede verificar que esta contribución se cancela con el término logartítmico dado por (\ref{loga}). Aislamos el término con $n=0$ en esta segunda sumatoria y extraemos el término logarítmico para $mL\ll 1$
\be
\int_{m L}^{\infty}dy\frac{C\left(y\right)}{y}\simeq -C(0)\log\left(m L\right)+\gamma  \,,\label{tutu}
\ee
donde
\be
\gamma=\lim_{y_0 \rightarrow 0}\left(\int_{y_0}^{\infty}dy\frac{C\left(y\right)}{y}+C(0) \log(y_0)\right)\,.\label{gam}
\ee
\begin{figure}[h]
\begin{center}
\includegraphics[width=10cm]{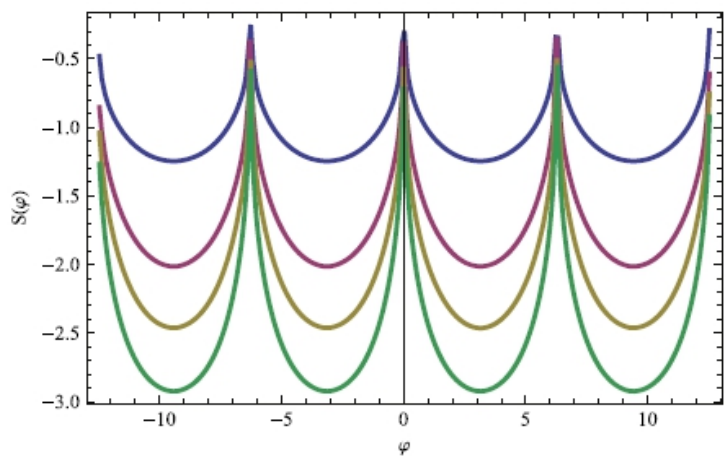}
\end{center}
\caption{\textbf{$S(\varphi)$ para el campo escalar no masivo y distintos valores de $L/D$.} De arriba hacia abajo: $L/D=1/10, 1/2, 1, 2$. En esta figura, tomamos el valor de la constante divergente infrarroja como $\gamma=2$. Notar que la forma de las fluctuaciones no tiende a anularse para valores pequeños de $L/D$, pero se hace lentamente cada vez `más plana' cuando $L/D\rightarrow 0$.} \label{figu2}
\end{figure}
Podemos evaluar (\ref{gam}) numéricamente utilizando los resultados para $C(y)$ dados en \cite{twoD} y, para el campo de Dirac, se tiene $\gamma \simeq-0.528$. Para el campo escalar, $\gamma$ está controlado por la física a bajas energías y puede ser grande. Si el regulador infrarrojo para el modo cero se ajusta con una masa pequeña se tiene $\gamma\sim -\log(-\log(m L))$. Esto es debido al segundo término más relevante en la expansión para $mL$ pequeño de la función $C(mL)$ (ecuación (\ref{dixi})). Si algún otro mecanismo fijara el regulador infrarrojo esto podría cambiar enormemente. Por ejemplo, imponiendo una condición de contorno antiperiódica en la dirección $x^2$ se tendría $\gamma\sim -\log(R/L)$, con $R$ el radio de compactificación en la dirección $x^2$.
\begin{figure}[h]
\begin{center}
\includegraphics[width=10cm]{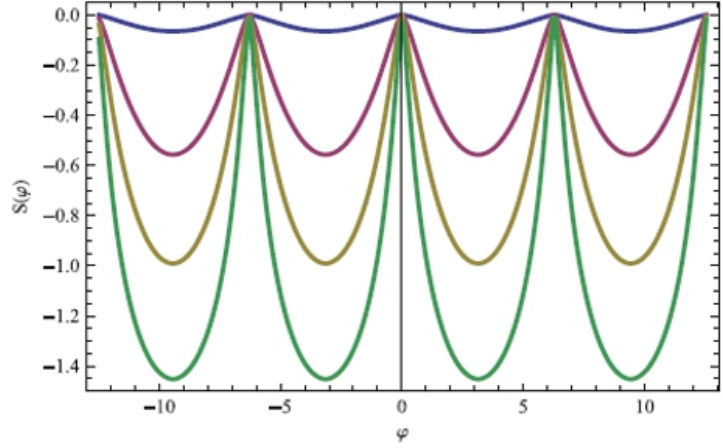}
\end{center}
\caption{\textbf{$S(\varphi)$ para el campo de Dirac no masivo y varios valores de $L/D$.} De arriba hacia abajo: $L/D=1/10, 1/2, 1, 2$. Para $L/D$ grande, las curvas son similares en forma a las del campo escalar, pero su tamaño se reduce a la mitad (comparar con figura \ref{figu2} y fórmula (\ref{contri})). Para $L/D$ pequeño, en contraste con el caso escalar, la función $S(\varphi)$ decae a cero.} \label{figu3}
\end{figure}
Al escribir la expresión completa para $S\left(\varphi\right)$, los términos logarítmicos que involucran a la masa $m$ en (\ref{loga}) y (\ref{tutu}) se cancelan. De este modo, la expresión para la entropía del campo no masivo queda
\begin{eqnarray}
S(\varphi)&=&\sum_{n\neq 0} \int^{\left|\frac{\left(2\pi n+\varphi\right)L}{D}\right|}_{\left|\frac{2\pi n L}{D}\right|} dy \, \frac{C(y)}{y}-\int_{\left|\frac{L\varphi}{D}\right|}^{\infty} dy\, \frac{C(y)}{y}+\\
&-&\frac{C(0)}{2}\log\left(2-2\cos\left(\varphi\right)\right)+\gamma-C(0)\log\left(\frac{L}{D}\right)\,.\nonumber
\end{eqnarray}
Naturalmente, $S(\varphi)$ es una función de  $L/D$. En las figuras (\ref{figu2}) y (\ref{figu3}) se grafica $S(\varphi)$ para algunos valores de $L/D$ para el campo escalar y el campo de Dirac, respectivamente.

Cuando $L/D\gg 1$ y $|L/D\,\varphi|\gg 1$ los primeros dos términos son exponencialmente pequeños (tomando $\varphi\in (-\pi,\pi)$), y la forma de las oscilaciones está dada por
\be
S(\varphi)=-\frac{C(0)}{2}\log\left(2-2\cos\left(\varphi\right)\right)+\gamma-C(0)\log\left(\frac{L}{D}\right)\,.\label{contri}
\ee
Exceptuando un factor $2$ y una constante global aditiva, la amplitud de estas oscilaciones es la misma para el campo de Dirac y el escalar.

El tamaño máximo $|S(\pi)|$ de las oscilaciones, en este caso, es
\be
|S(\pi)|=\frac{C(0)}{2}\log(4)-\gamma+C(0)\log\left(\frac{L}{D}\right)\,.
\ee
Este tamaño puede ser arbitrariamente largo para $L$ grande y $D$ fijo. Esta gran amplitud en la oscilación se debe a que el modo unidimensional no masivo $n=0$ tiene entropía que crece logarítmicamente con $L$, y esto está regulado por la masa efectiva que provee el campo magnético. Nótese que la dependencia dada por el primer término de (\ref{contri}) con $\varphi$ se produce por una contribución coherente de todos los modos a través del término de saturación $-C(0) \log(M)$ en la entropía.

La divergencia infrarroja (\ref{vuelta}) vuelve a aparecer en $S(\varphi)$, para el campo escalar no masivo, a través de la constante infrarroja divergente $\gamma$, y las grandes variaciones de la entropía cerca de $\varphi\rightarrow 0$. La variación de entropía para el campo escalar no masivo con y sin campo magnétio es divergente. Esta gran ``susceptibilidad'' al campo magnético no se presenta en el caso fermiónico dado que se debe al modo $n=0$ (clásico) del campo escalar. De todos modos, el valor de $\gamma$ no afecta las variaciones finitas de $S(\varphi)$ entre los diferentes valores de $\varphi\neq 0$. No modifica la forma de las curvas lejos de $\varphi=0$ sino que las desplaza a valores negativos grandes (ver figura \ref{figu2}).

En el límite opuesto, para $L/D\ll 1$, los modos se suman en forma no coherente y esto disminuye el tamaño de las oscilaciones. En este límite es mejor estudiar directamente la derivada $S^\prime(\varphi)$. Por (\ref{masses}) y (\ref{sum}), en el límite no masivo, esta derivada es
\be
S^\prime(\varphi)=\frac{L}{D} \sum_{n=-\infty}^\infty f\left(\frac{L}{D} (2 \pi n+\varphi )\right)\,, \label{sss}
\ee
donde
\be
f(x)=\frac{C(|x|)-C(0)}{x}\,.
\ee
La función $f(x)$ es antisimétrica y decae a cero exponencialmente rápido hacia el infinito. Si $f(x)$ fuese analítica podríamos aplicar la fórmula de Euler MacLaurin a (\ref{sss}) para concluir que $S^\prime(\varphi)$, y por lo tanto $S(\varphi)$, se anulan exponencialmente rápido con $L/D$ para $L/D$ pequeño. Sin embargo, $f(x)$ no es analítica en el origen; las expansiones dan $f(x)\sim -1/3 x \log(|x|)^2$ para fermiones y $f(x)\sim(x \log(|x|))^{-1}$ para escalares (ver (\ref{dixi}), (\ref{dixit})). En consecuencia, la amplitud de las oscilaciones cae como $(L/D)^2 \log(L/D)$ para fermiones en el límite de $L/D$ pequeño, mientras que $S^\prime(\varphi)$ decae sólo logarítmicamente, como $(\log(L/D))^{-1}$, para el campo escalar ($\varphi$ se mantiene fijo mientras $L/D\rightarrow 0$). Esta diferencia puede ser apreciada en las figuras \ref{figu2} y \ref{figu3}.

\begin{figure}[h]
\begin{center}
\includegraphics[width=10cm]{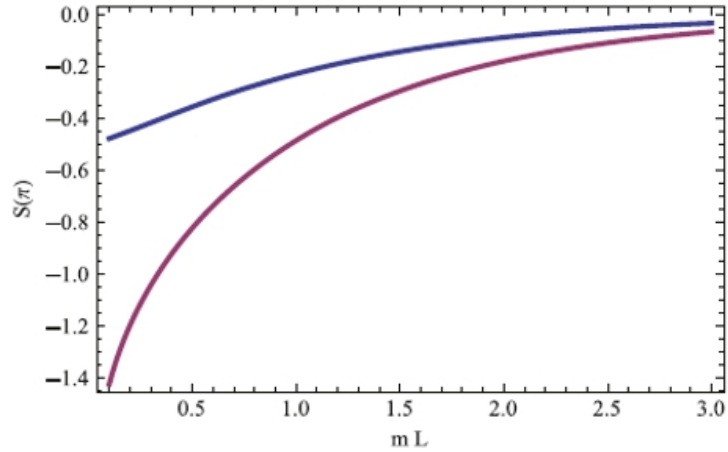}
\end{center}
\caption{\textbf{$S(\pi)$ (tamaño máximo del cambio de entropía con el flujo magnético) como función de la masa para $L/D=1$.} Para valores grandes de la masa $S(\pi)$ cae exponencialmente a cero. Para $m\rightarrow 0$ converge a un valor finito para el campo de Dirac (curva superior) y tiende a $-\infty$ para el campo escalar (curva inferior).} \label{figu4}
\end{figure}

Para campos masivos el efecto que produce el campo magnético en la entropía es menos intenso que el observado para el caso no masivo. Para $m L\gg 1$, el primer término de (\ref{complete}) da un número exponencialmente pequeño $\sim 1/2 (m L) K_1(2 m L)\sim (mL)^{1/2} e^{-2 m L}$. En este régimen, el tercer término de (\ref{complete}) da la contribución principal a la entropía
\be
S(\varphi)\sim -\frac{C(0)}{2}\log\left(\frac{\cosh(m D)-\cos(\varphi)}{\cosh(m D)-1}\right)\,.
\ee
Cuando adicionalmente se tiene $m D\gg 1$, este último término también da un número exponencialmente pequeño
\be
S(\varphi)\sim -C(0)\,e^{-m D}(1-\cos\left(\varphi\right))\,,
\ee
cuya forma ahora es puramente sinusoidal. En la figura (\ref{figu4}) se muestra el menor valor de la entropía $S(\pi)$, para el valor particular $L/D=1$, como función de la masa.

\section{Efecto AB e información mutua}

En las secciones anteriores mostramos que el efecto AB se manifiesta en la entropía de entrelazamiento de la región anular de tamaño $L$. Alternativamente, podemos estudiar la información mutua (dada por la ecuación \reef{infomutua}) entre dos cilindros semi-infinitos $A$ y $B$ separados por una distancia $L$ en la dirección $x^2$. Aquí $A$ y $B$ son dos regiones disjuntas que constituyen el complemento de la región tipo faja de la figura (\ref{figu1}). Como ya se ha mencionado en la sección \ref{secmutua}, esta cantidad mide la información compartida por los dos subsistemas y tiene la ventaja de ser independiente de la regularización (por su propia definición).

El cálculo es similar al realizado para la entropía; se hace reducción dimensional, y se utiliza la información mutua unidimensional $I_1(ML)$ para un campo de masa $M$ entre dos semirrectas separadas por una distancia $L$ en una dimensión. Esto da
\be
I(L,m,\varphi)=\sum_n I_1(L \sqrt{m^2+(2 n\pi+\varphi)^2/D^2} )\,.
\ee
En (\ref{infomutua}) sólo $S(A\cup B)$ cambia con $L$. Consecuentemente, $dI_1(LM)/dL=-dS_1(A\cup B)/dL$. Al igual que para la entropía $I_1(ML)$ también se anula para $L$ grande, dado que los estados en $A$ y $B$ se factorizan en este límite (recordar que $S_1(A\cup B)$ es igual a la entropía de su complemento $S_1(L)$, dado que el estado global es puro). Teniendo en cuenta todo esto, se obtiene
\be
I_1(L M)=\int_{L M}^\infty dy \, \frac{C(y)}{y}\,.
\ee
Esto es opuesto a lo obtenido para la entropía en (\ref{pp}), excepto por la ausencia del término de borde $\log(M)$ que es independiente de $L$ y se cancela en la información mutua (\ref{infomutua}). Luego, se obtiene una fórmula similar a (\ref{complete}) para la información mutua, pero sin el último término
\be
I(\varphi)=I(L,m,\varphi)-I(L,m,0)=\sum_{n=-\infty}^\infty  \int_{L M(n,\varphi)}^\infty dy \, \frac{C(y)}{y}
-\sum_{n=-\infty}^\infty  \int_{L M(n,0)}^\infty dy \, \frac{C(y)}{y}\,.\label{nunu}
\ee
La concavidad de la entropía unidimensional da $I^{\prime\prime}_1(ML)>0$; esto implica que la suma en (\ref{nunu}) es decreciente para $\varphi\in(0,\pi)$. En consecuencia, $I(\varphi)$ es siempre negativa, y alcanza su valor mínimo para $\varphi=\pi$. Esto muestra que el efecto AB siempre hace decrecer la información mutua.

La amplitud de las oscilaciones de la información mutua $I(\varphi)$ diverge en el límite no masivo $m\rightarrow 0$. Esto se debe a que la información mutua para el modo $n=0$ diverge en una dimensión para regiones semi-infinitas, mientras que esto no sucede para el caso en que $\varphi$ no es cero.

Cerramos este capítulo resumiendo que los resultados anteriores demuestran que la entropía de entrelazamiento mide el efecto Aharonov-Bohm. En nuestro caso, estudiamos un ejemplo sencillo en dos dimensiones donde el entrelazamiento en el vacío siempre decrece con el flujo magnético para holonomías no nulas. Esto puede atribuirse a la interferencia de modos, donde la holomonía induce una masa efectiva para los campos, reduciendo las correlaciones.

Otros escenarios donde el efecto AB puede calcularse son casos análogos a nuestro cálculo para campos libres, donde pueda hacerse reducción dimensional, pero en dimensiones más altas. En dimensiones más altas es necesario utilizar la información mutua para eliminar las divergencias espúreas, que aparecen en el cambio de la entropía con el flujo magnético, debidas al cambio en los términos de área inducidos por la masa \cite{wil,hertzberg}. Para dos regiones que constituyan el complemento de una región anular en un plano, las variaciones con el flujo magnético no deberían divergir en el caso no masivo, a diferencia de lo que sucede para la geometría que consideramos (ya que esta cantidad es finita para el caso en que el campo magnético se anula).

También sería interesante estudiar este efecto utilizando la entropía holográfica de entrelazamiento, presentada en la sección \ref{ch:relativa}.
\chapter{\label{ch:modular}Hamiltoniano modular}

En esta sección estudiaremos con detalle un objeto importante que caracteriza al estado de nuestro sistema. Cualquier operador densidad $\rho$ puede escribirse de la forma
\begin{equation}
\rho = \frac{e^{-H}}{\tr\left(e^{-H}\right)}\,,
\end{equation}
siendo $H$ un operador autoadjunto. En el contexto de la teoría algebraica de campos $H$ se conoce como \textbf{hamiltoniano modular} \cite{haag}\footnote{En la literatura de materia condensada este objeto se suele llamar hamiltoniano de entrelazamiento.}. Para un estado $\rho_V= \frac{e^{-H_V}}{\tr\left(e^{-H_V}\right)}$ asociado al dominio de dependencia causal de $V$, $H_V$ es un operador autoadjunto cuyo espectro se extiende en general desde $-\infty$ a $+\infty$ (y tiene al vacío como autovector de autovalor cero). La conjugación por $H_V$ de un operador $\cal{O}$
\begin{equation}
{\cal O}\left(\tau\right)= e^{i H_V\tau}{\cal O} e^{-i H_V\tau}\,,
\end{equation}
es un automorfismo del álgebra de operadores acotados en el dominio de dependencia causal de $V$. El grupo monoparamétrico de transformaciones unitarias dado por estos automorfismos, $U\left(\tau\right)=e^{-i H_V \tau}$, se conoce como \textbf{grupo modular} \cite{haag}. Estos operadores unitarios son generadores de simetría del sistema ya que
\begin{equation}
\textrm{tr}\left( \rho _{V}U\left( \tau \right) \mathcal{O}U\left( -\tau \right)
\right) =\textrm{tr}\left( \rho _{V}\mathcal{O}\right)\,,
\end{equation}
para cualquier operador $\mathcal{O}$ localizado en $V$. Extendiendo $\tau$ al plano complejo, encontramos que el grupo modular satisface la relación de periodicidad KMS (Kubo - Martin - Schwinger) \footnote{La condición KMS, para estados de teorías de campos cuyos correladores están determinados por las funciones de dos puntos, es $\textrm{tr}\left( \rho\mathcal{O}_{1}\left( i\beta\right) \mathcal{O}_{2}\right)
=\textrm{tr}\left( \rho\mathcal{O}_{2}\mathcal{O}%
_{1}\right)$. Si esta condición es válida, $\rho$ es invariante ante el grupo monoparamétrico de traslaciones temporales y define entonces un estado térmico a temperatura $T=1/\beta$.} en tiempo imaginario
\begin{equation}
\textrm{tr}\left( \rho _{V}\mathcal{O}_{1}\left( i\right) \mathcal{O}_{2}\right)
=\textrm{tr}\left( \rho _{V}U\left( i\right) \mathcal{O}_{1}U\left( -i\right) 
\mathcal{O}_{2}\right) =\textrm{tr}\left( \rho _{V}\mathcal{O}_{2}\mathcal{O}%
_{1}\right) \text{.}
\end{equation}
Esto surge de utilizar que $U\left( i\right) =\rho _{V}^{-1}$, $U\left( -i\right)=\rho _{V}$, y de la propiedad cíclica de la traza. Pensando a $\tau$ como un tiempo interno, $\rho_V$ se interpreta como un estado térmico a $T=1$, cuya evolución unitaria está dada por $U\left(\tau\right)=\rho_V^{i\tau}$. Esta evolución interna se conoce en la literatura como \textbf{flujo modular} \cite{haag}.

En general, el hamiltoniano modular es un operador no local, por lo que $U\left(\tau\right)$ no genera un flujo local en el dominio de dependencia causal de $V$ (si ${\cal O}$ es por ejemplo el operador de campo en un punto dado, ${\cal O}\left(\tau\right)$ no tendrá en general una expresión sencilla donde se involucre sólo un punto). Sin embargo, hay algunos casos especiales en donde se conoce que el flujo modular es local. En este capítulo, estudiamos algunos de esos ejemplos y presentamos los conceptos básicos que se utilizan en los capítulos siguientes. La importancia de los hamiltonianos modulares queda plasmada en el desarrollo de los capítulos siguientes y por ello, el estudio de hamiltonianos modulares se retoma en el capítulo \ref{ch:local}.

Comenzamos presentando el efecto Unruh y el hamiltoniano de Rindler. Luego, nos centramos en estudiar hamiltonianos modulares para estados térmicos. En particular, en la sección \ref{thermal} obtenemos los primeros resultados explícitos del hamiltoniano para un estado térmico en un intervalo.

\section{Hamiltoniano de Rindler y efecto Unruh}

\begin{figure}[h]
\begin{center}
\includegraphics[width=10cm]{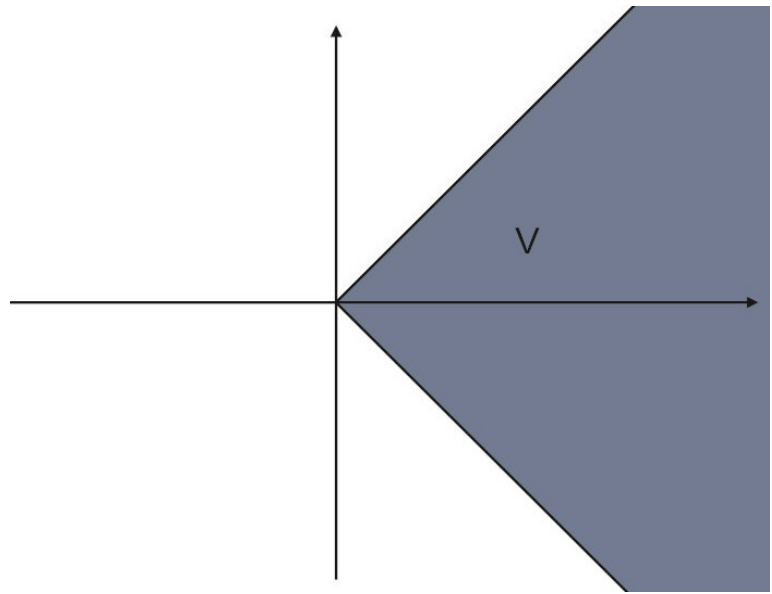}
\end{center}
\caption{\textbf{Wedge de Rindler}. En la figura se representa sombreado el wedge de Rindler, la mitad de la región causalmente espacial respecto del origen de coordenadas. Para cualquier teoría de campos, el hamiltoniano modular asociado a un estado reducido al wedge de Rindler tiene la forma universal $\rho_V=c e^{-2\pi K}$, donde $K$ es el generador de boosts en la dirección espacial en la cual se extiende el wegde de Rindler.} \label{wedge}
\end{figure}

Un observador con aceleración propia constante, en el vacío del espacio de Minkowski, percibe como si estuviera inmerso en un baño térmico a temperatura $T=a/2\pi$ \cite{unruh}. Este es el denominado efecto Unruh. La aceleración constante del observador le impide estar en contacto con todo el espacio de Minkowski y, de hecho, hay una porción del espacio tiempo desde la que las señales de luz no pueden alcanzar al observador. En consecuencia, los grados de libertad que se encuentran en esta región no son relevantes para la descripción de toda la física relativa al observador. Si trazamos sobre esa región, el estado de vacío se convierte en el estado mixto
\begin{equation}
\rho_V=c e^{-2\pi K}\,,\label{hamrind}
\end{equation}
donde $K=\int_{x>0} d^{d-1}x\, x \,T_{00}(\vec{x})$ es el operador de boosts, que mantiene fijo al wedge de Rindler $V$ (una de las regiones causalmente desconectadas del origen de coordenadas, ver figura \ref{wedge}).

\begin{figure}
\begin{center}
\includegraphics[width=13.5cm]{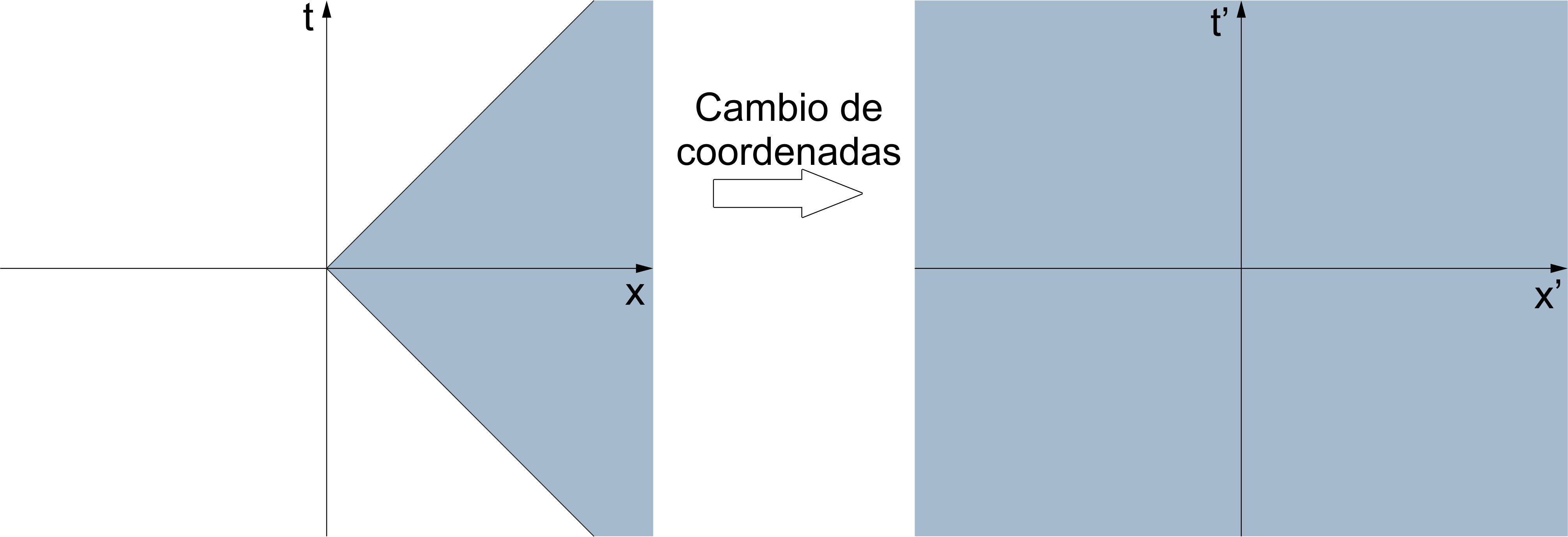}
\end{center}
\caption{\textbf{Transformación de coordenadas}. El wedge de Rindler dado por $u_{\pm}\geq 0$ se mapea en las nuevas coordenadas a todo el plano $(t',x')$.} \label{tr1}
\end{figure}

La expresión \reef{hamrind} es válida para cualquier teoría de campos cuando $V$ es el wedge de Rindler \cite{biso}. La evolución interna es en este caso local y causal, y está dada por transformaciones puntuales a lo largo de las trayectorias de aceleración constante, es decir, las curvas integrales del operador de boost $K$
\begin{equation}
\begin{pmatrix}
x\left( \tau \right) \\ 
t\left( \tau \right)%
\end{pmatrix}%
=%
\begin{pmatrix}
\cosh \left( 2\pi \tau \right) & \sinh \left( 2\pi \tau \right) \\ 
\sinh \left( 2\pi \tau \right) & \cosh \left( 2\pi \tau \right)%
\end{pmatrix}%
\begin{pmatrix}
x\left( 0\right) \\ 
t\left( 0\right)%
\end{pmatrix}%
\text{,}
\end{equation}
que en coordenadas nulas $u_{\pm }=x\pm t$ \footnote{Con esta definición, $u_{\pm }>0$ en el wedge de Rindler.} se escribe como
\begin{equation}
u_{\pm }\left( \tau \right) =e^{\pm 2\pi \tau }u_{\pm }\left( 0\right) \text{%
.}  \label{rel1}
\end{equation}
Interpretado en el sentido de Unruh, el estado en el espacio de Rindler es térmico con respecto a la noción de traslaciones temporales a lo largo de las órbitas del operador de boost. Definimos nuevas coordenadas
\begin{equation}
u_{\pm }^{\prime } =\ell _{1}\log \frac{u_{\pm } }{\ell _{2}}\,,
\label{nulasnuevas}
\end{equation}
donde hemos introducido las constantes $\ell_{1}$ y $\ell_{2}$ para que las dimensiones sean correctas; en estas nuevas coordenadas, el flujo modular se verá de una forma diferente. La relación \reef{nulasnuevas}, junto con la ecuación $\left( \ref{rel1}\right) $ nos permite encontrar el flujo modular en las coordenadas primadas escribiendo $u_{\pm}^{\prime }\left( \tau \right) $ como función de $u_{\pm }^{\prime} \left(0\right) $
\begin{equation}
u_{\pm }^{\prime }\left( \tau \right) =u_{\pm }^{\prime }\left( 0\right) \pm
2\pi \tau \ell _{1}\text{.}
\end{equation}
En este caso, es sencillo ver que la transformación modular corresponde a traslaciones temporales en las nuevas coordenadas. De hecho, utilizando el grupo de relaciones
\begin{equation}
x^{\prime } =\dfrac{u_{+}^{\prime }
+u_{-}^{\prime } }{2}\text{, }t^{\prime } =\dfrac{u_{+}^{\prime } -u_{-}^{\prime } }{2}\text{,} \label{equis}
\end{equation}
(donde ahora $x' \in (-\infty,+\infty)$ y $t' \in (-\infty,+\infty)$ dado que el wedge en las coordenadas originales está dado por $u_{\pm}\geq 0$ - ver figura \ref{tr1}) podemos encontrar una expresión para $t^{\prime }\left( \tau \right) $ y $x^{\prime }\left( \tau \right) $ en términos de $t^{\prime }\left( 0\right) $ y $x^{\prime
}\left( 0\right) $
\begin{eqnarray}
t^{\prime }\left( \tau \right) &=&t^{\prime }\left( 0\right) +2\pi \tau \ell
_{1}\text{,} \\
x^{\prime }\left( \tau \right) &=&x^{\prime }\left( 0\right) \text{.}
\end{eqnarray}
Sobre la superficie $t'\left(0\right)=0$, una deformación infinitesimal $\delta\tau$ nos da
\begin{equation}
\delta t'\left(\tau\right)=\left[ \frac{dt^{\prime }\left( \tau \right) }{d\tau }\right] _{\tau
=0,\,t^{\prime }\left( 0\right) =0} \delta\tau \,.
\end{equation}
El hamiltoniano modular es el generador que induce esta transformación infinitesimal en las coordenadas del campo
\begin{equation}
H=\int dx^{\prime }\left( 0\right) T'_{00}\left( x^{\prime }\left( 0\right)
\right) \left[ \frac{dt^{\prime }\left( \tau \right) }{d\tau }\right] _{\tau
=0,\,t^{\prime }\left( 0\right) =0}\,\text{.} 
\end{equation}
En este caso en particular, se tiene
\begin{equation}
\left[ \frac{dt^{\prime }\left( \tau \right) }{d\tau }\right] _{\tau
=0,\,t^{\prime }\left( 0\right) =0}=2\pi \ell _{1}\text{,}
\end{equation}
con lo que
\begin{equation}
H=\int dx^{\prime }\left( 0\right) T'_{00}\left( x^{\prime }\left( 0\right)
\right) 2\pi \ell _{1}\,\text{.}  \label{ham1}
\end{equation}
Esto es justamente el hamiltoniano modular correspondiente a un estado térmico
\begin{equation}
\rho =\frac{e^{-\beta H}}{\textrm{tr}\left( e^{-\beta H}\right) }\text{,}
\end{equation}
a temperatura $\beta=2\pi\ell_1$.

\section{Estados térmicos en teorías conformes bidimensionales}\label{thermal}

En esta sección consideramos teorías conformes en $1+1$. El objetivo es hallar el hamiltoniano modular de distintas regiones para estados térmicos, utilizando transformaciones modulares siguiendo la idea presentada en \cite{hmm}, que ha sido ilustrada en la sección anterior para el caso particular del efecto Unruh.

\subsection{Estado térmico en el espacio de Rindler}\label{semi}

Ahora consideramos el wedge dado por $x-\ell _{0}>\left\vert t\right\vert $. En esta sección, reproducimos el resultado hallado en \cite{yng} para el flujo modular de estados térmicos en la mitad del espacio de Rindler.

\begin{figure}[h]
\begin{center}
\includegraphics[width=4cm]{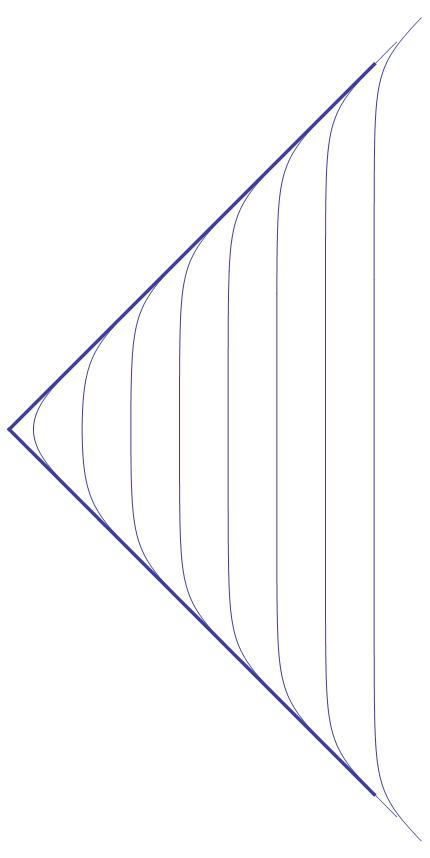}
\end{center}
\caption{\textbf{Flujo modular para un estado térmico en la mitad del espacio}. Cerca del borde del wedge de Rindler las curvas integrales tienden a las curvas integrales de los boosts de Lorentz. Lejos del borde del wegde, vemos que las curvas integrales son líneas verticales, lo que está asociado con traslaciones puras en el tiempo.} \label{ray}
\end{figure}

Comenzamos escribiendo el flujo modular en el wedge
\begin{equation}
\begin{pmatrix}
x\left( \tau \right) \\ 
t\left( \tau \right)%
\end{pmatrix}%
=%
\begin{pmatrix}
\cosh \left( 2\pi \tau \right) & \sinh \left( 2\pi \tau \right) \\ 
\sinh \left( 2\pi \tau \right) & \cosh \left( 2\pi \tau \right)%
\end{pmatrix}%
\begin{pmatrix}
x\left( 0\right) -\ell _{0} \\ 
t\left( 0\right)%
\end{pmatrix}%
+%
\begin{pmatrix}
\ell _{0} \\ 
0%
\end{pmatrix}%
\text{.}
\end{equation}
En coordenadas nulas, esta relación está dada por
\begin{equation}
u_{\pm }\left( \tau \right) =e^{\pm 2\pi \tau }\left( u_{\pm }\left(
0\right) -\ell _{0}\right) +\ell _{0}\text{.}
\end{equation}
\begin{figure}
\begin{center}
\includegraphics[width=13.5cm]{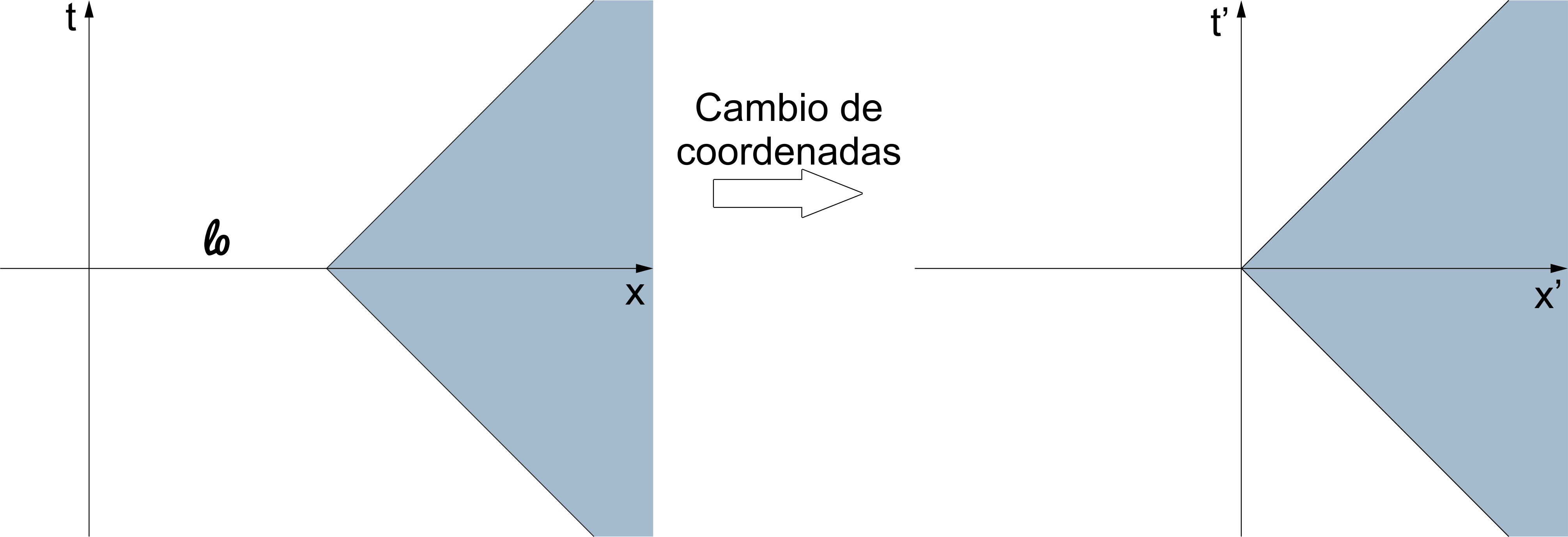}
\end{center}
\caption{\textbf{Transformación de coordenadas}. El wedge de Rindler original se mapea en las nuevas coordenadas a otro wedge en el plano $(t',x')$, cuyo vértice elegimos en el origen del sistema de coordenadas nuevo.} \label{tr2}
\end{figure}
Ahora podemos volver a utilizar la ecuación (\ref{nulasnuevas}) para hallar el flujo modular en las coordenadas primadas. $\ell _{1}$ y $\ell _{2}$ son nuevamente constantes que introdujimos para que las dimensiones sean correctas. En las nuevas coordenadas, el wedge de Rindler se transforma en un nuevo wedge (ver figura \ref{tr2}). Para que el vértice de este nuevo wedge se encuentre situado en $x'=t'=0$ debemos tomar $\ell _{2}=\ell _{0}$. Con esta elección, la relación entre $u_{\pm }^{\prime }\left( \tau \right) $ y $u_{\pm
}^{\prime }\left( 0\right) $ es
\begin{equation}
u_{\pm }^{\prime }\left( \tau \right) =\ell _{1}\log \left[ \exp \left( 
\frac{u_{\pm }^{\prime }\left( 0\right) }{\ell _{1}}\pm 2\pi \tau \right)
+\left( 1-e^{\pm 2\pi \tau }\right) \right] \text{,}
\end{equation}
y, utilizando las relaciones en (\ref{equis}), podemos encontrar expresiones para $t^{\prime
}\left( \tau \right) $ y $x^{\prime }\left( \tau \right) $ como funciones de $t^{\prime }\left( 0\right) $ y $x^{\prime }\left( 0\right) $
\footnotesize
\begin{eqnarray}
t^{\prime }\left( \tau \right) &=&\frac{\ell _{1}}{2}\log \left[ \frac{%
1+e^{2\pi \tau }\left( e^{\left( t^{\prime }\left( 0\right) +x^{\prime
}\left( 0\right) \right) /\ell _{1}}-1\right) }{1+e^{-2\pi \tau }\left(
e^{\left( -t^{\prime }\left( 0\right) +x^{\prime }\left( 0\right) \right)
/\ell _{1}}-1\right) }\right] \text{,} \\
x^{\prime }\left( \tau \right) &=&\\
&=&\frac{\ell _{1}}{2}\log \left\{ \left[
1+e^{2\pi \tau }\left( e^{\left( t^{\prime }\left( 0\right) +x^{\prime
}\left( 0\right) \right) /\ell _{1}}-1\right) \right] \left[ 1+e^{-2\pi \tau
}\left( e^{\left( -t^{\prime }\left( 0\right) +x^{\prime }\left( 0\right)
\right) /\ell _{1}}-1\right) \right] \right\} \text{.}\nonumber
\end{eqnarray}
\normalsize
En este caso, tenemos
\begin{equation}
\left[ \frac{dt^{\prime }\left( \tau \right) }{d\tau }\right] _{\tau
=0,\,t^{\prime }\left( 0\right) =0}=2\pi \ell _{1}\left( 1-e^{-\tfrac{%
x^{\prime }\left( 0\right) }{\ell _{1}}}\right) \text{.}
\end{equation}
Considerando que el wedge en el espacio $t'$, $x'$ está inmerso en un estado de temperatura inversa $\beta=2\pi \ell _{1}$, el hamiltoniano modular es entonces
\begin{equation}
H=\int dx^{\prime }\left( 0\right) T'_{00}\left( x^{\prime }\left( 0\right)
\right) \beta \left( 1-e^{-\tfrac{2\pi x^{\prime }\left( 0\right) }{\beta }%
}\right) \text{,}\label{hamtermrin}
\end{equation}
En la figura \ref{ray} se grafica el flujo modular asociado a este hamiltoniano modular. Podemos ver que cerca del borde del wedge, las curvas integrales son prácticamente boosts de Lorentz (hamiltoniano modular a temperatura igual a cero), mientras que lejos del borde la acción modular da traslaciones temporales.

\subsection{Estado térmico en un intervalo}\label{interval}

\begin{figure}[h]
\begin{center}
\includegraphics[width=14cm]{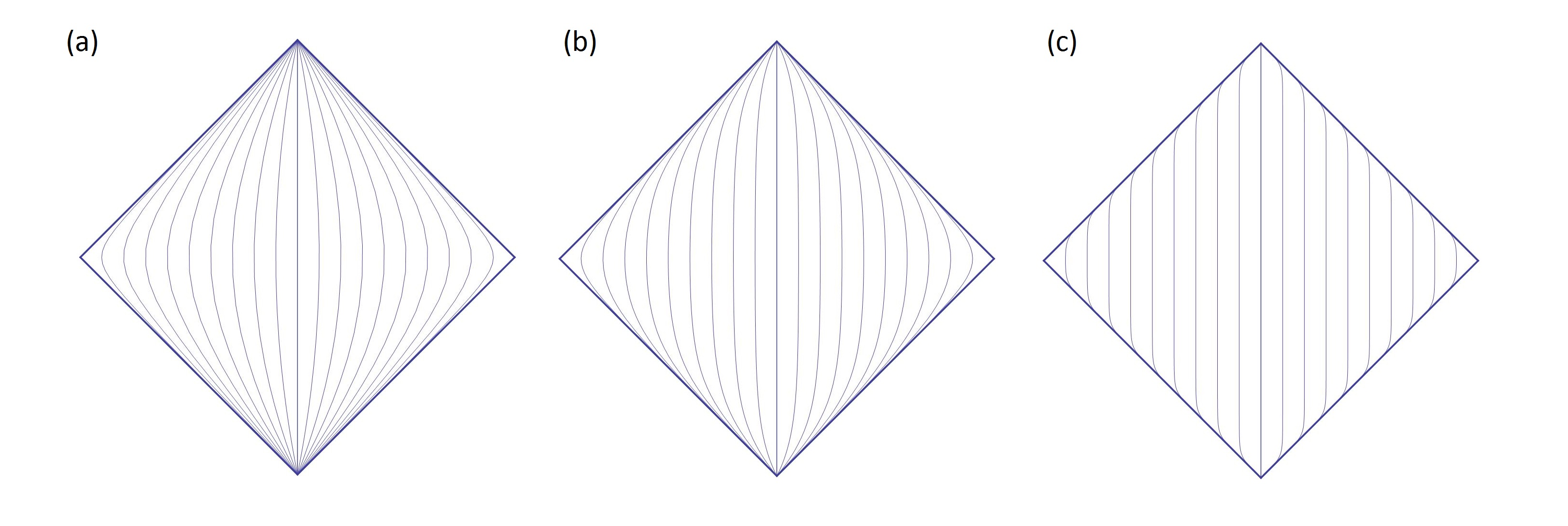}
\end{center}
\caption{\textbf{Flujo modular de un estado térmico en un intervalo finito}. Las figuras (a), (b) y (c) representan el flujo modular en el diamante causal asociado al intervalo para valores de temperatura creciente con respecto a la longitud $L$ del intervalo.} \label{segm}
\end{figure}

En esta subsección, aplicaremos el mismo razonamiento que en la sección anterior, pero ahora concentrándonos en el caso de un intervalo finito $\left(\ell _{0},\ell _{0}+\ell _{3}\right) $, con $\ell _{0}>0$. El flujo modular para este caso ya ha sido hallado en \cite{hmm} y está dado por \footnote{Para uniformar la notación con la de las secciones anteriores debemos tomar $R=\ell
_{3}/2$, $s=\tau $ y $x^{\pm }=u_{\pm }-(\ell _{0}+\ell _{3}/2)$ en la ecuación $\left( 2.13\right) $ de \cite{hmm}.}
\small
\begin{eqnarray}
u_{\pm }\left( \tau \right) &-&\left( \ell _{0}+\ell _{3}/2\right) =\\ &=&\frac{\ell
_{3}}{2}\frac{\left( \ell _{3}/2+u_{\pm }\left( 0\right) -\left( \ell
_{0}+\ell _{3}/2\right) \right) -e^{\mp 2\pi \tau }\left( \ell _{3}/2-u_{\pm
}\left( 0\right) +\left( \ell _{0}+\ell _{3}/2\right) \right) }{\left( \ell
_{3}/2+u_{\pm }\left( 0\right) -\left( \ell _{0}+\ell _{3}/2\right) \right)
+e^{\mp 2\pi \tau }\left( \ell _{3}/2-u_{\pm }\left( 0\right) +\left( \ell
_{0}+\ell _{3}/2\right) \right) }\text{.}\nonumber
\end{eqnarray}
\normalsize
\begin{figure}
\begin{center}
\includegraphics[width=13.5cm]{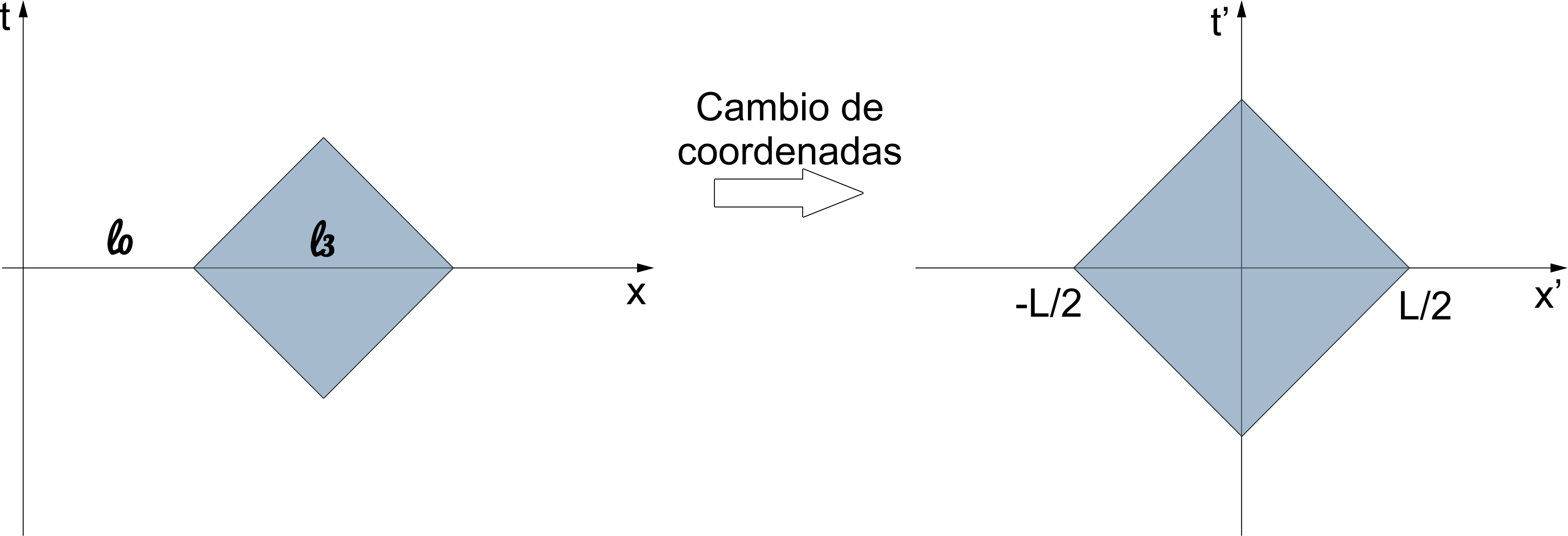}
\end{center}
\caption{\textbf{Transformación de coordenadas}. El intervalo original en el espacio de Rindler $u_{\pm}\geq 0$ se mapea en las nuevas coordenadas $(t',x')$ a un intervalo de longitud $L$ en un espacio con temperatura inversa $\beta=2\pi \ell _{1}$.} \label{tr3}
\end{figure}
Definiendo nuevamente $u_{\pm }^{\prime }$ y $u_{\pm}^{\prime }$, por medio de la ecuación (\ref{nulasnuevas}, encontramos el flujo modular en las coordenadas primadas (ver figura \ref{tr3})
\small
\begin{eqnarray}
u_{\pm }^{\prime }&\left( \tau \right)& =\\
&=&\ell _{1}\log \left\{ \frac{\ell _{0}%
}{\ell _{2}}+\frac{\ell _{3}}{2\ell _{2}}\left[ 1+\frac{\ell _{2}e^{u_{\pm
}^{\prime }\left( 0\right) /\ell _{1}}-\ell _{0}-e^{\mp 2\pi \tau }\left(
\ell _{3}-\ell _{2}e^{u_{\pm }^{\prime }\left( 0\right) /\ell _{1}}+\ell
_{0}\right) }{\ell _{2}e^{u_{\pm }^{\prime }\left( 0\right) /\ell _{1}}-\ell
_{0}+e^{\mp 2\pi \tau }\left( \ell _{3}-\ell _{2}e^{u_{\pm }^{\prime }\left(
0\right) /\ell _{1}}+\ell _{0}\right) }\right] \right\} \text{.}\nonumber
\end{eqnarray}
\normalsize
Las expresiones para $x^{\prime }\left( \tau \right) $ y $%
t^{\prime }\left( \tau \right) $ en términos $x^{\prime }\left( 0\right) $ y $t^{\prime }\left( 0\right) $ se pueden obtener de esta última ecuación, utilizando las relaciones de la ecuación (\ref{equis}). El hamiltoniano modular está nuevamente dado por
\begin{equation}
H=\int dx^{\prime }\left( 0\right) T_{00}\left( x^{\prime }\left( 0\right)
\right) F\left( x^{\prime }\left( 0\right) \right) \text{,}
\end{equation}
donde ahora
\small
\begin{equation}
F\left( x^{\prime }\left( 0\right) \right) =\left[ \frac{dt^{\prime }\left(
\tau \right) }{d\tau }\right] _{\tau =0,\,t^{\prime }\left( 0\right) =0}=-%
\frac{2\pi \ell _{1}e^{-x/\ell _{1}}}{\ell _{2}\ell _{3}}\left( \ell
_{0}-\ell _{2}e^{x/\ell _{1}}\right) \left( \ell _{0}-\ell _{2}e^{x/\ell
_{1}}+\ell _{3}\right) \text{;}
\end{equation}
\normalsize
notemos que hemos escrito $x$ en lugar $x^{\prime }\left( 0\right) $ para simplificar la notación. Para que el intervalo en el plano $x'$, $t'$ quede centrado en el nuevo origen $x'=t'=0$, debemos tener
\begin{equation}
F\left( L/2\right) =F\left( -L/2\right) =0\text{,}
\end{equation}
siendo $L$ la longitud del intervalo. Estas condiciones nos permiten poner $\ell _{0}$ y $\ell _{2}$ en términos de $\ell _{1}$ y $\ell _{3}$
\begin{equation}
\begin{array}{c}
\ell _{0}=\dfrac{\ell _{3}}{e^{L/\ell _{1}}-1}\text{,} \\ 
\ell _{2}=\dfrac{\ell _{3}e^{L/\left( 2\ell _{1}\right) }}{e^{L/\ell _{1}}-1}%
\text{.}%
\end{array}%
\end{equation}
Con esta elección tenemos
\begin{equation}
F\left( x\right) =2\pi \ell _{1}\frac{\cosh \left( L/\left( 2\ell
_{1}\right) \right) -\cosh \left( x/\ell _{1}\right) }{\sinh \left( L/\left(
2\ell _{1}\right) \right) }\text{.}
\end{equation}
En el límite $L/\ell _{1}\longrightarrow \infty $, $F$ se reduce a $F\left( x\right) =2\pi \ell _{1}$, con lo que nuevamente identificamos $\beta=2\pi \ell _{1}$ y entonces el hamiltoniano modular para un estado térmico en un intervalo de longitud $L$ queda escrito como
\begin{equation}
H=\int dx \, T'_{00}\left(x\right) \beta \frac{\cosh \left(\frac{\pi L}{\beta}\right)-\cosh \left(\frac{2\pi x}{\beta}\right)}{\sinh \left(\frac{\pi L}{\beta}\right)}\,.
\end{equation}
En la figura \ref{segm}, graficamos el flujo modular para distintas temperaturas, reobteniendo los casos conocidos para baja y al alta temperatura con relación al tamaño del intervalo.

\chapter{\label{ch:relativa}Entropía relativa y holografía}

En esta sección presentamos una cantidad fundamental en la teoría de la información cuántica: la entropía relativa. El objetivo es utilizar esta cantidad como herramienta para estudiar el entrelazamiento en el contexto de la dualidad de Maldacena, que también presentamos posteriormente en este capítulo. Existe un ansatz que permite calcular la entropía de entrelazamiento a partir de un elegante cálculo geométrico utilizando la dualidad de Maldacena. A la entropía calculada de esa forma se la conoce como entropía holográfica de entrelazamiento, y en el capítulo siguiente mostraremos que satisface una prueba de consistencia no trivial bastante importante.

\section{Entropía relativa}

Comenzamos introduciendo el concepto de entropía relativa y algunas de sus propiedades generales, así como también su relación con la segunda ley de la termodinámica y la cota de Bekenstein.

\subsection{Definición y propiedades generales}

La entropía relativa entre dos estados que actúan sobre el mismo espacio de Hilbert $\cal{H}$ es una cantidad estadística fundamental que mide la distancia entre estos dos estados. Dados dos estados $\rho_1$ y $\rho_0$, la entropía relativa $S(\rho_1|\rho_0)$ se define como
 \begin{equation}
S(\rho_1|\rho_0)=\textrm{tr}(\rho_1 \log \rho_1)-\textrm{tr}(\rho_1\log \rho_0)\,.
\labell{RSdef}
 \end{equation}
En general, se tiene $S(\rho_1|\rho_0)\ge0$. La entropía relativa se anula si y sólo si los dos estados son iguales. Nótese que según la definición, la entropía relativa no es simétrica respecto de sus argumentos (por lo que no es una distancia en el sentido matemático estricto).

Como dijimos, la entropía relativa da una medida de la distancia estadística entre los dos estados. Dado el estado $\rho_1$, la probabilidad de confundirlo con $\rho_0$ después de realizar $n$ mediciones de algún observable del sistema, es exponencialmente decreciente para $n$ grande
\begin{equation}
e^{-n S(\rho_1|\rho_0)}\,.\labell{sam}
\end{equation}
Es en este sentido en que se piensa a la entropía relativa como una cantidad que sirve para distinguir estados \cite{vedral}.

Para estados reducidos a alguna región $V$, la entropía relativa, además de ser positiva, es creciente con el tamaño del sistema
\begin{eqnarray}
&& S(\rho^V_1|\rho^V_0)\ge 0\,,\\
&& S(\rho_1^V|\rho_0^V)\le S(\rho_1^W|\rho_0^W)\,,\hspace{2cm} V\subseteq W\,.\labell{mon}
\end{eqnarray}
Esta propiedad de monotonicidad (\ref{mon}) es en realidad un caso particular de monotonicidad ante las llamadas transformaciones completamente positivas que preservan la traza (CPTP, por las siglas en inglés de Completely Positive Trace Preserving maps). Ejemplos de CPTP son por ejemplo los operadores de evolución unitarios, la traza parcial sobre un subsistema, y la evolución de un subsistema inicialmente desacoplado del resto en el que la dinámica global del sistema completo es unitaria. La expresión general de una transformación CPTP entre dos matrices densidad es \cite{nielsen}
\begin{equation}
\rho^\prime=\sum_i M_i \rho M_i^\dagger\,,\hspace{2cm} \sum_i M_i^\dagger M_i=I\,,
\end{equation}
para operadores $M_i$ de dimensión arbitraria, es decir, no necesariamente operadores cuadrados. La entropía relativa satisface esta propiedad de monotonicidad ante estas transformaciones más generales

\begin{equation}
S(\rho_1|\rho_0)\ge S(\rho_1^\prime|\rho_0^\prime)\,.\label{cptp}
\end{equation}

Estas transformaciones CPTP generalmente implican un aumento en la indistiguibilidad entre los dos estados (ejemplos en este sentido se obtienen mirando a la traza parcial sobre un sistema y a la propiedad (\ref{mon}).

\subsubsection{Subaditividad fuerte}\label{subcapsub}

La información mutua entre dos subsistemas $A$ y $B$: $I(A,B)=S(A)+S(B)-S(AB)$, introducida en \reef{infomutua}, es una medida de la información compartida por los dos subsistemas. Esta cantidad puede escribirse en términos de la entropía relativa como
\be
I(A,B)=S(\rho_{AB}|\rho_A\otimes \rho_B)\,
\ee
Las propiedades de positividad y monotonicidad con el tamaño de la región de la información mutua, surgen entonces como consecuencia de las propiedades de positividad y monotonicidad de la entropía relativa. En particular, la monotonicidad de la información mutua da
\be
I(A,BC)-I(A,B)=S(AB)+S(BC)-S(ABC)-S(B)\ge 0\,.
\ee
Por lo tanto, la propiedad de subaditividad fuerte de la entropía \ref{subfuerte} (la última desigualdad en la ecuación), es consecuencia de la monotonicidad de la entropía relativa. Destacamos que la propiedad de subaditividad fuerte en conjunto con otras propiedades de la entropía permiten demostrar en sentido contrario la validez de la monotonicidad de la entropía relativa \cite{wehrl}.

\subsection{Segunda ley de la termodinámica}\label{secsegunda}

Las primeras interpretaciones físicas de la positividad de la entropía relativa $S(\rho_1|\rho_0)$ están relacionadas con la termodinámica. Si $\rho_0$ representa un estado de equilibro térmico a temperatura $T$, entonces, la entropía relativa toma la forma $S(\rho_1|\rho_0)=(F(\rho_1)-F(\rho_0)/T)$, donde $F(\rho)=\textrm{tr} (\rho E )- T S(\rho)$ es la energía libre de Helmholtz evaluada para un estado general $\rho$ pero a temperatura fija $T$. Por lo tanto, la positividad de la entropía relativa nos dice que la energía libre de Helmholtz es mínima para el estado de equilibrio térmico.

Recordemos que la versión termodinámica de esta desigualdad es una consecuencia de la segunda ley de la termodinámica. En general, para un sistema en contacto con un baño térmico a temperatura $T$, la segunda ley implica que en cualquier proceso se tiene
\begin{equation}
\delta F\le \delta W\,,
\end{equation}
donde $\delta W$ es el trabajo hecho sobre el sistema. Por lo tanto, para un proceso espontáneano, en cual no se realiza trabajo, se debe tener $\delta F\le 0$. Es decir, la energía libre decrece a medida que el sistema evoluciona hacia el equilibrio térmico.

La segunda ley puede deducirse de las propiedades de la entropía relativa bajo ciertas hipótesis para la evolución temporal cuántica \cite{vedral,paper}. La entropía relativa también ha encontrado aplicaciones en la demostración de la segunda ley generalizada de la termodinámica, en el contexto de la evaporación de un agujero negro \cite{sorkin1,sorkin2,wall2,wall3}.

En estas demostraciones, la segunda ley se deduce a partir de la propiedad de monotonicidad ante transformaciones CPTP en la ecuación (\ref{cptp}).

La segunda ley de la termodinámica establece que la entropía de un sistema aislado no puede disminuir. En mecánica cuántica, la entropía de un sistema aislado que evoluciona en forma unitaria no cambia; por ello, debemos relajar la hipótesis de que el sistema esté completamente aislado y permitir algún intercambio de información con el ambiente. Como modelo para esta evolución consideremos el caso de un sistema cuyo estado $\rho(t)$ evoluciona ante transformaciones CPTP. Asumimos además, en consonancia con la idea de sistema ``aislado'', que la energía total $E$ se conserva. Además, asumimos que la evolución temporal deja invariante al estado de equilibrio térmico\footnote{En realidad, sólo es necesario pedir la existencia de un estado cuya entropía y energía sean preservadas.} $\rho_T=e^{-H/T}/\textrm{tr}(e^{-H/T})$ a cierta temperatura $T$, que corresponde a la energía conservada $E$, $\textrm{tr}(\rho_T H)=E$.

La entropía relativa $S(\rho(t),\rho_T)$ disminuye por la evolución a través de la transformación CPTP, y entonces para $t_1<t_2$ se tiene
\begin{equation}
F(\rho(t_2))-F(\rho_T)<F(\rho(t_1))-F(\rho_T)\,,
\end{equation}
donde utilizamos el hecho de que el estado térmico es invariante ante la evolución temporal. Expresando esta relación en términos de entropías y energías, y considerando que la energía se conserva, se tiene
\begin{equation}
S(t_2)>S(t_1)\,,
\end{equation}
que es la segunda ley de la termodinámica. Notar que la diferencia entre las energías libres entre el estado y el estado de equilibrio térmico es positiva y decrece en el tiempo. Consecuentemente, el estado se acerca al estado de equilibrio térmico durante la evolución. Eventualmente, si se alcanza el equilibro térmico, la diferencia en la energía libre se anula.

Otro caso en el cual es posible utilizar la entropía relativa para demostrar la segunda ley es el siguiente. Suponiendo que una transformación CPTP mantiene constante al estado de máxima entropía $\rho_0=I/n$, donde $n$ es la dimensión del espacio de Hilbert (este estado se entiende como la distribución microcanónica donde todos los estados son equiprobables), se puede demostrar que la entropía no decrece. Esto surge de que la entropía relativa en este caso se escribe como
\begin{equation}
S(\rho(t)|\rho_0)=\log(n)-S(\rho(t))\,.
\end{equation}
El aumento de la entropía surge entonces del decrecimiento de la entropía relativa ante transformaciones CPTP.

\subsection{Cota de Bekenstein}\label{secbeke}

La cota de Bekenstein \cite{beke0,beke00} es una propuesta que dice que todos los sitemas de la naturaleza deben verificar una desigualdad de la forma
\begin{equation}
S\le 2\pi R\, E \,,\labell{bek}
\end{equation}
donde $S$ y $E$ son la entropía y energía de un sistema confinado a una región de tamaño $R$. La propuesta de esta desigualdad surgió a partir de experimentos pensados que involucran agujeros negros. El hecho de que en \reef{bek} no aparezca la constante de Newton indica que la desigualdad debería expresar una propiedad que es válida aún cuando no hay gravedad. En particular, debería ser posible entender la ecuación \reef{bek} en espacio puramente plano. Si bien la forma de esta desigualdad parece simple, diversas discusiones sobre su validez \cite{marolf,marolf2,boussoa,bent,donrad} revelaron muchas sutilezas al interpretar las distintas cantidades que aparecen en la ecuación \reef{bek}. Eventualmente, se ha compredido que una versión bien definida de esta desigualdad en teoría cuántica de campos surge a partir de la positividad de la entropía relativa entre dos estados reducidos a una cierta región \cite{beke1,marolf,marolf2}.

El argumento original de Bekenstein para deducir la desigualdad (\ref{bek}) es el siguiente. Imaginemos que desde una distancia $R$ del horizonte de un agujero negro se deja caer libremente un pequeño objeto de prueba de tamaño $\lambda \lesssim R$. El objeto atraviesa el horizonte, llevando una entropía $S$ y cierta energía $E$, de acuerdo a lo que mide un observador ubicado en el punto desde el cual el objeto fue liberado. La energía absorbida por el agujero negro medida asintóticamente sufre un corrimiento al rojo $E T_{BH}/T_{rel}\simeq 2\pi R E T_{BH}$, donde $T_{BH}$ y $T_{rel}$ son las temperaturas de Hawking medidas en el infinito y la temperatura local medida en el punto en que se liberó al objeto, respectivamente. La variación en la masa del agujero negro está dada entonces por $\delta M=2\pi R E T_{BH}$ y la correspondiente variación de su entropía por $\delta S=\delta M/T_{BH}=2\pi R E$. La segunda ley de la termodinámica generalizada indica que el incremento en la entropía del agujero negro debe al menos compensar la pérdida de entropía en el exterior, es decir: $\delta S_{BH}\geq S$, de donde se deduce la desigualdad (\ref{bek}).

Un problema de la ecuación (\ref{bek}), es que la entropía y la energía de una región finita en teoría de campos no son cantidades bien definidas. Una propuesta para eliminar las ambigüedades en la definición de la entropía y la energía es considerar que las cantidades relevantes que intervienen en (\ref{bek}) son la diferencia entre la entropía del estado en la región relevante $V$ y la entropía del vacío en la misma región, y la diferencia entre los valores de espectación del hamiltoniano modular en cada estado. En la siguiente sección, mostramos una forma precisa de reinterpretar la ecuación (\ref{bek}) y relacionamos esta desigualdad con la positividad de la entropía relativa.

\subsubsection{Entropía relativa y la cota de Bekenstein}\label{secercb}

Recapitulando lo comentado en la sección \ref{secsegunda}, si un estado $\rho_0$ es térmico con respecto a un hamiltoniano $H$ (es decir, si $\rho_0=\frac{e{-H/T}}{\tr\left(e^{-H/T}\right)}$) entonces la entropía relativa con otro estado $\rho_1$ se expresa como
\begin{equation}
S\left(\rho_1 | \rho_0\right)=\frac{1}{T}\left(F\left(\rho_1\right)-F\left(\rho_0\right)\right)\,,
\label{helmrel}
\end{equation}
donde $F\left(\rho\right)$ es la energía libre dada por
\begin{equation}
F\left(\rho\right)=\tr\left(\rho H\right)-T S\left(\rho\right).\label{elibre}
\end{equation}
Es importante remarcar que $\rho_1$ no debe ser necesariamente un estado térmico; la temperatura que aparece en la definición de la energía libre $F\left(\rho_1\right)$ es la del estado $\rho_0$. Ahora, dada la expresión en la ecuación (\ref{helmrel}), la positividad de la entropía relativa es equivalente al hecho de que la energía libre a una temperatura $T$ fija es mínima para el estado de equilibrio térmico, como se comentó en la sección \ref{secsegunda}.

Para establecer la conexión de la entropía relativa con la cota de Bekenstein, consideremos ahora operadores describiendo estados de una teoría cuántica de campos en una región $V$.

Estos estados pueden escribirse en términos del hamiltoniano modular $H$ (presentado en la sección \ref{ch:modular}); en particular, para un estado $\rho_0$
\begin{equation}
\rho_0=\frac{e^{-H}}{\tr\left(e^{-H}\right)}\,.
\label{rhoz}
\end{equation}
La entropía relativa, a partir de las ecuaciones (\ref{helmrel}), (\ref{elibre}) y (\ref{rhoz}), se escribe entonces como
\begin{equation}
S(\rho_1 \vert \rho_0)= \Delta\langle H\rangle -\Delta S \labell{S1}
\end{equation}
donde
\begin{equation}
\Delta\langle H\rangle=\textrm{tr} (\rho_1\, H )- \textrm{tr} (\rho_0\, H )
\qquad{\rm and}\qquad
\Delta S=S(\rho_1)- S(\rho_0)\,.
 \labell{S2}
\end{equation}
Luego, la positividad de la entropía relativa se traduce en la siguiente desigualdad
\begin{equation}
\Delta\langle H\rangle\ge\Delta S\,.\labell{123}
\end{equation}
De este modo, al comparar dos estados, la variación de la entropía de entrelazamiento está acotada por la variación del valor de expectación del hamiltoniano modular. El capítulo siguiente de esta tesis se concentra en examinar la desigualdad (\ref{123}) en el contexto holográfico.

Volvamos a intentar establecer la conexión entre la cota de Bekenstein y la positividad de la entropía relativa. En el experimento original de Bekeinstein, $V$ sería la región exterior al agujero negro, cercana a su horizonte. Tanto para el objeto localizado fuera del agujero negro como para el estado de vacío en esta misma región $V$ la entropía de entrelazamiento es muy grande. Sólo la diferencia entre estas entropías (correspondientes a los estados inicial y final del proceso de absorción del objeto por el agujero negro) intervienen en la desigualdad (\ref{123}), de modo que la cantidad $S$ en la ecuación \ref{bek} debe interpretarse en realidad como $\Delta S$.

La cantidad $2\pi R\,E$ que aparece en \ref{bek} también sufre de ambigüedades similares en teoría de campos. Sin embargo, también se le puede dar un sentido preciso al interpretarla como $\Delta \langle H\rangle$, es decir, como la diferencia en los valores de expectación del hamiltoniano modular (para $\rho_V^0$) entre los dos estados \cite{beke1}.

Para hacer más precisa esta interpretación, empecemos notando que en el experimento pensado de Bekenstein la física relevante para la región cercana al horizonte de un agujero negro grande es prácticamente la misma que la del espacio de Rindler. Recordamos que el hamiltoniano modular en espacio de Rindler, como discutimos en el capítulo anterior, está dado por
\begin{equation}
H=2 \pi K=2 \pi \int_{x>0} d^{d-1}x\, x \,T_{00}(\vec{x})\,.
\labell{siete}
\end{equation}
Por lo tanto, evaluando $\Delta \langle H\rangle$ entre el estado inicial (con el objeto de prueba cerca del horizonte) y el estado final (el vacío), encontramos
 \be
\Delta \langle H\rangle =2\pi \int_{x>0} d^{d-1}x \ x \ \langle
T_{00}(x)\rangle_{\rho_V} \simeq 2 \pi R\,E  \,.
 \ee
De este modo, $\Delta \langle H\rangle$ reproduce la expresión que aparece a la derecha de la ecuación (\ref{bek}). Es claro también que $\Delta \langle H\rangle$ también permite dar una definición no ambigüa para el producto de la energía y el tamaño cuando se intenta aplicar la cota de Bekenstein a sistemas y regiones más generales. También resaltamos que esta versión cuántica de la cota soluciona el problema de que la energía puede ser localmente negativa, mientras que la entropía es siempre positiva; con esta nueva formulación, la diferencia de entropías puede ser negativa.

En resumen, el experimento pensado de Bekenstein y la ecuación \ref{bek} que se deduce del mismo no expresan más que la positividad de la entropía relativa, dada por la desigualdad (\ref{123}) y el hecho de que para el estado reducido para cualquier teoría de campos en el espacio de Rindler depende del operador de boosts como se expresa en la ecuación \reef{siete}. Esto implica que la física que hay detrás de la cota de Bekenstein es simplemente mecánica cuántica y relatividad especial. También es importante notar que la ecuación (\ref{123}) generaliza la cota de Bekenstein a regiones arbitrarias, mientras que la versión original de Bekenstein está limitada al espacio de Rindler.

\subsubsection{Entropía relativa y el problema de las especies}

Otro de los inconvenientes que se presenta en la cota original de Bekenstein (\ref{bek}) es el denominado problema de las especies \cite{marolf,marolf2}: si el número de especies de partículas independientes es grande, entonces la entropía también puede hacerse arbitrariamente grande manteniendo la energía fija. Interesantemente, la formulación cuántica de la cota de Bekenstein (ecuación \reef{123}) soluciona este problema. Es decir, al considerar teorías con un número grande de especies de campos cuánticos la desigualdad \reef{123} sigue siendo válida. Esto es debido a que, si bien al aumentar el número de grados de libertad la entropía de una excitación localizada puede hacerse grande manteniendo fija la energía, también la entropía de las fluctuaciones de vacío se hace grande, y la diferencia $\Delta S$ satura para un número arbitrariamente grande de especies. Al ir agregando más especies de partículas, el estado del objeto localizado y el vacío localizado resultan cada vez menos distinguibles, por lo que la entropía relativa entre ambos disminuye progresivamente (manteniéndose siempre positiva).

Para ver cómo se resuelve el problema de las especies con más detalle, comenzamos describiendo cómo fue originalmente propuesto. Consideremos una teoría de $\N$ copias desacopladas de alguna teoría cuántica de campos. Inicialmente, trabajaremos con estados globales en lugar de estados localizados en alguna región acotada. Sea $\hat{\rho}_0=|0\rangle \langle 0|$ el estado de vacío para una cierta especie, y $\hat{\rho}_1=|\psi \rangle \langle
\psi|$ cualquier otro estado ortogonal puro (por ejemplo, un estado de una partícula). El estado de vacío global
es $|\Omega\rangle=|0\rangle\otimes \cdots\otimes
|0\rangle$ y su correspondiente operador densidad
\be
\rho_0=|\Omega\rangle\langle \Omega| =\hr_{0} \otimes
\cdots\otimes \hr_{0}\,.
\labell{density}
\ee
Ahora, reemplacemos al vacío por el estado excitado $|\psi \rangle$ en la copia $i$-ésima de la teoría de campos, es decir $|\Psi_i\rangle=|0\rangle\otimes
\cdots\otimes|\psi\rangle \otimes \cdots\otimes|0\rangle$. En este caso, el operador densidad correspondiente es
 \be
\rho_{i}=|\Psi_i\rangle\langle \Psi_i| =\hr_{0} \otimes
\cdots\otimes \hr_{1} \otimes \cdots\otimes\hr_{0}
 \labell{den3s}
 \ee
Los estados $\rho_i$ son puros y ortogonales, $\langle \Psi_i|\Psi_j \rangle=0$ si $i\neq j$. Por lo tanto, la siguiente matriz densidad obtenida al combinar estas excitaciones de partículas para las diferentes especies
 \begin{equation}
\rho_{\textrm{mix}}=\frac{1}{{\cal N}}
\sum \rho_i=\frac{1}{{\cal N}} \sum |\Psi_i\rangle \langle\Psi_i|
\labell{upon}
 \end{equation}
ya se encuentra diagonalizada en la base de los $|\Psi_i\rangle$. Tiene ${\cal N}$ autovalores no nulos cuyo valor es $1/{\cal N}$.  Luego, $S(\rho_{\textrm{mix}})=\log({\cal N})$ y
 \begin{equation}
 \Delta S_{\textrm{tot}}=S(\rho_{\textrm{mix}})-S(\rho_0)=\log({\cal N})\,.
 \labell{oncex}
 \end{equation}
 %
%
%
Aquí $\Delta S$ aumenta indefinidamente cuando crece ${\cal N}$, mientras que la energía en $\rho_{\textrm{mix}}$ es independiente de ${\cal N}$. Recordemos de todos modos que estamos considerando estados globales, con lo que uno diría que $R \rightarrow \infty$ y entonces no habría violación de la ecuación $\reef{bek}$. Los estados puros $\hat{\rho}_0$ y $\hat{\rho_1}$ son ortogonales y perfectamente distinguibles entre sí, por lo que la entropía relativa entre ellos es infinita. Esto mismo vale para los estados dados por las ecuaciones \reef{den3s} y \reef{upon}, es decir, $S(\rho_{\textrm{mix}}|\rho_0)=\infty$. La diferencia de entropías $\Delta S_{\textrm{tot}}$ puede en este caso crecer sin límites debido a que en esta situación $\Delta \langle H\rangle$ es divergente. Para ver esto, pensemos en nuestro estado $\rho_0$ como el límite de un estado térmico $\rho_T\sim e^{-{\cal H}/T}$ ($H={\cal H}/T$ y $\cal H$ representa al hamiltoniano dinámico) cuando $T \rightarrow 0$: como $|\psi \rangle$ tiene energía fija finita, debe resultar $\Delta H\to\infty$ cuando $T\to0$.

Como mencionamos, estos estados son globales y para formular la cota de Bekenstein necesitamos tratar con estados localizados en una región de tamaño finito $R$. Consideremos, por simplicidad, el caso de estados reducidos al interior de una esfera $V$ de radio $R$. El estado reducido del vacío es
 \be
\rho_0=\Tr_{\bar V}[|\Omega\rangle\langle \Omega|] =\hr_{0} \otimes
\cdots\otimes \hr_{0}
 \labell{density1}
 \ee
donde ahora $\hr_{0}= \tr_{\bar V}[|0\rangle\langle 0|]$ es el operador densidad del vacío en cada copia individual de la teoría de campos. Nótese que introdujimos $\Tr$ para designar la traza en el espacio de Hilbert total, es decir, sobre todas las copias de la teoría de campos, mientras que $\tr$ denota la traza en una copia de la teoría de campos. Construímos ahora los operadores densidad análogos para los estados excitados \reef{upon}
 \be
\rho_{i}=\Tr_{\bar V}[|\Psi_i\rangle\langle \Psi_i|] =\hr_{0} \otimes
\cdots\otimes \hr_{1} \otimes \cdots\otimes\hr_{0}
 \labell{den3s1}
 \ee
donde $\hr_{1}= \tr_{\bar V}[|\psi\rangle\langle \psi|]$. El estado mezcla que combina las excitaciones de partículas para distintas especies es
 \begin{equation}
 \rho_{\textrm{mix}}=\frac{1}{{\cal N}} \sum \rho_i \labell{mix}\,.
 \end{equation}
Como las diferentes copias se encuentran todas desacopladas entre sí, el hamiltoniano modular toma la forma $H_{\textrm{tot}}=\sum H_i$, donde
 \begin{equation}
H_i = \mathbf{1}_1 \otimes \mathbf{1}_2 \otimes \cdots \otimes H \otimes
\cdots\otimes \mathbf{1}_\N\,.
 \labell{Hi}
 \end{equation}
En esta expresión, $H$ es precisamente el hamiltoniano modular para una copia de la teoría de campos.

Consideremos una situación análoga a la anterior, donde teníamos una excitación pura que es lo más distinguible posible del vacío, de modo que para estados globales la distinguibilidad con el vacío es infinita. Dentro de la esfera, esta distinguibilidad debe estar acotada. Para que el estado excitado sea lo más `diferente' posible del vacío dentro de la esfera, deberíamos construir un paquete de ondas con una longitud de onda $\lambda$ muy pequeña lejos del borde de la esfera (donde la temperatura efectiva es pequeña). Si nos limitamos por ejemplos a las teorías conformes, el hamiltoniano modular está dado por la ecuación \reef{sphereH} y podemos hacer la construcción en forma más explícita. Si situamos al paquete de ondas en el centro de la esfera, se tiene
\begin{equation}
\Delta H= \pi \frac{R}{\lambda}\gg 1\,.
\end{equation}
Este resultado puede ser ciertamente grande y cuando $\frac{R}{\lambda}\gg 1$ nos acercamos a la situación en la que el estado excitado es maximalmente distinguible de $\rho_0$. Notemos sin embargo que si bien puede ser grande, $\Delta H$ no puede ser divergente en la región acotada. En este régimen, el cálculo de entropía es similar al descripto para los estados globales y se tiene
\begin{equation}
\Delta \langle H\rangle-\Delta S=\pi \frac{R}{\lambda}-\log ({\cal N})\,.\label{asa}
\end{equation}
Cuando ${\cal N}$ aumenta, la entropía relativa decrece (la cota se hace más restrictiva) en acuerdo con la propiedad de mezclado de la entropía relativa \cite{wehrl}
\begin{equation}
S(\sum p_i \rho^{(1)}_i|\sum p_i \rho^{(2)}_i)\le \sum p_i S( \rho^{(1)}_i|
 \rho^{(2)}_i)\,,
\end{equation}
para $p_i>0$ y $\sum p_i=1$. Sin embargo, dado que $\Delta \langle H\rangle $ es independiente de ${\cal N}$ y la entropía relativa es siempre positiva, $\Delta S$ no puede tener un comportamiento tipo $\log({\cal N})$ para un número muy grande (${\cal N}\gtrsim e^{R/\lambda}$) de especies. La ecuación \reef{123} debe ser finalmente saturada, $\Delta S=\Delta\langle H\rangle$. Claramente debe haber un cambio en el comportamiento de $\Delta S$ respecto del crecimiento logarítmico hallado en la ecuación \reef{oncex} cuando ${\cal N}\gtrsim e^{R/\lambda}$. Intuitivamente, la probabilidad de encontrar un paquete de ondas excitado de la copia $i$-ésima de la teoría de campos conforme en el operador densidad del vacío (que tiene una temperatura efectiva del orden $1/R$ en el lugar donde se encuentra el paquete de ondas) es $e^{-R/\lambda}/Z$, independientemente de ${\cal N}$. Para el estado excitado en $\rho_{\textrm{mix}}$, esta probabilidad es $\frac1{\cal N}+\frac{e^{-R/\lambda}}{Z}$. Por lo tanto, cuando ${\cal N}\gtrsim e^{R/\lambda}$, el vacío y el estado mezcla son mucho menos distinguibles y nos encontramos en una situación donde $\Delta S\simeq \Delta \langle H\rangle$.

Es evidente la importancia que tiene expresar el producto original $2 \pi R\, E$ en el lado derecho de la ecuación \reef{bek} como el cambio de `energía modular' $\Delta\langle H\rangle$, y de considerar la diferencia de entropías $\Delta S$ en lugar de sólo la entropía $S$. Esto último asegura la saturación de la cota en el caso de un número grande de especies. Cuando el número de especies es suficientemente grande, la excitación de partícula cuya probabilidad está distribuida entre las varias copias en el estado excitado queda inmersa en la nube de excitaciones que se producen en el proceso de localización del vacío a una región finita. De este modo, $\rho_{\textrm{mix}}$ y $\rho_0$ dejan de ser fácilmente distinguibles.

En general, la transición de la forma \reef{asa} a cero para ${\cal N}$ será una función complicada. En el apéndice A.4 de \cite{blanco2} presentamos un estudio de las primeras correcciones no triviales para el caso de pequeñas desviaciones del vacío. Encontramos que a primer orden el nuevo estado mixto satura la desigualdad \reef{123}, de modo que no hay distinguibilidad con respecto al vacío; para ello, es necesario ir hasta segundo orden en la perturbación respecto al vacío. Se encuentra que, a segundo orden, la diferencia $\Delta \langle H \rangle - \Delta S$ es distinta de cero (de modo que hay distinguibilidad entre los estados) y explícitamente positiva (como es requerido por la positividad de la entropía relativa).

Como comentario final de esta sección, notemos que el experimento pensado de Bekenstein involuca un proceso dinámico con intercambio de entropía y energía entre dos sistemas. Interpretando la cota de Bekenstein en términos de la entropía relativa, el mismo razonamiento puede ser aplicado en espacio sin gravedad y para cualquier región, en particular, sin involucrar agujeros negros. El experimento en espacio plano involucraría una excitación con una diferencia de energía modular $\Delta\langle H \rangle$ con respecto al vacío en la región $V$. Bajo cierta evolución, se asume que esta energía modular (la energía de Rindler en el experimento pensado de Bekenstein) es conservada y al mismo tiempo pasa a un reservorio térmico (representado por el agujero negro en el experimento de Bekenstein) -en lenguaje termodinámico, se convierte en `calor'-. Esto da $\Delta
S_{\textrm{res}}=\Delta\langle H \rangle$ para el reservorio, dado que en el caso de un número grande de grados de libertad nos encontramos en la situación de pequeñas desviaciones (notar que la temperatura es $T=1$). El aumento de entropía bajo esta evolución requiere $\Delta S_{\textrm{res}}- \Delta S=\Delta\langle
H \rangle-\Delta S\ge 0$. De hecho, como mencionamos en la sección \ref{secsegunda}, la positividad de la entropía relativa puede siempre ser interpretada como una consecuencia de la segunda ley para evoluciones temporales específicas que son CPTP pero no unitarias en la región. Un ejemplo simple para el caso presente está dado por una evolución que agrega especies de campos idénticas e independientes y mezcla el estado en espacios de Hilbert de la forma descripta anteriormente en esta sección. Este proceso puede representar, para nuestros propósitos, la evolución del sistema inicial que finalmente es absorbido por el reservorio. Implícitamente, la discusión anterior muestra que esta `evolución' preserva el valor de $\Delta\langle H \rangle$. Además, en el límite de un número grande de especies, deberíamos tener $\Delta
S_{\textrm{res}}=\Delta\langle H \rangle$. $\Delta S_{\textrm{res}}$ es aquí la variación de entropía del baño térmico debida a la absorción del objeto de prueba. La cota dada por la entropía relativa puede considerarse entonces como una consecuencia de una segunda ley bajo la evolución ante transformaciones CPTP, en analogía con la derivación de la cota de Bekenstein utilizando la segunda ley de la termodinámica generalizada.

\section{Entropía holográfica de entrelazamiento}

Muchas veces, los nuevos hallazgos de la física téorica aparecen al entender que dos conceptos diferentes se encuentran relacionados entre sí a un nivel más fundamental. Ejemplos de estas relaciones son las dualidades que vinculan dos aparentemente diferentes teorías cuánticas; en tales casos, los espacios de Hilbert y la dinámica de las dos teorías coinciden, a pesar de que los lagrangianos sean distintos. La dualidad Anti-de Sitter/Conformal Field Theory (AdS/CFT), es otro tipo de dualidad que relaciona una teoría de campos conforme en espacio plano con una teoría de cuerdas. Intuitivamente, la teoría de campos cuántica en espacio plano no aparenta ser una teoría cuántica de la gravedad. Sin embargo, la correspondencia AdS/CFT establece que las dos teorías son equivalentes.

En un contexto más general, la dualidad AdS/CFT es una realización precisa del principio holográfico. Este principio establece que en una teoría gravitatoria, el número de grados de libertad en cierto volumen $V$ escalea como el área de la superficie $\partial V$ del volumen dado. La teoría cuántica de la gravedad involucrada en la correspondencia está definida en general sobre una variedad de la forma AdS$\times X$, donde $X$ es un espacio compacto, y la teoría de campos conforme queda definida sobre el borde del espacio AdS.

En este capítulo introducimos a grandes rasgos las ideas generales que hay detrás de esta dualidad y, en particular, estudiamos una conjetura que permite calcular la entropía de entrelazamiento en la teoría de campos a partir de un cálculo geométrico en la teoría gravitatoria.

\subsection{Correspondencia AdS/CFT}

Como mencionamos en la introducción al capítulo, la correspondencia AdS/CFT relaciona teorías gravitatorias en espacios asintóticamente Anti-de Sitter con teorías de campos conformes. Una realización específica de esta dualidad se da para las teorías de Super Yang-Mills (SYM) con ${\cal N}=4$ en $3+1$ dimensiones y la teoría de supercuerdas tipo $IIB$ en AdS$_5 \times S^5$. La correspondencia AdS$_5$/CFT$_4$ (en su versión más fuerte) establece que la teoría de SYM ${\cal N}=4$, con grupo de simetría $SU\left(N\right)$ y constante de acoplamiento $g_{YM}$ es dinámicamente equivalente a la teoría de supercuerdas tipo $IIB$, donde la longitud de las cuerdas es $l_s=\sqrt{\alpha'}$ y la constante de acoplamiento es $g_s$ en AdS$_5 \times S^5$ (el radio de curvatura del espacio AdS y de la esfera $L$ son iguales). Los parámetros libres de la teoría de campos ($g_{YM}$ y $N$) se relacionan con los parámetros libres de la teoría de supercuerdas ($g_s$ y $L/\sqrt{\alpha'}$ por medio de las ecuaciones
\begin{equation}
g_{YM}^2=2\pi g_s\,,\,\,\,\,2g_{YM}^2 N=\frac{L^4}{\alpha'^2\,}\,.
\end{equation}
La realización de esta dualidad es bastante peculiar, dado que permite relacionar una (posible) teoría cuántica de la gravedad (la teoría de supercuerdas $IIB$) con una teoría de campos que no contiene grados de libertad gravitacionales.

A pesar de que la forma más fuerte de la correspondencia AdS$_5$/CFT$_4$ presentada es muy interesante, es bastante complicado realizar cálculos específicos para valores genéricos de los parámetros libres. Por este motivo, es natural preguntarse si en ciertos límites existe una versión más débil de la correspondencia, con la cual hacer cálculos explícitos sea más factible. Dado que la teoría de cuerdas se entiende mejor en el régimen perturbativo, una propuesta es especificar la dualidad para el caso en $g_s \ll 1$, manteniendo $L/\sqrt{\alpha'}$ fijo. A primer orden no trivial en $g_s$, la parte AdS de la correspondencia queda reducida a una teoría clásica de cuerdas (en el sentido de que sólo intervienen diagramas tipo árbol). La longitud de la cuerda $l_s$ medida en unidades de $L$ se mantiene constante. En este límite (versión fuerte de la correspondencia), la teoría de campos involucrada en la correspondencia tiene $g_{YM}\ll 1$ y $\lambda\doteq g_{YM}^2 N$ (constante de 't Hooft) finito. En otras palabras, en la teoría de campos tenemos que tomar el límite de $N\longrightarrow \infty$ manteniendo $\lambda$ fijo, lo que se conoce como límite de 't Hooft; esto corresponde al límite planar de la teoría de gauge.  En este régimen, la correspondencia AdS/CFT es una realización concreta de la idea de 't Hooft de que el límite planar de una teoría de campos cuántica es una teoría de cuerdas.

En el límite de 't Hooft, sólo hay un parámetro libre en cada lado de la correspondencia: el parámetro de 't Hooft $\lambda$ en la teoría de campos y $L/\sqrt{\alpha'}$ en la teoría de cuerdas. Estos dos parámetros se relacionan por $L^4/\alpha'^2=2\lambda$. Si estamos interesados en estudiar teorías de campos fuertemente interactuantes (para estudiar, por ejemplo, fenómenos no perturbativos), podemos tomar el límite $\lambda \longrightarrow \infty$ en la teoría de campos, lo que da $\sqrt{\alpha'}/L \longrightarrow 0$ en la teoría de cuerdas. En este caso, la longitud de la cuerda es muy pequeña comparada con el radio de curvatura, lo que indica que este es el límite de partícula puntual de la teoría $IIB$ (que está dado por la teoría de supergravedad $IIB$ en AdS$_5\times$S$^5$). Este límite de la correspondencia se conoce como forma débil de la dualidad AdS/CFT.

\subsection{Entropía holográfica de entrelazamiento}

En este contexto holográfico, se ha propuesto un ansatz para calcular la entropía de entrelazamiento usando la dualidad. La entropía de entrelazamiento en la teoría de campos del borde se determina mediante un elegante cálculo geométrico en el dual gravitatorio \cite{rt1,rt2,nishioka,taka}. En particular, la entropía de entrelazamiento entre una región espacial $V$ y su complemento $\bar V$ en el borde se obtiene a partir de
 \be
S(V) = \frac{2\pi}{\lp^{d-1}}\ \mathrel{\mathop {\rm
ext}_{\scriptscriptstyle{\mv\sim V}} {}\!\!} \left[A(\mv)\right]
 \labell{definep}
 \ee
donde hay que extremar el valor del área sobre todas las superficies $\mv$ en el bulk homólogas a $V$. Hemos adoptado la convención $\lp^{d-1}=8\pi G_\mt{N}$, donde $d$ es la dimensión espacio temporal del borde. Para casos estáticos, la ecuación \reef{definep} se conoce como conjetura de Ryu-Takayanagi. En geometrías dependientes del tiempo, la ecuación \reef{definep} se suele conocer como fórmula HRT (Hubeny, Rangamani, Takayanagi). En geometrías estáticas hay una coordenada tiempo $t$ natural y por simetría la superficie extremal siempre estará contenida en un corte del espacio tiempo a tiempo fijo. Si el tiempo está fijo, la superficie extremal es entonces la que tiene área mínimal, de modo que para el caso estático, la fórmula \reef{definep} se reduce a encontrar la superficie que minimiza el funcional área. En cambio, para deformaciones en la dirección de $t$ las superficies extremales son las que tienen área máxima (dado que el área puede hacerse tan pequeña como se desee haciendo que la superficie se aproxime a una superficie nula).

La prescripción \reef{definep} ha pasado muchas pruebas de consistencia \cite{rt1,rt2,nishioka,taka,head,hms}; algunas de ellas se describen en la sección siguiente. Esta fórmula también ha sido derivada para el caso especial de regiones de entrelazamiento esféricas \cite{hmm} y, recientemente en \cite{lewko} la derivación ha sido extendida para el caso de superficies de entrelazamiento suaves generales.

La entropía de entrelazamiento es una medida de cómo se encuentra oganizada espacialmente la información cuántica en un estado cuántico. Para una teoría de campos general (no conforme), esperamos que la fórmula para calcular la entropía de entrelazamiento utilizando holografía sea extremadamente complicada. El hecho de que se simplifique a un sencillo cálculo geométrico en CFT holográficas revela una profunda característica de los sistemas fuertemente interactuantes; en el régimen de acoplamiento fuerte, la organización de la información cuántica tiene una estructura sencilla y universal. Cómo es que esto sucede y qué relación tiene con la geometría emergente es un importante problema abierto y de gran interés actual.

La teoría de la información cuántica provee también otras herramientas que permitirían refinar nuestro entendimiento sobre el entrelazamiento en teorías holográficas. Las entropías de Rényi, por ejemplo, son una familia infinita de medidas de entrelazamiento \cite{renyi0,renyi1,karol1} que proveen una descripción completa del espectro del operador densidad \cite{cala}. Desafortunadamente, el progreso en el entendimiento de las entropías de Rényi en teorías holográficas es aún bastante limitado \cite{head,twod1,twod2,yale} Una buena comprensión de las mismas se ha logrado para el caso de teorías conformes bidimensionales y también para el caso especial de regiones de entrelazamiento esféricas en cualquier dimensión \cite{yale}. Sin embargo, aún no se tiene una propuesta eficiente para calcular la entropías de Rényi en holografías para situaciones más generales. En el capítulo \ref{ch:hee} estudiamos otra medida de información, la entropía relativa (presentada en el capítulo anterior), en el contexto holográfico.

Volvamos a la fórmula \reef{definep}. Es importante destacar que la misma da una contribución proporcional al área para la entropía, como se espera en general para el orden no trivial más importante. Además, es sencillo ver que si el estado es puro (esto se traduce en la ausencia de horizontes en el bulk), la entropía holográfica de una región y su complemento resultan idénticas, como se espera que suceda para la entropía de entrelazamiento.

A continuación, para entender un poco mejor cómo se utiliza la fórmula de Ryu-Takayanagi, presentamos un par de ejemplos en donde se calcula la entropía de entrelazamiento utilizando la ecuación \reef{definep}, y que también son indicadores de que esta prescripción sirve en efecto para calcular la entropía de entrelazamiento.

\subsubsection{Estado de vacío en 1+1 CFT}

Consideremos una teoría conforme en $1+1$ dimensiones y sea $V$ un intervalo de longitud $L$ dado por los puntos
\begin{equation}
x \in \left[-\frac{L}{2},\frac{L}{2}\right]\,.
\end{equation}
La geometría dual a esta teoría conforme es $AdS_3$, cuya métrica está dada por
\begin{equation}
ds^2=\frac{\ell^2}{z^2}\left(-dt^2+dx^2+dz^2\right)\,;
\end{equation}
$z$ es la dirección radial, y la teoría de campos se encuentra en la superficie dada por $z=0$.

Consideremos el estado de vacío en la teoría de campos, para el cual queremos hallar la entropía. Este estado es estático, por lo que podemos tomar $t=0$. Una superficie extremal de codimensión $2$ en $AdS_3$ es unidimensional, es decir, es una geodésica. Luego, la fórmula \reef{definep} nos indica que debemos hallar una geodésica espacial en la geometría
\begin{equation}
ds^2=\frac{\ell^2}{z^2}\left(dx^2+dz^2\right)\,,
\end{equation}
que conecte los puntos $P_1=\left(z_1,x_1\right)=\left(0,-\frac{L}{2}\right)$ y $P_2=\left(z_2,x_2\right)=\left(0,\frac{L}{2}\right)$. Sin embargo, cualquier geodésica que termine en $z=0$ va a tener longitud infinita, dado que $\int \frac{dz}{z}=+\infty$. Esto es el reflejo, en el dual gravitatorio, de las divergencias ultravioletas que presenta la entropía de entrelazamiento en teoría de campos. Para regular estas divergencias, introducimos un cutoff en $z=\epsilon$.

Parametrizando la curva en la forma $\left(z(\lambda),x(\lambda)\right)$, tenemos que la longitud (área) de la geodésica regularizada está dada por
\begin{equation}
A=\int ds = 2\ell \int_{\epsilon}^{z_{max}}\frac{dz}{z}\sqrt{x'(z)^2+1}\,.\label{aresp}
\end{equation}
Es sencillo mostrar que la geodésica está dada por una porción de círculo (casi un semicírculo)
\begin{equation}
x=\frac{L}{2}\cos\lambda\,,\,\,\,z=\frac{L}{2}\sin\lambda\,,\,\,\,\textrm{con}\,\,\lambda \in \left(\frac{\epsilon}{L},\pi-\frac{\epsilon}{L}\right)\,.
\end{equation}
Sustituyendo en \reef{aresp}, encontramos que el área de la superficie mínima es
\begin{equation}
A=2\ell\log\left(\frac{L}{\epsilon}\right)\,.
\end{equation}
Aplicando entonces la fórmula \reef{definep}, la entropía de entrelazamiento resulta
\begin{equation}
S=\frac{4\pi \ell}{\lp}\log\left(\frac{L}{\epsilon}\right)=\frac{c}{3}\log\left(\frac{L}{\epsilon}\right)\,,
\end{equation}
donde hemos utilizado que los parámetros de la teoría gravitatoria y la teoría conforme están relacionados por $c=\frac{12\pi \ell}{\lp}$. Este resultado está en perfecta concordancia con los cálculos de la entropía de entrelazamiento para teorías conformes que presentamos en la ecuación \reef{paracft}.

\subsubsection{Subaditividad fuerte de la entropía holográfica}

\begin{figure}[h]
\begin{center}
\includegraphics[width=13cm]{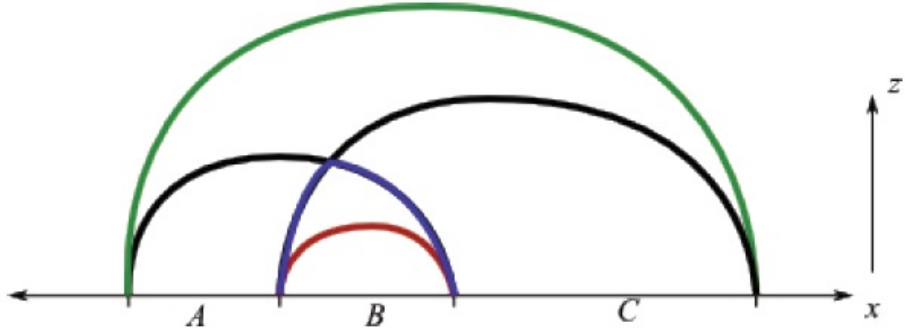}
\end{center}
\caption{\textbf{Subaditividad fuerte para la entropía holográfica}. Las curvas suaves homólogas a cada región, representan las superficies de área mínima. Las dos curvas asociadas a las superficies minimales homólogas a $AB$ y $BC$ pueden pensarse como dos curvas, en general no suaves, asociadas a superficies homólogas a las regiones $ABC$ y $B$ (representadas en negro y azul, respectivamente). Como las curvas en verde y rojo representan las superficies de área mínima homólogas a las regiones $ABC$ y $B$, respectivamente, se deduce que el área de la curva roja es menor a la de la curva azul, y el área de la curva negra es menor al de la verde. La suma de esas dos desigualdades demuestra la subaditividad fuerte para la entropía holográfica. } \label{holossa}
\end{figure}

En esta sección, derivamos una de las propiedades que satisface la entropía holográfica de entrelazamiento. La validez de esta propiedad, que se sabe que satisface la entropía de entrelazamiento, constituye una prueba de consistencia para la fórmula \reef{definep}. La propiedad de subaditividad fuerte fue presentada en la sección \ref{sec:propentro}; recordamos que para un sistema tripartito, la subaditividad fuerte dice que
\begin{equation}
S_{ABC}+S_B \leq S_{AB} + S_{BC}\,.\label{subanueva}
\end{equation}
La demostración de la validez de esta propiedad para la entropía holográfica es bastante sencilla para estados estáticos. En la figura \ref{holossa} se grafican las superficies minimales correspondientes a las regiones $ABC$, $B$, $AB$ y $BC$. Las dos superficies minimales homólogas a $AB$ y $BC$ pueden pensarse como dos superficies, en general no suaves, asociadas a superficies homólogas a las regiones $ABC$ y $B$, representadas en la figura en negro y azul, respectivamente. Como las superficies coloreadas en verde y rojo con las de área mínima homólogas a las regiones $ABC$ y $B$, respectivamente, deducimos que el área de la superficie roja es menor al de la superficie azul, y el área de la superficie negra es menor al de la verde. Sumando estas dos desigualdades y multiplicando globalmente por $\frac{2\pi}{\ell_p^{d-1}}$ obtenemos la propiedad de subaditividad fuerte \reef{subanueva}.

Este argumento ha sido también extendido al caso de estados no estáticos. Naturalmente, la demostración es un poco más complicada, dado que las superficies extremales no se encuentran necesariamente en un corte a tiempo constante.
%
%
%

\section{Entropía relativa y holografía}\label{secdeltaSH}

En el capítulo siguiente, estudiamos la validez de la ecuación \reef{123} para la entropía holográfica de entrelazamiento dada por \reef{define}. Interesantemente, para dos estados infinitesimalmente próximos, la desigualdad \reef{123} se transforma en una igualdad para la variación a primer orden de la entropía de entrelazamiento $\Delta S$ y la variación a primer orden del valor de expectación del hamiltoniano modular $\Delta \langle H\rangle$. La igualdad a primer orden puede entenderse debido a que estamos examinando la entropía relativa entre dos estados muy cercanos. 

\begin{figure}[h]
\begin{center}
\includegraphics[width=10cm]{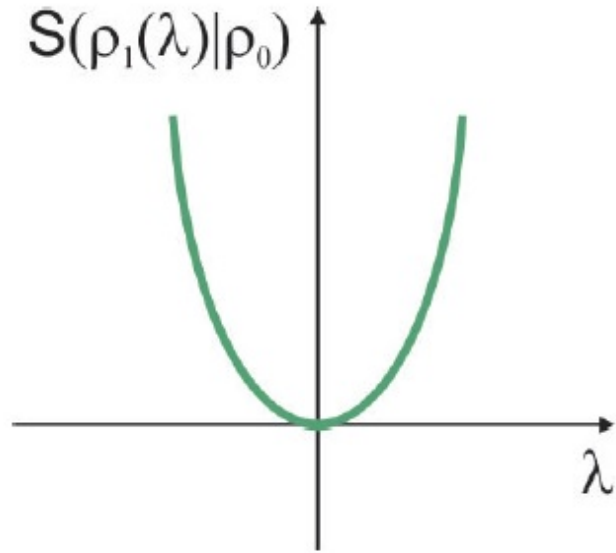}
\end{center}
\caption{\textbf{$\Delta\langle H\rangle =\Delta S$}. La entropía relativa $S(\rho_1(\lambda)|\rho_0)$ es positiva para valores positivos y negativos de $\lambda$, y se anula para $\lambda=0$ donde se tiene $\rho_1(\lambda=0)=\rho_0$. Si $S(\rho_1(\lambda)|\rho_0)$ es una función suave de $\lambda$, entonces su primera derivada debe anularse en $\lambda=0$, lo que da la ecuación $\Delta\langle H\rangle =\Delta S$ a primer orden en $\lambda$.} \label{relativita}
\end{figure}

Consideremos un estado fijo $\rho_0$ en el espacio de los estados y parametricemos con $\lambda$ al resto de los estados en la forma $\rho_1(\lambda)$, de modo que $\rho_1(\lambda=0)=\rho_0$. Como los dos estados coinciden para $\lambda=0$, tenemos que $S(\rho_1(0)|\rho_0)=0$, pero $S(\rho_1(\lambda)|\rho_0)>0$ tanto para $\lambda$ positivo y negativo. Luego, si $S(\rho_1(\lambda)|\rho_0)$ es una función suave de $\lambda$, su primera derivada debe anularse en $\lambda=0$ (ver figura \ref{relativita}). Esto implica que
\begin{equation}
\Delta\langle H\rangle =\Delta S \labell{eeex}
\end{equation}
En términos termodinámicos, esta es la conocida ecuación $dE=T dS$ válida para estados cerca del equilibrio térmico.

Una forma más precisa de obtener la igualdad \reef{eeex} es evaluando la perturbación de primer orden de $S(\rho)$ y $H(\rho)$ para la matrix densidad
 \be
\rho=\frac{e^{-(H+\delta H)}}{\textrm{tr}(e^{-(H+\delta H)})}\,.
 \ee
Luego, a orden lineal en $\delta H$, tenemos que ambas coinciden
\begin{equation}
\Delta S=\Delta \langle H\rangle=\frac{\textrm{tr}(e^{-H} H )\textrm{tr}(e^{-H}
\delta H )}{(\textrm{tr}(e^{-H} ))^2}-\frac{\textrm{tr}(e^{-H} H \delta H )}{
\textrm{tr}(e^{-H} )}=\langle H\rangle\langle\delta H\rangle-\langle H \delta
H\rangle\,,\labell{deviation}
\end{equation}
donde en la última expresión los valores de expectación se calculan utilizando la matriz densidad sin perturbar. Al derivar la ecuación \ref{deviation} hemos tratado a $\delta H$ como una perturbación numérica y no como un operador. Esto está justificado en este caso dado que estamos manipulando operadores dentro de la traza y tomando sólo términos que son funciones de $H$ con sólo un único operador $\delta H$ (por lo que no es necesario prestar atención al ordenamiento de los operadores). Esta fórmula asume que la perturbación de $\rho$ es pequeña con respecto a $\rho$. En este punto, tenemos que ser cuidadosos dado que en teoría cuántica de campos las matrices densidad tienen un número infinito de autovalores, que deben tener desviaciones pequeñas. Por ejemplo, insertando una excitación de partícula, que esté bien localizada en el interior de una región grande $A$, no debería cambiar mucho la entropía con respecto a la del estado de vacío. En particular, si la partícula se encuentra lejos del borde $\partial A$, de donde proviene la mayor contribución al entrelazamiento, la entropía debería ser aproximadamente la misma que la del vacío. Sin embargo, $\Delta \langle H\rangle$ va a medir la energía del paquete de ondas de la partícula. La razón de la discrepancia entre $\Delta S$ y $\Delta \langle H\rangle$ en este caso es debida a que el estado de partícula nunca se acerca al de vacío mientras que la distancia $R$ entre el paquete de ondas y el borde de la región sea más grande que la longitud de onda $\lambda$ del paquete de ondas. De hecho, el estado global con la excitación de partícula es siempre ortogonal al vacío global y esperamos que la entropía relativa tienda a infinito en el límite de $R/\lambda$ grande, lo que corresponde a una perfecta distinguibilidad entre los estados. Además, debido a las relaciones de incerteza, la energía de la partícula escalea como $1/\lambda$ y $\Delta \langle H\rangle \sim R/\lambda$.

Podemos hacernos una idea intuitiva de cuándo la igualdad (\ref{eeex}) será válida. Cerca del borde de una región, las matrices densidad tendrán la forma de Rindler \reef{siete}, lo que sugiere una interpretación térmica en el sentido de Unruh \cite{unruh}. En particular, la temperatura decrece con $1/x$ a medida que nos alejamos del borde una distancia $x$ hacia dentro de la región. Para una región de tamaño finito $R$ tenemos entonces una temperatura mínima $T\sim 1/R$ \cite{rovelli}. Queremos cambiar el estado agregándole una perturbación. Supongamos entonces que tenemos un estado térmico y lo mezclamos con el estado$|E\rangle \langle E|$ de energía $E$ con una probabilidad $p$ pequeña. Para que el cambio en los autovalores sea pequeño, debemos tomar $p\ll e^{-\beta E}/Z=p_E$, es decir, $p$ debe ser más pequeño que la probabilidad con la que el mismo estado aparece en el ensamble térmico. Esto último siempre se cumple si el cambio en energía es menor que la energía promedio típica para el mismo estado en el baño térmico. Por lo tanto, en nuestro problema original requerimos que la densidad de energía depositada en algún lugar, donde la temperatura local es $\sim T(x)$, sea mucho menor que $T^{d}$. Luego, el cambio en la entropía satisface $\Delta S\sim \Delta E/T\ll 1$ y estamos perturbando el baño térmico. De otro modo, las excitaciones en la región generan estados lejos del equilibrio. La conclusión es que la igualdad \reef{eeex} será válidad para regiones compactas y estados que den un valor de expectación del tensor de energía impulso pequeño. Para tener perturbaciones de baja densidad de energía dentro de $A$ podemos tomar una combinación del vacío con un paquete de ondas; por ejemplo, $\vert 0 \rangle+\epsilon \vert \phi\rangle$ para $\epsilon$ pequeño. En este caso, podemos hacer que la densidad de energía del estado sea tan baja como deseemos sin requerir que el estado tenga una longitud de onda grande.

Evaluando la entropía relativa entre el vacío y otros estados para regiones esféricas utilizando la dualidad AdS/CFT, en el capítulo siguiente verificamos que las ecuaciones y desigualdades relevantes se cumplen para una amplia clase de estados, lo que le brinda un gran soporte a la fórmula holográfica para el cálculo de la entropía de entrelazamiento. En el capítulo 7, discutimos sobre algunos usos potenciales de la ecuación \reef{eeex}, y en el capítulo 8 desarrollamos una de estas aplicaciones, que es la obtención de una nueva cota tipo Bekenstein.
\chapter{\label{ch:hee}Una prueba de consistencia para la entropía holográfica}

Como comentamos en el capítulo anterior, en el contexto de la dualidad AdS/CFT, la entropía de entrelazamiento en la teoría de campos del borde se determina a partir de un cálculo elegante en la teoría dual gravitatoria \cite{rt1,rt2,nishioka,taka}. En particular, la entropía de entrelazamiento entre una región espacial $V$ y su complemento $\bar V$ en el borde está dada por
 \be
S(V) = \frac{2\pi}{\lp^{d-1}}\ \mathrel{\mathop {\rm
ext}_{\scriptscriptstyle{\mv\sim V}} {}\!\!} \left[A(\mv)\right]
 \labell{define}
 \ee
donde la extremización se realiza sobre todas las superficies $\mv$ en el bulk que tienen a $V$ como su borde.

En este capítulo, nos concentraremos en estudiar la validez de la desigualdad
\begin{equation}
\Delta\langle H\rangle\ge\Delta S\,.\labell{1123}
\end{equation}
dada por la positividad de la entropía relativa. Recordamos que
\begin{equation}
\Delta\langle H\rangle=\textrm{tr} (\rho_1\, H )- \textrm{tr} (\rho_0\, H )
\qquad{\rm and}\qquad
\Delta S=S(\rho_1)- S(\rho_0)\,.
 \labell{S2}
\end{equation}
La prescripción holográfica \reef{define} nos permite calcular las entropías de entrelazamiento necesarias para hallar $\Delta S$. Para evaluar la otra parte de la desigualdad es necesario conocer el hamiltoniano modular. Como ya se ha comentado en el capítulo \ref{ch:modular}, la forma precisa del hamiltoniano modular se conoce en pocas situaciones:

\begin{enumerate}
\item Para el estado de vacío de cualquier teoría de campos restringida al semiespacio $x>0$. En este caso, el hamiltoniano modular es proporcional a $K$, el generador de transformaciones de boosts en la dirección $x$ \cite{bisognano,bisognano2}
\begin{equation}
H=2 \pi K=2 \pi \int_{x>0} d^{d-1}x\, x \,T_{00}(\vec{x})\,.
\labell{iete}
\end{equation}
\item Para el estado de vacío de una teoría conforme de campos, para una región de entrelazamiento esférica
\begin{equation}
H=2 \pi \int_{|x|<R} d^{d-1}x\, \frac{R^2-r^2}{2 R}\ T_{00}(\vec{x})\,.
\labell{sphereH}
\end{equation}
Este resultado se deriva sencillamente de la ecuación \reef{iete}, dado que hay una transformación conforme especial (y una traslación) que mapea el wedge de Rindler a la esfera $|x|<R$ \cite{hmm,haag}.
\item Otra situación donde se conoce un hamiltoniano modular local, y que hemos derivado, es el caso de una teoría conforme bidimensional en un estado térmico (a temperatura $T$) en el wedge de Rindler \cite{yng}
\begin{equation}
H=\frac1T \int_{x>0}dx\ \left(1-e^{-2\pi T x}\right) \,T_{00}(\vec{x})\,.
 \labell{2dcftRinT}
\end{equation}
\end{enumerate}
A continuación, nos concentraremos principalmente en el caso de una superficie de entrelazamiento esférica, donde el hamiltoniano modular está dado por \reef{sphereH}.

En los capítulos siguientes mostraremos que la desigualdad \reef{123} se satisface. En particular, veremos que se satura para perturbaciones lineales del estado, como es esperado según la discusión de la sección \ref{secdeltaSH}.

Además de utilizar la desigualdad \reef{1123}, que sale de la positividad de la entropía relativa, también podemos examinar la relación de monotonicidad de la entropía relativa. Es decir, deberíamos poder comprobar que la entropía relativa aumenta con el radio de la región esférica de entrelazamiento
 \begin{equation}
\partial_R S(\rho_1|\rho_0)=\partial_R
\left[\Delta\langle H\rangle- \Delta S \right] \ge0\,;
 \labell{include}
 \end{equation}
por supuesto, esto podremos verlo sólo para los casos en que $\Delta\langle H\rangle\ne\Delta S$.

En la sección \ref{simple} comenzamos presentando un par de ejemplos donde evaluamos estas desigualdades y la igualdad para perturbaciones lineales (\ref{eeex}). En la sección \ref{general} analizamos el caso general de una perturbación lineal del vacío y demostramos la validez de la ecuación (\ref{eeex}). También analizamos el caso de las perturbaciones cuadráticas y encontramos que en todos los casos la entropía relativa es positiva y monótona con las regiones. En la sección \ref{two} analizamos algunos ejemplos en  $d=2$ en donde puede calcularse la entropía de entrelazamiento en forma exacta. En la sección \ref{puzzle} del próximo capítulo discutimos aspectos relacionados con la localización de las contribuciones a $\Delta \langle H\rangle$ y, también en el próximo capítulo, en la sección \ref{discuss} realizamos una discusión de los resultados y aplicaciones más relevantes de este trabajo. En particular, discutimos sobre el potencial uso de la ecuación (\ref{eeex}) para realizar tomografías del estado de vacío utilizando la entropía de entrelazamiento, y demostramos que los resultados de la sección \ref{general} permiten reconstruir el operador densidad en una región esférica a partir de la prescripción de área extremal para la entropía, obteniéndose concordancia con el resultado de la teoría de campos conforme. El trabajo expuesto en este capítulo y el siguiente fue realizado en colaboración con Horacio Casini, Robert C. Myers del Perimeter Institute y Ling-Yan Hung de Harvard University.

\section{Algunos ejemplos sencillos}
\labell{simple}

Como ya comentamos, la estrategia será evaluar la desigualdad \reef{123} en el contexto holográfico para una superficie de entrelazamiento esférica, para la cual se conoce el hamiltoniano modular. La prescripción de Ryu-Takayanagi \cite{rt1,rt2,nishioka,taka} nos permite calcular las entropías de entrelazamiento y consecuentemente también $\Delta S$.

En entos casos, podemos evaluar $\Delta \langle H \rangle$ si conocemos el valor de expectación del tensor de energía-impulso $\la T_{ab}\ra$. En esta sección, nuestro estado de referencia $\rho_0$ es el vacío de la teoría de campos conforme, mientras que $\rho_1$ será el dual holográfico de un agujero negro.

El bulk dual al estado de vacío de una teoría de campos conforme $d$-dimensional es el espacio AdS$_{d+1}$, cuya métrica en coordenadas de Poincaré tiene la forma
\begin{equation}
ds^2=\frac{L^2}{z^2}\left(-dt^2
+d\vec{x}^2_{d-1}+dz^2\right)\,.
\labell{adsvac}
\end{equation}
Elegimos nuestra región $V$ como una esfera en la teoría de campos del borde, es decir, $V$ es la esfera $\lbrace t=0, r\le R\rbrace$.

El tensor de energía-impulso tiene valor de expectación nulo en el vacío, de modo que valor de expectación del hamiltoniano modular \reef{sphereH} se anula para este estado: $\langle H\rangle_0=\tr(\rho_0\,H)=0$.

La entropía de entrelazamiento para la región $V$ en este estado puede calcularse utilizando la prescripción holográfica \reef{define}. En este caso, la superficie de área mínima $v$ está dada por \cite{rt1,rt2,nishioka,taka}
\begin{equation}
 z=z_0(r)\equiv\sqrt{R^2-r^2}\,. \labell{sphereb}
\end{equation}
Luego, la entropía de entrelazamiento toma la forma
\begin{equation}
S_0=2\pi\, \frac{A(v)}{\lp^{d-1}}
=2\pi\frac{L^{d-1}}{\lp^{d-1}}\Omega_{d-2}\int_0^R dr\, \frac{r^{d-2}}{z^{d-1}}
\sqrt{1+\partial_r z^{\,2}}\,,
\labell{sphereEE}
\end{equation}
donde $\Omega_{d-2}$ denota el área de una esfera $(d-2)$-dimensional
 \be
\Omega_{d-2}=\frac{2\,\pi^{(d-1)/2}}{\Gamma((d-1)/2)}\,.
 \labell{units}
 \ee
Para nuestros propósitos no será necesario evaluar explícitamente la integral de la ecuación \reef{sphereEE} (el resultado puede encontrarse por ejemplo en \cite{rt1,rt2,nishioka,taka,hmm}.

Nuestro segundo estado $\rho_1$, será el dual holográfico de una brana negra, es decir, un agujero negro planar en AdS. En general, el valor de expectación del tensor de energía-impulso dual a una brana negra estacionaria tiene la forma del de un fluido ideal
 \be
\langle T_{\mu\nu}\rangle= (\veps +P)\, u_\mu u_\nu + P\,\eta_{\mu\nu}\,,
 \labell{idealf}
 \ee
donde $\veps$, $P$ y $u_\mu$ corresponden a la densidad de energía, presión y velocidad $d$-dimensional del fluido, respectivamente. Como la teoría del borde es conforme, la traza del tensor de energía-impulso es nula $\langle T^\mu{}_\mu\rangle=0$, lo que impone $P=\veps/(d-1)$.

\subsection{Brana negra estática}

Como primer ejemplo, consideremos una brana negra AdS planar estática, para la cual la métrica se escribe
\begin{equation}
ds^2=\frac{L^2}{z^2}\left(-f(z)\,dt^2+d\vec{x}^2_{d-1}
+\frac{dz^2}{f(z)}\right)\quad{\rm con}\ \ f(z)=1-\frac{z^d}{\zh^d}
\,. \labell{static}
\end{equation}
En este caso, el plasma dual se encuentra en reposo, es decir, $u^\mu=(1,\vec{0}_{d-1})$, y entonces la ecuación \reef{idealf} se reduce a
 \be
\langle T_{\mu\nu}\rangle= \veps\ {\rm diag} (1,1/(d-1),1/(d-1),\cdots)\,.
 \labell{enerden1a}
 \ee
El diccionario holográfico usual \cite{construct,construct2,nice} nos dice que la densidad de energía es
 \be
\veps =  \frac{d-1}{2}\,\frac{L^{d-1}}{\lp^{d-1}}\,\frac{1}{\zh^{d}}\,.
 \labell{enerden1}
 \ee
Esta última ecuación puede interpretarse como $\veps= c\, T^d$, utilizando la expresión para la temperatura del agujero negro
\begin{equation}
T=\frac{d}{4 \pi \zh}\,.
 \labell{temper1}
\end{equation}
Con estas expresiones, la evaluación del valor de expectación del hamiltoniano modular en este estado es inmediata
 \bea
\langle H\rangle_1&=&\pi \Omega_{d-2}\frac{\veps}{R}\,\int_0^R
dr\,r^{d-2}\left(R^2-r^2\right)
 \nonumber\\
 &=& \frac{2 \pi \Omega_{d-2}}{d^2-1}\  R^d\,\veps\,.
 \labell{modH1}
 \eea
De este modo, obtenemos que
 \be
\Delta\langle H\rangle=\langle H\rangle_1-\langle H\rangle_0= \frac{\pi
\Omega_{d-2}}{d+1} \,\frac{L^{d-1}}{\lp^{d-1}}\,\frac{R^d}{\zh^{d}}
 \labell{dmodH1}
 \ee
habiendo reemplazado $\veps$ utilizando la ecuación \reef{enerden1}.

Para completar la evaluación de la ecuación \reef{123} sólo resta calcular la entropía de entrelazamiento para una superficie esférica en el espacio cuya métrica está dada por la ecuación (\ref{static}). Aplicando la prescripción holográfica \reef{define}, la entropía resulta
\begin{equation}
S_1=2\pi\frac{L^{d-1}}{\lp^{d-1}}\Omega_{d-2}\int_0^R dr\, \frac{r^{d-2}}{z^{d-1}}
\sqrt{1+\frac{(\partial_r z)^{2}}{f(z)}}\,,
\labell{sphereEEBH}
\end{equation}
donde $f(z)$ es la función dada en la ecuación \reef{static}. En principio, podríamos extremar la expresión anterior y evaluar el funcional entropía para ese caso extremo en forma numérica \footnote{En \cite{finiteT} pueden encontrarse varias aproximaciones analíticas interesantes.}. Para obtener un resultado analítico realizaremos un cálculo perturbativo para esferas (o temperaturas) "pequeñas", en el que consideraremos el límite $R/\zh\ll1$ (o alternativamente,
$RT\ll1$). En este caso, la superficie mínima siente principalmente la región asintótica de la geometría \reef{static} y entonces la solución es muy parecida a la del caso AdS puro \reef{sphereb}, es decir, $z(r)=z_0(r)+\delta
z(r)$. Ahora, dado que $z_0(r)$ minimiza el funcional entropía para el espacio AdS \reef{sphereEE}, la desviación $\delta z(r)$ no modificará el resultado a primer orden en nuestro cálculo perturbativo. Por lo tanto, la diferencia a primer orden se obtiene al evaluar \reef{sphereEEBH} con $z=z_0(r)$ y determinando la contribución principal en $R/\zh$. Expandiendo la ecuación \reef{sphereEEBH} al primer orden no trivial en $1/\zh^d$ obtenemos
 \begin{eqnarray}
\Delta S&=&\pi\frac{L^{d-1}}{\lp^{d-1}}\Omega_{d-2}\int_0^R dr\,\left.
\frac{r^{d-2}\,z\,(\partial_r z)^{2}}{\zh^d\,\sqrt{1+(\partial_r z)^{2}}}
\right|_{z=z_0(r)}=
\nonumber\\
&=&\pi\frac{L^{d-1}}{\lp^{d-1}}\Omega_{d-2}\int_0^R dr\,
\frac{r^{d}}{\zh^d\,R}=\frac{\pi \Omega_{d-2}}{d+1} \, \frac{L^{d-1}}{\lp^{d-1}}
\,\frac{R^d}{\zh^{d}}
 \labell{temp}
 \end{eqnarray}
Luego, al comparar esta expresión con \reef{dmodH1}, vemos que a primer orden
\begin{equation}
\Delta \langle H \rangle=\Delta S\,,
\labell{agree}
\end{equation}
es decir, se satura la desigualdad \reef{123}.
%

\subsection{Brana negra `boosteada'} \labell{boost}

Ahora repetimos los cálculos de la sección anterior para el caso en que la brana no es estática. Es decir, el estado $\rho_1$ es un plasma térmico uniformemente boosteado en alguna dirección. Este estado $\rho_1$ se encuentra entonces caracterizado por la temperatura $T$ y la velocidad $v$. Nuestros cálculos estarán hechos al orden más bajo no trivial en la temperatura y a todo orden en la velocidad.

El tensor de energía-impulso toma la forma dada en la ecuación \reef{idealf} ahora con $u^\mu =
(\gamma, \gamma v, \vec{0}_{d-2})$ donde $\gamma=1/\sqrt{1-v^2}$, y $P=\veps/(d-1)$. En particular, tenemos
\begin{equation}
\langle T_{00}\rangle=\veps\left( 1+\frac{d}{d-1}\,\gamma^2\,v^2
\right)\,.
\labell{newstress}
\end{equation}
La correspondiente solución gravitatoria se deriva simplemente aplicando un boost en la dirección de $x^1\equiv x$ a la métrica dada por \reef{static}. Es conveniente escribir la métrica resultante en la forma
\begin{equation}
ds^{2}=\frac{L^{2}}{z^{2}}\left[ -dt^2 + dx^2 + \gamma^2\frac{z^{d}}{
\zh^{d}}\left( dt+
v dx\right) ^{2}+d\vec{x}_{d-2}^{2} +\frac{dz^{2}}{1-\frac{z^{d}}{\zh^{d}} }
\right] \,.
 \labell{metrica}
\end{equation}
Con las herramientas holográficas usuales \cite{construct,construct2,nice}, se ve que la ecuación \reef{newstress} se verifica con $\veps$ dado nuevamente por la ecuación \reef{enerden1}.

Ahora evaluamos el cambio en el valor de expectación del hamiltoniano modular. Como la densidad de energía sigue siendo uniforme, el cálculo de $\langle H\rangle_1$ es igual al hecho anteriormente, salvo por el factor global adicional de la ecuación \reef{newstress}. Por lo tanto
\begin{equation}
\Delta \langle H^{\prime }\rangle=\Delta \langle H\rangle
\left( 1+\frac{d}{d-1}\,\gamma^2\,v^2
\right) \,,  \labell{energy}
\end{equation}
donde $\Delta \langle H\rangle$ es la variación del hamiltoniano modular dado en \reef{dmodH1}.

Ahora, como el fondo gravitatorio \reef{metrica} es en principio estacionario (pero no estático),para evaluar la entropía holográfica de entrelazamiento deberíamos aplicar la prescripción covariante sugerida por \cite{station}. En efecto, la prescripción holográfica dada por la ecuación \reef{define} se ajusta a la situación. En este nuevo fondo gravitatorio, necesitaríamos hallar la superficie extremal con un perfil definido por $z=z(x,y)$ y $t=t(x,y)$ donde
$y^2\equiv\sum_{i=2}^{d-1} (x^{i})^2$ --- notar que, en particular, la superficie extremal no permanecerá en un corte a tiempo fijo en el bulk. De todos modos, nuestro objetivo es evaluar el cambio en la entropía de entrelazamiento $\Delta S'$; un razonamiento análogo al de la sección anterior nos permite deducir que la contribución principal a $\Delta S'$ se obtiene evaluando el funcional área para la geometría \reef{metrica} con el perfil de temperatura cero \reef{sphereb}. Por lo tanto, a partir de ahora ignoramos las desviaciones de la superficie extremal respecto del plano a tiempo constante.

Para un perfil $z=z(x,y)$, es inmediato demostrar que la entropía en el fondo gravitatorio boosteado \reef{metrica} es
\footnotesize
\begin{equation}
S^\prime_1=2\pi\frac{L^{d-1}}{\lp^{d-1}}\Omega_{d-3}\int_{-R}^R dx
\int_0^{\sqrt{R^2-x^2}}\!\!\!dy\
\frac{y^{d-3}}{z^{d-1}}\,
\left[\left(1+\gamma^2v^2\frac{z^d}{\zh^d}\right)
\left(1+\frac{\partial_y z^{\,2}}{f(z)}\right)+\frac{\partial_x z^{\,2}}{f(z)}
\right]^{1/2}\,,
\labell{sphereEEboost}
\end{equation}
\normalsize
donde nuevamente $f(z)$ está dada en \reef{static}.

En la evaluación de $\Delta \langle H^{\prime }\rangle$ en la ecuación \reef{energy} no fue necesario realizar ninguna aproximación (al igual que en la sección anterior). Para el cambio de entropía trabajamos al orden más bajo no trivial en el límite $R/\zh\ll1$. Nuevamente, aplicando el mismo razonamiento que anteriormente, concluimos que el orden más bajo no trivial en el cambio de entropía se obtiene evaluando \reef{sphereEEboost} con el perfil de temperatura cero \reef{sphereb}, es decir, $z=z_0(r)=\sqrt{R^2-x^2-y^2}$. Primero expandimos la expresión anterior a primer orden en $1/\zh^d$ y luego sustraemos la contribución de orden cero \reef{sphereEE}, obteniendo
\begin{equation}
\Delta S'=\pi\frac{L^{d-1}}{\lp^{d-1}}\Omega_{d-3}\int_{-R}^R dx
\int_0^{\sqrt{R^2-x^2}}\!\!\!dy\
\frac{y^{d-3}\,z}{\zh^d\,\sqrt{1+\partial_r z^{\,2}}}
\left[\partial_r z^{\,2}+\gamma^2v^2\left(1+\partial_y z^{\,2}\right)
\right]\,,
\labell{sphereEEboost1}
\end{equation}
donde hemos simplificado $\partial_x z^{2}+\partial_y z^{2}=\partial_r z^{2}$ antes de sustituir $z=z_0(r)$. Con esta sustitución, el primero de los términos dentro del corchete dará el resultado sin boostear $\Delta
S$ de la ecuación \reef{temp}. Por lo tanto
 \begin{eqnarray}
\Delta S'&=&\Delta S\,+\,\pi\frac{L^{d-1}}{\lp^{d-1}}\Omega_{d-3}\,\gamma^2v^2
\int_{-R}^R dx \int_0^{\sqrt{R^2-x^2}}\!\!\!dy\ \left. \frac{y^{d-3}\,z
\left(1+\partial_y z^{\,2}\right) }{\zh^d\,\sqrt{1+\partial_r z^{\,2}}}
\right|_{z=z_0(r)}
\nonumber\\
&=&\Delta S\,+\,\pi\frac{L^{d-1}}{\lp^{d-1}}\Omega_{d-3}\,\gamma^2v^2
\int_{-R}^R dx \int_0^{\sqrt{R^2-x^2}}\!\!\!dy\ \frac{y^{d-3}
\left(R^2-x^2\right) }{\zh^d\,R}
 \nonumber\\
&=&\Delta S\,+\,\pi\frac{L^{d-1}}{\lp^{d-1}}\Omega_{d-3}\,\gamma^2v^2
\frac{R^d}{\zh^d}\,\frac{\sqrt{\pi}}{d-2}\,\frac{\Gamma\left(d/2+1\right)}{
\Gamma\left(d/2+3/2\right)}
 \nonumber\\
&=&\Delta S
 \left( 1+\frac{d}{d-1}\,\gamma^2\,v^2
\right)\,,
 \labell{sphereEEboost2}
 \end{eqnarray}
donde hemos utilizado las ecuaciones \reef{units} y \reef{temp} en la última línea, logrando una expresión final bastante sencilla.

Recordamos que en la sección anterior teníamos $\Delta \langle H \rangle=\Delta S$; luego, comparando las ecuaciones \reef{energy} y \reef{sphereEEboost2}, se obtiene nuevamente que al orden más bajo no trivial resulta
\begin{equation}
\Delta \langle H' \rangle=\Delta S'
\labell{agree1}
\end{equation}
para el plasma boosteado. Este resultado podría haber sido anticipado a partir de la discusión relativa a la ecuación \reef{eeex}. En este caso, estamos considerando una familia de operadores densidad caracterizados por la temperatura $T$ y la velocidad $v$. Si bien nuestros cálculos son válidos para todo orden en la velocidad, $\Delta\langle H' \rangle$ y $\Delta S'$ ha sido evaluado al orden más bajo no trivial en $(RT)^d$.

\subsection{Brana negra cargada} \labell{charge}

Continuando con el análisis de la sección \ref{simple}, otro fondo gravitatorio interesante a considerar se obtiene tomando a $\rho_1$ como el estado dual a una brana negra AdS cargada. En este caso, el estado en la teoría del borde está caracterizado por el potencial químico, $\mu$, así como por la temperatura $T$. En este caso, nuestros cálculos serán al orden más bajo no trivial en $RT$, y a primer orden en $\mu/T$.

Consideramos la acción gravitatoria en el bulk
\begin{equation}
I=\frac{1}{2 \lp^{d-1}}\int d^{d+1}x\, \sqrt{-g}\,
\left(\frac{d(d-1)}{L^2}+R-\frac{L^2}{4}
F_{\mu\nu} F^{\mu\nu}\right)
\end{equation}
con $d\ge3$.\footnote{La normalización del término del campo de gauge queda determinada por los detalles microscópicos de la construcción holográfica --- véase, por ejemplo, la discusión en \cite{chemical}. Aquí elegimos el factor de $L^2$ por pura conveniencia. Nótese que en el caso $d=2$, aparecen en la solución términos logarítmicos.} La métrica para un agujero negro planar con carga se puede escribir como
\begin{equation}
ds^{2}=\frac{L^{2}}{z^{2}}\left( -h(z)\, dt^{2}+d\vec{x}
^{2}_{d-1}+\frac{dz^{2}}{h}\right)
\labell{chargemet}
\end{equation}
donde
\begin{equation}
h=1-\left(
1+\zh^2\,q^{2}\right) \frac{z^d}{\zh^d}
+q^{2} \frac{z^{2d-2}}{\zh^{2d-4}} \,,
 \labell{mimi}
\end{equation}
y el potencial electromagnético correspondiente tiene sólo una componente no nula
 \begin{equation}
A_0(z)=\sqrt{\frac{2(d-1)}{d-2}} \,q\,
\left(1-\frac{z^{d-2}}{\zh^{d-2}}\right)\,.
 \labell{pot}
 \end{equation}
Aquí, $z=\zh$ corresponde a la posición del horizonte y $q$ está relacionado con la densidad de carga arrastrada por el mismo. La temperatura del plasma dual está dada por
 \begin{equation}
T=\frac{d}{4 \pi \zh}\left(1-\frac{d-2}{d}\zh^2\, q^{2}\right)
\labell{temper2}
\end{equation}
y el potencial químico está dado por el valor asintótico del potencial de gauge
 \be
\mu=\lim_{z\to0} A_0 = \sqrt{\frac{2(d-1)}{d-2}} \,q\,.
 \labell{chemu}
 \ee
Dado que el plasma en la teoría conforme está en reposo, la ecuación \reef{idealf} se reduce a $\langle
T_{\mu\nu}\rangle= \veps\ {\rm diag} (1,1/(d-1),1/(d-1),\cdots)$ y la prescripción holográfica usual da \cite{construct,construct2,nice}
 \be
\veps =  \frac{d-1}{2}\,\frac{L^{d-1}}{\lp^{d-1}}\,\frac{1}{\zh^{d}}\left(
1+\zh^2\,q^{2}\right)\,.
 \labell{enerden2}
 \ee
Empezamos evaluando el cambio en el valor de expectación del hamiltoniano modular entre este nuevo estado y el vacío. Dado que la densidad de energía es aquí también uniforme, la evaluación de $\langle H\rangle_1$ se realiza precisamente del mismo modo que el cálculo en la ecuación \reef{modH1}, difiriendo sólo en el factor adicional constante que aparece en la ecuación \reef{enerden2}. Por lo tanto, llegamos a que
\begin{equation}
\Delta \langle H^{\prime\prime }\rangle=\Delta \langle H\rangle\
\left(1+\zh^2\,q^{2}\right) \,,  \labell{energy2}
\end{equation}
donde $\Delta \langle H\rangle$ es el resultado dado en la ecuación \reef{dmodH1}.

Como la brana negra es estática, la superficie extremal cuya área da la entropía holográfica tiene un perfil con simetría esférica $z=z(r)$, para una superficie de entrelazamiento esférica. Consecuentemente, dada la métrica \reef{chargemet}, el funcional entropía queda
\begin{equation}
S''_1=2\pi\frac{L^{d-1}}{\lp^{d-1}}\Omega_{d-2}\int_0^R dr\, \frac{r^{d-2}}{z^{d-1}}
\sqrt{1+\frac{(\partial_r z)^{2}}{h(z)}}\,,
\labell{sphereEEBH2}
\end{equation}
donde $h(z)$ está dado en la ecuación \reef{mimi}. En lo que sigue, limitamos nuestro análisis a un cálculo perturbativo con $R/\zh\ll1$, pero consideramos $\zh q =
O(1)$. Como en los casos anteriores, la contribución al cambio de entropía se obtiene simplemente evaluando \reef{sphereEEBH2} con el perfil solución del vacío $z=z_0(r)$ y expandiendo en $R/\zh$. En este caso, es conveniente hacer una aclaración al respecto. Tanto aquí, como en los ejemplos anteriores, el orden más bajo no trivial al cambio en la métrica es $O(z^d/\zh^d)$, por lo que el orden más bajo no trivial al cambio de la entropía es $\Delta S''=O(R^d/\zh^d)$. El orden más bajo no trivial al cambio del perfil de la superficie extremal, $\delta z$, también está controlado por los cambios en la métrica. Sin embargo, como se argumentó anteriormente, la entropía sólo cambia recién a orden cuadrático en $\delta z$, con lo que esta contribución produce un cambio en la entropía de orden que es $\Delta S''(\delta
z^2)=O(R^{2d}/\zh^{2d})$ --- ver la sección \ref{quad} para un cálculo explícito. Vemos entonces que el segundo orden más bajo no trivial en el cambio de la métrica \ref{chargemet} es $O(z^{2d-2}/\zh^{2d-2})$, dado que consideramos que $\zh q$ es del orden de 1. Si hacemos la cuenta con estos cambios en la métrica y usando el perfil original de la solución de vacío, aparece una contribución adicional al cambio de entropía que es $O(R^{2d-2}/\zh^{2d-2})$ --- esto se verá explícitamente en el cálculo que sigue a continuación. Esta contribución sigue siendo de orden menor en la expansión en $R/\zh$ que las contribuciones provenientes de cambios en el perfil. Consecuentemente, en este caso es también legítimo no considerar los cambios en el perfil de la superficie extremal. Expandimos entonces la ecuación\reef{sphereEEBH2} como
 \begin{eqnarray}
\Delta S''&=&\pi\frac{L^{d-1}}{\lp^{d-1}}\Omega_{d-2}\int_0^R dr\,\left.
\frac{r^{d-2}\,z\,(\partial_r z)^{2}}{\zh^d\,\sqrt{1+(\partial_r z)^{2}}}
\left[\left(
1+\zh^2\,q^{2}\right)-q^2\frac{z^{d-2}}{\zh^{d-4}}\right]\right|_{z=z_0(r)}
\nonumber\\
&=&\Delta S\,\left( 1+\zh^2\,q^{2}\right)-
\pi\frac{L^{d-1}}{\lp^{d-1}}\Omega_{d-2} \frac{q^2}{\zh^{2d-4}R} \int_0^R
dr\,r^{d}\left(R^2-r^2\right)^{\frac{d-2}{2}}
\nonumber\\
&=&\Delta S\,\left( 1+\zh^2\,q^{2}\right)-
\frac{d-1}{2}\pi^{\frac{d+1}{2}}\frac{\Gamma(d/2)
}{\Gamma\left(d+\frac12\right)} \frac{L^{d-1}}{\lp^{d-1}} (\zh q)^2
\frac{R^{2d-2}}{\zh^{2d-2}}
 \labell{temp2}
 \end{eqnarray}
donde $\Delta S$ corresponde a la variación dada por la ecuación \reef{temp}.
Recordamos que en la ecuación \reef{agree} encontramos $\Delta \langle H \rangle=\Delta S$. Por lo tanto, al comparar \reef{energy2} y \reef{temp2}, encontramos nuevamente que los términos al orden no trivial más bajo son nuevamente iguales. Sin embargo, al incluir la contribución de $O(R^{2d-2}/z_{0}^{2d-2})$, se obtiene
\begin{equation}
\Delta \langle H'' \rangle>\Delta S''\,.
\labell{agree2}
\end{equation}
De este modo, al agregar un potencial químico se introduce una contribución a un orden superior al más bajo no trivial y que asegura la positividad de la entropía relativa. A partir de las expresiones anteriores se tiene
 \be
S(\rho_1|\rho_0)\simeq\frac{\pi}{2}\frac{L^{d-1}}{\lp^{d-1}}\Omega_{d-2}\frac{\Gamma(d/2)
\Gamma((d+1)/2)}{\Gamma\left(d+\frac12\right)} (\zh q)^2
\frac{R^{2d-2}}{\zh^{2d-2}}
 \labell{relate2}
 \ee
Dado que $S(\rho_1|\rho_0)\propto R^{2d-2}$, podemos verificar sencillamente que la entropía relativa también satisface la propiedad de monotonicidad \reef{include}, es decir,
$\partial_RS(\rho_1|\rho_0)>0$. Utilizando las ecuaciones \reef{temper2} y \reef{chemu}, es posible reescribir el lado derecho como función de la temperatura y el potencial químico. Si bien la expresión completa no es particularmente reveladora, en la situación $1\gg\mu/T\gg RT$, tenemos $S(\rho_1|\rho_0)\sim (RT)^{2d-2}(\mu/T)^2$ y entonces, en particular, observamos que esta contribución no nula comienza a orden cuadrático en el potencial químico.

Para terminar con esta sección, remarcamos que el resultado en la ecuación \reef{temp2} se obtuvo a partir de una deformación a primer orden de la métrica asintótica, mientras que el último resultado es causa de una \textit{back-reaction} del campo de gauge en la geometría con lo que el cambio en la entropía relativa es cuadrático en el coeficiente $q$ correspondiente.

\section{Análisis general} \labell{general}

En esta sección generalizamos los resultados obtenidos anteriormente para estados holográficos más generales. Para superficies de entrelazamiento esféricas, es muy sencillo evaluar $\Delta\langle H\rangle$ utilizando la ecuación \reef{sphereH}, dado que las prescripciones holográficas usuales nos permiten determinar $\langle T_{\mu\nu}\rangle$
\cite{construct,construct2,nice}. En principio, el cálculo de la entropía de entrelazamiento utilizando la ecuación \reef{define} es más complicado, dado que involucra determinar la superficie de área extremal en la geometría del bulk, que está asociada al estado $\rho_1$. De todos modos, en los ejemplos anteriores vimos que si $\rho_1$ describe una "pequeña" perturbación del estado inicial de vacío $\rho_0$, los cálculos se restringen a considerar perturbaciones asintóticas de la geometría $AdS$. En este contexto perturbativo, nuestro análisis de la entropía holográfica de entrelazamiento se simplifica considerablemente. También resulta natural pensar en formular estos cálculos utilizando la expansión asintótica de Fefferman-Graham (FG) \cite{feffer,feffer2} --- ver también \cite{construct,construct2}. En particular, este acercamiento nos permitirá considerar una clase más general de estados perturbados sin necesidad de preocuparse por los detalles que tenga la geometría del bulk en el infrarrojo lejano.

A continuación, utilizando la expansión de FG, apuntaremos a realizar tres cálculos diferentes: comenzaremos considerando estados descriptos puramente por excitaciones gravitatorias en el bulk AdS. Es decir, el tensor de energía impulso es el único operador que tiene valores de expectación no nulos en estos estados. Introducimos un parámetro perturbativo pequeño $\vep$ que controla la magnitud de los valores de expectación $\langle T_{\mu\nu}\rangle$. Nuestro primer resultado es demostrar que la desigualdad \reef{123} se satura siempre, es decir, que $\Delta\langle H\rangle=\Delta S$ a primer orden en $\vep$. Remarcamos que esta igualdad es válida incluso cuando $\langle T_{\mu\nu}\rangle$ varía en escalas comparables a $R$, el radio de la superficie de entrelazamiento. En segunda instancia, en la sección \ref{quad} extenderemos estos cálculos a segundo orden en $\vep$. Allí demostraremos que, si bien $\Delta\langle H\rangle$ no cambia, las contribuciones adicionales a la entropía tienen un signo definido que asegura la validez de la ecuación $\Delta\langle H\rangle>\Delta S$. El tercer punto de análisis, desarrollado en la sección \ref{matter}, considera el caso de estados para los cuales aparecen otros operadores distintos del $T_{\mu\nu}$ que adquieren valores de expectación no nulos. Como vimos en la sección \ref{charge}, es relativamente sencillo determinar las correcciones cuadráticas a la entropía provenientes de estas perturbaciones. Extendemos este análisis a una clase mucho más amplia de estados y verificamos que las contribuciones cuadráticas nuevamente aseguran que se cumpla la desigualdad $\Delta\langle H\rangle>\Delta S$.

Como ya hemos comentado, nuestro análisis general estará formulado en el contexto de la expansión de Fefferman-Graham para las soluciones asintóticas del bulk \cite{construct,construct2,feffer,feffer2}. Por lo tanto, comenzamos considerando una métrica general para el bulk, escrita en coordenadas de FG
 \be
ds^2 =\frac{L^2}{z^2} \left( dz^2 + g_{\mu\nu}(z,x^\mu) dx^\mu dx^\nu\right)\,.
 \labell{bulkG}
 \ee
Estamos considerando la geometría asintótica cuando $z\simeq0$. Siempre elegiremos la métrica asintótica (es decir, la métrica sobre la cual se define la teoría de campos conforme) como plana, por lo que
 \be
g_{\mu\nu}(z,x^\mu) = \eta_{\mu\nu} + \delta g_{\mu\nu}(z,x^\mu)
 \labell{expand}
 \ee
donde la expansión de $\delta g_{\mu\nu}$ comienza con términos de orden $z^d$. Estamos interesados en calcular la entropía holográfica de entrelazamiento \reef{define} y, por lo tanto, necesitaremos evaluar el área de diferentes superficies extremales en el bulk. En principio, en situaciones para las que la geometría de fondo no es estática, el perfil de estas superficies $(d-1)$-dimensionales quedará especificado al dar la posición radial y el tiempo en el bulk como funciones de las restantes coordenadas espaciales, es decir, $z=z(x^i)$ y $t=t(x^i)$. Sin embargo, nuestro objetivo es evaluar el cambio de entropía de entrelazamiento $\Delta S$ y, como se ha discutido en la sección \ref{boost}, será suficiente considerar superficies en el bulk a tiempo constante. Por lo tanto, con un perfil radial de la forma $z=z(x^i)$, la métrica inducida $h_{ij}$ en esta superficie está dada por
 \be
h_{ij}dx^i dx^j = \frac{L^2}{z^2}\left(g_{ij} + \partial_i z\partial_j z
\right) dx^i dx^j
 \labell{induced}
 \ee
y el área correspondiente es entonces
 \be
A= L^{d-1}\int d^{d-1}x  \sqrt{h} = L^{d-1}\int d^{d-1}x \sqrt{{\rm
det}g_{ij}}\,\sqrt{1+ g^{ij}\,\partial_i z\,\partial_j z}\,.
 \labell{action}
 \ee
En principio, la ecuación \reef{action} puede ser utilizada como una acción efectiva para hallar el perfil extremal $z=z(x^i)$. Sin embargo, como hemos visto en los ejemplos anteriores, para determinar $\Delta S$ al orden más bajo no trivial, alcanza con evaluar el área en la nueva geometría utilizando el perfil original \reef{sphereb}.

\subsection{Correcciones lineales a la entropía relativa} \labell{line}

Comenzamos considerando estados $\rho_1$ cuya pequeña diferencia con el vacío está caracterizada por el valor de expectación del tensor de energía impulso $T^0_{\mu\nu}$ en la teoría de campos conforme en el borde.\footnote{Para simplificar la notación, a partir de ahora omitimos los corchetes angulares que denotan el valor de expectación de $T^0_{\mu\nu}$ .}
Suponemos que este valor de expectación es "muy pequeño" y está caracterizado por un parámetro adimensional $\vep \ll 1$. Como hicimos anteriormente, limitaremos nuestra atención al caso de superficies de entrelazamiento esféricas, para las cuales el hamiltoniano modular (del vacío) \reef{sphereH} es lineal en el tensor de energía impulso y entonces $\Delta\langle H\rangle$ es lineal en $\vep$. Sin embargo, en la ecuación \reef{123}, el cambio de entropía recibirá contribuciones a todo orden en $\vep$. En esta sección sólo evaluaremos $\Delta S$ a orden lineal en $\vep$.

En general, utilizando la expansión de FG, la desviación de la métrica del bulk respecto del caso $AdS$ puro planteada en la ecuación \reef{expand}, toma la forma
 \be
\delta g_{\mu\nu} = \frac{2}d\frac{\lp^{d-1}}{L^{d-1}} z^d\sum_{n=0} z^{2n}\,
T^{(n)}_{\mu\nu}\,.
 \labell{FGexp}
 \ee
Las ecuaciones de Einstein en el bulk determinan $T^{(n)}_{\mu\nu}$ para $n>0$ en término de los valores de expectación $T^{(0)}_{\mu\nu}$. Como comentamos, la estrategia será encontrar $T^{(n)}_{\mu\nu}$ al orden más bajo en $\vep$ (o, a orden lineal en $T^{{(0)}}_{\mu\nu}$).

Antes de pasar a resolver las ecuaciones de Einstein, recordamos que el objetivo es evaluar el cambio de la entropía holográfica de entrelazamiento en la métrica perturbada. En este caso, podemos aplicar el mismo razonamiento que el realizado en la sección \ref{simple}. En particular, en el $AdS$ puro, existe una solución analítica \reef{sphereb} para la superficie extremal en el bulk correspondiente a una superficie de entrelazamiento esférica de radio $R$ en el borde
 \be
z_0^2 + r^2 = R^2, \qquad {\rm donde}\ \ r^2 = \sum_{i=1}^{d-1} x_i^2\,.
\labell{wox}
 \ee
La superficie de entrelazamiento extremal para la métrica perturbada también admite una expansión en $\vep$, es decir, tendremos $z(x^i)= z_0(x^i) + \vep z_1(x^i) + \cdots$. De todos modos, como describimos en la sección anterior, dado que el perfil $z_0$ es extremal al orden más bajo, la perturbación $z_1$ sólo contribuirá al orden $\vep^2$. Por lo tanto, podemos obtener el cambio lineal en el área simplemente evaluando el área \reef{action} en la métrica perturbada con el perfil original $z_0$. Consecuentemente, dado (\ref{FGexp}), encontramos que a orden lineal en $\vep$
 \be
\Delta S=2\pi\frac{\Delta A}{\lp^{d-1}} = \frac{2\pi R}{d} \int_{|{x}|\le
R}\!\!\!d^{d-1}x\ \sum_{n=0} z_0^{2n}\left(T^{(n)}{}_{i}{}^{i} -
T^{(n)}{}_{\,ij}\, \frac{x^i\,x^j}{R^2}\right)\,.
  \labell{linearA}
  \ee
Ahora resolvemos las ecuaciones de Einstein, que se escriben como
 \be
\hat{R}_{AB} - \frac{1}{2}G_{AB}\left(\hat{R} + \frac{d(d-1)}{L^2}\right)=0\,,
 \ee
donde $\hat{R}_{AB}$ es el tensor de Ricci del bulk evaluado en la métrica $G_{AB}$ dada como en la ecuación (\ref{bulkG}). Utilizando los resultados de \cite{useful}, podemos escribir a orden lineal en  $\vep$,
\begin{eqnarray}
\hat{R}_{\rho\rho} &=&  -\frac{d}{4\rho^2}-\frac{1}{2} \partial^2_\rho
\delta g^{\mu}{}_{\mu} \,,\nonumber \\
\hat{R}_{\mu\rho}  &=& \frac{1}{2}\left(\partial_\rho
\partial_{\nu}\delta g^{\nu}{}_{\mu}-
\partial_{\mu}\partial_\rho\delta g^{\nu}{}_{\nu}\right) \nonumber \,,\\
\hat{R}_{\mu\nu} &=& R_{\mu\nu} -2\rho\partial^2_{\rho}\delta g_{\mu\nu}
+ (d-2)\partial_{\rho}\delta g_{\mu\nu}
+ \eta_{\mu\nu}\partial_{\rho}\delta g_{\gamma}{}^{\gamma} -
\frac{d}{\rho} (\eta_{\mu\nu} + \delta g_{\mu\nu}) \nonumber \,,\\
\hat{R} &=& -d(d+1) + \rho R + 2(d-1) \rho \partial_\rho \delta g_{\mu}{}^{\mu} -
4\rho^2 \partial^2_\rho\delta g_{\mu}{}^{\mu}\,,
\end{eqnarray}
donde elegimos una coordenada radial (adimensional) $\rho = z^2/L^2$. Además, $R_{\mu\nu}$ y $R$ son tensores de curvatura evaluados en $g_{\mu\nu}$, tratando a $z$ (o $\rho$) como un parámetro externo. Explícitamente
 \be
R_{\mu\nu} = \frac{1}{2}\left(\partial_{\nu}\partial_{\gamma}\delta
g^{\gamma}{}_{\mu} +
\partial_\mu\partial_\gamma\delta g^{\gamma}{}_{\nu} - \Box \delta g_{\mu\nu} -
\partial_\mu\partial_\nu\delta g^{\gamma}{}_{\gamma}\right)\,.
 \ee
Sustituyendo la ecuación (\ref{FGexp}) y la expresión anterior en las ecuaciones de Einstein, obtenemos las siguientes relaciones para $T^{(n)}$ utilizando las componentes $\rho\rho$ y $\mu\rho$, respectivamente
\footnotesize
\begin{eqnarray}
&& \partial^\mu\partial^\nu T^{(n)}_{\mu\nu} - \Box T^{(n)\,\mu}{}_{\mu} +
(d-1)(d+2n+2)\,T^{(n+1)\,\mu}{}_{\mu} =0\,,
 \quad T^{(0)\,\mu}{}_{\mu}=0\,,
 \\
&& \partial_{\nu}T^{(n)}{}_\mu{}^{\nu} -\partial_\nu
T^{(n)\,\mu}{}_{\mu} =0 \,.
\end{eqnarray}
\normalsize
Estas dos ecuaciones implican que
 \be
T^{(n)\,\mu}{}_{\mu} =0\,,\qquad
\partial_{\nu}{T^{(n)}}_{\mu}{}^{\nu} =0\,,
 \labell{boat}
 \ee
para todo $n$. Vemos entonces que las ecuaciones de Einstein aseguran que $T^{(n)}$ tenga traza nula y sea conservado para \emph{todo} $n$. Finalmente, las componentes $\mu\nu$ de las ecuaciones de Einstein se reducen a
 \be
T^{(n)}_{\mu\nu} = -\frac{\Box T^{{(n-1)}}_{\mu\nu}}{2n(d+2n)}\,,
 \ee
lo que implica
 \be
T^{(n)}_{\mu\nu} =    \frac{ (-1)^n \Gamma[d/2+1]}{2^{2n} n! \Gamma[d/2+n+1]}
\,\Box^n T^{(0)}_{\mu\nu}\,.
 \labell{box}
 \ee
Desde ya, es posible sustituir estos resultados en (\ref{linearA}) y así expresar $\Delta S$ enteramente en términos de $T^{(0)}_{\mu\nu}$.

Para lo que sigue, será más conveniente escribir al tensor de energía impulso en términos de su expansión en serie de Fourier
 \be
T^{(0)}_{\mu\nu}(x) = \int d^{d}p \,\, \exp(-i p\cdot x)\
\widehat{T}_{\mu\nu}(p)\,.
 \labell{fourier}
 \ee
Utilizando los resultados previos, el cambio en la entropía de entrelazamiento \reef{linearA} resulta
\begin{eqnarray}
&&\Delta S = \frac{2\pi R}{d}\int d^{d-1}x \, \int d^dp \, \exp(-i p\cdot x)\,\times
 \labell{main} \\
&&\frac{\Gamma[d/2+1]}{(z_0|p|/2)^{d/2}}
\sum_{n=0} \left[\frac{1}{n!\Gamma[d/2+n+1]}\left(\frac{|p|z_0}{2}\right)^{2n+
d/2}\right]\left(\widehat{T}_{i}{}^{i}(p)- \widehat{T}_{ij}(p)\frac{x^ix^j}{R^2}
\right)\,,
 \nonumber
\end{eqnarray}
donde $|p|= |\sqrt{p_\mu p^\mu}|$. Vemos que la suma entre corchetes da precisamente
 \be
\sum_{n=0} \left[\frac{1}{n!\Gamma[d/2+n+1]}\left(\frac{|p|z_0}{2} \right)^{2n+
d/2}\right] = I_{d/2}(|p|z_0)\,.
 \labell{sumI}
 \ee
Para $p$ temporal en signatura Lorentziana, se obtiene en cambio $J_{d/2}(|p|z_0)$. Es decir, obtenemos una expresión que es precisamente proporcional a la función de Green del gravitón en AdS$_{d+1}$. Sin embargo, notamos que la condición de contorno asintótica se toma de modo que el término constante más importante sea cero --- ver, por ejemplo, \cite{vijay0}. Esto último puede contrastarse con la función de Green del bulk al borde usual , que resulta proporcional a $K_{d/2}(|p|z_0)$, donde la condición de contorno se elige de modo que el término más relevante cerca del borde de $AdS$ sea constante.

\subsection*{Saturando la desigualdad \reef{123}:}

Volvemos a la ecuación \reef{123} para demostrar que la misma se satura a primer orden en $\vep$, para la clase general de estados considerados. Dado el hamiltoniano modular \reef{sphereH} (para una superficie de entrelazamiento esférica), podemos escribir
\begin{equation}
\Delta\langle H\rangle=\frac{\pi}{R} \int_{|x|\le R} d^{d-1}x\, z_0^2\
T^{(0)}_{00} \,.
\labell{spheredH}
\end{equation}
donde $z_0$ es el perfil extremal dado en la ecuación \reef{wox}. En principio, esta expresión no resulta parecida a la de $\Delta S$ en la ecuación \reef{linearA}, incluso aún después de sustituir los resultados de la ecuación \reef{box}.

Para demostrar que la desigualdad \reef{123} se satura, comenzamos examinando la ecuación (\ref{main}) para una sola componente de momento bien definido correspondiente al $\delta H$ de la última ecuación. Elegimos al momento en la dirección espacial de $x^1$, es decir,
\begin{equation}
 T^{(0)}_{\mu\nu}(x)=\widehat{T}_{\mu\nu}\, e^{-i p\cdot x}\,.
 \labell{2four}
\end{equation}
Para ilustrar la cuenta suponemos que el momento es tipo tiempo. La cuenta para un momento tipo espacio es análoga.

La conservación de $T^{(0)}_{\mu\nu}$ y el hecho de que tiene traza nula, implican que
\be
 \widehat{T}_i{}^i=\widehat{T}_{00}\,,\quad
\widehat{T}_{10}=-\frac{p^0}{p^1}\,\widehat{T}_{00}\quad{\rm and}\quad
\widehat{T}_{11}=\frac{(p^0)^2}{(p^1)^2}\, \widehat{T}_{00}\,.
 \labell{polar}
 \ee

Luego, notamos que dado el tensor de energía impulso elegido en la ecuación \reef{2four}, la integral de la ecuación (\ref{main}) es simétrica ante rotaciones que dejan a $x^1$ fijo. Esto implica que la integral que contiene el término $\widehat{T}_{ij}\, x^i x^j$ se anula para $i\neq j$. Además, para $i=j=2,\cdots,(d-2)$, todas las integrales son iguales. Por lo tanto, podemos reemplazar dentro de la integral
\begin{eqnarray}
&& \widehat{T}_i{}^i-\widehat{T}_{ij} \frac{x^i x^j}{R^2}\rightarrow
\widehat{T}_{00}-\widehat{T}_{11}\, \frac{(x^1)^2}{R^2}-
\sum_{i=2}^{d-2}\widehat{T}_{ii}\, \frac{(x^i)^2}{R^2}\\
 && \rightarrow \widehat{T}_{00}-\widehat{T}_{11} \frac{(x^1)^2}{R^2}- \sum_{i=2}^{d-2}
 \widehat{T}_{ii}\frac{\sum_{j=2}^{d-2} (x^j)^2}{(d-2) R^2}
\rightarrow
 \widehat{T}_{00}-\widehat{T}_{11} \frac{(x^1)^2}{R^2}- \sum_{i=2}^{d-2}
 \widehat{T}_{ii} \frac{r^2-(x^1)^2}{(d-2) R^2}
 \nonumber \\
 &&\rightarrow \widehat{T}_{00}-\widehat{T}_{11} \frac{(x^1)^2}{R^2}-
 (\widehat{T}_i{}^i-\widehat{T}_{11})\,
 \frac{r^2-(x^1)^2}{(d-2) R^2}
\\&&\hspace{2cm}\rightarrow
 \widehat{T}_{00} \left(1-\frac{(p^0)^2}{(p^1)^2}\frac{(x^1)^2}{R^2}-
\frac{\left(1-\frac{(p^0)^2}{(p^1)^2}\right)
(r^2-(x^1)^2)}{(d-2) R^2}\right)\,.\nonumber
\end{eqnarray}
En la última transformación utilizamos (\ref{polar}). Esta expresión final depende sólo de $\widehat{T}_{00}$, que es necesario para obtener una igualdad con $\Delta\langle H\rangle$.

$\Delta S$ en coordenadas polares queda entonces
\begin{eqnarray}
\Delta S&=& \frac{2^{(d+2)/2} \pi R }{d |p|^{d/2}}\,\Gamma[d/2+1]
\Omega_{d-3}\,\widehat{T}_{00}e^{ip^0 t}\,\int_0^R dr\, r^{d-2}\int_0^\pi d \theta\,
\sin^{d-3}\!\theta\, e^{-i p^1 r \cos(\theta)} \nonumber\\
 && \times   \frac{J_{d/2}(|p|\sqrt{R^2-r^2})}{(R^2-r^2)^{d/4}}
\left(1-\frac{(p^0)^2}{(p^1)^2}\frac{r^2 \cos^2\theta}{R^2}-
\frac{\left(1-\frac{(p^0)^2}{(p^1)^2}\right)r^2\sin^2\theta}{(d-2) R^2}\right)\,.\labell{dss}
 \end{eqnarray}
Las integrales sobre $\theta$ se pueden hacer explícitamente utilizando
\begin{equation}
\int_0^\pi d \theta\, \sin^q(\theta) e^{-i x \cos(\theta)}=2^{q/2}
\sqrt{\pi}\,\Gamma[(q+1)/2]\, \frac{J_{q/2}(| x|)}{|x |^{q/2}}\,.\labell{angle}
\end{equation}
Ahora, para la variación del hamiltoniano modular, sustituimos la ecuación \reef{2four} en \reef{spheredH}, lo que da
\small
\begin{eqnarray}
\Delta\langle H\rangle &=&  2\pi \Omega_{d-3}\,\widehat{T}_{00}e^{ip^0 t}\,
\int_0^R dr\,r^{d-2}
\int_0^\pi d\theta\, \sin^{d-3}\!\theta\, \frac{R^2-r^2}{2 R}
e^{-i p_1 r \cos(\theta)}\labell{modu2}\\
&=&
 2^{(d-1)/2}\pi^{3/2} \Omega_{d-3}\,
\Gamma[(d-2)/2]\,\widehat{T}_{00}e^{ip^0 t}\,
 \frac{R^{(d-1)/2}}{ |p^1|^{(d+1)/2}} J_{(d+1)/2}(|p^1| R)\,.\nonumber
\end{eqnarray}
\normalsize
Observamos que la integral para $\Delta S$ en la ecuación \reef{dss} depende de un parámetro adicional $|p|$ que no está presente en la integral de la ecuación \reef{modu2}. Esto implica que para que $\Delta S$ y $\Delta\langle H\rangle$ sean iguales, es necesario que la expresión (\ref{dss}) sea independiente de $p$ para un valor fijo de $p^1$. Esto afortunadamente sucede; es posible verlo haciendo una expansión en potencias de $p$ y $p^1$, y reemplazando $(p^0)^2=p^2+(p^1)^2$ en la integral de la ecuación (\ref{dss}). Juntando los términos con las mismas potencias de $p$ y $p^1$, se llega a expresiones que se pueden integrar en $\theta$ y $r$  de forma analítica. El resultado es que el coeficiente de $(p^1)^m p^n $ en la expansión de $\Delta S$ es cero para todo $n>0$. Consecuentemente, podemos tomar el límite de $p\rightarrow 0$ en el integrando para simplificar los cálculos y la ecuación (\ref{dss}) queda
\small
\begin{eqnarray}
 \delta S &=& \frac{4\pi}{d}\Omega_{d-3}\,\widehat{T}_{00}e^{ip^0 t}\,
 \int_0^R
 dr\,r^{d-2}\int_0^\pi d\theta\, \sin^{d-3}\!\theta\,
\frac{R^2-r^2 \cos(\theta)^2}{2 R} e^{-i p_1 r \cos(\theta)} \labell{bango}\nonumber \\
&=&  2^{(d-1)/2}\pi^{3/2} \Omega_{d-3}\,
\Gamma[(d-2)/2]\,\widehat{T}_{00}e^{ip^0 t}\,
 \frac{R^{(d-1)/2}}{ |p^1|^{(d+1)/2}} J_{(d+1)/2}(|p^1| R)\,.
\end{eqnarray}
\normalsize
Comparando ahora las ecuaciones (\ref{modu2}) y \reef{bango} vemos que
\begin{equation}
\Delta\langle H\rangle=\Delta S\,. \labell{equalz}
\end{equation}
Si bien este análisis se realizó para una sola onda plana \reef{2four}, como estamos considerando perturbaciones lineales, la misma igualdad vale para la expansión de Fourier general \reef{fourier}. De este modo, concluimos que la ecuación \reef{equalz} vale para cualquier perturbación a primer orden del tensor de energía impulso. En particular, esta igualdad es también válida cuando $T^{(0)}_{\mu\nu}$ varía en escalas comparables a $R$, el tamaño de la superficie esférica de entrelazamiento.

\subsection{Correcciones cuadráticas a la entropía relativa} \labell{quad}

Si bien demostrar la igualdad \reef{equalz} requirió de bastantes cálculos técnicos, era razonable esperar su validez en vistas de la discusión presentada en la sección \ref{secdeltaSH}. Similarmente, al extender nuestros cálculos de $\Delta S$ a segundo orden en $\vep$, esperamos que las nuevas contribuciones a este orden sean tales que aseguren la validez de la desigualdad \reef{123}. En esta sección, verificamos que efectivamente esto sucede. Por simplicidad, restringiremos nuestro cálculo al caso de tensores de energía impulso constantes.

Para hallar las correcciones cuadráticas a la entropía relativa procedemos del siguiente modo: primero, expandimos la métrica del bulk a orden cuadrático en el tensor de energía impulso. Luego, expandimos el funcional área \reef{action} a orden cuadrático en el parámetro perturbativo $\vep$. En particular, obtenemos las ecuaciones de movimiento que gobiernan la deformación de la superficie extremal a orden lineal en el tensor de energía impulso. Luego, resolviendo las ecuaciones de movimiento, sustituimos los resultados en el funcional área y separamos la corrección cuadrática a la entropía relativa.

\subsubsection*{Paso 1: Métrica del bulk}

En la ecuación (\ref{FGexp}) la métrica del bulk está expandida a primer orden. A orden cuadrático, la expansión toma la forma
 \be
\delta g_{\mu\nu} = \eta_{\mu\nu} + a\,z^d\, T_{\mu\nu} + a^2 \,z^{2d}  \left(
n_1\, T_{\mu\alpha}T^{\alpha}{}_{\nu} + n_2\, \eta_{\mu\nu}T_{\alpha\beta}
T^{\alpha\beta}\right)+\cdots\,,
 \labell{secbulk}
 \ee
donde
 \be
a= \frac{2}{d}\, \frac{\lp^{d-1}}{L^{d-1}}\,.
 \labell{aaax}
 \ee
El término cuadrático en el tensor de energía impulso tiene la forma más general permitida por la invariancia de Lorentz, la simetría entre $\mu$ y $\nu$ y que respeta la nulidad de la traza de $T_{\mu\nu}$\footnote{Recordamos que nos estamos limitando al caso en que $T_{\mu\nu}$ es constante y por lo tanto los términos de derivadas \reef{box}, que aparecían a orden lineal en $\vep$ anteriormente, se anulan.}. La potencia de $z^{2d}$ en este término se determina simplemente por análisis dimensional. Sólo resta fijar los coeficientes $n_{1,2}$, lo que puede hacerse comparando esta expresión con la de la métrica de la brana negra \reef{static} cuando esta última se reescribe utilizando las coordenadas de FG \reef{bulkG}. Esto requiere transformar a una nueva coordenada radial en la geometría asintótica $AdS$
 \be
\tz = z\,\left(1+\frac{1}{2d}\frac{z^d}{\zh^d}+\frac{2+3d}{16d^2}
\frac{z^{2d}}{\zh^{2d}}+\cdots\right)\,.
 \labell{newz}
 \ee
Esta nueva coordenada se elige para para obtener $G_{zz}=L^2/\tz^2$, como es requerido en la ecuación \reef{bulkG}. En términos de esta coordenada radial, las componentes restantes de la métrica en la expansión asintótica toman la forma
 \bea
g_{00} &=& -1 + \frac{d-1}{d}\frac{\tz^d}{\zh^d}
-\frac{4d^2-9d+4}{8d^2}\frac{\tz^{2d}}{\zh^{2d}}
+\cdots \nonumber\\
g_{ij}&=&\delta_{ij}\left(1 +
\frac{1}{d}\frac{\tz^d}{\zh^d}-\frac{d-4}{8d^2}\frac{\tz^{2d}}{\zh^{2d}}+\cdots
\right)\,.
 \labell{asmetric}
 \eea
Recordamos que el tensor de energía impulso toma la forma dada en la ecuación \reef{enerden1a} y \reef{enerden1}, como puede leerse de los términos más relevantes en la expansión de arriba. Luego, comparando las ecuaciones \reef{secbulk} y \reef{asmetric}, podemos obtener $n_1$ y $n_2$
 \be
n_1 = \frac{1}{2}\qquad{\rm and}\qquad n_2 = -\frac{1}{8(d-1)}\,.
 \labell{coeff3}
 \ee

\subsubsection*{Paso 2: Expansión del funcional área y ecuaciones de movimiento}

El perfil para la superficie extremal recibe correcciones debidas al cambio en la métrica del bulk. Recordamos de las secciones anteriores que la superficie minimal en un gauge estático se puede describir por $z(x^i)$, es decir, la coordenada radial del bulk queda especificada por una función de las coordenadas espaciales $x^i$. En la presentación construcción perturbativa, podemos expandir
 \be z(x^i) = z_0(x_i) + \vep\, z_1(x_i) +
\vep^2 z_2(x_i) + \cdots\,,
 \labell{profler}
 \ee
donde $z_0$ está dado en la ecuación \reef{wox}. Notar que como sólo estamos interesados en correcciones cuadráticas a la entropía, $z_2$ es irrelevante dado que aparecería linealmente en el funcional área y consecuentemente se anularía en virtud de las ecuaciones de movimiento.

La corrección a orden $\vep^2$ del funcional área \reef{action} se puede escribir como
 \be
A_{(2)} = A_{2,0} + A_{2,1} + A_{2,2}\,,
 \ee
donde estamos separando las contribuciones en tres términos, de acuerdo a la potencia de $z_1$ que aparece en las expresiones (lo que denotamos con el segundo subíndice). Sólo $A_{2,1}$ y $A_{2,2}$ contribuyen a las ecuaciones de movimiento linearizadas para $z_1$.

Expandiendo, encontramos
\small
\begin{eqnarray}
&& A_{2,0}=L^{d-1} a^2 \int d^{d-1}x\,  R z_0^{d} \bigg(-\frac{1}{16}
\left(1-\frac{r^2}{(d-1)R^2}\right)
(T_{00}^2 + T_{ij}T^{ij})
\nonumber \\
&&
+\frac{T_{i0}T^{i0}}{8}\left(1+ \frac{r^2}{(d-1)R^2}\right)+
\frac{x^i x^k}{4R^2}T_{i\alpha}T^{\alpha}{}_{k} +
\frac{1}{8}(T^2 - T_x^2 - 2T T_x)\bigg)
\end{eqnarray}
\normalsize
donde
 \be
T\equiv T_i{}^i\qquad{\rm and}\qquad T_x \equiv T_{ij}\frac{x^ix^j}{R^2}\,.
 \ee
Notar que utilizamos $z_0^2 = R^2- r^2$ para simplificar la expresión anterior\footnote{Recordamos que los índices griegos $\mu,\nu,\cdots$ corren entre todos los índices correspondientes a las direcciones planas del borde, mientras que los índices latinos $i,j,\cdots$ se restringen a las direcciones espaciales.}. Adicionalmente, tenemos
\begin{eqnarray}
A_{2,1} &=&L^{d-1} a \int d^{d-1}x\,\frac{R}{2z_0}\bigg(
T\big(z_1 - \frac{z_0^2}{R^2} x^i\partial_i z_1\big)  \nonumber \\
&&\qquad
+T_{ij}\left(\frac{2z_0^2 x^i \partial^j z_1}{R^2} -\frac{z_1 x^ix^j}{R^2}
 - \frac{z_0^2 x^ix^j x^k\partial_k z_1}{R^4}\right)\bigg)\,,
\end{eqnarray}
y
\begin{eqnarray}
A_{2,2}&=& L^{d-1}\int d^{d-1}x\, \frac{R}{z_0^d}\bigg( \frac{d(d-1)z_1^2}{2z_0^2} +
\frac{z_0^2(\partial z_1)^2}{2R^2}\nonumber\\
&&\qquad -\frac{z_0^2(x^i\partial_i z_1)^2}{2R^4} +
\frac{(d-1)}{2}\frac{x^i\partial_i z_1^2}{R^2}
\bigg)\,.
\end{eqnarray}
Remarcamos que en $A_{2,1}$ hemos omitido términos que se anulan al ser evaluados en la superficie minimal $z_0$. También recordamos que los términos de borde no contribuyen. Las ecuaciones de movimiento para $z_1$ se derivan variando $A_{2,1} + A_{2,2}$ y pueden escribirse del siguiente modo
 \be
 \frac{1}{z_0^{d-1}R}\left(\partial^2 (z_0\,z_1) -
\frac{x^ix^j}{R^2}\partial_i\partial_j(z_0\,z_1)\right) = \frac{z_0}{2R}
\left((d-2) T + (d+2) T_x \right)\,. \labell{bingo}
 \ee
Esperamos que la perturbación $z_1$ tenga la forma $T f_1(r) + T_{ij}x^i x^j f_2(r)$. Después de varios intentos de proponer soluciones para $f_2$, y determinando las condiciones de contorno apropiadas sumando soluciones apropiadas de la ecuación homogénea, llegamos a la siguiente forma sencilla para la solución para $d$ general
 \be
 z_1 = -\frac{a R^2z_0^{d-1}}{2(d+1)}(T + T_x)\,.
 \labell{solz1}
 \ee

\subsubsection*{Paso 3: Sustitución en el funcional área}

Después de haber realizado los cálculos anteriores, estamos ahora listos para sustituir todo en el funcional área. Después de hacer un poco de álgebra llegamos a
\small
\begin{eqnarray}
A_{(2)} = L^{d-1}a^2 &\int & d^{d-1}x \,\,\bigg(c_1 T^2 + c_2 T_x^2 + c_3 T_{ij}^2 + \nonumber \\
&+& c_4 T_{i0}T^{i0} + c_5 \frac{x^i T_{ij}T^{j}{}_{k}x^k}{R^2}
+ c_6  \frac{x^i T_{i0}T^{0}{}_{j}x^j}{R^2} + c_7 T T_x\bigg)\,,
\end{eqnarray}
\normalsize
con los coeficientes dados por
\begin{eqnarray}
c_1 &=& \frac{z_0^{d-4}}{16(d+1)^2(d-1)R} \bigg((d+1)^2 r^6 + (3 + d (3d^2+d-15)) r^4 R^2
 \nonumber \\
&\qquad+ &(d^2(13 - 8 d)+ 2d) -3) r^2 R^4 +(3 d^3- 7 d^2+ d +3) R^6 \bigg)\,,\\
c_2 &=& \frac{ z_0^{d-4}}{8(d+1)^2} \bigg((1 - 5 d^2) r^2 R^3 + ( d(4d+3)-3) R^5\bigg)\,,  \\
c_3 &=& \frac{(\frac{r^2}{d-1}- R^2) z_0^d}{16 R}\,,\\
c_4 &=& \frac{(\frac{r^2}{d-1}+ R^2) z_0^d}{8 R}\,,\\
c_5 &=& R z_0^{d}\frac{d(d-2)-1}{4(d+1)^2}\,,\\
c_6 &=& \frac{R}{4} z_0^d \,, \\
c_7 &=& z_0^{d-4}\frac{R^3(d-1)}{4(d+1)^2}\bigg((1 - 3 d) r^2 + (2d+1) R^2\bigg)\,.
\end{eqnarray}
Al proceder con el cálculo de las integrales restantes es útil observar que, por simetría, cuando una integral tiene la forma $\int d^{d-1}x\,\, (x^i x^j x^k x^l \cdots) f(r)$, es decir cuando hay $n$ pares de $x^i$'s en el integrando, podemos simplemente reemplazarla por
 \be
N(\delta_{ij}\delta_{kl}\cdots + \textrm{permutations})\int d^{d-1}\!x\
r^{2n}\, f(r)\,,
 \labell{gummy}
 \ee
siendo $N$ una constante de normalización apropiada. Utilizando esto, llegamos a un resultado final de la forma
 \be
A_{(2)} =a^2 L^{d-1} \Omega_{d-2}\left(C_1 T^2 + C_2 T_{ij}^2 + C_3 T_{0i}^2\right)\,,
 \labell{final}
 \ee
donde
\begin{eqnarray}
C_1 &&= -\frac{d\sqrt{\pi}R^{2d}\Gamma[d+1]}{2^{d+4}(d+1)\Gamma[d+\frac{3}{2}]}
\,,\nonumber\\
C_2 &&= C_1\,, \labell{corel}\\
C_3 &&= -\frac{(d+2)\sqrt{\pi}R^{2d}\Gamma[d+1]}{2^{d+3}(d-1) \Gamma[d+\frac{3}{2}]}\,.
 \nonumber
\end{eqnarray}
Observar que en la expresión anterior $T_{0i}^2 \equiv T_{0i}
T_{0j}\delta^{ij}\ge0$. Por lo tanto, como los tres coeficientes son negativos, queda demostrado que la perturbación al área a segundo orden es negativa y, consecuentemente, la contribución a segundo orden a la entropía holográfica de entrelazamiento asegura la validez de la desigualdad \reef{123}. Como a segundo orden tenemos $\Delta\langle H\rangle\ne\Delta S$, la validez de la propiedad de monotonicidad \reef{include} no es trivial. Sin embargo, a partir del resultado de arriba, encontramos que $S(\rho_1|\rho_0)\propto R^{2d}$ y entonces la desigualdad se satisface siempre: $\partial_RS(\rho_1|\rho_0)>0$.

A modo de ejemplo, podemos aplicar estos resultados generales al gas térmico estático descripto por el agujero negro planar AdS. El tensor de energía impulso correspondiente está dado por las ecuaciones \reef{enerden1a} y \reef{enerden1}\footnote{Para mantener el mismo lenguaje que en el desarrollo anterior, podríamos introducir un parámetro $\vep$ en estas expresiones. Sin embargo, seguiremos el procedimiento más sencillo de elegir formalmente $\vep=1$ en las expresiones anteriores. De nuestro estudio anterior del gas térmico, así como también de los resultados aquí, podemos inferir que $\Delta S$ aparece como una expansión en el parámetro pequeño $aR^d\veps$.}, es decir, tenemos $T_{00} = \veps$ y $T_{ij} = \delta_{ij} \veps /(d-1)$. La solución de \reef{bingo} puede escribirse como
 \be
z_1(r) = \frac{k_1}{\sqrt{R^2-r^2}} + a\veps \left( \frac{ ((d-1 ) R^{d+2} -
(R^2-r^2)^{d/2} (r^2 + (d-1) R^2))}{
 2 (d^2-1 ) \sqrt{R^2-r^2 }}\right)\,,
 \ee
donde $k_1$ es una constante de integración. Para asegurar que $r\to R$ a medida que $z\to 0$, que de por sí se satisface para $z_0$, debemos elegir 
 \be
 k_1 = -\frac{a \veps\, R^{d+2}}{2 (d+1)}\,.
 \ee
Esta elección arroja precisamente la solución para $z_1$ dada en la ecuación (\ref{solz1}). Sustituyendo esta solución en el funcional área, encontramos
 \be \Delta S_{(2)}
= - \frac{\pi^{3/2}d\,\Omega_{d-2}\,
   \Gamma[d-1]}{2^{d+1}(d+1)\, \Gamma[d+\frac32]}\,\frac{L^{d-1}}{\lp^{d-1}}
   \,R^{2 d}\veps^2\,.
\ee
que es negativo, como era necesario. Es destacable que aunque el integrando involucra una serie complicada de polinomios en $d$, el resultado final se reduzca a la simple forma arriba.

\subsection{Correcciones por operadores adicionales}\labell{matter}

Hasta ahora hemos considerado una clase especial de estados para la cual sólo el tensor de energía impulso tiene valores de expectación no nulos. Para perturbaciones genéricas del vacío se espera que otros operadores adquieran valores de expectación no triviales. Consecuentemente, en esta sección consideramos estados para los cuales otros operadores distintos al tensor de energía impulso adquieren valores de expectación no nulos. La descripción dual involucrará soluciones gravitatorias en el bulk en las cuales se excitarán campos de materia adicionales. Como vimos en el ejemplo de la brana negra cargada en la sección \ref{charge}, es relativamente sencillo determinar las correcciones cuadráticas a la entropía provenientes de tales campos de materia. A continuación, evaluamos las contribuciones análogas a $\Delta S$ para dos tipos de estados: Los primeros darán un valor de expectación no nulo para un operador escalar. La descripción dual requerirá agregar un campo escalar masivo a la teoría gravitatoria. La segunda clase de estados involucrará perturbaciones por una corriente conservada en la teoría del borde, o un campo de gauge en el bulk. Analizar estas configuraciones brindará una simple generalización del caso de la brana negra cargada. Para ambas familias de estados, encontramos que las contribuciones cuadráticas nuevamente aseguran la validez de $\Delta\langle H\rangle>\Delta S$.

\subsubsection*{Perturbaciones por un condensado escalar}

En nuestra primera clase de estados, un operador escalar $\mathcal{O}$ de dimensión $\Delta$ adquiere un valor de expectación no trivial en el vacío (en ausencia de fuentes). La descripción dual corresponde a pensar que ha aparecido un campo escalar que interactúa con la geometría y cambia la entropía de entrelazamiento. Aquí nos limitaremos a calcular la contribución más relevante de esta ``backreaction'' con la geometría. La acción del bulk que consideramos está dada por
 \be
I = \frac{1}{2\lp^{d-1}}\int d^{d+1}x\, \sqrt{G} \left[R -\frac{1}{2}(\partial
\phi)^2 - V(\phi) \right]\,.
 \ee
Como apuntamos a hacer una cuenta perturbativa en $\phi$ sólo es necesario mantener los términos cuadráticos en el campo escalar, de modo que el potencial se aproxima por
 \be V(\phi)=-\frac{d(d-1)}{L^2} + \frac{1}{2}m^2 \phi^2\,, \ee
donde el primer término es lo que da la constante cosmológica negativa.

Un resultado estándar \cite{revue} de la correspondencia AdS/CFT es que al orden más bajo en el condensado, el campo escalar $\phi$ de masa $m = \Delta (d-\Delta)$ se comporta asintóticamente como
 \be
\phi = \gamma\, \mathcal{O}\, z^\Delta + \cdots\,,
 \ee
con alguna constante de normalización  $\gamma$. Esto puede sustituirse en la ecuación de Einstein, que en presencia del campo escalar se escribe como
 \be
\hat{R}_{AB} = \frac{1}{2}\partial_A \phi
\partial_B \phi + \frac{1}{d-1} G_{AB} V(\phi)\,. \labell{albert}
 \ee
En presencia del campo escalar, esperamos que la expansión de la métrica en el borde se vea alterada \cite{construct,construct2}. Sin embargo, dado que sólo estamos interesados en las contribuciones más relevantes de la perturbación, no es necesario incluir los términos que contienen conjuntamente al tensor de energía impulso y al condensado escalar. A orden lineal en el tensor de energía impulso del borde y a orden cuadrático en el operador escalar, la expansión de la métrica $\delta g_{\mu\nu}$ en la ecuación \reef{expand} toma la forma
 \be
\delta g_{\mu\nu} = a z^d \sum_{n=0}z^{2n} T_{\mu\nu}^{(n)}+z^{2\Delta}
\sum_{n=0} z^{2n} \, \sigma^{(n)}_{\mu\nu}+ \cdots\,,\labell{poww}
 \ee
y recordamos que los términos en la primera suma fueron analizados en la sección \ref{line}. En ambas sumas el supraíndice $(n)$ indica que el operador correspondiente contiene un total de $2n$ derivadas; ver por ejemplo la ecuación \reef{box}. Entonces, para $n=0$, la única contribución posible del escalar es $\sigma^{(0)}_{\mu\nu}= \alpha_0\,\eta_{\mu\nu} \mathcal{O}^2$ donde $\alpha_0$ es alguna constante. Esta constante se determina sencillamente sustituyendo la expansión de la métrica y la del campo escalar en las ecuaciones de Einstein \reef{albert}, con lo que se termina obteniendo
 \be \sigma^{(0)}_{\mu\nu}
= -\frac{\gamma^2}{4(d-1)}\,\eta_{\mu\nu}\,\mathcal{O}^2\,.
 \labell{sigma0}
 \ee
Notamos que el coeficiente es definido negativo, hecho que luego será crucial al evaluar el cambio en la entropía de entrelazamiento.

Si bien no utilizaremos este resultado, mostramos por completitud cómo calcular el próximo término $\sigma^{(1)}_{\mu\nu}$ (que tiene dos derivadas actuando sobre el condensado $\mathcal{O}$). Pidiendo invariancia de Lorentz y simetría en $\mu,\nu$ podemos escribir la forma general
 \be
\sigma^{(1)}_{\mu\nu}=\alpha_1
\partial_\mu\mathcal{O}\partial_\nu\mathcal{O} +\alpha_2
\mathcal{O}\partial_{\mu}\partial_{\nu}\mathcal{O} +\alpha_3
\eta_{\mu\nu}\mathcal{O}\Box \mathcal{O}+ \alpha_4 \eta_{\mu\nu}(\partial
\mathcal{O})^2\,,
 \ee
con ciertas coeficientes $\alpha_i$ a definir. Nuevamente, utilizando las ecuaciones de movimiento \reef{albert} llegamos a
\begin{eqnarray}
\sigma^{(1)}_{\mu\nu}&&= \frac{\gamma^2}{4(d-1)(\Delta+1)(2\Delta+2-d)}\bigg(
\big((d-2)\,\mathcal{O}\partial_{\mu}\partial_{\nu}\mathcal{O} + \Delta \,
\eta_{\mu\nu}\mathcal{O}\Box\mathcal{O}\big)\nonumber \\
&&\qquad-\big(d\,\partial_\mu\mathcal{O}\partial_\nu\mathcal{O}
- \eta_{\mu\nu} (\partial\mathcal{O})^2\big)\bigg)\,.
\end{eqnarray}
Para un operador general $\mathcal{O}(x)$, tendríamos que considerar las sumas en la ecuación \reef{poww} a todos los órdenes en las derivadas. Sin embargo, si $\mathcal{O}$ varía poco en la escala de $R$, $\sigma^{(0)}$ da la contribución más importante al cambio en la entropía de entrelazamiento; esto es lo que supondremos aquí. Como en nuestros cálculos previos, determinamos esta contribución evaluando el funcional área \reef{action} de la métrica perturbada con el perfil original \reef{wox}. El cambio de la entropía de entrelazamiento resulta
\small
\begin{eqnarray}
\Delta S(\mathcal{O})&=& \frac{\pi L^{d-1 }R}{\lp^{d-1}}
\int \frac{d^{d-1}x}{z_0^{d-2\Delta}} (\sigma^{(0)\,i}{}_{i} -
 \sigma^{0}_{ij}\frac{x^ix^j}{R^2}) \nonumber \\
&=& -\frac{\pi\gamma^2 L^{d-1}R}{4\lp^{d-1}}\, \mathcal{O}^2\,
\int \frac{d^{d-1}x}{z_0^{d-2\Delta}}
 \left(1-\frac{r^2}{(d-1)R^2}\right)
 \nonumber\\
&=& -\frac{\gamma^2 L^{d-1}}{\lp^{d-1}}\, \frac{\pi^{3/2} \(
\Delta-\frac{(d-2)^2}{2(d-1)}   \) \Gamma[\Delta - \frac{d}{2} + 1]}{
8 \Gamma[\Delta - \frac{d}{2} + \frac52]}\,\Omega_{d-2}\,R^{2\Delta}
\mathcal{O}^2\,.
 \labell{apple}
\end{eqnarray}
\normalsize
Notar que la cota unitaria $\Delta>\frac{d}2-1$ asegura que el prefactor numérico de la última ecuación es positivo y, por lo tanto, $\Delta S$ resulta negativo. Remarcamos el hecho de que este signo negativo proviene directamente de la ecuación (\ref{sigma0}). Esto es muy interesante ya que sugiere que, al nivel de la expansión de FG, la métrica parece contener información sobre la positividad de la entropía relativa.

Es interesante comparar la contribución del condensado escalar $\mathcal O$ con la contribución más importante proveniente del tensor de energía impulso. En particular, podemos considerar un escenario en el cual los valores de expectación de ambos operadores estuviera determinado por una única escala $\mu$ (por ejemplo, la temperatura), en cuyo caso tendríamos $\mathcal{O}\sim \mu^\Delta$ y $T_{\mu\nu}\sim \mu^d$. Luego, las contribuciones correspondientes a la escala escalarían como $\Delta S(\mathcal{O})\sim (R\mu)^{2\Delta}$ y $\Delta S(T_{\mu\nu})\sim (R\mu)^{d}$ (válido en el régimen en que $R\mu\ll1$). Por lo tanto, si $\mathcal O$ es suficientemente relevante, es decir si $\frac{d}2-1<\Delta<\frac{d}2$, su contribución sería la dominante. En el caso $\frac{d}2<\Delta<d$ el tensor de energía impulso daría la contribución dominante, mientras que para el caso especial en que $\Delta=\frac{d}2$ el escaleo de ambas contribuciones sería el mismo.

De la ecuación \reef{apple} se ve que la entropía relativa es proporcional a $R^{2\Delta}$, por lo que deducimos que también se satisface la desigualdad de monotonicidad \reef{include}.

\subsubsection*{Perturbaciones por una corriente} \labell{river}

En esta sección realizamos una breve descripción de la extensión del análisis de la sección \ref{charge} para un estado con una corriente general $J_\mu$ en el borde. Recordamos que la idea es construir una métrica en la forma de FG, como en las ecuaciones \reef{bulkG} y
\reef{expand}. Por simplicidad, asumiremos que el valor de expectación de la corriente es constante y entonces, al orden más importante, la perturbación de la métrica toma la forma
\begin{equation}
\delta g_{\mu\nu}=a \, z^d \, T^{(0)}_{\mu\nu}+
z^{2d-2}\,(b \, J_\mu J_\nu+c\, \eta_{\mu\nu} J^2)\,,\labell{ff}
\end{equation}
donde las constantes $a$, $b$ y $c$ son adimensionales. Como estamos trabajamos a orden lineal en la perturbación de la métrica, podemos considerar las contribuciones de cada uno de los términos en la ecuación \reef{ff} independientemente, tal como hicimos para el caso del condensado escalar. Sabemos que la contribución de $T^{(0)}_{\mu\nu}$ satura la desigualdad \reef{123}, por lo que las perturbaciones por la corriente deben producir una contribución negativa al cambio de entropía de entrelazamiento.

Recordamos que $a$ está dado en la ecuación \reef{aaax}. Para determinar las constantes restantes, comparamos con el caso de la métrica para la brana negra cargada \reef{chargemet}. Es conveniente escribir la función en la métrica \reef{mimi} como
\begin{equation}
h=1-\gamma \tilde{z}^d+\beta \tilde{z}^{2d-2}\,,
\end{equation}
con $\gamma$ y $\beta$ constantes positivas. Para llevar la métrica a la forma de FG \reef{bulkG} debemos cambiar la coordenada radial $z$. Con esto hecho, encontramos que al orden más importante las componentes restantes de la métrica toman la forma
\begin{eqnarray}
g_{00}&=&-\left(1-\gamma \left(1-\frac{1}{d}\right)z^d+\beta
\left(1-\frac{1}{2d-2}\right)z^{2d-2}\right)\,,
\nonumber\\
g_{ij}&=&\delta_{ij}\left(1+\gamma \frac{z^d}{d}-\beta \frac{z^{2d-2}}{2d-2}\right)\,.
\labell{popper}
\end{eqnarray}
Poniendo $J_i = 0$ en la ecuación \reef{ff}, podemos comparar la expresión resultante con la de arriba y llegar a
\begin{eqnarray}
b = - 2(d-1)\,c \,,\qquad
c =\frac{\beta}{2(d - 1)J_0^2} \,.\labell{si}
\end{eqnarray}
Además, identicando $J_0\equiv\lim_{z\to0}z^{d-3}\partial_zA_0$ en la solución de la brana negra cargada, encontramos que $c$ es una constante independiente de la corriente y positiva
\begin{equation}
c =\frac{1}{4(d-1)^2(d-2)}\,.
\end{equation}
Ahora, la parte relevante de la perturbación de la métrica es
\begin{equation}
\delta g_{\mu\nu}= c\, z^{2d-2}(-2(d - 1)J_{\mu}J_\nu + \eta_{\mu\nu}J^2) \,.
\end{equation}
Insertando esta expresión en el funcional área \reef{action} arribamos a
\begin{equation}
\Delta S =\frac{\pi R L^{d-1}}{\lp^{d-1}}
\int d^{d-1}x \, \frac{1}{z_0^d}\left(\delta g_i{}^i -\delta g_{ij}
 \frac{x^i x^j}{R^2}\right)\,.
\end{equation}
Para una corriente constante la integral da
\begin{equation}
\Delta S= -\frac{\pi^{3/2} (d-3)!\, \Omega_{d-2} }{2^{d+1} \Gamma[d+\frac12]}
\,\frac{L^{d-1} R^{2d-2}}{\lp^{d-1}}\,( \vec{J}^2 + (J^0)^2) \,.\labell{contri}
\end{equation}
Por lo tanto, de la ecuación (\ref{contri}) se ve que la entropía relativa $\Delta \langle H\rangle-\Delta S$ es positiva, y además aumenta con $R^{2d-2}$, satisfaciendo así la relación de monotonicidad \reef{include}.

\subsection{Correcciones para superficies generales} \labell{generic}

En esta sección extendemos nuestro análisis al caso en que tenemos superficies de entrelazamiento que no son necesariamente esferas. Comenzamos considerando el funcional área \reef{action} para una superficie de entrelazamiento general en el borde y una perturbación del estado de vacío donde se excita al tensor de energía impulso. A orden lineal, la perturbación de la geometría del bulk sigue teniendo la forma presentada en la ecuación \reef{FGexp}, donde los coeficientes $T^{(n)}_{\mu\nu}$ están dados por \reef{box}. Como en nuestros ejemplos anteriores, el cálculo holográfico de la entropía de entrelazamiento en AdS involucrará algún perfil extremal $z_0(x^i)$, que dependerá de la geometría de la superficie de entrelazamiento en el borde. Si bien este perfil se modifica en la métrica perturbada, esta perturbación sólo contribuirá al cambio de área a segundo orden. Por lo tanto, podemos hallar el cambio lineal del área evaluando el área \reef{action} en la geometría perturbada con el perfil original $z_0$. Para una superficie de entrelazamiento genérica, la perturbación lineal de la entropía de enetrelazamiento es entonces
 \be
\Delta S=2\pi\frac{\Delta A}{\lp^{d-1}} = \frac{2\pi}{d} \int d^{d-1}x\,
\sqrt{1+(\partial z_0)^2}\,
 \sum_{n=0} z_0^{2n+1}\left(T^{(n)}{}_{i}{}^{i} -
T^{(n)}{}_{\,ij}\, \frac{\partial^iz_0\,\partial^jz_0}{1+(\partial
z_0)^2}\right)\,,
  \labell{linearA2}
  \ee
donde $(\partial z_0)^2 = \delta^{ij}\partial_i z_0\partial_j z_0$ y
la geometría del borde es el espacio de Minkowski. Previamente, concluimos en la ecuación \reef{boat} que todos los tensores $T^{(n)}_{\mu\nu}$ tienen traza nula, por lo que podemos reemplazar $T^{(n)}{}_{i}{}^{i}=T^{(n)}_{00}$ (están relacionados con la densidad local de energía $T^{(0)}_{00}$ por la ecuación \reef{box}). Por lo tanto, el primer término de arriba está controlado enteramente por la densidad de energía. Sin embargo, la conexión con la densidad de energía no es clara en el segundo término. En la sección \reef{line}, la simetría rotacional de la superficie esférica de entrelazamiento y el correspondiente perfil en el bulk \reef{wox} resultó esencial para reducir la expresión a una controlada puramente por $T^{(0)}_{00}$. Observamos entonces que en principio es lógico esperar contribuciones de otras componentes del tensor de energía impulso a $\Delta S$, incluso a orden lineal, para superficies de entrelazado con menos simetría.

Para ilustrar explícitamente este comportamiento, consideramos el caso bien estudiado de una geometría tipo \textit{slab} (``faja''), donde la superficie de entrelazamiento se compone de dos planos a $x=\pm\ell/2$ \cite{rt1,rt2,nishioka,taka}. La superficie extremal en AdS vacío tiene un perfil $z(x)$ y el área es
 \be
A=L^{d-1} B^{d-2}\int_{-\ell/2}^{\ell/2} \frac{dx}{z^{d-1}}\sqrt{1+z'^2}\,,
 \labell{actions}
 \ee
donde $B$ es un regulador infrarrojo para el tamaño de los dos planos ($B^{d-2}$ es el área de un plano). Mirando este área como una acción para $z(x)$, sacamos un vínculo para el perfil dado por una cantidad conservada \cite{rt1,rt2,nishioka,taka}
 \be
 z^{d-1} \,\sqrt{1+z'^2} = z_*^{d-1}\,.
 \labell{conserve}
 \ee
$z_*$ es el máximo valor de $z$ que la superficie extremal alcanza en el bulk a $x=0$,
 \be
z_* = \frac{ \Gamma[\frac{1}{2 (d-1)}]}{2 \sqrt{\pi} \, \Gamma[\frac{d}{2 (d-1
)}]}\ \ell\,.
 \labell{zmax}
 \ee
El cambio en la entropía \reef{linearA2} queda
 \be
\Delta S = \frac{2\pi}{d} B^{d-2} z_*^{d-1} \int_{-\ell/2}^{\ell/2} dx
\sum_{n=0} z^{2n+2-d}\left[T^{(n)}_{\,00} - T^{(n)}_{\,xx}\,
\left(1-\frac{z^{2(d-1)}}{z_*^{2(d-1)}}\right)\right]\,.
 \labell{change8}
 \ee
Vemos entonces que tanto la densidad de energía como la presión en la dirección $x$ contribuyen al resultado. Para dar un resultado más explícito, podemos simplificar el cálculo suponiendo que el valor de expectación del tensor de energía impulso es uniforme, es decir: $T^{(n)}{}_{\mu\nu}=0$ for $n\ge1$. De este modo, la ecuación \reef{change8} queda
 \bea
\Delta S&=& \frac{2\pi}{d} B^{d-2} z_*^{d-1} \int_{-\ell/2}^{\ell/2}
\frac{dx}{z^{d-2}}\left[T_{00} - T_{xx}\,
\left(1-\frac{z^{2(d-1)}}{z_*^{2(d-1)}}\right)\right]
 \nonumber\\
 &=&  \frac{\pi^{1/2}
  \Gamma[\frac{d}{d-1}] \Gamma[\frac{1}{2 (d-1)}]^2}{8d \Gamma[
\frac{3d-1 }{2 (d-1)}] \Gamma[\frac{d}{2 (d-1)}]^2}\, B^{d-2} \ell^{2}\,
\left[\left(\frac{d+1}{d-1}\right)\,T_{00} - T_{xx}\right] \,,
  \labell{linearA3}
  \eea
donde hemos utilizamos las ecuaciones \reef{conserve} y \reef{zmax}. Vemos nuevamente que el resultado contiene un término proporcional a $T_{xx}$.

De los cálculos a primer orden realizados aquí, esperamos que la desigualdad \reef{123} sea saturada, o sea $\Delta\langle H\rangle=\Delta S$. Por lo tanto, de este resultado inferimos que el hamiltoniano modular para la geometría tipo slab contenga términos lineales en el operador $T_{xx}$. Consecuentemente, para regiones con superficies de entrelazamiento generales, aparecen en el hamiltoniano modular otras componentes del tensor de energía además de $T_{00}$.

Agregamos unas observaciones más acerca de $\Delta S$ para superficies de entrelazamiento generales. Primero, notamos que si realizamos una transformada de Fourier del tensor de energía impulso, como en la ecuación \reef{fourier}, la ecuación \reef{linearA2} se reescribe utilizando \reef{sumI} como
 \bea
\Delta S &=& \pi\,\Gamma[d/2] \int d^{d-1}x \, \int d^dp \, \exp(-i p\cdot x)
\, \sqrt{1+(\partial z_0)^2}
 \labell{linearA2x}\\
&&\qquad\qquad\qquad\qquad
\frac{I_{d/2}(|p|z_0)}{(z_0|p|/2)^{d/2}}\,\left(\widehat{T}_{00}(p) -
\widehat{T}_{ij}(p)\, \frac{\partial^iz_0\,\partial^jz_0}{1+(\partial
z_0)^2}\right)\,,
  \nonumber
  \eea
donde $|p|= |\sqrt{p_\mu p^\mu}|$. Vemos que aparece la misma función de Green $I_{d/2}(|p|z_0)$ al evaluar la contribución principal a $\Delta S$ para superficies de entrelazamiento generales. Desafortunamente, sin la simetría que tienen las superficies esféricas, esta expresión no se simplifica.

En la ecuación \reef{linearA2} se realiza una suposición importante acerca de la superficie extremal en el bulk: se asume que es univaluada como función de las coordenadas del borde $x^i$ o, alternativamente, que la superficie extremal no se extiende para valores de $x^i$ más allá de la región $V$. Desafortunadamente, esto no siempre sucede. Por ejemplo, una expansión estándar tipo FG de la superficie extremal describe la superficie del bulk como $X^\mu(y^a,z)$, donde $y^a$ son coordenadas a lo largo de la superficie  de entrelazamiento y $z$ es la coordenada radial usual en el bulk \cite{adam,calc}. Cerca del borde de AdS, encontramos entonces que
 \be
X^i = X^{i}_0(y^a) - \frac{1}{2(d-2)}K^i(y^a)\, z^2 +\cdots
 \labell{expansion}
 \ee
donde $X^{i}_0(y^a)$ describe la posición de la superficie de entrelazamiento en el borde y $K^i$ es la traza de la curvatura extrínsica para la normal espacial a la superficie de entrelazamiento. Nuestras convenciones son tales que $X^i < X^{i}_0(y^a)$ corresponde a la región dentro de la superficie de entrelazamiento y $K^i=+(d-2)X^i/R^2$ para una esfera de radio $R$, centrada en $X^i=0$. Por lo tanto, para una superficie de entrelazamiento esférica, la expresión de arriba muestra cómo la superficie extremal comienza a extenderse `hacia el interior' de $V$ a medida que se extiende hacia el bulk. Sin embargo, si la geometría del bulk es tal que $K^i<0$ en alguna parte de la superficie de entrelazamiento, la superficie extremal se extiende entonces a $X^i > X^{i}_0(y^a)$. Claramente, la ecuación \reef{linearA2} no da esta situación en donde la integración incluiría contribuciones provenientes del exterior de $V$ -- ver sección \ref{puzzle} del siguiente capítulo para una discusión más detallada.

También podemos utilizar la expansión \reef{expansion} para realizar observaciones interesantes acerca de las contribuciones a $\Delta S$ de las cercanías a la superficie de entrelazamiento. Asumimos que $K^i$ es positivo en todos lados y utilizamos la ecuación \reef{expansion} para evaluar $\partial_i z$ al orden más relevante en $z$ pequeño, o equivalentemente al orden principal en $X^i- X^i_0(y^a)$,
\be
\partial_i z=-\frac{d-2}{z}\left(\frac{1}{K^i(y^a)}-\frac{ \frac{\partial X^{i}_0}{\partial y^b}\frac{\partial y^b}{\partial X^i}}{K^i(y^a) }\right) +\cdots\,.
 \ee
Podemos elegir coordenadas $y^a$ que coincidan con $d-2$ de las coordenadas $Xi$ a primer orden en la vecindad de un punto en el borde, y llamamos $r$ a la coordenada $X$ restante, ortogonal al borde. Sustituyendo en la ecuación (\ref{linearA2}), encontramos al orden más importante
 \be
\Delta S = 2\pi\frac{d-2}{d}  \int d^{d-1}x \,K^{-1}\(T_{00} -
T_{rr}\) + \cdots\,,
  \label{generalK}
  \ee
donde $K=\sqrt{\sum(K^i)^2}$. En esta expresión, hemos descartado las contribuciones de derivadas superiores con $T^{(n)}_{\mu\nu}$. Notar además que el integrando está bien aproximado sólo para regiones cercanas a la superficie de entrelazamiento. Como sabemos, en el caso especial de una esfera de radio $R$, tenemos $K^r=+(d-2)/R$; en ese caso, la expresión general (\ref{generalK}) se reduce a
\be
\Delta S = \frac{2\pi R}{d}  \int d^{d-1}x \,\bigg(T_{00} - T_{rr}\bigg) +
\cdots\,,
 \labell{roundS}
 \ee
que coincide con la expansión \reef{linearA} a primer orden en $(R-r)$. Notamos sin embargo que esto no da una buena aproximación de $\Delta H$ según \reef{spheredH}, incluso para $(R-r)$ pequeño. Esto sugiere que la expansión en infinitas derivadas de la ecuación \reef{linearA} es crucial si queremos introducir fuentes localizadas en la vecindad del borde de la región.

Como explicamos en la sección \ref{drind}, si $T_{\mu\nu}$ estuviera localizado suficientemente cerca de la superficie de entrelazamiento esperaríamos que $\Delta \langle H\rangle$ se redujera al resultado obtenido para el hamiltoniano modular de Rindler \reef{siete}. Además, en el régimen en que $\Delta S=\Delta \langle H\rangle$, se espera que esta forma se refleje en el resultado para $\Delta S$. Sin embargo, como demostramos anteriormente, esto no puede obtenerse holográficamente expandiendo sólo al orden más importante en $z$ cerca del borde, independientemente de qué tan localizado y cerca esté $T_{\mu\nu}$ de la superficie de borde. De hecho, mientras más localizado esté $T_{\mu\nu}$, más relevantes pasan a ser los términos de derivadas altas, lo que deriva en una corrección significativa al orden principal en $z$. Como concluimos arriba entonces, el conocimiento del infrarrojo de la superficie minimal en el bulk es siempre relevante.

\section{Teorías bidimensionales} \labell{two}

Para teorías de dos dimensiones en el borde, podemos describir un estado térmico mediante un agujero negro BTZ \cite{btz}. Aún en este caso, la geometría del bulk es todavía AdS$_3$ localmente. Además, para los cálculos de entropía holográfica, las superficies extremales son simplemente geodésicas. Combinando estas dos observaciones, podemos determinar las superficies extremales en forma analítica y de este modo extender nuestro análisis previo más allá de la teoría de perturbaciones. En contraste con los resultados de la sección \ref{simple}, en esta sección evaluamos $\Delta \langle H\rangle$ y $\Delta S$ para valores arbitrarios de $R T$. El presente análisis además nos permite ver el efecto de compactificar el borde de AdS como también verificar la validez de la desigualdad \reef{123} en una situación donde la superficie extremal exhibe una "transición de fase".

La ecuación \reef{static} ya describe el agujero negro tridimensional apropiado. Sin embargo, como queremos considerar compacta a la dimensión espacial, escribimos la métrica euclídea BTZ en coordenadas más familiares como
\begin{equation}
ds^2_\mt{E}=\frac{r^2-r_+^2}{R^2} d\tau^2+ \frac{L^2\,dr^2}{r^2-r_+^2} +r^2\,
d\phi^2\,,
 \labell{btzmet}
\end{equation}
donde, como es usual, $L$ es el radio de AdS y el período de $\phi$ es $2\pi$. Esta geometría es suave siempre que se elija a $\tau$ con período $\beta=2\pi
LR/r_+$, de modo que la temperatura está dada simplemente por $T = 1/\beta= r_+/(2\pi LR)$.

Las coordenadas en la ecuación \reef{btzmet} están normalizadas para que la métrica del borde sea
 \be
ds^2_{boundary} = d\tau^2 + R^2\, d\phi^2\,.
 \labell{bound9}
 \ee
Consecuentemente, la periodicidad de la dirección espacial es $2\pi R$ y el borde es un cilindro de área total $2\pi R\beta$. Remarcamos que, como la dimensión espacial es compacta, existe una transición de fase de Hawking-Page \cite{HP,HP2}. La geometría de agujero negro de arriba es la dominante para la integral de caminos gravitatoria en el régimen $T>1/(2\pi R)$, mientras que para $T<1/(2\pi R)$ la geometría dominante es la AdS$_3$ térmico. La métrica de esta última geometría se puede escribir como
\begin{equation}
ds^2_\mt{E}=\frac{r^2+L^2}{R^2} d\tau^2+ \frac{L^2\,dr^2}{r^2+L^2} +r^2\,
d\phi^2\,.
 \labell{thermalmet}
\end{equation}
Implícitamente, $\tau$ y $\phi$ quedan elegidos con la misma periodicidad como en el caso anterior, y la métrica del borde está dada nuevamente por la ecuación \reef{bound9}.

Comenzamos por centrarnos en la fase de alta temperatura, para la cual la geometría correcta se describe por la ecuación \reef{btzmet}. Es relativamente sencillo evaluar la entropía de entrelazamiento de un intervalo de tamaño angular $\Delta\phi$ (en una superficie de $\tau$ constante). Siguiendo \reef{define}, la entropía está dada por la longitud de la geodésica que conecta los extremos del intervalo $V$ en el borde \cite{rt1,rt2,nishioka,taka},
\begin{equation}
S(V)=\frac{c}{3}\log\left[\frac{\beta}{\pi \epsilon}\sinh\(\frac{\pi R
\Delta \phi}{\beta}\)\right]\,,
 \labell{finiteT}
\end{equation}
donde $c= 3L/(2 G_3)= 12\pi L/\lp$ es la carga central de la teoría conforme del borde y $\epsilon$ es un regulador de alta energía en la teoría conforme\footnote{Este cutoff aparece en el cálculo holográfico cuando se elijen como extremos de la geodésica dos puntos ubicados en la superficie $r=r_\mt{UV}=LR \epsilon$ y no sobre el borde del bulk.}.

Esta expresión coincide con el resultado previamente derivado en \cite{korepin,cacardy} para teorías conformes bidimensionales a temperatura finita. Remarcamos sin embargo que este resultado previo fue derivado para el caso en que la dirección espacial es no-compacta. Es decir, esta misma expresión \reef{finiteT}, fue derivada para cualquier teoría conforme de campos bidimensional pero sólo en el límite $R\to\infty$, manteniendo $\Delta x=R\Delta\phi$ fijo. Del cálculo holográfico vemos entonces que compactificar la dimensión espacial no afecta esta entropía de entrelazamiento a temperatura finita \reef{finiteT}. Naturalmente, esta afirmación es válida cuando la física del bulk está bien descripta por la relatividad general de Einstein, y en cuyo caso \reef{finiteT} representa la contribución más importante de una expansión para $c$ grande.

\begin{figure}
\centering
\leavevmode
\epsfysize=7cm
 \begin{tabular}{ccc}
\includegraphics[width=0.4\textwidth]{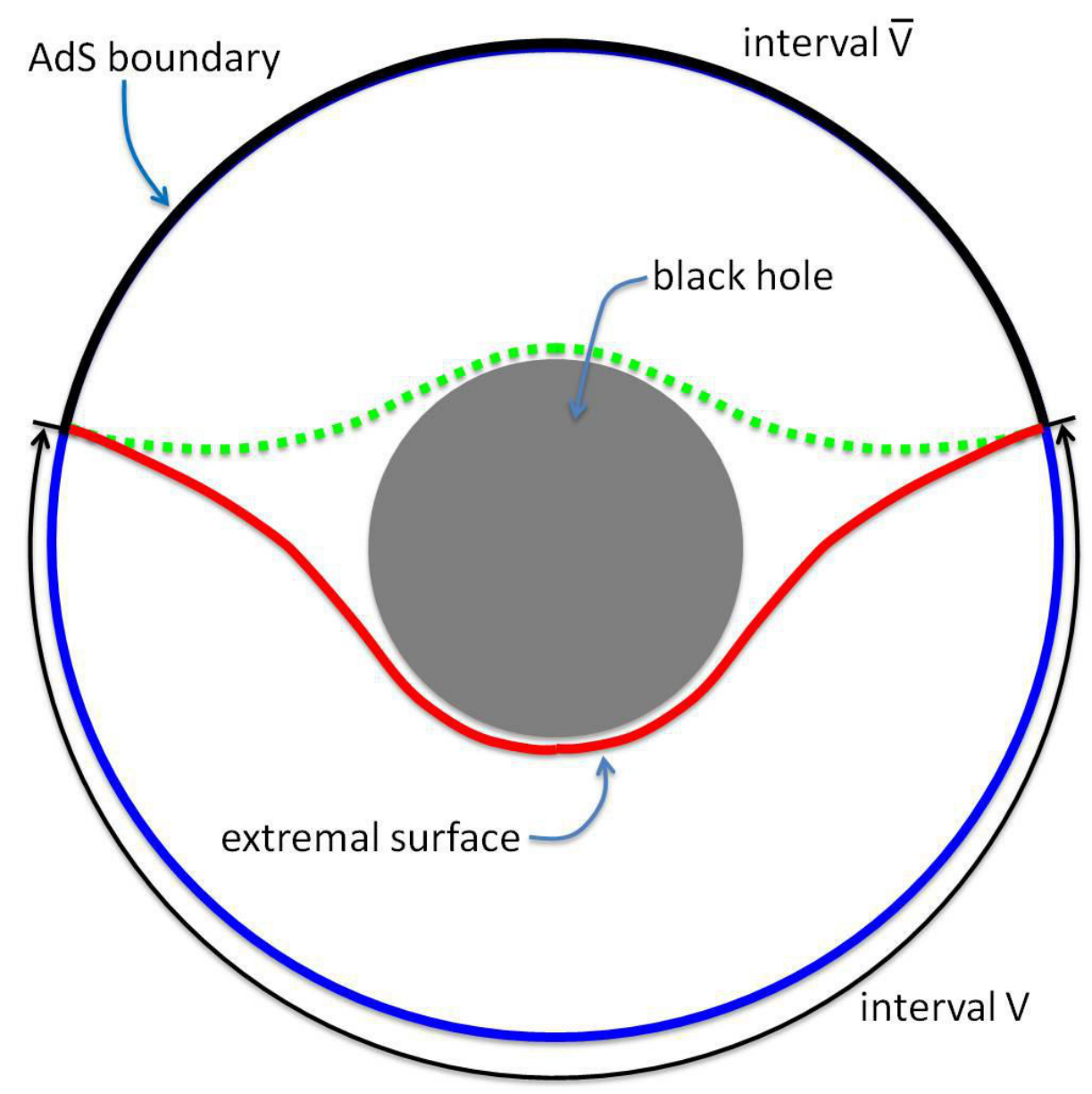}
&$\quad$&
\includegraphics[width=0.4\textwidth]{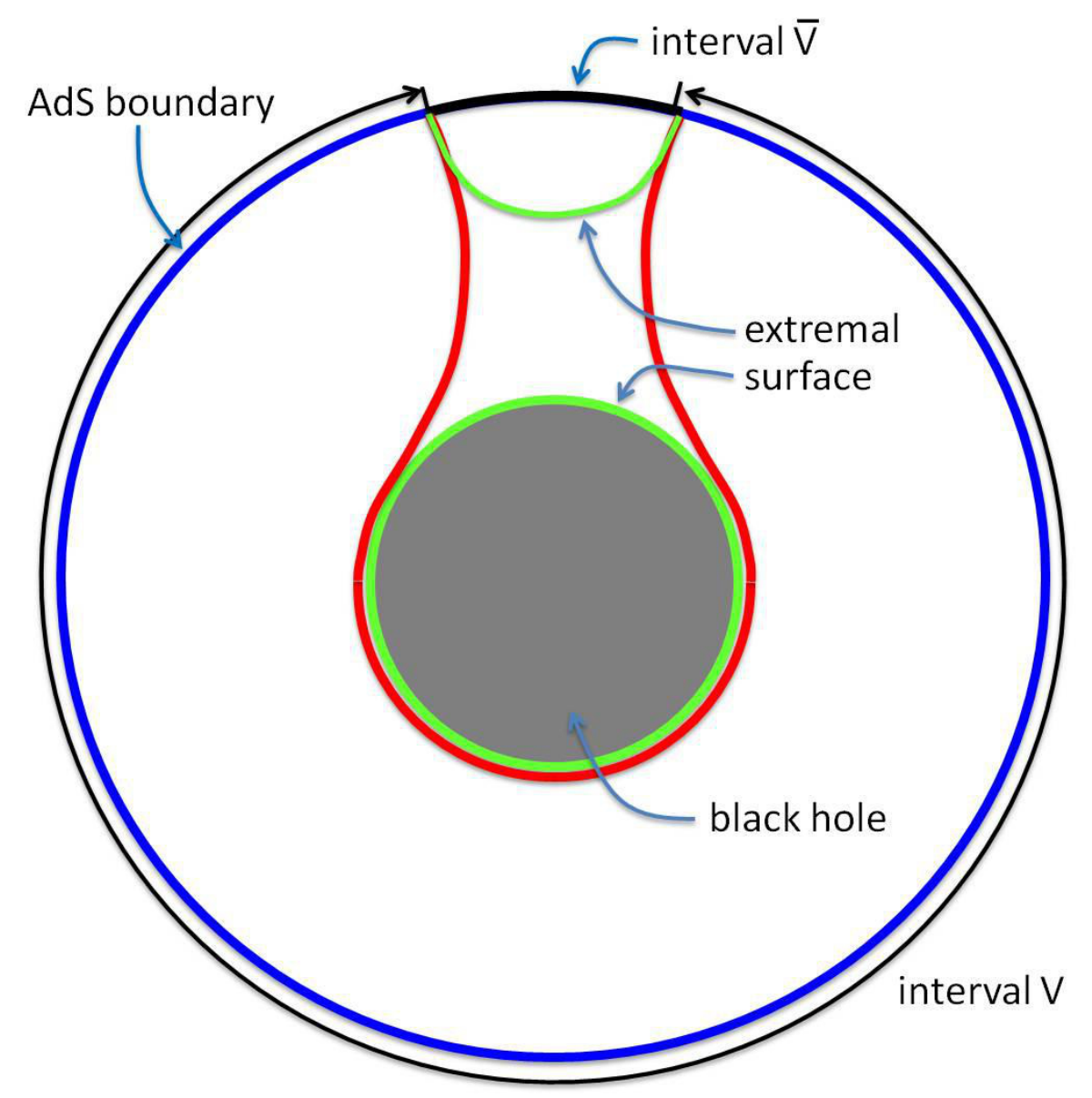}
\\
\ \ (a) & &\ (b)
\end{tabular}
\caption{\textbf{Superficies extremales en la fase de alta temperatura.} Las figuras muestran una sección de la geometría de agujero negro AdS$_3$ a tiempo constante. (a) Para $\Delta\phi$ suficientemente pequeño, la entropía holográfica de entrelazamiento \reef{define} se evalúa con la geodésica roja. La geodésica punteada de color verde, al otro lado del agujero negro, no es homóloga al intervalo $V$, sin embargo, da la entropía para el intervalo complementario $\bar V$. (b) Para $\Delta\phi$ grande, la solución relevante (en verde) está formada por dos componentes disconexas: la geodésica homóloga a $\bar V$ y la geodésica que se envuelve en el horizonte del agujero negro.}
\labell{disconnect}
\end{figure}
El resultado anterior asume implícitamente que $\Delta\phi$ es suficientemente pequeño. En esta fase de temperatura alta, para $\Delta \phi$ suficientemente grande se encuentra que la entropía de entrelazamiento experimenta una ``transición de fase'', como se describe en la figura \ref{disconnect}. Para valores grandes de $\Delta\phi$, hay dos geodésicas que conectan los puntos extremos del intervalo en el borde, y que están separados por el agujero negro, como se observa en la figura \ref{disconnect}a. Sin embargo, sólo una de ellas (la geodésica de color rojo en la figura) es homólogo al intervalo $V$ del borde y es entonces el que debe elegirse para evaluar la entropía holográfica. La otra geodésica (punteada en color verde en la figura) se puede utilizar para evaluar la entropía holográfica de la región complementaria $\bar V$, cuyo resultado es
\begin{equation}
S(\bar{V})=\frac{c}{3}\log\left[\frac{\beta}{\pi \epsilon}\sinh\(\frac{\pi R
(2\pi-\Delta \phi)}{\beta}\)\right]\,.
 \labell{finiteT2}
\end{equation}
Naturalmente, para $\Delta\phi>\pi$, $S(\bar{V})$ es menor que $S(V)$ (dada en la ecuación \reef{finiteT}). Aunque esta geodésica no es homóloga a la región de interés, puede ser utilizada para contruir otra superficie extremal que consta de dos componentes disconexas, como se ilustra en la figura \ref{disconnect}b, que es homóloga a $V$. La segunda componente es una geodésica cerrada que se cierra alrededor del horizonte del agujero negro. Esta última, contribuye a la entropía usual del horizonte
 \be
S_\mt{BH}=\frac{2\pi}{\lp}\, A(r_+)=\frac{2\pi^2
r_+}{\lp}=\frac{2\pi^2c}{3}\,\frac{R}{\beta}\,.
 \labell{finiteT3}
 \ee
Combinando estos resultados, la entropía holográfica para un intervalo arbitrario está dada por
\small
\begin{equation}
S=\frac{c}{3}\min\left[\,\log\left(\frac{\beta}{\pi \epsilon}\sinh\left(\frac{\pi R
\Delta \phi}{\beta}\right)\right)\, ,\,\log\left(\frac{\beta}{\pi \epsilon}\sinh\left(
\frac{\pi R(2\pi-\Delta \phi)}{\beta}\right)\right)+\frac{2 \pi^2 R}{\beta} \,\right]\,.
\labell{finiteT4}
\end{equation}
\normalsize
Para valores arbitrarios de $R/\beta$, el valor preciso de $\Delta\phi$ para el cual se produce la transición de fase podría calcularse numéricamente. En el límite de alta temperatura $R/\beta\gg1$, se demuestra sencillamente que la transición de fase ocurre para
 \be
\Delta\phi\simeq 2\pi - \log 2\,\frac{\beta}{2\pi R} + \cdots\,,
 \labell{ouch}
 \ee
donde los puntos supensivos indican correcciones que se suprimen exponencialmente por $e^{-2\pi^2 R/\beta}$.

Recordamos que en la fase de temperatura baja con $R/\beta<1/(2\pi)$, la geometría del bulk es simplemente la geometría de AdS$_3$ térmico \reef{thermalmet}. En este caso, existe siempre una única geodésica que une los extremos del intervalo del borde y se tiene
\begin{equation}
S=\frac{c}{3}\log\left(\frac{2R}{\epsilon}\sin(\Delta\phi/2)\right)\,.
\labell{lowT}
\end{equation}
Esta expresión coincide nuevamente con un resultado previamente derivado para teorías conformes bidimensionales \cite{wilcCFT,cacardy}. En este caso, la expresión \reef{lowT} vale para cualquier teoría conforme de campos bidimensional pero sólo en el límite $T=0$. Vemos entonces que para el cálculo holográfico no hay modificaciones en la entropía de entrelazamiento \reef{finiteT} (al orden más relevante en la expansión $c$ grande) si el sistema se encuentra a una temperatura $T$ pequeña distinta de cero.

Comparando entonces la entropía del estado a baja temperatura con la del estado vacío (es decir, a $T=0$) para un intervalo, encontramos $\Delta S=0$ (las contribuciones de orden $c$ se cancelan y por lo tanto $\Delta S$ tiene orden uno). Si en cambio comparamos la entropía de un estado de alta temperatura con el vacío, encontramos
 \bea
\Delta S&=& \frac{c}3\,\log\(\frac{1}{2\pi R T} \frac{\sinh\(\pi
RT\Delta\phi\)}{\sin\(\Delta\phi/2\)} \)
 \labell{deltaST1}\\
 &=& \frac{\pi^2}{18}\,c\,\(R^2T^2+\frac{1}{4\pi^2}\)\Delta\phi^2+O\(\Delta\phi^4\)\,.
 \eea
En la primera línea, asumimos que $\Delta\phi$ es suficientemente pequeño de modo que la entropía a temperatura finita está dada por la ecuación \reef{finiteT}. En la segunda línea, expandimos el resultado para $\Delta\phi\ll1$. Notamos que las dos expresiones \reef{finiteT} y \reef{lowT} son aproximadamente iguales en este límite $\Delta \phi\ll 1$ donde los efectos de compactificación y temperatura finita son despreciables.

El hamiltoniano modular correspondiente a una teoría conforme de campos $d$-dimensional, para el estado de vacío en una geometría cilíndrica $R\times S^{d-1}$, puede obtenerse haciendo transformaciones conformes del resultado \reef{sphereH} que se tiene para la esfera en espacio de Minkowski \cite{hmm}. Aplicando esta transformación para el caso presente con $d=2$, tenemos
\begin{equation}
H=2 \pi R^2\int_{-\Delta \phi/2}^{\Delta\phi/2}d\phi\,
\frac{\cos(\phi)-\cos(\Delta \phi/2)}{\sin(\Delta \phi/2)}\, T_{00}   \,.
\labell{cylinH}
\end{equation}
Para el vacío, en el cilindro, la densidad de energía está dada por $T_{00}=-\frac{c}{24\pi R^2}$ \cite{difra}. En general, a temperatura finita, la expresión para la densidad de energía será bastante complicada, pero al orden más importante en la carga central la densidad de energía no cambia hasta que se alcanza fase de alta temperatura $RT>(2 \pi)^{-1}$ \cite{bh3d}. En esta fase de alta temperatura tenemos $T_{00}=\frac{\pi}{6}\,c\,  T^2$, que es el resultado usual para cualquier teoría conforme en el límite de temperatura alta \cite{difra}.

Combinando estos dos resultados
 \bea
\Delta \langle H\rangle&=&\frac{2 \pi^2c}{3} \
\left[1-\frac{\Delta\phi/2}{\tan(\Delta\phi/2)} \right]\,
\(R^2T^2+\frac1{4\pi^2}\)
 \labell{dhtx8}\\
&=& \frac{\pi^2}{18}\,c\,
\(R^2T^2+\frac{1}{4\pi^2}\)\Delta\phi^2+O\(\Delta\phi^4\)\,.
 \nonumber
 \eea
La segunda línea da una expansión del resultado para  $\Delta\phi\ll1$. Comparando con la expansión de la ecuación \reef{deltaST1}, vemos que el término más importante en ambos casos coincide y entonces para $\Delta\phi$ pequeño se satura la desigualdad \reef{123}.
 
Los resultados anteriores son válidos para cualquier valor de $\Delta\phi$, por lo que podemos examinar la desigualdad \reef{123} para valores finitos. La figura \ref{paralelas1}a muestra la diferencia $\Delta \langle H\rangle-\Delta S$ como función de $\Delta\phi$ para la fase de alta temperatura. Vemos que esta diferencia es positiva y creciente para todos los ángulos. Por lo tanto, las desigualdades en las ecuaciones \reef{123} y \reef{include} se satisfacen para todo el rango de $\Delta\phi$. Notar que la transición de fase para tamaños angulares grandes, que fue discutida anteriormente, contribuye muy poco a la pendiente de las curvas. La figura \ref{paralelas1}b muestra el cociente $\Delta S/ \Delta \langle H\rangle$ como función de $\Delta\phi$. Este cociente decrece con el tamaño y se ve que $\Delta S\simeq \Delta \langle H\rangle$ para intervalos pequeños, como se discutió anteriormente.

\begin{figure}[h!]
\centering
\leavevmode
\epsfysize=4.8cm
 \begin{tabular}{ccc}
\includegraphics[width=0.45\textwidth]{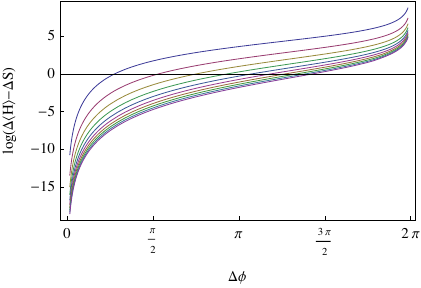}
&$\quad$&
\includegraphics[width=0.44\textwidth]{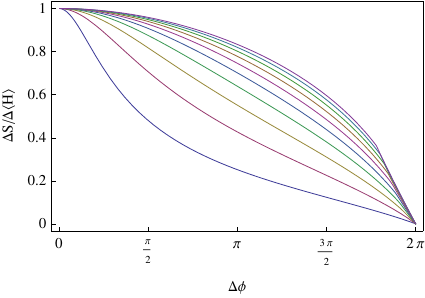}
\\
\ \ (a) & &\ (b)
\end{tabular}
\caption{\textbf{Comparación de $\Delta \langle H\rangle$ y $\Delta S$ en la fase de alta temperatura.} En (a) se grafica el logaritmo de la entropía relativa y en (b) el cociente $\Delta S/\Delta \langle H\rangle$, ambos en función de $\Delta\phi\in (0,2\pi)$. Las distintas curvas son para $\beta/R=2\pi \frac{i}{10}$, con $i=1,...,10$. Las curvas correspondientes a temperatura alta ($\beta$ pequeño) tienen mayor entropía relativa en (a) y menor cociente $\Delta S/\Delta \langle H\rangle$ en (b).}
\labell{paralelas1}
\end{figure}

\subsection{Espacio de Rindler térmico} \labell{RindlerT2}

En esta sección consideramos un estado térmico en el espacio de Rindler para una teoría conforme bidimensional. El hamiltoniano modular para este caso está dado en la ecuación \reef{2dcftRinT}. Utilizaremos esa expresión para calcular la entropía relativa entre dos estados a distintas temperaturas, es decir, tanto $\rho_0$ como $\rho_1$ describirán estados térmicos de temperatura $T_0$ y $T_1$, respectivamente. El valor de expectación del tensor de energía impulso para ambos estados es $T_{00}(x)= \frac{\pi}6\, c T_i^2$, donde $T_i$ es la temperatura correspondiente a cada estado.

Como el espacio de Rindler es infinito, es necesario introducir un regulador infrarrojo $\Lambda$, es decir, integraremos sólo en $0\le x\le \Lambda$ que caen a cero para $\Lambda$ grande. Dada la ecuación \reef{2dcftRinT}, fijamos el hamiltoniano modular $H_0=H(T=T_0)$ correspondiente a $\rho_0$. Luego, el cambio en el valor de expectación del hamiltoniano modular entre $\rho_1$ y $\rho_0$ está dado por
\begin{equation}
\Delta \langle H \rangle=\Tr\(\rho_1 H_0\)-\Tr\(\rho_0 H_0\)=
 \frac{\pi }{6}\,c\Lambda\left(\frac{T_1^2}{T_0}
-T_0\right)-\frac{c}{12}\left(\frac{T_1^2}{T_0^2}-1\right)\,.\label{dooce}
\end{equation}

En esta última expresión hemos omitido los términos proporcionales a $\exp(-2\pi T_0\Lambda)$. El primer término del lado derecho es la contribución puramente térmica y extensiva ($\propto \Lambda$), proveniente de la parte del espacio de Rindler a distancias de $x=0$ mayores a $T^{-1}$. El segundo término se atribuye a la contribución del entrelazamiento en la superficie $x=0$.

Pasamos ahora al cálculo de la entropía holográfica, para lo cual utilizamos la métrica original (\ref{static}) con $d=2$ para describir la geometría del agujero negro. La superficie extremal apropiada con la que se debe evaluar \reef{define} es la geodésica que comienza en $x=0$ sobre el borde del espacio AdS ($z=0$) y se extiende a lo largo del horizonte de eventos ($z=\zh$) para $x$ grande. Esta geodésica está dada por
\begin{eqnarray}
x(s)&=&\frac{1}{2} \zh \log\left(4 e^{2 s/L}+1\right)\,,
\labell{geodesic9}\\
z(s)&=&\frac{\zh}{\left(\frac{1}{4}e^{-2 s/L}+1\right)^{1/2}}\,,
\nonumber
 \end{eqnarray}
siendo $s$ el parámetro afín a lo largo de la geodésica. Notar que la geodésica va hacia el borde del espacio AdS cuando $s\to-\infty$ y se extiende hacia el horizonte para $s\to+\infty$. Para $d=2$, la ecuación \reef{temper1} da $T=1/(2\pi \zh)$, y recordamos que $c=12\pi L/\lp$. Imponemos un cutoff ultravioleta en $z=\epsilon$ y uno infrarrojo en $x=\Lambda$. Con esto, la entropía a una temperatura arbitraria $T$ da
\begin{equation}
S(T)=\frac{2\pi}{\lp}\(s(x=\Lambda)-s(z=\epsilon)\)=\frac{c}{12}\log
\left(\frac{e^{4 \pi T \Lambda}-1}{4 \pi^2 \epsilon^2 T^2}\right)\,.
 \labell{entropy87x}
\end{equation}
La variación en la entropía es entonces
\begin{equation}
\Delta S=S(T_1)-S(T_0)=\frac{\pi}{3}\,c\Lambda  (T_1-T_0)
-\frac{c}{6}\log\left(\frac{T_1}{T_0}\right)\,,
 \labell{entropy87y}
\end{equation}
donde nuevamente hemos omitido los términos que caen exponencialmente con $\Lambda$.

Combinando las ecuaciones \reef{dooce} y \reef{entropy87y}, la entropía relativa resulta
\begin{equation}
S(\rho_1|\rho_0)=\Delta \langle H \rangle-\Delta S=\frac{\pi }{6}\,c\Lambda T_0
  \(\frac{T_1}{T_0}-1\)^2+\frac{c}{12}\left(1
+2\log\left(\frac{T_1}{T_0}\right)-\frac{T_1^2}{T_0^2}\right)\,.
 \labell{relent98}
\end{equation}
Para $T_1$ arbitrario, este resultado es siempre positivo ya que está dominado por el primer término (dado que $\Lambda T_{0,1}\gg
1$). La región $T_1\sim T_0$ debe tratarse con más cuidado. Para
$T_1=T_0$, tanto $S(\rho_1|\rho_0)$ y la derivada primera
$\partial_{T_1}S(\rho_1|\rho_0)$ se anulan. La derivada segunda es
 \be
\partial^2_{T_1}S(\rho_1|\rho_0)=\frac{c}{6}\(2\pi\frac{\Lambda}{T_0}
-\frac1{T_0^2} -\frac1{T_1^2}\)\,.
 \labell{secondX}
 \ee
Esta cantidad es nuevamente positiva dado que $\Lambda
T_{0,1}\gg 1$, por lo que la entropía relativa es entonces positiva alrededor de la región $T_1=T_0$. Como la derivada primera se anula, tenemos nuevamente la igualdad para pequeñas perturbaciones $\delta T=T_1-T_0$
\begin{equation}
\Delta S=\Delta \langle H\rangle=c\,\left(\frac{\pi }{3}\ \Lambda
-\frac{1}{6 T_0}\right)\, \delta T\,.
\end{equation}
Para comparar el estado térmico con el vacío en el espacio de Rindler podemos tomar $T_1=0$. $\Delta \langle H \rangle$ según la ecuación (\ref{dooce}) da
\begin{equation}
\Delta \langle H \rangle=-\frac{ \pi }{6}\,c  \Lambda T_0+\frac{c}{12}\,.
 \labell{fireX}
\end{equation}
El vacío en espacio de Rindler tiene entropía logarítmica $S\sim c/6 \log(\Lambda/\epsilon)$. De este modo, la diferencia de entropía es
\begin{equation}
\Delta S=-\frac{ \pi }{3}\,c  \Lambda T_0 +\frac{c}{6}\log(\Lambda T_0)
 + {\cal O}(\Lambda^0)\,.
\end{equation}
En consecuencia, la desigualdad $\Delta \langle H\rangle>\Delta S$ es siempre válida. Notar que es incorrecto decir que la entropía relativa\footnote{Notar que cuando evaluamos $S(\rho_1|\rho_0)=\Delta \langle H\rangle-\Delta S$,
$\rho_1$ corresponde al estado vacío mientras que $\rho_0$ es el estado térmico. En los cálculos previos los roles están invertidos.} se aproxima a cero ($\Delta
\langle H\rangle\to\Delta S$) para temperaturas pequeñas dado que debemos tener siempre $\Lambda T_0 \gg 1$. De hecho, el estado de vacío en espacio de Rindler siempre se encuentra a una distancia estadística infinita de cualquier estado térmico dado que suficientemente lejos del origen, es decir para $x\gg 1/T_0$, los modos termalmente excitados sienten una temperatura de Unruh casi nula. Esto no sucede al comparar el vacío y un estado térmico sobre un intervalo finito de tamaño $\ell$. Para temperaturas suficientemente pequeñas ($T_0\lesssim 1/\ell$) la diferencia de hamiltonianos modulares compensa el cambio en la entropía de entrelazamiento. En particular, en la sección anterior, vimos que $\Delta \langle H\rangle$ y $\Delta S$ eran siempre prácticamente iguales para $\Delta\phi$ suficientemente pequeño, independientemente de la temperatura.
\include{aplicaciones}
\chapter{\label{ch:negativa}Entropía y energía negativa}

En la secciones \ref{ch:relativa} y \ref{ch:hee} nos centramos en el estudio de la propiedad de positividad de la entropía relativa y de sus consecuencias, en particular, en el marco holográfico. En esta sección estudiamos otra propiedad de la entropía relativa, la \textbf{monotonicidad} definida por la ecuación (\ref{mon}). Demostramos que de esta desigualdad se desprende una relación reminiscente a la cota de Bekenstein (\ref{bek}), y en particular, que esta nueva relación impone severas restricciones a la localización espacial de energía negativa en las teorías conformes.

Este trabajo fue realizado en colaboración con Horacio Casini y parte de los resultados de este capítulo ha sido publicado en \cite{blanco3}.

\section{Energía en teoría de campos}

En las teorías clásicas para la materia, se requiere que el tensor de energía impulso $T_{\mu\nu}$ satisfaga una serie de relaciones que se conocen como \textit{condiciones clásicas de energía} \cite{review2}. Estas relaciones se imponen para que la teoría satisfaga ciertos supuestos físicos que se entiende que deben respetarse. Por ejemplo, la condición de energía nula (NEC, por Null Energy Condition)
\begin{equation}
T_{\mu\nu}u^{\mu}u^{\nu} \geq 0\,,\,\,\textrm{para}\,\,\textrm{todo}\,\,u\,\,\textrm{nulo}\,,\label{nec1}
\end{equation}
está relacionada con el teorema de Hawking que implica que el área de los agujeros negros crece con el tiempo \cite{area}. Otra de ellas, la condición de energía débil (WEC, por Weak Energy Condition)
\begin{equation}
T_{\mu\nu}u^{\mu}u^{\nu} \geq 0\,,\,\,\textrm{para}\,\,\textrm{todo}\,\,u\,\,\textrm{temporal}\,,\label{wec1}
\end{equation}
está asociada a que desde el punto de vista de cualquier observador la \textit{densidad de energía es no negativa}.

Debido a la simetría de Lorentz y a la existencia de un estado fundamental, la energía es siempre positiva en teoria de campos cuántica. Sin embargo, \textit{la densidad de energía puede tomar valores negativos} si es compensada por la presencia de energía positiva en otras regiones del espacio. De hecho, en toda teoría de campos siempre hay estados con densidad de energía negativa \cite{nega-necesaria}. Esto es un fenómeno puramente cuántico y no se espera que persista en el límite clásico. Por ejemplo, la densidad de energía para un campo escalar libre clásico $T^{00}(x)=\frac{1}{2}\left(\dot{\phi}^2+(\nabla \phi)^2+m^2 \phi^2\right)$ es definida positiva. En el proceso de cuantización, la sustracción de la energía de punto cero hace que el operador densidad de energía tenga signo no definido.

La violación cuántica de la condición de energía nula \reef{nec1} es necesaria para la evaporación de agujeros negros. También se entiende que la existencia de agujeros de gusano atravesables y máquinas del tiempo requieren la presencia de una cantidad suficiente de energía negativa \cite{wormhole}. Más recientemente, en el contexto de modelos holográficos, las condiciones de energía en el espacio tiempo del bulk han sido relacionadas con propiedades de la teoría de campos del borde, tales como la subaditividad fuerte de la entropía de entrelazamiento \cite{wall} y la irreversibilidad del grupo de renormalización \cite{msinha1,cteo}.

En conexión con estas aplicaciones, es de gran interés conocer qué tanta energía negativa permite la mecánica cuántica en violación de las condiciones clásicas de energía. A pesar de que la respuesta a esta pregunta en espacio curvo está lejos de ser respondida, en espacio de Minkowski se ha realizado un progreso importante \cite{review2}. Una desigualdad de energía cuántica (QEI, por quantum energy inequality) es en forma genérica una cota para una combinación de los valores de expectación de componentes del tensor de energía impulso, pesadas con alguna función del espacio tiempo. En la actualidad se conocen varias de estas cotas \cite{varios,varios2,varios3} (ver también \cite{review2} y los trabajos a los que hace referencia). Sin embargo, la mayor parte de los ejemplos conocidos sólo se aplican a campos libres y las desigualdades típicas no restringen la distribución \textit{espacial} (ver sin embargo \cite{oso}), sino que establecen una cota para la posible duración en el tiempo de la densidad de energía negativa en un punto específico del espacio.

En las siguientes secciones, mediante un sencillo argumento mostraremos que, en una teoría de campos conforme, la distribución espacial de energía negativa se encuentra severamente restringida. Más precisamente, la energía negativa parece estar confinada a ocupar regiones cercanas a las zonas de energía positiva y tiene que estar menos dispersa que ella. Veremos también que es posible obtener una restricción más fuerte a partir de propiedades de la entropía relativa y analizaremos la relación de las desigualdades obtenidas con la cota de Bekenstein.

\section{Cotas para la distribución espacial de energía negativa}

La existencia de cotas para la presencia y manipulación de energía negativa es necesaria para la validez de la segunda ley de la termodinámica. Por ejemplo, si se dejara caer una cantidad de energía negativa en un agujero negro podría reducir el tamaño del mismo y su entropía sin una compensación en la entropía emitida mediante el aumento de la radiación de Hawking.

Interesantemente, como comentamos en la sección \ref{secbeke}, un experimento pensado que involucra agujeros negros y la segunda ley generalizada de la termodinámica da lugar a la cota de Bekenstein
 \begin{equation}
S_A\le 2 \pi \,R \,E_A\,,\label{bek1}
\end{equation}
donde recordamos que $S_A$ y $E_A$ son la entropía y energía de cualquier objeto que se deja caer al agujero negro. Dado que la entropía en mecánica cuántica es siempre positiva, esto parecería implicar que la energía de una región no puede ser negativa. Vimos que esto no es estrictamente correcto y que el problema es que las cantidades que aparecen en la ecuación \reef{bek} deben definirse con cuidado en teoría cuántica de campos. También explicamos que una versión bien definida de la cota de Bekenstein requiere escribir el lado derecho de (\ref{bek}) como una substracción $\Delta S_A=S_A^1-S_A^0$ entre la entropía $S_A^1=-\textrm{tr} \rho_A^1 \log \rho_A^1$ del estado del objeto $\rho_A^1$ reducida a la región $A$ y la entropía del vacío $S_A^0=-\textrm{tr}\rho_A^0 \log \rho_A^0$ en la misma región \cite{beke0,marolf}. Esto elimina las divergencias ultravioletas artificialmente producidas por la localización que aparecen en la entropía. Adicionalmente, el producto $2\pi E R$ en el lado derecho de la ecuación (\ref{bek}) debe reescribirse en términos del hamiltoniano modular $H_A=-\log (\rho_A^0)$ correspondiente a la matriz densidad reducida del vacío en $A$. La relación entre $H_A$ con la energía y el tamaño se clarifica si elegimos $A$ como el espacio de Rindler $x^1>0$. En este caso, $H_A$ está dado por el generador de la simetría de boost dentro de $A$ (como vimos en el capítulo \ref{ch:modular}) para cualquier teoría de campos en dimensión espacio temporal $d$
\begin{equation}
H_A=2\pi \int_{x^1>0} d^{d-1}x \,\,x^1 \,T^{00}(x)\,.\label{boost}  
\end{equation}
En el capítulo anterior, vimos que teniendo en cuenta (\ref{boost}), una interpretación cuántica natural de (\ref{bek}) está dada por
\begin{equation}
\Delta S_A\le \Delta \langle H_A\rangle\,,\label{ine}
\end{equation}
donde $\Delta \langle H_A\rangle= \textrm{tr}(\rho_A^1 H_A)-\textrm{tr}(\rho_A^0 H_A)$ es la variación del valor de expectación del hamiltoniano modular entre el estado del objeto y el vacío. Ya comentamos que (\ref{ine}) es válida para cualquier región $A$ (no necesariamente la mitad del espacio) y para cualquier estado $\rho^1$, debido a la positividad de la entropía relativa
 \begin{equation}
 S(\rho_A^1|\rho_A^0)=\textrm{tr}(\rho^1_A \log \rho^1_A-\rho^1_A \log \rho^0_A)=\Delta \langle H_A\rangle-\Delta S_A\ge 0\,.
 \end{equation}
La ecuación (\ref{ine}) permite valores negativos a ambos lados de la desigualdad, a diferencia de la ecuación (\ref{bek}), ya que el valor de expectación de (\ref{boost}) puede ser negativo para algunos estados. En este caso, la energía negativa en $A$ debe ser acompañada por una reducción de la entropía de entrelazamiento del estado con respecto a la del vacío.

La consistencia de la desigualdad (\ref{ine}) con la correspondiente a la región complementaria $\bar{A}$, requiere la existencia de alguna restricción para la distribución de energía negativa, como hemos mencionado al final de la sección \ref{ch:mencion}. Siguiendo esa idea, en la sección \ref{ch:nuevobeke} presentamos una versión cuántica de la cota de Bekenstein, que resulta universalmente válida e involucra sólo cantidades positivas a ambos lados de la desigualdad. Esta nueva desigualdad mejora la cota para la localización de energía negativa que se deduce utilizando puramente argumentos de simetría conforme y cuya deducción presentamos en la siguiente subsección.

\subsection{Un generador de simetría positivo}

En una teoría covariante de Lorentz es posible transformar el hamiltoniano $H=P^0$ con un boost a cualquier operador de la forma $P_\mu a^\mu$, con $a^\mu$ un vector en el cono de luz futuro. Como estos operadores tienen el mismo espectro que el hamiltoniano esto nos dice que son definidos positivos. En una teoría conforme, el grupo de Lorentz es un subgrupo del grupo de transformaciones conformes, que permiten transformar al hamiltoniano dentro de un cono más grande de operadores positivos.

Vamos a hacer entonces una transformación conforme del hamiltoniano. Para que el resultado sea lo más simétrico posible, consideramos primero la transformación conforme $\hat{I}=R.I$, donde $R$ es una reflexión espacial e $I$ es la inversión de coordenadas $x^{\mu\prime}=\frac{x^\mu}{x^2}$. La reflexión espacial $R$ es necesaria para que $R.I$ pertenezca a la parte del grupo conforme conectada con la identidad. La transformación de coordenadas compuesta $\hat{I}^{-1}. \delta t . \hat{I}$, donde $\delta t$ es una traslación temporal en un tiempo pequeño  $\delta t^\mu\equiv (\delta t,0,0,0)$ es, a primer orden en $\delta t$,
\begin{equation}
x^{\mu\prime}\simeq x^\mu+  \, x^2 \, \delta t^\mu - 2  x^\mu (\delta t^\alpha x_\alpha)\,.
\end{equation}
El generador $G$ que implementa esta transformación conforme se obtiene mirando el efecto en los puntos de la superficie $x^0=0$,
 \begin{equation}
G=\int d^{d-1}x\, |\vec{x}|^2 T^{00}(x)\,.\label{this}
\end{equation}
Por lo tanto, se tiene para los operadores cuánticos $G=\hat{I}^\dagger.H.\hat{I}$, y $G$ es definido positivo.

Mirando la forma general que tienen los generadores conformes se deduce que una transformación de $P^0$ (por una de estas transformaciones conformes generales) resulta en una combinación lineal con coeficientes positivos del hamiltoniano y las traslaciones de $G$.

\subsection{Localización de energía negativa}

La positividad de $G$ parece decir que el ``momento de inercia'' de la densidad de energía es positivo. Para los valores de expectación en cualquier estado se tiene
\begin{equation}
\int d^{d-1}x\,\, |\vec{x}-\vec{x}_0|^2 \langle T^{00}(x)\rangle \ge 0\,.\label{30}
\end{equation}
La cota se hace óptima al minimizar (\ref{30}) sobre la posición $\vec{x}_0$.

Para clarificar qué dice (\ref{30}) sobre la distribución de energía, llamemos $E_+$ a la cantidad total de energía positiva y $E_-$ al valor absoluto de la cantidad total de energía negativa
\begin{equation}
E_\pm=\int d^{d-1}x\,\,\theta(\pm\langle T^{00}(x)\rangle)\,\,|\langle T^{00}(x)\rangle|\,.
\end{equation}
La energía total es entonces $E=E_+-E_-\ge 0$. Definimos también los centros de energía positiva y negativa $\vec{x}_\pm$ en la forma natural
\begin{equation}
\vec{x}_\pm E_\pm=\int d^{d-1}x\, \vec{x} \,\theta(\pm\langle T^{00}(x)\rangle)\,\,|\langle T^{00}(x)\rangle|\,,
\end{equation}
y las dispersiones $r_\pm$ de las distribuciones de energía positiva y negativa como
\begin{equation}
(r_\pm)^2 E_\pm =\int d^{d-1}x\, |\vec{x}-\vec{x}_{\pm}|^2 \,\theta(\pm\langle T^{00}(x)\rangle)\,\,|\langle T^{00}(x)\rangle|\,.
\end{equation}  
Teniendo en cuenta que el valor de $\vec{x}_0$ que minimiza la ecuación (\ref{30}) es
\begin{equation}
\vec{x}_0=\frac{E_+ \vec{x}_+-E_- \vec{x}_-}{E}\,,\label{punto}
\end{equation}
obtenemos la siguiente desigualdad
\begin{equation}
|\vec{x}_+-\vec{x}_-|^2\le E\frac{E_+ r_+^2-E_- r_-^2}{E_+ E_-}\,.\label{bi}
\end{equation}
En particular, el tamaño intrínseco del ``momento de inercia'' de energía negativa está acotado por lo de energía positiva
\begin{equation}
 E_- r_-^2\le E_+ r_+^2 \,.
\end{equation}
De la ecuación (\ref{bi}) vemos que la energía positiva y negativa no pueden separarse demasiado. Por ejemplo, si tenemos una región de densidad de energía negativa pequeña $r_-\ll r_+$, y una pequeña cantidad de energía , $E_-\ll E_+$, resulta $|\vec{x}_+-\vec{x}_-|\lesssim \sqrt{\frac{E}{E_-}} r_+$. 

\section{Una nueva cota tipo Bekenstein}\label{ch:nuevobeke}

Consideramos una teoría cuántica de campos y una región $A$ arbitraria. Recordamos la definición del operador hamiltoniano modular completo
\begin{equation}
\hat{H}_A=H_A-H_{\bar{A}}=-\log(\rho^0_A)\otimes 1+1\otimes \log{\rho^0_{\bar{A}}}\,,\label{full}
\end{equation}
Escribimos al estado vacío en su descomposición de Schmidt sobre el espacio producto ${\cal H}_A\otimes{\cal H}_{\bar{A}}$, con lo que se tiene que
\begin{equation}
\hat{H}_A|0\rangle=(H_A-H_{\bar{A}})|0\rangle=0\,.\label{refi}
\end{equation}
Consideramos ahora un estado excitado $\rho^1$ y cierta región $B$ tal que $B\subseteq A$. De la monotonicidad \reef{mon} de la entropía relativa, se obtiene que
\begin{equation}
\Delta \langle H_A\rangle- \Delta S_A\ge \Delta \langle H_B\rangle- \Delta S_B\,,\label{ii}
\end{equation}
y para las regiones complementarias
\begin{equation}
\Delta \langle H_{\bar{B}}\rangle- \Delta S_{\bar{B}}\ge \Delta \langle H_{\bar{A}}\rangle- \Delta S_{\bar{A}}\,.\label{jj}
\end{equation}
De la propiedad (\ref{refi}) se desprende que $\langle H_A\rangle^0 =\langle H_{\bar{A}}\rangle^0$ y $\langle H_B\rangle^0 =\langle H_{\bar{B}}\rangle^0$. Además, como el vacío es un estado puro, se tiene $ S_A^0=S_{\bar{A}}^0$ y $S_B^0= S_{\bar{B}}^0$. Luego, sumando las ecuaciones (\ref{ii}) y (\ref{jj}) obtenemos 
\begin{equation}
\langle \hat{H}_A-\hat{H}_B\rangle^1 \ge S_A^1-S_B^1+S_{\bar{B}}^{1}-S_{\bar{A}}^1\equiv 2\, S_f(A,B)\,.\label{perio}
\end{equation}
En esta desigualdad el estado vacío está involucrado sólo a través de la definición de los hamiltonianos modulares. La combinación de entropías que aparece en la ecuación (\ref{perio}), que denotaremos como $2\, S_f(A,B)$ por conveniencia futura, es siempre positiva como consecuencia de la propiedad de monotonicidad débil de la entropía $S(X)+S(Y)\ge S(X-Y)+S(Y-X)$ aplicada a las regiones $X=A$, $Y=\bar{B}$. Como la relación (\ref{perio}) es válida para cualquier estado $\rho^1$, la diferencia $\hat{H}_{A}-\hat{H}_B$ es un operador positivo si $B\subseteq A$ (sin relacionarlo con la entropía relativa, esto ha sido discutido en el contexto de la formulación algebraica de la teoría de campos en \cite{1}).

La desigualdad (\ref{perio}) es nuestra propuesta de una nueva y universalmente válida cota de Bekenstein cuántica. Para ver la relación de esta nueva cota con la desigualdad original (\ref{bek}), aplicamos (\ref{perio}) al caso en que las regiones son dos semiespacios, uno incluído dentro del otro, es decir, $A$ es la región dada por $x^1>0$, $B$ está dada por $x^1>L>0$, y $A-B$ es entonces una región tipo faja (strip) de ancho $L$. Utilizando la forma conocida para el hamiltoniano modular (\ref{boost}) obtenemos
\begin{eqnarray}
 \pi\, L\,  E\ge S_f(A,B)=\label{piso}\frac{1}{2}( S(x^1>0)-S(x^1>L)\nonumber \\
 +S(x^1<L)-S(x^1<0))\,.
\end{eqnarray}
En el límite clásico, esta fórmula es reminescente a la formulación original de Bekenstein (\ref{bek}).

Para comprender un poco mejor el significado de $S_f(A,B)$ se puede pensar en calcularla para el caso de un gas en equilibrio térmico a alta temperatura. En este caso, la particular combinación de entropías involucradas en $S_f(A,B)$ causa que la contribución del entrelazamiento en los bordes de las regiones se cancele; $S_f(A,B)$ retiene exactamente la entropía térmica extensiva del gas dentro de $A$ pero fuera de $B$. Por este preciso motivo, insertamos un factor $2$ en la definición de $S_f(A,B)$. Observamos también que para un estado global puro se tiene $S(X)=S(\bar{X})$, y entonces $S_f(A,B)\equiv 0$. Por lo tanto, $S_f(A,B)$ no mide la entropía de la faja $A-B$ producida por pares de partículas entrelazadas (una en $A-B$ y la otra afuera). Esto, en cierto grado, nos lleva a interpretar a $S_f(A,B)$ como una {entropía \sl ``libre'' (o ``global'') localizada entre los bordes de $A$ y $B$}. 
 
Es interesante notar que, interpretando la entropía del estado global como si proviniera de la traza parcial sobre un sector oculto $\aleph$ utilizado para purificar el estado ($\rho^1=\textrm{tr}_{\aleph} |\psi\rangle \langle\psi|$, para algún vector $|\psi\rangle$ en ${\cal H}\otimes \aleph$), se puede escribir 
 \begin{equation}
 S_f(A,B)=\frac{I(A,\aleph)-I(B,\aleph)}{2}\,,
 \end{equation}           
donde $I(X,Y)=S(X)+S(Y)-S(X\cup Y)$ es la información mutua entre $X$ e $Y$, anteriormente introducida en la ecuación \reef{infomutua}. Por lo tanto, $S_f(A,B)$ es aproximadamente extensiva en $A-B$ y depende sólo de $A-B$ en la medida en que $I(X,\aleph)$ sea aproximadamente extensiva para regiones espaciales $X$. En esta representación, se ve también que $S_f(A,B)$ es efectivamente monótonamente creciente con el tamaño de $A-B$, ya que la información mutua crece monotónamente. Es más, satisface en forma trivial cierta extensividad parcial: Si $C\subset B$ tenemos $S_f(A,C)=S_f(A,B)+S_f(B,C)$.
 
Si bien el aspecto general de (\ref{piso}) es similar a la formulación original de Bekenstein, se diferencian en algunos aspectos interesantes. Por ejemplo, $L$ ahora es la dimensión más pequeña de la región tipo faja en lugar de ser el diámetro que circunscribe a la región como se plantea en la formulación original de Bekenstein (en este sentido, nuestra cota es similar a la propuesta en \cite{boussoa}). Además, en el lado izquierdo tenemos ahora la energía global $E$ y no una medida de la energía en la región.

La nueva formulación (\ref{piso}) (o más generalmente (\ref{perio})) también presenta diferencias respecto de la primera versión cuántica de la cota de Bekenstein (\ref{ine}). Por empezar, (\ref{piso}) involucra a la energía y entropía en un estado, en contraste con la diferencia entre dos estados en (\ref{ine}). Matemáticamente, (\ref{piso}) proviene de una propiedad diferente: la monotonicidad de la entropía relativa y no su positividad como en el caso de (\ref{ine}).  Finalmente, en contraste con (\ref{ine}), ambos lados de la nueva desigualdad son positivos.

\subsection{Entropía y energía negativa}

La conexión entre la positivad de $G$ y la entropía relativa proviene del hecho de que, en una teoría conforme, el hamiltoniano modular para una región esférica en el vacío es proporcional al generador de transformaciones conformes que mantiene la esfera invariante \cite{modular}. El hamiltoniano modular para una esfera $A$ de radio $R$ es
\begin{equation}
H_A=2 \pi \int_{|\vec{x}|\le R} d^{d-1}x\, \frac{R^2-|\vec{x}|^2}{2 R} T^{00}(x)\,,\label{tera}
\end{equation}
mientras que para la región complementaria $\bar{A}$ es
\begin{equation}
H_{\bar{A}}=2 \pi \int_{|\vec{x}|\ge R} d^{d-1}x\, \frac{|\vec{x}|^2-R^2}{2 R} T^{00}(x)\,.\label{tira}
\end{equation}
Si definimos $A$ como una esfera de radio $R_1$ y $B$ como una esfera concéntrica de radio $R_2$, con $R_2<R_1$, podemos utilizar las ecuaciones (\ref{tera}) y (\ref{tira}) para obtener 
\begin{equation}
\frac{\pi}{2} (R_1-R_2)\left(\langle P^0\rangle +  \langle G\rangle/(R_1 R_2)\right)\ge \,S_f(A,B)>0\,.\label{poro}
\end{equation}
Tomando el límite en que $R_2\rightarrow 0$ se reobtiene la positividad de $G$ expresada en la ecuación (\ref{this}).

\begin{figure}[h]
\begin{center}
\includegraphics[width=6cm]{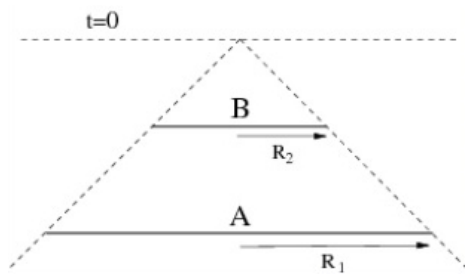}
\end{center}
\caption{\textbf{Construcción geométrica para obtener una cota sólo en términos de $G$}. Dos esferas espaciales, $A$ de radio $R_1$ y $B$ de radio $R_2$ localizadas en el cono de luz pasado. $S_f(A,B)$ es una medida de la entropía que atraviesa el cono nulo entre las fronteras de $A$ y $B$.} \label{ffff}
\end{figure}

También es posible utilizar la desigualdad (\ref{perio}), aplicada a dos esferas en una teoría conforme, para obtener una cota sólo en términos del operador $G$, sin que aparezcan contribuciones del hamiltoniano como en la cota obtenida en la ecuación (\ref{poro}). Para esto, debemos utilizar esferas situadas en tiempos diferentes. Tomamos $A$ como una esfera de radio $R_1$ a un tiempo $t=-R_1$, centrada en el origen de coordenadas espacial, y $B$ otra esfera de radio $R_2<R_1$ centrada en el origen de coordenadas pero a un tiempo $t=-R_2$ (ver figura \ref{ffff}). $A$ y $B$ se encuentran en el cono de luz pasado.

La corriente conforme que da el generador conforme correspondiente a $\hat{H}_A$ es
\begin{equation}
J_A^\mu
= 2 \pi T^{\mu\beta} x_\beta+T^{\mu\beta}\left(c^\alpha x_\alpha x_\beta-\frac{1}{2} c_\beta x^\alpha x_\alpha\right)
\end{equation} 
con $c_\beta\equiv(2 \pi/R_1,0,...,0)$. La corriente $J_B^\mu$ se obtiene de $J^\mu_A$ reemplazando $R_1$ por $R_2$. Por lo tanto, en este caso la cota se escribe
\begin{eqnarray}
&&\langle \hat{H}_A-\hat{H}_B\rangle=\int d^{d-1}x\, (J^0_A-J^0_B)\\&&=\pi \left(\frac{1}{R_2}-\frac{1}{R_1}\right) \int |\vec{x}|^2\,\langle  T^{00}(x)\rangle \ge 2\,S_f(A,B)\,.\nonumber
\end{eqnarray} 
En particular, eligiendo el centro de las coordenadas espaciales en el punto $\vec{x}_0$ dado por la ecuación (\ref{punto}), obtenemos
\begin{eqnarray}
&&(E_+ r_+^2-E_- r_-^2)-\frac{E_+ E_-}{E}|\vec{x}_+-\vec{x}_-|^2\nonumber\\ &&\hspace{.4cm}\ge\max_{R_1,R_2,R_1>R_2}\left(\frac{2 R_1 R_2}{(R_1-R_2)\pi}\,S_f(R_1,R_2)\right)\,.
\end{eqnarray}
Por lo tanto, la entropía hace a la cota para la localización de energía negativa más restrictiva. Esto es bastante natural desde el punto de vista de la motivación original a la búsqueda de cotas para la energía negativa basada en la segunda ley \cite{ford}: un estado puro con energía negativa que se mezcla con un estado térmico disminuye su energía, y consecuentemente el espacio de fases disponible, posiblemente reduciendo su entropía y violando la segunda ley. En este sentido, la materia con energía negativa y entropía positiva sólo podría empeorar el problema, dado que la segunda ley se violaría aún más. Sin embargo, la entropía de entrelazamiento entre la fuente de energía negativa y el reservorio de energía positiva no representa un problema adicional, dado que esta entropía desaparece una vez que estos estados se mezclan; en otros términos, esta entropía de entrelazamiento no es considerada en el balance de entropía para los estados globales inicial y final en la segunda ley. Esto se refleja en el hecho de que la cantidad entrópica específica que aparece en la cota `no siente' el entrelazamiento espacial.

Como comentario final de esta sección, queremos mencionar que es posible que la existencia de dos formas cuánticas de la cota de Bekenstein quizás sugiere que hay otra forma de entender a la segunda ley cuando el entrelazamiento está involucrado debido a la presencia de agujeros negros.

\section{Cotas para estados no puros}

En la sección \ref{ch:nuevobeke} derivamos la desigualdad (\ref{perio}) a partir de la propiedad de monotonicidad de la entropía relativa. Es importante notar que en esta ecuación, el estado de referencia $\rho_0$ era el vacío, un estado puro, lo que aseguraba que $S_A^0=S_{\bar{A}}^0$ y $S_B^0=S_{\bar{B}}^0$, y que $\langle H_A\rangle^0 =\langle H_{\bar{A}}\rangle^0$ y $\langle H_B\rangle^0 =\langle H_{\bar{B}}\rangle^0$. En esta sección vamos a permitir que $\rho_0$ sea un estado no puro. El objetivo es encontrar una desigualdad general entre la entropía y la energía modular para un estado $\rho_1$ arbitrario.

Partimos entonces de las ecuaciones que expresan la monotonicidad de la entropía relativa. Si consideramos dos regiones $A$ y $B$, con $B \subseteq A$, y $\rho_0$ y $\rho_1$ dos estados arbitrarios, la monotonicidad de la entropía relativa implica que
\begin{equation}
\Delta \langle H_A \rangle - \Delta S_A \geq \Delta \langle H_B \rangle - \Delta S_B \,,
\label{primera}
\end{equation}
y
\begin{equation}
\Delta \langle H_{\bar{B}} \rangle - \Delta S_{\bar{B}} \geq \Delta \langle H_{\bar{A}} \rangle - \Delta S_{\bar{A}} \,,
\label{segunda}
\end{equation}
dado que $\bar{A} \subseteq \bar{B}$. Sumando las ecuaciones (\ref{primera}) y (\ref{segunda}) obtenemos la siguiente desigualdad
\begin{equation}
\langle \hat{H}_A-\hat{H}_B \rangle _1 - \langle \hat{H}_A-\hat{H}_B \rangle _0 \geq \left(S_A^1-S_B^1+S_{\bar{B}}^1-S_{\bar{A}}^1 \right) - \left(S_A^0-S_B^0+S_{\bar{B}}^0-S_{\bar{A}}^0 \right)\,,
\label{desi}
\end{equation}
donde $\hat{H}_X$ representa al hamiltoniano modular completo de la región $X$
\begin{equation}
\hat{H}_X = H_X-H_{\bar{X}}\,.
\end{equation}
La entropía libre $S_f$ (definida en la ecuación \reef{perio}) es siempre positiva como consecuencia de la subaditividad fuerte de la entropía aplicada a $A$ y $\bar{B}$
\begin{equation}
S_A+S_{\bar{B}} \geq S_{A-\bar{B}}+S_{\bar{B}-A} = S_B+S_{\bar{A}}\,.
\end{equation}
Por lo tanto, se tiene $S_A^1-S_B^1+S_{\bar{B}}^1-S_{\bar{A}}^1 \geq 0$, y entonces podemos obtener la desigualdad más débil (\ref{desi})
\begin{equation}
\langle \hat{H}_A-\hat{H}_B \rangle _1 + \chi \geq 0 \,,
\label{desiweak}
\end{equation}
donde hemos definido
\begin{equation}
\chi \doteq \left(S_A^0-S_B^0+S_{\bar{B}}^0-S_{\bar{A}}^0 \right) - \langle \hat{H}_A-\hat{H}_B \rangle _0\,.
\label{chi}
\end{equation}
Remarcamos que $\chi$ es una cantidad que sólo involucra al estado $\rho_0$ y es, en principio, calculable si el estado $\rho_0$ es conocido. En tal caso, la ecuación (\ref{desiweak}) nos da una cota para el valor de expectación del hamiltoniano modular (definido por el estado $\rho_0$) en un estado arbitrario $\rho_1$. Recordamos nuevamente que $\rho_0$ puede ser cualquier estado, no necesariamente puro.
%
%
Analizando la desigualdad (\ref{desiweak}) utilizando algún estado de referencia $\rho_0$ para el cual $\chi$ sea calculable, será posible obtener otra desigualdad cuántica de energía para el estado $\rho_1$. En el capítulo \ref{ch:modular} vimos que para un estado térmico en la mitad del espacio de Rindler, el hamiltoniano modular es conocido y está dado por la ecuación \reef{hamtermrin}. En tal caso, será posible calcular $\chi$ explícitamente. Este trabajo se encuentra actualmente en desarrollo.
\chapter{\label{ch:local}Términos locales en el hamiltoniano modular de campos libres}

En el capítulo \ref{ch:modular} presentamos algunos ejemplos de hamiltonianos modulares locales conocidos, que posteriormente utilizamos en los capítulos siguientes. Vimos que el hamiltoniano modular es un objeto fundamental que, entre otras cosas, resulta relevante al momento de buscar relaciones tipo Bekenstein en la teoría cuántica. En este capítulo, exploramos la forma que deben tener los términos locales que aparecen en el hamiltoniano modular de campos libres. En particular, probamos que, para campos libres, los términos locales que aparecen en el hamiltoniano modular de una región espacial $\Sigma$ están dados por el flujo sobre $\Sigma$ de una corriente (en general no conservada) construida con el tensor de energía impulso, $j^{\mu}(x,\partial\Sigma)=T^{\mu\nu}(x)a_\nu(x,\partial\Sigma)$. El vector temporal $a_\nu$ está dirigido hacia el futuro, depende del borde de la región, es independiente de la masa del campo y es el mismo para escalares y fermiones. El vector $a_\nu(x,\partial \Sigma)$ da una correspondencia entre regiones y vectores dirigidos hacia el futuro en $x$ que conecta el orden parcial dado por la inclusión de regiones en el ordenamiento temporal de vectores.

Este trabajo ha sido realizado en colaboración con Raúl E. Arias, Horacio Casini y Marina Huerta, y una versión extendida y corregida de este estudio puede encontrarse en \cite{blanco5}.

\section{Un primer ejemplo}

En general, los hamiltonianos modulares son objetos no locales. El primer ejemplo explícito de hamiltoniano modular no local fue presentado en \cite{fermion} y corresponde al campo de Dirac no masivo en $1+1$ sobre una región $V$ formada por $n$ intervalos de la forma $\left(a_i,b_i\right)$. Para el campo de Dirac no masivo en $1+1$, la matriz densidad reducida sobre cualquier región $V$ se factoriza en las coordenadas nulas $u_{\pm}=t\pm x$
\begin{eqnarray}
\rho_V &=& \rho(u_+)\otimes \rho(u_-)= c e^{-H_+} \otimes e^{-H_-}\,,\\
H_{\pm}&=&\int_{V_{\pm}} du_{\pm}^{1} \int_{V_{\pm}} du_{\pm}^{2} \Psi_{\pm}^{\dagger}\left(u_{\pm}^1\right) K_{\pm}\left(u_{\pm}^1,u_{\pm}^2\right)\Psi_{\pm}\left(u_{\pm}^2\right)\,.\label{enojado}
\end{eqnarray}
$V_{\pm}$ son las proyecciones de $V$ sobre los ejes nulos y $\Psi_{\pm}$ las componentes quirales del campo de Dirac dadas por $\Psi_{\pm}=Q_{\mp}\Psi$, donde $Q_{\pm}$ son los proyectores de quiralidad
\be
Q^+=\frac{1+\gamma^3}{2}\hspace{2cm}Q^-=\frac{1-\gamma^3}{2}\,.
\ee
El problema se reduce entonces a dos problemas unidimensionales sobre las coordenadas nulas y el hamiltoniano modular es la suma $H=H_+ + H_-$. 

Si la región está formada por $n$ intervalos disjuntos de la forma $V= \cup_{i=1}^{n} \left(a_i,b_i\right)$ cada núcleo $K_{\pm}$ (utilizamos $K$ para designar a cualquiera de los núcleos integrales que aparecen en la ecuación \reef{enojado}; no confundir con el operador de boosts) está dado por \cite{fermion} 
\begin{equation}
K\left(x,y\right)=K_{\textrm{loc}}\left(x,y\right)+K_{\textrm{noloc}}\left(x,y\right)\,,
\end{equation}
con
\begin{equation}
K_{\textrm{loc}}\left(x,y\right)= \pi i \left[2 \left(\frac{dz(x)}{dx}\right)^{-1}\partial _x + \frac{d}{dx} \left(\frac{dz(x)}{dx}\right)^{-1}\right]\delta\left(x-y\right)\,\,\textrm{y}
\end{equation}
\begin{equation}
K_{\textrm{noloc}}\left(x,y\right)=-2\pi i \sum_{l, x_l\left(z(x)\right)\neq x}{\frac{1}{x-y} \left(\frac{dz(y)}{dy}\right)^{-1} \delta \left(y-x_l\left(z(x)\right)\right)}\,,
\end{equation}
donde
\begin{equation}
z(x)=\log \left[-\dfrac{\prod_{i=1}^{n}{\left(x-a_i\right)}}{\prod_{i=1}^{n}{\left(x-b_i\right)}}\right]\,.\label{zeta}
\end{equation}
La función $z(x)$ es estrictamente monótona en cada intervalo $\left(a_i,b_i\right)$ y su rango es $\left(-\infty,+\infty\right)$, por lo que dado un $z \in \left(-\infty,+\infty\right)$ existe una única solución $x_l$ de \reef{zeta} para cada intervalo $\left(a_l,b_l\right)$. Estas soluciones están dadas por las raíces de un polinomio. La parte no local del hamiltoniano modular mezcla entonces un número finito de puntos, los $x_l(z)$ que tienen el mismo $z$, uno por cada intervalo.

La parte local del hamiltoniano modular
\footnotesize
\be
H_\textrm{loc}= \int_{V_{+}} dx \int_{V_{+}} dy \Psi_{+}^{\dagger}\left(x\right) K_\textrm{loc}\left(x,y\right)\Psi_{+}\left(y\right) + \int_{V_{-}} dx \int_{V_{-}} dy \Psi_{-}^{\dagger}\left(x\right) K_\textrm{loc}\left(x,y\right)\Psi_{-}\left(y\right)
\ee
\normalsize
puede reescribirse del siguiente modo
\be
H_{\textrm{loc}}=\int_{V_+} dx\, 2 \pi f(x) T_{++}(x)+\int_{V_-} dx\, 2 \pi f(x) T_{--}(x)\,,\label{hlocalfer}
\ee
siendo
\be
f(x)=\left(\sum_{i=1}^n \frac{1}{x-a_i}+\sum_{i=1}^n\frac{1}{b_i-x}\right)^{-1}\,,\label{ffunc}
\ee
y $T_{\pm \pm}$ las componentes nulas del tensor de energía impulso. La expresión \reef{hlocalfer} manifiesta explícitamente que esa contribución al hamiltoniano modular es local.

La función $f$ en \reef{ffunc} es positiva y además crece con el tamaño de la región ya que
\bea
\frac{d f(x)}{d b_i}&=&\frac{f(x)^2}{(b_i-x)^2}>0\,,\\
\frac{d f(x)}{d a_i}&=&-\frac{f(x)^2}{(x-a_i)^2}<0\,.
\eea
Al comparar regiones con distintos números de componentes, esta desigualdad surge de las ecuaciones anteriores y del hecho de que la función $f(x)$ para $n$ componentes (para $x$ fijo) tiende a la función de $n-1$ componentes cuando $b_i-a_i\rightarrow 0$ (y $x\notin (a_i,b_i)$) o cuando $b_i\rightarrow a_{i+1}$ (en este último caso, dos intervalos se juntan y se transforman en uno).

\section{Desigualdades para los términos locales en campos libres}

En esta sección daremos argumentos que nos permiten obtener ciertas consecuencias generales para los términos locales de campos libres. Para evaluar la forma que tienen estos términos, será relevante poder inducir excitaciones localizadas en el tensor de energía impulso. Para campos escalares libres, estos estados con tensor de energía impulso altamente localizado en una región pequeña se pueden generar utilizando estados coherentes.

\subsection{Estados coherentes}

Dentro de una teoría para un campo escalar libre $\phi$, consideremos el siguiente estado coherente normalizado
\be
|\alpha\rangle=e^{i\int dx \alpha(x) \phi(x)}|0\rangle\,,
\ee
donde $\alpha(x)$ es una función real. Se tiene que
\be
[\phi(x),e^{i\int dx \alpha(y) \phi(y)}]=e^{i\int dx \alpha(y) \phi(y)} \int dy\,\alpha(y) i [\phi(x),\phi(y)]=e^{i\int dx \alpha(y) \phi(y)} f(x,\alpha)\,,\label{conmu}
\ee
donde hemos definido
\be
f(x,\alpha)=-\int dy\,\Delta(x-y)\alpha(y)\,,\hspace{2cm} \Delta(x-y)=-i [\phi(x),\phi(y)]\,.
\ee
El conmutador $\Delta(x-y)$ es una solución de las ecuaciones de movimiento homogéneas y se anula fuera del cono de luz. Por lo tanto, $f(x,\alpha)$ es también una solución a las ecuaciones de movimiento $(\partial^2 +m^2)f(x,\alpha)=0$ y se anula fuera del soporte (pasado y futuro) de la función $\alpha(x)$.

Utilizando (\ref{conmu}) tenemos que\footnote{Notar que hemos omitido escribir la dependencia de $f$ con $\alpha$, para simplificar la notación.}
\be
\langle\alpha |\phi(x)|\alpha\rangle=f(x)\,.
\ee
Para el correlador de dos puntos se obtiene análogamente
\be
\langle\alpha |\phi(x)\phi(y)|\alpha\rangle=\langle 0|\phi(x)\phi(y)|0\rangle + f(x)f(y)\,.
\ee
Vemos entonces que si sustraemos la contribución del vacío a los correladores de dos puntos tenemos una expresión puramente clásica, que se obtiene reemplazado al operador $\phi(x)$ (y sus derivadas) por la función clásica $f(x)$. El hamiltoniano modular de cualquier región para un campo escalar libre es cuadrático en el campo y el momento, con lo que el estado de expectación del mismo en el estado coherente considerado (definiendo a $H$ de modo que tenga valor de expectación nulo en el vacío) está dado por la expresión clásica
\be
\langle\alpha |H|\alpha\rangle=\int_V dx\,dy\, (M(x,y)f(x)f(y)+N(x,y)\dot{f}(x)\dot{f}(y) )\,.
\ee
De esta última ecuación, vemos que si elegimos $\alpha$ con soporte en una esfera de radio $\epsilon$ pequeño alrededor de algún punto $p$ en la región espacial $V$, las únicas contribuciones no nulas a $\langle\alpha |H|\alpha\rangle$ provienen de $|x-p|\le \epsilon$, $|y-p|\le \epsilon$. En el límite en que consideramos una sucesión de funciones $f_{\epsilon}(x)$ con soporte $\epsilon$ aproximándose a cero, estos estados sirven para evaluar las contribuciones locales de los términos cuadráticos en el núcleo integral. Con este procedimiento, la parte local queda definida como la parte del núcleo con soporte en $x=y$, y deberían ser múltiplos de la delta de Dirac o sus derivadas.

\subsection{Términos locales en el hamiltoniano de Rindler}

Como ejemplo de un término local tenemos al hamiltoniano modular de Rindler
\be
H=2 \pi\int_{x^1>0} d^{d-1}x\, x^1 T_{00}(x)\,,
\ee
que, particularmente, es completamente local. Escrito para un \textit{wedge} completamente general y como la integral de una superficie espacial genérica, el hamiltoniano modular de Rindler tiene la siguiente forma
\be
H=2\pi \int_\Sigma  d\sigma\, \eta^\mu T_{\mu\nu} \omega^{\nu\delta} x_\delta \,,
\ee
donde $\eta$ es un vector causalmente temporal, normal a la superficie espacial $\Sigma$ (cuyo dominio de dependencia es el espacio de Rindler), y $\omega^{\nu\delta}=j^\nu t^\delta-j^\delta t^\nu$, con $j^\nu$ y $t^\delta$ dos vectores espacial y temporal respectivamente, ortogonales al horizonte del \textit{wedge}.

Recordamos que en el análisis posterior a la ecuación \reef{perio}, mencionamos que
\be
\hat{H}_A-\hat{H}_B\,\, \textrm{es un operador positivo si}\,\, B \subseteq A\,.\label{denoche}
\ee
Consideremos dos regiones tipo \textit{wedge de Rindler}, con una de ellas incluida en la otra; sus horizontes deben ser paralelos y estar desplazados un cierto vector espacial $a^\mu$. La desigualdad dada por \reef{denoche} aplicada a estas dos regiones, expresadas en dos superficies con el mismo $\eta$ en el mismo punto $x$ de las superficies de Cauchy, implica que para una perturbación local alrededor de $x$ se tiene
\be
\eta^\mu \langle T_{\mu\nu}(x)\rangle \omega^{\nu\delta} a_\delta\ge 0\,,
\ee
donde debe entenderse al valor de expectación como la integral sobre la pequeña región donde el tensor de energía impulso no se anula. Podemos tomar $j^\mu$ proporcional a $a^\mu$ y perpendicular a $t^\mu$, y obtenemos
\be
\eta^\mu \langle T_{\mu\nu}(x)\rangle t^\nu\ge 0\,.
\ee
Esta última desigualdad debe ser válida para el vector temporal $t$ en presencia de la excitación localizada. En este caso, utilizando otro argumento podemos verificar su validez: $\eta^\mu \langle T_{\mu\nu}(x)\rangle=P_\nu$ para la perturbación localizada, y $P_\nu t^\nu\ge 0$ es la condición espectral para estados, uno de los axiomas de la teoría de campos axiomática de Wightman \cite{wightman}.

Este resultado es general para cualquier teoría. En particular, para los campos escalares libres los estados coherentes que presentamos sirven naturalmente para generar la perturbación localizada al tensor de energía impulso y tenemos
\be
\langle\alpha |T_{\mu\nu}|\alpha\rangle-\langle 0 |T_{\mu\nu}|0\rangle=\partial_\mu f(x) \partial_\nu f(x)-g_{\mu\nu}\frac{1}{2}(\partial_\beta f(x) \partial^\beta f(x)+m^2 f(x)^2 )\,.
\ee

\subsection{Consecuencias generales para los términos locales}

Basándonos en el ejemplo del espacio de Rindler y en la ecuación \reef{denoche}, consideremos el caso de un campo escalar no masivo. Supongamos que una región $V$ de forma arbitraria se encuentra dentro de una esfera $S_2$ y contiene en su interior a otra esfera $S_1$. Tomemos un punto $x$ en una superficie de Cauchy $\Sigma$ común a las tres regiones y sea $\eta$ un versor unitario normal a $\Sigma$ en $x$. La contribución al hamiltoniano modular de $S_2$ en $x$ es de la forma $\eta^{\nu} T_{\mu\nu} v_2^{\nu}\left( x\right)$ y la de la esfera $S_1$ es $\eta^{\nu} T_{\mu\nu} v_2^{\nu}\left( x\right)$. Por la desigualdad que derivamos en la sección anterior, $v_1$ y $v_2$ deben ser vectores causalmente temporales dirigidos hacia el futuro, así como también lo debe ser su diferencia $v_2-v_1$.

El valor de expectación del hamiltoniano modular de $V$ para una excitación localizada en $x$ tiene que estar entre los valores de expectación correspondientes a los hamiltonianos modulares para $S_1$ y $S_2$; por este motivo, $H_V$ debe tener un término local. Este término local puede ser de la forma $\eta^{\mu} T_{\mu\nu} w^{\nu}\left(x\right)$ con $v_2-w$ y $w-v_1$ vectores causalmente temporales dirigidos hacia el futuro. Veremos que este es el único tipo de términos locales permitidos.

Recordemos que para campos libres, el hamiltoniano modular asume una forma cuadrática en los campos. Si como estado localizado consideramos un estado coherente, el término local será una combinación del campo y sus derivadas. Debe además ser positivo, dado que la entropía relativa es positiva, y la entropía de entrelazamiento para este tipo de estados localizados es igual a la del vacío. Podemos imaginar diferencias combinaciones positivas que contribuyan al valor de expectación del hamiltoniano modular, por ejemplo
\be
f(x)^2\,\hspace{1cm}(\partial_x f(x))^2\,,\hspace{1cm} (\partial_y\partial_x f(x))^2\,,
\ee
etc. La positividad de algunos de estos términos será consecuencia directa de su expresión. Para otros, será necesario utilizar las ecuaciones de movimiento. Sin embargo, el punto crucial es que los valores de estos términos serán, en general, independientes entre sí. Si consideramos por ejemplo una excitación localizada, $\int dx \,f(x)^2$ y $\int dx \,(\partial_x f(x))^2$ pueden ser muy diferentes, dependiendo de la elección de la función localizada $f(x)$. Por ejemplo, para $f(x)=e^{-b^2 x^2/c} \cos(b x)$ en el límite $c\rightarrow \infty$ (excitación muy localizada), tenemos
\be
\frac{\int dx \,f(x)^2}{\int dx \,(\partial_x f(x))^2}=\frac{1}{b^2}\,,
\ee
que puede ser muy grande o muy pequeño, dependiendo de la elección de $b$. Con esto dicho, parece poco probable que para un dado operador local no podamos encontrar una $f(x)$ adecuada que evite que el resultado viole la cota dada por la combinación de los tensores de energía impulso de las dos esferas.

Nuestra conclusión preliminar es entonces que el hamiltoniano modular de cualquier región $V$ tiene que tener un término local en $x$ de la forma $\eta^\mu T_{\mu\nu} w^\nu(x)$, con $w$ dirigido hacia el futuro. Este $w$ es `mayor' (en términos de la relación de orden parcial definida por el ordenamiento de vectores en el cono de luz futuro) que todo $v_1$ de cualquier esfera que contenga $x$ que esté contenida en $V$, y `menor' que cualquier $v_2$ en $x$ de cualquier esfera que contenga a $V$.

Examinemos con más detalle la estructura de estos vectores dirigidos hacia el futuro. El ordenamiento trasciende la comparación con esferas; cualquier región puede compararse con cualquier otra ahora que la estructura de los términos locales es más clara. Dado un punto $x$, hay un vector dirigido hacia el futuro $w(x,V)$ para cualquier $V$ que contenga (causalmente) al punto $x$. Notemos que, en principio, $w$ depende de $x$ pero no de $\eta$. El conjunto de los $w(x,V)$ establece un mapeo monótono (es decir, que respeta el ordenamiento) entre el conjunto de todas las regiones que contienen a $x$ con los vectores en el cono de luz futuro; $w(x,V)$ relaciona entonces el ordenamiento de inclusión de conjuntos con el ordenamiento en el cono de luz. Esto a simple vista constituye un hecho interesante, dado lo vasto que parece el conjunto de regiones posibles a considerarse en comparación con la estructura más restringida de los vectores en el cono de luz.

\section{Fermiones libres}

En esta sección volvemos a considerar como ejemplo el caso del campo de Dirac en $1+1$, pero ahora permitiremos que el campo tenga una masa $m$. En este caso, es posible obtener las funciones de correlación expandiendo en el orden de la masa y se pueden hallar por ejemplo las entropías para distintas regiones a un orden determinado en la masa \cite{fermion,blanco1}. Siguiendo estas ideas, en esta sección hallaremos la corrección a primer orden en la masa $m$ del hamiltoniano modular para una región compuesta por un intervalo.

El campo de Dirac libre satisface la ecuación de movimiento
\be
D \psi(x)=(i\gamma^\mu\partial_\mu-m)\psi(x)=0\,,
\ee
y las relaciones de anticonmutación a tiempos iguales
\be
\left.\{\psi(x),\psi^\dagger(y)\}\right|_{x^0=y^0}=\delta^{d-1}(x-y)\,.
\ee

El hamiltoniano modular para un fermión de Dirac libre en $d=2$ para un intervalo se escribe como \cite{fermion}
\be
H=\int_0^L dx\, dy\, \bar{\psi}(x) K(x,y) \psi(y)\,.
\ee
con $K(x,y)$ dado por
\be
K=-\int_{1/2}^\infty d\beta\, (R(\beta)+R(-\beta))\,,
\ee
en términos de la resolvente
\be
R(\beta)=(C-1/2+\beta)^{-1}\,,
\ee
donde $C(x,y)=\langle 0|\psi(x)\psi^\dagger(y)|0\rangle$ es el correlador en el intervalo de longitud $L$. Expandiendo $C$ a primer orden en la masa se tiene
\be
R(\beta)=R^0(\beta)-R^0(\beta)\,\delta C \,R^0(\beta)+...\label{23}
\ee
con
\bea
R^0(\beta)(x,y)&=&\int_{-\infty}^\infty ds\, \psi_s(x) M(\beta,s) \psi^*_s(y)\,,\\
M(\beta,s)&=&\left(\beta {\bf 1}-\tanh(\pi s)\frac{\gamma^3}{2}\right)^{-1}\\
\psi_s(x)&=& \frac{L^{1/2}}{(2\pi)^{1/2}\sqrt{x (L-x)}}e^{-i s z(x)},~~z(x)=\log\left(\frac{x}{(L-x)}\right)\,,\\
\delta C(x,y)&=&-\frac{m}{2\pi}\left(\gamma_E-\log(2)+\log(m|x-y|)\right)\gamma^0\,.
\eea

\subsubsection{Orden cero en la masa}
A orden cero en la expansión masiva, tenemos
\be
\int_{1/2}^\infty d\beta\, (M(\beta,s)+M(-\beta,s)=2 \pi s \gamma^3\,.
\ee
y luego
\bea
H_0&=&-\gamma^3\int_{-\infty}^\infty ds\,\frac{L}{(x(L-x)y(L-y))^{1/2}} s \,e^{-i s (z(x)-z(y))}= \nonumber \\
&=& 2\pi i\gamma^3\frac{L}{(x(L-x)y(L-y))^{1/2}} \delta^{\prime}(z(y)-z(x))= \nonumber \\
&=& i 2\pi \gamma^3 (\frac{x(L-x)}{L}\delta^{\prime}(x-y)+\frac{(L-2 x)}{2L} \delta(x-y))\,.
\eea
El término proporcional a la delta de Dirac hace que el núcleo entero sea hermítico. Teniendo en cuenta que el hamiltoniano (dinámico) de Dirac se escribe como
\be
i\alpha_x \partial_x=i\gamma^3\partial_x \,,
\ee
esto da el término a orden cero en la masa del hamiltoniano modular
\be
H_0=\int_0^L dx\, 2\pi \frac{x(L-x)}{L} T_{00}(x)\,,
\ee
con $T_{00}=(i/2) \psi^\dagger \gamma^3 \stackrel{\leftrightarrow}{\partial}_x \psi$.

\subsubsection{Primer orden en la masa}
Veamos ahora cómo son las correcciones a primer orden en la masa. Para ello, necesitamos calcular
\bea
\int_{1/2}^\infty d\beta\, (M(\beta,s)\gamma^0 M(\beta,s^\prime)&+& M(-\beta,s)\gamma^0 M(-\beta,s^\prime))=\\
&=& 4\pi \gamma^0 (s+s^\prime) \frac{\cosh(\pi s)\cosh(\pi s^\prime)}{\sinh(\pi(s+s^\prime))}\nonumber \,.
\eea
Para hacer la integral intermedia en $x$ en (\ref{23}), separamos la contribución de $\delta C$ en dos partes: una parte independiente de las coordenadas y otra proporcional a $\log|x-y|$. Escribamos al término constante como
\be
k=-\frac{m}{2\pi}\left(\gamma_E-\log(2)+\log(m)\right)\,,
\ee
y llamemos $H_{1,0}$ a su contribución al hamiltoniano modular. Utilizando
\be
\int_0^L dx\, \psi_s(x)=\left(\frac{\pi}{2}\right)^{1/2} L^{1/2} \textrm{sech}(\pi s)
\ee
obtenemos
\be
H_{1,0}=k L^2 \pi \gamma^0 \int_{-\infty}^\infty ds\,\int_{-\infty}^\infty ds^\prime\,\frac{e^{-i (s z(x)- s^\prime z(y))}}{(x(L-x)y(L-y))^{1/2}}   \frac{(s+s^\prime)}{\sinh(\pi(s+s^\prime))}\,.
\ee
y cambiando variables a $u=s+s^\prime$ y $v=s-s^\prime$ llegamos a
\small
\bea
H_{1,0}&=& k L^2 \pi^2  \gamma^0 \frac{1}{(x(L-x)y(L-y))^{1/2}} \delta(z(x)+z(y)) \int_{-\infty}^\infty du\, \frac{u e^{-i \frac{u}{2}(z(x)-z(y))}}{\sinh(\pi u)}
=\nonumber \\
&=& 4 k  \pi^2  \gamma^0 \frac{x y}{L}\,\,\delta(x+y-L)\,.\label{h10}
\eea
\normalsize
Este término es bastante interesante. Vemos que es no local; diríamos que es `cuasilocal' dado que sólo mezcla $x$ con $L-x$ (similar al caso de muchos intervalos para el campo no masivo que presentamos al comienzo de este capítulo). Vemos también que cerca del borde modifica la contribución tipo Rindler agregando una contribución no local que mezcla la contribución proveniente del otro borde.

Veamos ahora la contribución proporcional a $m \log|x-y|$, a la que llamaremos $H_{1,1}$. En este caso, tenemos que calcular
\bea
H_{1,1}&=& -2 m \gamma^0 \int ds \int ds'\,(s+s^\prime) \frac{\cosh(\pi s)\cosh(\pi s^\prime)}{\sinh(\pi(s+s^\prime))}\psi_s(x)\psi^*_{s'}(y)\times \nonumber\\
&\times &\int dx'\int dy'\psi_s^*(x')\log|x'-y'|\psi_{s'}(y').\label{h11}
\eea
Para realizar las integrales definimos la función
\be
F(x')=\int dy'\log|x'-y'|\psi_{s'}(y')\,,
\ee
y utilizamos que
\be
F'(x')=A+\int dy\,\frac{1}{x-y}\psi_s(y)=A+i\pi\tanh{\pi s}\psi_s(x),
\ee
donde $A$ es una constante de integración que podemos fijar utilizando el valor de la integral en $y'$ para $x'=0$ en \reef{h11}
\be
A=\int dy'\log|-y'|\psi_{s'}(y')=\sqrt{\frac{L\pi }{2}}\,\text{sech}\left(\pi s'\right) \left(\log (L)+H_{-i s'-\frac{1}{2}}\right).
\ee
$H_{-i s'-\frac{1}{2}}$ representa la función número armónico. Así, llegamos a
\bea
F(x')&=&\sqrt{\frac{L\pi }{2}}\,\text{sech}\left(\pi s'\right) \left(\log (L)+H_{-i s'-\frac{1}{2}}\right)+\\
&-&\frac{\sqrt{2 \pi } \tanh \left(\pi  s'\right) \left(x'\right)^{\frac{1}{2}-i s'} \left(L-x'\right)^{\frac{1}{2}+i s'} \,
   _2F_1\left(1,1;\frac{3}{2}-i s';\frac{x'}{L}\right)}{\sqrt{L} \left(2 s'+i\right)}\nonumber
\eea
y la corrección al hamiltoniano modular puede escribirse como
\small
\be
H_{1,1}=-2m\gamma^0 \int ds \int ds'\,(s+s^\prime)\frac{\cosh(\pi s)\cosh(\pi s^\prime)}{\sinh(\pi(s+s^\prime))}\psi_s(x)\psi^*_{s'}(y)\int dx'F(x')\psi_s^*(x').
\label{h11dos}
\ee
\normalsize
Hacemos la integral en $x'$ y obtenemos la expresión antisimetrizada en las variables $s, s'$ 
\footnotesize
\bea
\int dx'F(x')\psi^*_s(x')&=&\frac{1}{4} \pi\,  L\, \text{csch}\left(\pi  \left(s-s'\right)\right) \left[\tanh (\pi  s) \left(H_{-i s'-\frac{1}{2}}+H_{i s'-\frac{1}{2}}+2 \log
   (L)\right)\right. +\nonumber \\
   &&\left.-\tanh \left(\pi  s'\right) \left(H_{-i s-\frac{1}{2}}+H_{i s-\frac{1}{2}}+2 \log (L)\right)\right].
   \eea
   \normalsize
Notemos que los términos proporcionales a $\log(L)$ dan integrales como las que aparecen en la ecuación \reef{h10}, por lo que el resultado será proporcional a $\delta(x+y-L)$,
\small
\be
-m L \pi \log(L)\int ds ds'\psi_s(x)\psi_s'^*(y)\frac{s+s'}{\sinh(\pi(s+s'))}=-4\pi \,m \log(L)\frac{x y}{L}\delta(x+y-L).
\ee
\normalsize
Los términos que involucran las funciones armónicas pueden integrarse realizando el cambio de variables $u=s+s', v=s-s'$. De este modo, la expresión a estudiar es
\small
\bea
&&\frac{m L^2}{16 \sqrt{x y (L-x) (L-y)}}\int du\,dv\, u\, \text{csch}(\pi  u) \text{csch}(\pi  v) \times \\
&&\times \left[\left(H_{-\frac{1}{2} i (u+v-i)}+H_{\frac{1}{2} i (u+v+i)}\right) (\sinh (\pi  u)-\sinh (\pi
   v))+ \right.\nonumber \\
&&\left.-\left(H_{-\frac{1}{2} i (u-v-i)}+H_{\frac{1}{2} i (u-v+i)}\right) (\sinh (\pi  u)+\sinh (\pi  v))\right]\times \nonumber \\
&&\times e^{-\frac{1}{2} i (u
   (z(x)-z(y))+v (z(x)+z(y)))}\nonumber \label{integrand}.
\eea
\normalsize
Sustrayendo los términos singulares de \reef{integrand} obtenemos una integral en $u, v$ que da un resultado finito pero es complicada de calcular. Sin embargo, nuestro interés reside en hallar la contribución de estos términos singulares al hamiltoniano modular. Para extraer el término proporcional a $\delta(x-y)$ utilizamos que la expansión del integrando cuando $u\rightarrow\infty$ va como $4\,v\cosh(v)$. Realizando las integrales en $u$ y $v$ llegamos a que el resultado final para \reef{integrand} es
\be
\frac{2 \pi  m\, y (L-y)}{L}\delta(x-y).
\ee
Ahora, podemos analizar el límite $v\rightarrow\infty$  del integrando en \reef{integrand} para extraer otra contribución singular al hamiltoniano modular proveniente de este límite. La expresión resultante es
\bea
&&\frac{m L^2}{4 \sqrt{x y (L-x) (L-y)}} \int { du\,dv\, u\, \text{csch}(\pi  u) \left(\log \left(\frac{1}{|v|}\right)-\gamma +\log (2)\right)}\times \nonumber \\
&&\times e^{-\frac{1}{2} i (u
   (z(x)-z(y))+v (z(x)+z(y)))},
\eea
donde $\gamma$ es el número de Euler. El término proporcional a $(-\gamma+\log(2))$ puede ser integrado y da
\be
4\pi m (-\gamma+\log(2))\frac{x y}{L}\delta(x+y-L).
\ee
La integral del término que involucra a $\log\left(\frac{1}{|v|}\right)$ da
\be
\frac{4\pi\,m \,L^2 }{\left(\sqrt{x} \sqrt{L-y}+\sqrt{y} \sqrt{L-x}\right)^2 |\log \left(\frac{x y}{(L-x)(L-y)}\right)|}+4\pi\,m\,\gamma\frac{x\,y}{L}\delta(x+y-L).
\ee
Notemos que en el límite $x\rightarrow L-y$ el primer sumando diverge como $\frac{m}{x+y-L}$.

Finalmente, el resultado final para los términos singulares en la corrección a primer orden en la masa del hamiltoniano modular es
\bea
H_{1,0}+H_{1,1}&=& 4\pi \left[3 k  \pi+ m\left(\gamma+\log\left(\frac{m}{L}\right)\right)\right] \frac{x y}{L}\,\,\delta(x+y-L)+ \nonumber \\
&+& \frac{2 \pi  m\, y (L-y)}{L}\delta(x-y)+ \\
&+& \frac{4\pi\,m \,L^2 }{\left(\sqrt{x} \sqrt{L-y}+\sqrt{y} \sqrt{L-x}\right)^2 |\log \left(\frac{x y}{(L-x)(L-y)}\right)|}\gamma^0\,.\nonumber
\eea

Estos resultados fueron verificados con cálculos realizados en la red, utilizando el formalismo presentado en la sección \ref{discreto} y la forma que tienen los correladores fermiónicos en la red en $1+1$ dimensiones, dada por \cite{review}
\begin{equation}
C_{jk}=\frac{1}{2}\delta_{jk}-\int_{-\pi}^{\pi} dx \frac{m\gamma^0+\sin(x)\gamma^0 \gamma^1}{4\pi\sqrt{m^2+\sin^2(x)}}e^{-i x (j-k)}\,.
\end{equation}

\section{Propagación de los términos locales}

En esta sección analizaremos cómo evolucionan los términos locales en el hamiltoniano modular. Nos concentraremos primeramente en el campo de Dirac en $d=2$. El hamiltoniano modular en una superficie espacial $\Sigma$ es
\be
H=\int_{\Sigma} ds_1\,ds_2\, \psi^\dagger(s_1) K(s_1,s_2) \psi(s_2)\,,
\ee
donde $s_1,s_2$ son parámetros de distancia sobre la superficie. Los términos locales tienen la forma
\be
\int_\Sigma ds\, j^\mu(x(s)) \eta_\mu(s)\,,
\ee
donde $j^{\mu}(x)$ es independiente de la superficie $\Sigma$ que pasa por $x$.

El hamiltoniano modular puede reescribirse sobre cualquier otra superficie que tenga el mismo dominio de dependencia causal $V$ utilizando la ecuación de propagación para los campos
\bea
\psi(x)=\int_{\Sigma^\prime} ds^\prime\, S(x-x^\prime)\gamma^\mu\eta^\prime_\mu(s^\prime) \psi(s^\prime)\,,\label{54}\\
\bar{\psi}(x)=\int_{\Sigma^\prime} ds^\prime\, \bar{\psi}(s^\prime) \gamma^\mu\eta^\prime_\mu(s^\prime) S(x^\prime-x)\,,
\eea
donde
\be
S(x-y)=\{\psi(x),\bar{\psi}(y)\}=  (i\gamma^\mu \partial_\mu+m) i\Delta(x-y)  \,.
\ee
Miremos entonces cómo se propagan los términos locales. Un término local en $\Sigma^\prime$ surgirá del término local en $\Sigma$ convolucionado con los términos singulares de la función de Green $S(x-x^\prime)$. Como las integrales que dan lugar a un término local en $\Sigma^\prime$ son sobre regiones compactas en la coordenada $s$ no tendremos ningún otro término singular proveniente de los términos finitos. Miremos entonces la estructura de singularidades de la función de Green. Tenemos
\small
\be
i\Delta(x)=[\phi(x),\phi(0)]=\int \frac{d^2p}{2\pi}\,\epsilon(p^0) \delta(p^2-m^2) e^{-i p x} =\epsilon(x^0) \theta(x^2) i \frac{\textrm{Im}(K_0(i m \sqrt{x^2}))}{\pi}\,,
\ee
\normalsize
Dado que
\be
i\frac{\textrm{Im}(K_0(i y))}{\pi}\sim -\frac{i}{2} + {\cal O}(y^2)
\ee
para $y$ positivo pequeño, tenemos que $i\Delta(x)$ es suave en todos lados a excepción del cono de luz nulo, donde tiene un salto como el de la función
\be
-\frac{i}{4} (\epsilon(x^+)+\epsilon(x^-)).
\ee
En el desarrollo utilizaremos coordenadas nulas
\bea
x^+&=&x^0+x^1\,,\hspace{2cm}\,\,\,\,\,\,\,
x^- = x^0-x^1\\
x^0 &=&\frac{x^++x^-}{2}\,,\hspace{2cm}\,\,\,\,
x^1=\frac{x^+-x^-}{2}\\
\partial_+ &=&\frac{1}{2}(\partial_0+\partial_1)\hspace{2cm}\,\,\,\partial_- =\frac{1}{2}(\partial_0-\partial_1) \\
g^{\mu\nu} &=&\left(\begin{array}{cc} 0 & 2 \\ 2 & 0  \end{array}\right)\,,\hspace{2cm}  g_{\mu\nu}=\left(\begin{array}{cc} 0 & 1/2 \\ 1/2 & 0  \end{array}\right)
\eea
También utilizaremos la representación quiral para las matrices de Dirac, en la cual $\gamma^3=\gamma^0\gamma^1$ es diagonal
\be
\gamma^3=\left(\begin{array}{cc} 1 & 0 \\ 0 & -1  \end{array}\right)\,,\hspace{2cm}
\gamma^0=\left(\begin{array}{cc} 0 & 1\\ 1 & 0  \end{array}\right)
\ee
Los proyectores de quiralidad son
\be
Q^+=\frac{1+\gamma^3}{2}\hspace{2cm}Q^-=\frac{1-\gamma^3}{2}
\ee
y
\be
\gamma^\mu \partial_\mu=2\gamma^0(Q^+\partial_++ Q^-\partial_-)
\ee
La estructura de singularidades del anticonmutador en la representación quiral y en coordenadas nulas está dada entonces por
\bea
S(x)&\simeq &\frac{1}{4}(\gamma^\mu \partial_\mu-i m)(\epsilon(x^+)+\epsilon(x^-))=\nonumber \\
&=&\frac{1}{4}(2\gamma^0(Q^+\partial_++ Q^-\partial_-)-i m)(\epsilon(x^+)+\epsilon(x^-))= \nonumber\\
&=&\gamma^0  \left(\begin{array}{cc}
\delta(x^+) & -i \frac{m}{4} (\epsilon(x^+)+\epsilon(x^-))\\-i \frac{m}{4} (\epsilon(x^+)+\epsilon(x^-)) &  \delta(x^-)
\end{array}\right)\,,\label{gf}
\eea
donde hemos omitido otros términos menos singulares.

\subsection{Propagación en el caso no masivo}

Consideremos la propagación de un término local proporcional al tensor de energía impulso
\be
\int_{\Sigma} ds\, \eta_\mu T^{\mu\nu} a_\nu \label{intre}
\ee
El tensor de energía impulso se escribe como
\be
T^{\mu\nu}=\frac{i}{4}\bar{\psi}(\gamma^\mu \stackrel{\leftrightarrow}{\partial}^\nu+\gamma^\nu \stackrel{\leftrightarrow}{\partial}^\mu)\psi\label{ten}
\ee
Estamos interesados en la propagación en un dado punto $x$ de la superficie $\Sigma$ a otra superficie $\Sigma^\prime$. Sin pérdida de generalidad, podemos elegir el sistema de coordenadas de modo que el vector normal sea $\eta_\mu(x)=(1,0,\hdots)$. El operador en $x$ en la integral (\ref{intre}) se escribe entonces como
\bea
a^0 T_{00}+a^1 T_{01}=\bar{\psi}(x) (a^0 (-\frac{i}{2}) \gamma^1 \stackrel{\leftrightarrow}{\partial}_x + \frac{i}{2} a^1 \gamma^0 \stackrel{\leftrightarrow}{\partial}_x) ) \psi(x)\nonumber\\
=\frac{i}{2}\bar{\psi}(x)\gamma^0 (a^+ Q_- -a^- Q_+ ) \stackrel{\leftrightarrow}{\partial}_x \psi(x)\label{esa}
\eea
Las derivadas simetrizadas en este caso actúan sólo sobre los campos fermiónicos y no sobre las componentes de $a^\mu$. La propagación está dada por
\footnotesize
\be
\int_{\Sigma^\prime} ds_1^\prime \, ds_2^\prime\,\int_\Sigma dx\, \bar{\psi}(x_1^\prime)\eta_\beta(s_1^\prime)\gamma^\beta S(x_1^\prime-x) \frac{i}{2}\gamma^0 (a^+ Q_- -a^- Q_+ ) \stackrel{\leftrightarrow}{\partial}_x  S(x-x_2^\prime) \eta_\alpha(s_2^\prime)\gamma^\alpha\psi(x_2^\prime)\label{hjhk}
\ee
\normalsize
Utilizando que para las integrales en el parámetro de longitud a lo largo de la superficie
\be
\gamma^0 \eta_\alpha(s)\gamma^\alpha ds=\left(\begin{array}{cc} - dx^- & 0 \\ 0 & dx^+ \end{array}\right)
\ee
podemos reescribir (\ref{hjhk}) como
\small
\bea
&& \int_{\Sigma^\prime} \int_\Sigma dx\, \bar{\psi}(x_1^\prime)\gamma^0 \left(\begin{array}{cc} - dx_1^{\prime -} & 0 \\ 0 & dx_1^{\prime +} \end{array}\right) \gamma^0  \left(\begin{array}{cc}
\delta(x^{\prime+}_1-x^+) & 0\\0 &  \delta(x_1^{\prime -}-x^-)
\end{array}\right) \times \nonumber \\
&& \times \frac{i}{2}\gamma^0 (a^+ Q_- -a^- Q_+ ) \stackrel{\leftrightarrow}{\partial}_x \times \\
&&\times\,\,\,\,\,  \gamma^0  \left(\begin{array}{cc}
\delta(x^+-x^{\prime+}_2) & 0\\0 &  \delta(x^--x_2^{\prime -})
\end{array}\right) \gamma^0 \left(\begin{array}{cc} - dx_2^{\prime -} & 0 \\ 0 & dx_2^{\prime +} \end{array}\right)\psi(x_2^\prime)= \nonumber \\
&& =-\int_{\Sigma^\prime}dx^{-} a^- (x^-)\bar{\psi}(x^-)\gamma^0Q^+ \frac{i}{2}\stackrel{\leftrightarrow}{\partial}_{x^-}\psi(x^-)+\nonumber\\
&&\int_{\Sigma^\prime}dx^{+} a^+ (x^+)\bar{\psi}(x^+)\gamma^0Q^- \frac{i}{2}\stackrel{\leftrightarrow}{\partial}_{x^+}\psi(x^+)\,,\nonumber\label{ope}
\eea
\normalsize
donde $a^{\pm}$ está calculado sobre la superficie $\Sigma$. Luego, si $a^\mu$ está definido en todos lados por las componentes $a^-(x^-)$, independiente de $x^+$, y $a^+(x^+)$, independiente de $x^-$, dado por los valores en $\Sigma$, utilizando (\ref{ten}) para $\Sigma^\prime$ tenemos que el operador (\ref{ope}) queda
\small
\be
\int_{\Sigma^\prime} ds \,\eta^\mu a^\nu T_{\mu\nu}=\int_{\Sigma^\prime} ds \,(\eta^+ a^+ T_{++}+\eta^- a^- T_{--})=\int_{\Sigma^\prime} dx^+ \, a^+ T_{++}-\int_{\Sigma^\prime} dx^- \, a^- T_{--} \,.
\ee
\normalsize
donde hemos utilizado que
\be
ds (\eta^+,\eta^-)=(dx^+,-dx^-)
\ee
y, como anteriormente, hay un cambio de signo debido a los diferenciales de $x^-$ con respecto a $x$, aunque no hemos cambiado los límites de integración\footnote{Si bien esto es conveniente porque tenemos que mirar sólo el cambio en los diferenciales, también involucra la introducción de un signo cuando integramos una delta de Dirac donde aparece la variable $x^-$.}. Vemos entonces que la forma del término local se mantiene igual que en la superficie original y, en este caso, no se generan términos locales.

Para el campo no masivo, podemos entender que esto es una consecuencia del hecho de que $j^\mu=a_\nu T^{\mu\nu}$ es una corriente conservada y su flujo es independiente de la superficie. De hecho, cuando $\partial_+ a^-=\partial_- a^+=0$, utilizando que en el caso no masivo la traza es cero $T^{+-}=0$, y $\partial_+ T^{++}=0$, $\partial_- T^{--}=0$, por conservación,
\be
\partial_\mu T^{\mu\nu}a_\nu=T^{++}\partial_+ a_++T^{--}\partial_- a_-=0
\ee
Notemos que $\nabla^2 a^\mu=0$, aunque ninguna combinación de $\epsilon_{\mu\nu}\partial^\mu a^\nu$ y $\partial_\nu a^\nu$ se anula. Por lo tanto, no podemos escribir de este modo una ecuación de movimiento de primer orden para $a^\mu$, incluso aunque la propagación depende sólo del valor del campo en la superficie inicial y no de su derivada.

\subsection{Propagación en el caso masivo}

En el caso masivo tenemos ciertas diferencias con respecto al caso anterior. La primera de ellas, es el cambio en la ecuación (\ref{esa}) para incorporar el término de masa en el tensor de energía impulso
\be
a^0 T_{00}+a^1 T_{01}
=\frac{i}{2}\bar{\psi}(x)\gamma^0 (a^+ Q_- -a^- Q_+ ) \stackrel{\leftrightarrow}{\partial}_x \psi(x)\label{esa}+ a^0 \bar{\psi}(x) m \psi(x)
\ee
La segunda diferencia tiene que ver con los términos adicionales que aparecen en el propagador. Estos términos son no singulares, pero al ser contraídos con el término con derivadas del tensor de energía impulso producirán deltas de Dirac. Sólo los términos de primer orden en la masa darán deltas, mientras que todas las otras contribuciones de orden mayor darán lugar a términos no localizados. El término generado por la adición al tensor de energía impulso da
\small
\bea
&&\int_{\Sigma^\prime} \int_\Sigma dx\, \bar{\psi}(x_1^\prime)\gamma^0 \left(\begin{array}{cc} - dx_1^{\prime -} & 0 \\ 0 & dx_1^{\prime +} \end{array}\right) \gamma^0  \left(\begin{array}{cc}
\delta(x^{\prime+}_1-x^+) & 0\\0 &  \delta(x_1^{\prime -}-x^-)
\end{array}\right)  a^0 m \times \nonumber \\
&&\times\,\,\,\,\,  \gamma^0  \left(\begin{array}{cc}
\delta(x^+-x^{\prime+}_2) & 0\\0 &  \delta(x^--x_2^{\prime -})
\end{array}\right) \gamma^0 \left(\begin{array}{cc} - dx_2^{\prime -} & 0 \\ 0 & dx_2^{\prime +} \end{array}\right)\psi(x_2^\prime)\\
\label{ope1}
\eea
\normalsize
Notemos que esta contribución no genera términos localizados en la superficie $\Sigma^\prime$ dado que involucra una delta entre dos puntos en $\Sigma^\prime$ cuyas coordenadas $x^+$ y $x^-$ coinciden con las coordenadas correspondientes del punto $x$ en $\Sigma$. O sea que las perturbaciones singulares se propagan a la velocidad de la luz en diferentes direcciones generando muy posiblemente términos cuasilocales.

Veamos ahora las perturbaciones del propagador en el caso no masivo para el tensor de energía impulso. Tenemos
\bea
&&\int_{\Sigma^\prime} \int_\Sigma dx\, \bar{\psi}(x_1^\prime)\gamma^0 \left(\begin{array}{cc} - dx_1^{\prime -} & 0 \\ 0 & dx_1^{\prime +} \end{array}\right) \gamma^0  \left(\begin{array}{cc}
\delta(x^{\prime+}_1-x^+) & 0\\0 &  \delta(x_1^{\prime -}-x^-)
\end{array}\right)\times \nonumber \\
&&\times \frac{i}{2}\gamma^0 (a^+ Q_- -a^- Q_+ ) \stackrel{\leftrightarrow}{\partial}_x \times \\
&&\times\,\,\,\,\,  -i \frac{m}{4} (\epsilon(x^+-x^{\prime+}_2)+\epsilon(x^--x^{\prime-}_2)) \gamma^0 \left(\begin{array}{cc} - dx_2^{\prime -} & 0 \\ 0 & dx_2^{\prime +} \end{array}\right)\psi(x_2^\prime)+h.c.\nonumber\label{ope2}
\eea
Las dos derivadas pueden hacerse actuar en el mismo lado dado que al derivar las componentes de $a^\mu$ sólo se generan términos no singulares. Obtenemos
\footnotesize
\bea
&&\int_{\Sigma^\prime} \int_\Sigma dx\, \bar{\psi}(x_1^\prime)\gamma^0 \left(\begin{array}{cc} - dx_1^{\prime -} & 0 \\ 0 & dx_1^{\prime +} \end{array}\right) \gamma^0  \left(\begin{array}{cc}
\delta(x^{\prime+}_1-x^+) & 0\\0 &  \delta(x_1^{\prime -}-x^-)
\end{array}\right) \\
&&\times \gamma^0 \frac{1}{2}(a^+ Q_- -a^- Q_+ ) m (\delta(x^+-x^{\prime+}_2)-\delta(x^--x^{\prime-}_2)) \gamma^0 \left(\begin{array}{cc} - dx_2^{\prime -} & 0 \\ 0 & dx_2^{\prime +} \end{array}\right)\psi(x_2^\prime)+h.c.\nonumber\label{ope3}
\eea
\normalsize
Esta expresión tiene términos locales y cuasilocales. Los términos cuasilocales cancelan los producidos por (\ref{ope1}), mientras que los términos locales son
\bea
-\frac{m}{2}\int_{\Sigma^\prime} \psi^\dagger(x)  \gamma^0    \psi(x) (-a^-(x^-) dx^{+} +a^+(x^+) dx^-)\,.
\eea
Utilizando
\be
-a^- dx^++a^+ dx^-=-2 ds a_\mu \eta^\mu\,,
\ee
obtenemos
\be
m\int_{\Sigma^\prime} ds\, \psi^\dagger(x)  \gamma^0    \psi(x) \, a_\mu \eta^\mu\,.
\ee
Vemos entonces que el término local generado en $\Sigma^\prime$ a partir del término original en $\Sigma$ tiene nuevamente la forma de un flujo del tensor de energía impulso, donde el vector $a^\mu$ se ha propagado del mismo modo que en el caso no masivo.

\subsection{Un enfoque más general sobre la propagación}

En el caso no masivo podemos arribar a la ley de propagación para $a^\mu$ utilizando la conservación del infinito número de corrientes, mientras que en el caso masivo no tenemos todas estas simetrías. Sin embargo, podemos utilizar la localidad y linealidad de la propagación a lo largo de las líneas nulas, junto con la conservación de los generadores de boosts, para dar otra demostración que prueba que la propagación es la que encontramos anteriormente. Este método además tiene el potencial de ser útil para generalizaciones a más dimensiones.

Empezando de $\int_\Sigma ds \eta^\mu T_{\mu\nu} a^\nu$ en $\Sigma$, no es complicado ver para la parte singular del propagador que el término generado en $\Sigma^\prime$ es lineal en $a^\mu$ y no involucra derivadas de $a^\mu$ (si el núcleo integral se escribe con derivadas simetrizadas). Además, en el punto $x^\prime$ en $\Sigma^\prime$ puede haber contribuciones provenientes sólo de un punto $x_1$ en $\Sigma$ con el mismo $x_1^-=x^{\prime-}$ y de un punto $x_2$ con $x_2^+=x^{\prime+}$. Como la propagación se realiza a lo largo de las líneas nulas no puede haber dependencia alguna en las distancias, es decir, un dado operador en $x_1^-=x^{\prime-}$ producirá el mismo operador en $\Sigma^\prime$ independientemente de la coordenada $x^{+}$. Luego, en referencia al término local generado en $x^\prime$ nos podemos preguntar si es proporcional al tensor de energía impulso o si además involucra a otros campos locales. Y si es proporcional al tensor de energía impulso, entonces sería interesante entender cómo se propaga $a^\mu$. Estas preguntas pueden responderse utilizando el hecho de que hay un generador de boosts conservado, que se escribe en términos del tensor de energía impulso sobre cualquier superficie
\be
\int_\Sigma ds\, \eta_\mu T^{\mu\nu}(x(s)) \omega_{\nu\alpha}x^\alpha= \int_{\Sigma^\prime} ds^\prime\, \eta^\prime_\mu(s^\prime) T^{\mu\nu}(x^\prime(s^\prime)) \omega_{\nu\alpha}x^{\alpha\prime}\,,
\ee
donde el tensor antisimétrico $\omega_{\mu\nu}$ es arbitrario. Escrito en coordenadas nulas esto da
\bea
&&\int_\Sigma ds\, \eta_\mu T^{\mu\nu}(x(s)) \omega_{\nu\alpha}x^\alpha=\\
&&=\int_\Sigma dx^+ (T_{++} \omega^{+-}x_-+T_{+-} \omega^{-+}x_+)-\int_\Sigma dx^- (T_{--} \omega^{-+}x_++T_{-+} \omega^{+-}x_-)\,.\nonumber
\eea
Dado un punto fijo $x^\prime$ en $\Sigma^\prime$, podemos cambiar la superficie $\Sigma$ y la contribución de las cargas conservadas debe ser igual. Tenemos una contribución proveniente de un punto con $x^+=x^{\prime+}$ y $x^{-}$ arbitrario. Como esta contribución no puede cambiar con $x^-$, sumado al hecho de que las contribuciones se suman y sólo pueden provenir de la propagación a lo largo de las líneas nulas, concluimos que
\be
\int_\Sigma dx^+ T_{+-} \omega^{-+}x_+-\int_\sigma dx^- T_{--} \omega^{-+}x_+=\int_\Sigma ds\, \eta^\mu T_{\mu -} \omega^{-+}x_+
\ee
se propaga en sí mismo a lo largo de las líneas de $x^+$ constante. Esto significa que $ds \eta^\mu T_{\mu -} a^-$ se propaga en sí mismo a lo largo de las líneas con $x^+$ constante para cualquier $a^-(x^+)$. Del mismo modo, $ ds \eta^\mu T_{\mu +} a^+
$ se propaga en sí mismo a lo largo de las líneas con $x^-$ constante, para cualquier $a^+(x^-)$. Naturalmente, esta propagación involucrará en general a términos no locales. Sólo cuando $a^{\mu}=b^\mu+\omega^{\mu\nu}x_{\nu}$ los términos no locales se cancelan exactamente y la cantidad conservada será combinación del flujo de las tres corrientes conservadas correspondientes a traslaciones y la simetrías ante boosts. En este argumento es crucial la simetría ante boosts.

Este mismo argumento debería aplicarse a todos los campos libres y, posiblemente, también a campos interactuantes en $d=2$. Para más dimensiones, quizás resulte posible generalizar este argumento (redefiniendo apropiadamente las coordenadas nulas como $x^\pm=t\pm r$), aunque hay ciertos detalles propios de la presencia de más dimensiones que deben tenerse en cuenta. La investigación en este caso todavía se encuentra en desarrollo.
\chapter{\label{ch:conclusiones}Conclusiones y comentarios finales}

Cerramos este trabajo con un resumen de los resultados más importantes que hemos presentado y mencionando también algunos posibles caminos futuros de extensión de esta investigación.

Los últimos años, la entropía de entrelazamiento ha tenido una relevancia muy grande en distintos ámbitos de la física. En este trabajo, probamos que en particular, la entropía de entrelazamiento presenta una dependencia con la fase magnética, al estudiar el efecto Aharonov-Bohm sobre las fluctuaciones de vacío. Esto reafirma el rol fundamental de la entropía en el estudio de diversos fenómenos de la física y abre las puertas a futuras investigaciones de este efecto en teoría de campos; en particular, sería interesar buscar un contexto donde sea posible estudiar este fenómeno utilizando holografía.

Si bien el cálculo preciso de la entropía suele ser dificultoso, en los últimos años se ha logrado un gran progreso analítico y numérico en este sentido y es deseable seguir sumando escenarios donde sea posible calcular esta cantidad y otras relacionadas, como la información mutua, entropías de Renyi, entropía relativa, etc.

Otra de las cantidades de la teoría de la información que estudiamos fue la entropía relativa, cuya conexión con la entropía de entrelazamiento se da principalmente en el contexto de la formulación de la segunda ley de la termodinámica y en el de la derivación de una cota de Bekenstein cuántica. En este trabajo explotamos dos propiedades matemáticas de la entropía relativa: la positividad y la monotonicidad. Vimos que la propiedad de positividad establece la desigualdad $\Delta S \leq \Delta \langle H \rangle$, entre la diferencia de entropía entre dos estados y la diferencia en los valores de expectación en estos estados del hamiltoniano modular. Pudimos demostrar que para una clase de estados muy generales, la entropía holográfica de entrelazamiento (propuesta para calcular la entropía de campos utilizando la dualidad de Maldacena) satisface la desigualdad $\Delta S \leq \Delta \langle H \rangle$, lo que representa una prueba de consistencia muy fuerte para esta cantidad.

Con relación a la holografía, cuando se consideran modificaciones a la gravedad de Einstein, las prescripciones adecuadas para realizar el cálculo geométrico son diferentes a la usual (en general, se requiere alguna condición extra que permita determinar unívocamente la superficie extremal). Sería interesante explorar estos escenarios, extendiendo de ser posible los métodos presentados en este trabajo y encontrar generalizaciones adecuadas de la prescripción holográfica para teorías de gravedad construídas con invariantes de curvatura superiores al escalar de curvatura. 

También utilizando la desigualdad $\Delta S \leq \Delta \langle H \rangle$, encontramos que al querer localizar estos valores de expectación aparecen algunos vínculos no triviales para las propiedades de los estados físicos. Vimos por ejemplo que, para un estado puro `cercano' al vacío, no es posible inyectar una cantidad de energía positiva en una región $V$ sin hacerlo también en su complemento. Por otro lado, mostramos que utilizando la igualdad a primer orden $\Delta S = \Delta \langle H \rangle$ es posible reconstruir el estado de vacío reducido a cualquier región sólo a partir del conocimiento del funcional entropía (tomografía del estado vacío). Este trabajo fue pionero en el estudio de la igualdad a primer orden, conocida como la primera ley de la termodinámica para la entropía de entrelazamiento, en el contexto holográfico.

Basados en el estudio de estos problemas relacionados con la localización, entendimos que los estados de una teoría conforme no pueden soportar la presencia de una densidad de energía negativa distribuida en grandes regiones del espacio. Pudimos demostrar esto en forma precisa utilizando argumentos propios de la teoría de campos, obteniendo los primeros resultados que ponen severas restricciones a la localización espacial de energía negativa en una teoría de campos. También vimos que, desde otro punto de vista, esta desigualdad se desprende de la propiedad de monotonicidad de la entropía relativa. La formulación con este enfoque, lleva a una nueva relación entre la entropía y energía de un sistema, similar (aunque también diferente en algunos sentidos, como se ha discutido) a las versiones previas de la cota de Bekenstein. Esta nueva cota de Bekenstein cuántica que encontramos sugiere que quizás existe otra manera de entender a la segunda ley de la termodinámica.

A lo largo del desarrollo de este trabajo, notamos la importancia que tiene conocer detalles sobre la estructura de los hamiltonianos modulares. Pudimos obtener la forma precisa del hamiltoniano modular para un estado térmico en dos dimensiones sobre un intervalo finito y, en particular para estados térmicos, estudiamos un ejemplo en el que es posible establecer una desigualdad para los valores de expectación del mismo en un estado genérico.

En el último capítulo, presentamos una investigación que aún se encuentra en curso, sobre la forma que deben tener los términos locales que aparecen en los hamiltonianos modulares de campos libres. Para una región espacial, vimos que los términos locales están dados por el flujo de una corriente sobre la región. Esta corriente es la contracción del tensor de energía impulso con un vector temporal dirigido hacia al futuro independiente de la masa del campo y de su estadística. También estudiamos cómo se propagan estos términos locales para el caso de dos dimensiones y presentamos una idea que hace factible la generalización a más dimensiones. Sin dudas, este trabajo todavía se encuentra muy abierto y se presta a una exploración más profunda en diversos aspectos.

Finalmente, queremos destacar que si bien el progreso que se realizó al estudiar la entropía en teoría de campos es sin dudas notable, hay muchos elementos de la teoría de la información aún no han sido estudiados en un contexto relativista, principalmente por la dificultad involucrada en los cálculos. El cómputo de algunas medidas de entrelazamiento (como el entrelazamiento de formación y el entrelazamiento de destilado) constituye un desafío enorme aún en mecánica cuántica ordinaria, donde el progreso que se ha hecho ha sido principalmente numérico. A pesar de que se cuenta con pocos resultados concretos, se conocen ciertas relaciones de jerarquía entre estas medidas de entrelazamiento, que dicen por ejemplo que el entrelazamiento destilable es siempre menor al de formación, y este a su vez es menor a la entropía. En teoría de campos relativista, estas desigualdades así planteadas no son relevantes dado que todos los términos son divergentes; sería entonces interesante pensar en si es posible establecer alguna jerarquía análoga que tenga sentido en teoría de campos, redefiniendo o regularizando apropiadamente las cantidades involucradas. Otro aspecto interesante que sería de interés estudiar es la búsqueda de testigos del entrelazamiento en teoría de campos. Un testigo del entrelazamiento es un operador que tiene valores de expectación de distinto signo en estados separables y estados entrelazados. La existencia de testigos del entrelazamiento es una consecuencia del Teorema de Hahn-Banach y de la convexidad del conjunto de los estados separables. Es sabido que el problema de encontrar testigos del entrelazamiento es tan complicado como encontrar operadores positivos. Este problema tiene una dificultad elevada y cualquier progreso en este sentido (incluso en mecánica cuántica ordinaria) representaría un gran avance en el área.

Este trabajo nos deja como enseñanza general que las herramientas y conceptos de la teoría de la información cuántica tienen aplicaciones importantes en el estudio de la física de altas energías y que, además, en muchos casos brindan perspectivas sobre ciertos problemas que con los acercamientos usuales no se revelan fácilmente.

\begin{appendix}

\renewcommand{\chapnumber}{\nonumber}
\renewcommand{\partname}{}
\renewcommand{\thepart}{}

\end{appendix}

\backmatter
\pagestyle{empty}
\renewcommand\bibname{Referencias}
%



\onecolumn

\cleardoublepage

\thispagestyle{empty}
\thispagestyle{empty}

\end{document}